\documentclass[11pt, oneside,reqno]{article}
\usepackage[margin=2.54cm]{geometry}
\usepackage[utf8]{inputenc}
\usepackage[english]{babel}
\usepackage{amsmath, mathtools, amsthm, amssymb}
\usepackage{dsfont}
\usepackage{changepage}
\usepackage{caption, subcaption}
\usepackage{enumitem}
\usepackage{hyperref}
\usepackage{float}
\usepackage{titling}
\usepackage{cite}
\usepackage[titletoc,toc,title]{appendix}
\usepackage{color}
\usepackage{xcolor}
\usepackage{bbm}
\usepackage{algorithm}
\usepackage{algpseudocode}
\usepackage[export]{adjustbox}
\usepackage{pgfplots}
\pgfplotsset{compat=1.15}
\usepackage{mathrsfs}
\usetikzlibrary{arrows}

\DeclareMathAlphabet{\mathdutchcal}{U}{dutchcal}{m}{n}

\theoremstyle{plain}
\newtheorem{thm}{Theorem}[section] % reset theorem numbering for each section

\theoremstyle{definition}
\newtheorem{defn}[thm]{Definition} % definition numbers are dependent on theorem numbers

\makeatletter
\def\thmheadassumption#1#2#3{%
	\thmname{#1}\thmnumber{\@ifnotempty{#1}{ }\@upn{#2}}%
	\thmnote{ {\the\thm@notefont[#3]}}}
\makeatother

\newtheoremstyle{assumption}% Name
{}% space above
{}% space below
{\itshape}% body font
{}% indent
{\bfseries}% head font
{.}% punctuation after head
{ }% space after head (has to be space or dimension!)
{\thmheadassumption{#1}{(A#2)}{#3}}% head spec

\theoremstyle{assumption}

\theoremstyle{remark}
\newtheorem{rmk}{Remark}
\newtheorem*{notation}{Notation}

\newcommand{\R}{\mathbb{R}}
\newcommand{\N}{\mathbb{N}}

\newcommand{\dx}{\, \mathrm{d}x}

\newcommand{\dt}{\, \mathrm{d}t}
\newcommand{\di}{\, \mathrm{d}}

\newcommand{\probspace}{(\Omega, \mathcal{F}, \mathbb{P})}

\newcommand{\hilbert}{\mathcal H}
\newcommand{\iid}{\overset{\mathrm{iid}}{\sim}}
\newcommand{\bigzero}{\mbox{\normalfont\Large\bfseries 0}}

\makeatletter
\newcommand{\ALOOP}[1]{\ALC@it\algorithmicloop\ #1%
	\begin{ALC@loop}}
	\newcommand{\ENDALOOP}{\end{ALC@loop}\ALC@it\algorithmicendloop}

\newcommand{\algorithmicbreak}{\textbf{break}}
\newcommand{\BREAK}{\State \algorithmicbreak}
\makeatother

\makeatletter
\newcommand{\subalign}[1]{%
	\vcenter{%
		\Let@ \restore@math@cr \default@tag
		\baselineskip\fontdimen10 \scriptfont\tw@
		\advance\baselineskip\fontdimen12 \scriptfont\tw@
		\lineskip\thr@@\fontdimen8 \scriptfont\thr@@
		\lineskiplimit\lineskip
		\ialign{\hfil$\m@th\scriptstyle##$&$\m@th\scriptstyle{}##$\hfil\crcr
			#1\crcr
		}%
	}%
}
\makeatother

\begin{document}
	\definecolor{yqqqqq}{rgb}{0.5019607843137255,0,0}
	\definecolor{zzffff}{rgb}{0.6,1,1}
	\definecolor{qqwuqq}{rgb}{0,0.39215686274509803,0}
	\definecolor{ffffzz}{rgb}{1,1,0.6}
	\definecolor{ffzzqq}{rgb}{1,0.6,0}
	\definecolor{qqqqff}{rgb}{0,0,1}
	\definecolor{ffqqqq}{rgb}{1,0,0}
	\definecolor{wwccqq}{rgb}{0.4,0.8,0}
	\definecolor{wwwwww}{rgb}{0.4,0.4,0.4}
	\definecolor{wwqqcc}{rgb}{0.4,0,0.8}
	\definecolor{ffffww}{rgb}{1,1,0.4}
	\definecolor{zzffzz}{rgb}{0.6,1,0.6}
	\definecolor{ffqqtt}{rgb}{1,0,0.2}
	\definecolor{qqttcc}{rgb}{0,0.2,0.8}
	\definecolor{zzttff}{rgb}{0.6,0.2,1}
	
\title{A localized particle filter for geophysical data assimilation}

\author{Dan Crisan\thanks{Department of Mathematics, Imperial College London, {\tt d.crisan@imperial.ac.uk}.}
		\and
		Eliana Fausti\thanks{Department of Mathematics, Imperial College London, {\tt e.fausti@imperial.ac.uk}.}}
	
\date{\today}
	
\maketitle
	
\begin{abstract}
	
Particle filters are computational techniques for estimating the state of dynamical systems by integrating observational data with model predictions. This work introduces a class of Localized Particle Filters (LPFs) that exploit spatial localization to reduce computational costs and mitigate particle degeneracy in high-dimensional systems. By partitioning the state space into smaller regions and performing particle weight updates and resampling separately within each region, these filters leverage assumptions of limited spatial correlation to achieve substantial computational gains. This approach proves particularly valuable for geophysical data assimilation applications, including weather forecasting and ocean modeling, where system dimensions are vast, and complex interactions and nonlinearities demand efficient yet accurate state estimation methods. We demonstrate the methodology on a partially observed rotating shallow water system, achieving favourable performance in terms of algorithm stability and error estimates.
\end{abstract}
	
\begin{adjustwidth}{0.6cm}{0.5cm}
		\begin{flushleft}
			
		\end{flushleft}
\end{adjustwidth}

\section{Introduction}

Sequential data assimilation plays a critical role in geophysical sciences, providing a framework to integrate numerical models with observational data to estimate the state of dynamical systems describing weather, ocean circulation, and climate (for a recent survey, see Carrassi~et~al.~\cite{carrassi2018_reviewDA}). Among the most popular data assimilation (DA) methods are ensemble Kalman filters (EnKF, Evansen~\cite{evensen94}) and its variants. EnKF-like algorithms have been widely adopted in atmospheric and oceanographic applications due to their computational efficiency and simplicity of implementation. However, these methods rely on Gaussian assumptions that fail in the presence of strong nonlinear dynamics or non-Gaussian error distributions (Bocquet~et~al.~\cite{bocquet2010}, Poterjoy~\cite{poterjoy22}), which is a problem increasingly relevant as observational data improve in resolution and coverage. In particular, the \textit{Surface Water and Ocean Topography (SWOT)} mission by NASA and CNES~\cite{swot} has been collecting new high-resolution altimetric data, which poses significant difficulties for current DA methodologies due to its (swath) structure, frequency and nonlinearity.

Particle filters (PFs), on the other hand, provide a fully nonlinear, non-Gaussian approach to DA (see e.g.~Crisan~and~Doucet~\cite{crisan_doucet_2002}, Doucet~et~al.~\cite{Doucet2001}, Reich~and~Cotter~\cite{reich_cotter15}, van~Leeuwen~\cite{vanLeeuwen2009_review}). PFs represent the (posterior) probability distribution of the state of a system using a set of weighted particles, allowing them to approximate arbitrarily complex distributions and enabling theoretically exact inference in nonlinear systems (see Bain~and~Crisan~\cite{bain_crisan} for a background in stochastic filtering). Despite their theoretical appeal, the application of PFs in high-dimensional geophysical systems has been somewhat limited by the so-called \textit{curse of dimensionality}: as system dimension increases, particle weights tend to degenerate, with a single particle dominating the ensemble (Bickel~et~al.~\cite{bickel2008}, Snyder~et~al.~\cite{snyder2008}). These results, which apply to basic PFs (with or without resampling), have sometimes lead to discarding PFs as viable DA algorithms in realistic, large-scale applications. On the other hand, Beskos~et~al.~\cite{beskos_crisan_AAP14} demonstrated that, with the appropriate modifications, more advanced particle filters are stable even in high-dimensions, giving a strong theoretical backing for their development in practice.

The curse of dimensionality, therefore, is not insurmountable, and several methodologies have been developed to mitigate particle degeneracy. For example, \textit{jittering}, which involves slightly perturbing the resampled particles, improves diversity and maintains the ensemble spread (called \textit{roughening} for the bootstrap particle filter in Gordon~et~al.~\cite{gordon_bootstrap_93}; jittering MCMC techniques first proposed in Liu~and~Chen~\cite{Liu1998}; see also Andrieu~et~al.~\cite{andrieu10}). Jittering is a powerful tool if used in conjunction with \textit{tempering}, which reduces the variance among the weights by smoothing the transition from prior to posterior distribution, inserting intermediate steps to go from one to the other (which can be understood as a series of changes of measure, see Del~Moral~et~al.~\cite{delmoral2006}, also~\cite{beskos_crisan_AAP14} for convergence results). PFs with tempering and jittering in the context of fluid or geophysical fluid dynamics have been implemented successfully in Kantas~et~al.~\cite{kantas_14} and Llopis~et~al.~\cite{llopis_kantas_18_pf_NS} (for the Navier--Stokes equations), Cotter~et~al.~\cite{cotter_et_al_PF_2D_euler_20, cotter_et_al_quasigeo_20} (for the 2D Euler equations and QG equations) and Lang~et~al.~\cite{lang_22_PF_RSW} (for the stochastic RSW equations). More recently, Cotter~et~al.~\cite{cotter_crisan_singh_24} applied a PFs with tempering and jittering to the Camassa--Holm equation, and Crisan~et~al.~\cite{lobbe_RSW_23} and Lobbe~et~al.~\cite{lobbe_RSW_24} to the RSW model, in conjunction with noise calibration. Other techniques like \textit{nudging} (Akyildiz~and~Miguez~\cite{akyildiz_20_nudging}), variants like the space-time PF (Beskos~et~al.~\cite{beskos_crisan_AdvAP17}) or the Ensemble Transport PF (Reich~\cite{reich2013}), and hybridizations with variational and Ensemble Kalman methods have been suggested, and they can be used effectively to stabilize PF algorithms,  even in the face of high-dimensional challenges (for a recent review of high-dimensional PFs, see van~Leeuwen~et~al.~\cite{vanLeeuwen2019_review}).

One of the most promising strategies to mitigate weight degeneracy in PFs is \textit{localization}, a procedure originally developed in the context of the EnKF (Houtekamer~and~Mitchell~\cite{houtekamer2001locEnKF}). The premise behind localization is that, since most geophysical systems exhibit nearly independent short-term evolution in distant regions, it seems practical, and efficient, to ignore far away observations when updating the positions of the ensemble members. By localizing, one restricts the impact of observations to nearby state variables, thus reducing the effective dimensionality of the DA (or filtering) problem. A simple localization methodology for the EnKF relies on ``weighting" the entries of the sample covariance matrix to eliminate long-range spurious correlations; matters are considerably less simple for PFs, since modifying (read: resampling) particles locally and gluing them together in \textit{composite particles} can easily lead to discontinuities.

The survey paper by Farchi and Bocquet~\cite{farchi_18_locPF_comp} highlights these challenges, and compares (and sometimes generalizes) different localization algorithms for PFs. It groups them roughly in two classes, based on the type of localization developed: \textit{state-block-domain} localization or \textit{sequential-observations} localization. We will not discuss the second type of localization, in which observations are assimilated serially, here, since it is not relevant to our work, but refer the reader to Poterjoy~\cite{poterjoy_16_locPF}, Poterjoy~and~Anderson~\cite{poterjoy_16_b} and subsequent works.
%We also mention the Local Adaptive Particle Filter (LAPF) by Potthast~et~al.~\cite{potthast_19_locPF} as one of the few localized PF methods tested in operational numerical weather forecast.

State-block-domain localization relies on a block-decomposition of the domain, with the analysis step performed independently in each block by computing local weights. This is the case for the Local PF (LPF) proposed by Penny~and~Miyoshi~\cite{penny_miyoshi_16} (although the first application goes back to van~Leeuwen~\cite{vanLeeuwen2003}): they deal with the issue of discontinuities in composite particles by averaging over the grid-points within a smoothing radius, for each particle and at each grid-point. In general, some spatial smoothness for the weights and ``regularity" in the resampling scheme (in~\cite{penny_miyoshi_16} local weights are sorted in descending order before resampling; see also Robert~and~K{\"u}nsch\cite{robert2017}) can help mitigate discontinuities, but spatial averaging will always be necessary---without composite particles, after all, we would not have localization.

In this paper, we propose a state-block-domain localization method for PFs along the lines of \cite{penny_miyoshi_16, farchi_18_locPF_comp}. We suggest a domain decomposition in blocks that are not disjoint, but overlap at the boundaries, to facilitate interpolation at the end of each assimilation step. The local weights in each block depend on the observations on the whole domain, with the impact of observations located outside the block being weighted down by a smooth function of the distance of the observation locations from the boundary of the block, in the spirit of Gaspari~and~Cohn~\cite{gaspari_cohn}. Tempering is performed locally, before interpolation. As it is generally the case for state-block-domain localization algorithms, our method allows for parallelization, thus guaranteeing a significant speed-up in the computations.

We test our algorithm on twin experiments (or filtering problems) where the signal is given by a Stochastic Rotating Shallow Water (SRSW) model from the Stochastic Advection by Lie Transport (SALT) framework. The SALT framework, developed by Holm~\cite{Holm_SALT_15}, incorporates stochasticity in (geophysical) fluid dynamics equations in a physical way through Hamilton's principle (see Street~and~Crisan~\cite{street_21} for an introduction to SALT accessible to stochastic analysts). In the context of DA for geophysical systems, the use of stochastic models such as those derived in the SALT and LU (fluid dynamics under Location Uncertainty, developed by M{\'e}min~\cite{memin2014_LU}) frameworks gives a natural parametrization of the small scales, and avoids the need to introduce noise in the system in artificial ways. Finally, we remark that in our numerical experiments we \textit{only} observe the height field, and \textit{not} the associated velocities. This is to try to mimic (albeit in a greatly simplified setting) the satellite data collected through the SWOT mission, and increase the difficulty in our DA experiments.

\subsection{Structure of the paper}

This paper is organized as follows: in Section~\ref{sec:set-up} we set-up a filtering problem where the signal is given by an SPDE, and briefly introduce particle filters. In Section~\ref{sec:localization} we describe in detail our localization algorithm. In Section~\ref{sec:numerics} we present the results of the twin experiments, comparing our LPF with a PF with tempering and jittering, and show both stability of the algorithm and that the approximation errors due to the localization procedure do not impact the performance of the LPF. In particular, we are interested in providing numerical evidence towards supporting three claims:
\begin{enumerate}
	\item Particle filters (localized and not) are viable algorithms for high-dimensional geophysical data assimilation problems.
	\item Even when the PF is stable (and so localization is not strictly necessary), the error introduced by the localization procedure is negligible. Moreover, when working with PF algorithms with a relatively low number of particles, localization can even improve on the results of the standard PF.
	\item Given comparable performance, the LPF is substantially faster than the PF, owing to both parallelization capabilities and the dimensionality reduction achieved through localization.
\end{enumerate}
These are the main numerical experiments contained in Section~\ref{sec:numerics}:
\begin{itemize}
	\item Comparison between the performances of the PF and the LPF (both augmented with tempering and jittering), tested in twin experiments with observations located on equi-spaced grids of $2^n \times 2^n$ grid-points (for $n$ varying from 1 to 7) in a 128 $\times $ 128 domain, for 1000 assimilation steps (Figures~\ref{fig:PF_obs_grid_test_2_grids}, \ref{fig:PF_obs_grid_test_2}, \ref{fig:PF_obs_grid_test_rmse_2_grids} and  \ref{fig:PF_obs_grid_test_rmse_2}). Also comparison of the (average) number of tempering steps required by a PF and by a LPF (with 4 localization regions), given equivalent performance (Figure~\ref{fig:tempering_steps}).
	\item Sensitivity of the LPF to the experiment settings (to the observations frequency in Figure~\ref{fig:LPF_test_freq} and \ref{fig:LPF_freq_evol}, and to the observations noise in Figure~\ref{fig:LPF_test_obs_noise}) and to the number of localization regions (Figure \ref{fig:LPF_test_nregions}).
	\item Analysis of the performance of the LPF with observations located on a thin strip of the domain periodically changing position in time (Figures~\ref{fig:strip_subM} and \ref{fig:strip_M}).
\end{itemize}

Finally, this paper contains a few appendices: in Appendix~\ref{app:numerics_details} we give details about our numerical implementation;  Appendix~\ref{app:pseudo-codes} contains pseudocodes for all the particle filtering algorithms implemented for this work (in particular, in Appendix~\ref{app:basic_filt} we review resampling, tempering and jittering); and in Appendix~\ref{app:jittering_comp} we compare two jittering algorithms, one (more computationally expensive) that employs Metropolis--Hastings to sample from the posterior distribution, and one that simply perturbs the resampled particles with spatial noise.

\section{Geophysical state space models and filtering}\label{sec:set-up}

Geophysical systems used for weather-forecast and ocean-prediction are generally described by continuous-time state space models. The dynamics of the state of the system $u$, which comprises physical quantities such as wind or water velocity, temperature, wave height, etc., are modelled by systems of Partial Differential Equations (PDEs) or Stochastic Partial Differential Equations (SPDEs). Crucially, even when the dynamics are taken to be deterministic, some randomness is always incorporated into the system to account for model uncertainty (for example, by adding noisy perturbations to the PDE at discrete times). The state $u = (u_t)_{t \ge 0}$ is then a continuous-time stochastic process.  In most cases, $u$ is unknown and cannot be observed directly, but only $\textit{estimated}$ by collecting instantaneous, partial, noisy measurements at some given times $\{ t_k : k = 1, 2, \dots, M \}$. The relation between $u_{t_k}$ (the state at times $t_k$) and these measurements, which we denote by $Y_k$, is often nonlinear. For $t_{k} \le t < t_{k+1}$, the conditional distribution of $u_{t}$ given all the past data $Y_1 = y_1, Y_2 = y_2,..., Y_k = y_k$  is found by solving a filtering problem, where the state $u_t$ plays the role of the continuous-time \textit{hidden signal}, and $Y_k$ of the discrete \textit{observations}. The resulting \textit{filter} gives the optimal estimate for $u$.

We consider this framework and introduce some notation. Let $\probspace$ be a probability space and consider a finite time interval $[0,T]$. Consider a state space model where the state of the system $u$ is modelled by an SPDE of the form
\begin{equation}\label{eq:SPDE_general}
	\di u_t = \mathcal{A}(u_t) \dt + \mathcal{R}(u_t) \di W_t, \quad t \in [0,T],
\end{equation}
and some initial conditions $u_0 \in \mathcal{H}$, where $\mathcal{H}$ is a Hilbert space. We denote the spatial domain of \eqref{eq:SPDE_general} by $D \subset \R^{d}$, where  $d = 1, 2$ or $3$. The noise $W$ is assumed to be some infinite-dimensional Wiener-process taking values in a separable Hilbert space $\mathcal{K}$ and adapted to a filtration $(\mathcal{F}_t)_{t \ge 0}$ on $\probspace$. For geophysical models, we usually have $\mathcal{K} = L^2(D)$, and $W$ is taken to be a $\mathcal{K}$-cylindrical Brownian motion, which can be written as
\begin{equation}\label{eq:noise_SPDE}
	W_t(\varphi) = \sum_{n = 1}^{\infty} \langle \varphi, e_n \rangle W_t (e_n), \quad \forall \varphi \in \mathcal{K},
\end{equation}
where $\langle \cdot, \cdot \rangle$ denotes the inner product on $\mathcal{K}$ and $\{ e_n\}$ is any complete orthonormal system in $\mathcal{K}$. Here, $W_t^n  := W_t(e_n)$ is now a family of independent real-valued Brownian motion. The SPDE \eqref{eq:SPDE_general} needs to be understood in a suitable (variational) sense, and we assume enough regularity on the coefficients $\mathcal{A}$ and $\mathcal{R}$ such that \eqref{eq:SPDE_general} has a unique weak solution $u$ for $t \in [0,T]$, and $u$ is a $\hilbert$-valued Markov process. For general background on SPDEs, we refer to Lototsky~and~Rozovsky~\cite{rozovsky_spdes} or Liu~and~R\"{o}ckner~\cite{rockner_spdes}.

Let $\mathcal{P}(\hilbert)$ and  $\mathcal{B}(\hilbert)$ denote the probability measures and the Borel sets on $\hilbert$ respectively. We are interested in computing the conditional law of $u$ (given the observations) at the assimilation times $t_1, t_2, \dots, t_M$. Since $u$ is Markov by assumption, it follows that the sequence $U_k = u_{t_k}$ for $k = 0, 1, \dots, M$ is a $\hilbert$-valued Markov chain, with (time-dependent) transition kernel $P_k : (\hilbert, \mathcal{B}(\hilbert) ) \to [0,1]$ given by
\begin{equation}\label{eq:transition_kernel}
	P_{k+1} (u_{k}, \di u_{k+1}) : = \mathbb{P} (U_{k+1}\in \di u_{k+1} | U_{k}  = u_{k}), \quad k = 0, 1, \dots, M.
\end{equation}

Note that in practical data assimilation problems one rarely works with the SPDE \eqref{eq:SPDE_general}, but with a numerical approximation of its solution, and therefore of the transition kernel \eqref{eq:transition_kernel}, instead. We cast the filtering problem in this numerical setting. In view of developing a localization algorithm later, this also dispenses us from having to deal with the technical issue of pointwise evaluation of $u \in \hilbert$ at $x \in D$ when $\hilbert$ is a general Hilbert space.

We start by discretizing time and space. For $k = 1, \dots M$, let $\delta t_k = (t_k - t_{k-1})/L_{k}$, where $\{t_k\}_{k=1}^M$ are the observation times, $t_0 = 0$, $t_M = T$, and $L_k \in \N$ indicates the level of discretization of the time interval $[t_{k-1}, t_k]$. For the spatial domain $D$, we consider a 1-, 2-, or 3-dimensional grid (depending on whether $D \subset \R$, $\R^2$ or $\R^3$). The number of points in the grid gives us the dimension of the numerical solution of \eqref{eq:SPDE_general}: denote this quantity by $N_u$.

Slightly abusing notation, from now on let $u$ be the numerical solution to \eqref{eq:SPDE_general}, so $u_{t_{k-1} + m \delta t_k} \in \R^{N_u}$ for all $m = 0, \dots, L_k -1$ and all $k = 1, \dots, M$. Note that if $z$ is a grid-point in the discretized domain $D$, by $u_{t_{k-1} + m \delta t_k}(z)$ we mean the numerical solution of the SPDE \eqref{eq:SPDE_general} at time $t_{k-1} + m \delta t_k$ at the grid-point $z$. As before, define the Markov chain $U_k = u_{t_k}$ for $k = 0, \dots, M$ (but note that $(U_k)_{k}$ is now $\R^{N_u}$-valued), and the numerical transition kernel corresponding to \eqref{eq:transition_kernel} by $P_{k} : (\R^{N_u}, \mathcal{B}(\R^{N_u})) \to [0,1] $.

For the observations, let $h_k : \R^{N_{u}} \to \R^{N_y}$ be a family of (possibly nonlinear) measurable functions. Define
\begin{equation}\label{eq:observations}
	Y_k = h_k (U_{k}) + B_k, \quad \mathrm{for} \, k = 1, \dots, M,
\end{equation}
and $Y_0 = 0$, with $ B_k \iid N(0, \Sigma)$ and independent of $W$ and $\Sigma \in \R^{N_y \times N_y}$ positive-definite. Note that the \textit{sensor function} $h_k$ is allowed to vary in time (and similarly one could take the noise covariance matrix $\Sigma$ to be time-dependent, although we avoid this for clarity of exposition).

Then, if we denote the distribution of $U_k$ by ${\mu}_k(S) = \mathbb{P} (U_k \in S)$ for $S \in \mathcal{B}(\R^{N_u})$, we have
\begin{equation*}
\mu_{k+1}(S) = P_{k+1} \mu_{k}(S) = \int_{\R^{N_u}} P_{k+1} (u, S) \di \mu_{k} (u).
\end{equation*}
Let $(\mathcal{Y}_k)_{k \ge 1}$ be the $\sigma$-algebra generated by $Y$. The filtering distribution $\pi$ is a discrete-time $\mathcal{P}(\R^{N_u})$-valued random process defined as
\begin{equation}\label{eq:filter_def}
	\pi_k(S) = \mathbb{P}(U_k \in S | (\mathcal{Y}_j)_{j \le k}), \quad k = 1, \dots, M,
\end{equation}
for all $S \in \mathcal{B}(\R^{N_u})$, and  $\pi_{0} = \mu_0 = \mathrm{law}(U_0)$. One can prove that the following recurrence formula holds for all $S \in \mathcal{B}(\R^{N_u})$ and  $k \ge 0$
\begin{equation}\label{eq:filter_recurrence}
	\pi_{k+1}(S)
	= \frac{\int_S g_{k+1}(u) P_{k+1} \pi_k (\di u)}{\int_{\R^{N_u}}  g_{k+1}(u)  P_{k+1} \pi_k (\di u) },
\end{equation}
where for all $k\ge1$, $g_k : \R^{N_u} \to \R$ is given by
\begin{equation}\label{eq:likelihood}
	g_{k}(u) =  g (Y_{k} - h_k(u)),
\end{equation}
and $g$ is the probability density function of $Y_k$ given $U_k$, so in our case, since $B_k$ are i.i.d.  multivariate normal random variables,
\begin{equation}\label{eq:multivariate_normal}
	g (y) = \frac{1}{\sqrt{(2 \pi)^{N_y} \mathrm{det}(\Sigma)}} \exp \Big\{ -\frac{1}{2} y^{\top} \Sigma^{-1} y  \Big\}, \quad \forall y \in \R^{N_y}.
\end{equation}
From a Bayesian point of view, the function $g_k$ is referred to as the \textit{likelihood function}, while $q_{k} := P_{k} \pi_{k-1}$ is the \textit{prior} (or \textit{forecasted}) distribution of $U_{k}$ and $\pi_{k}$ is its \textit{posterior}. Finally, recall that while $U$ is a discrete-time process, we might be interested in estimating (or forecasting) the state of the system $u$ at all times in $[0,T]$ and not only at the assimilation times $\{ t_1, \dots, t_M\}$. From the recurrence relation \eqref{eq:filter_recurrence} and the Markov property of $u$, it is easy to see that, for $t_{k-1} + m\delta t_k \in (t_{k-1}, t_{k})$ (where $0<m<L_{k}$), the probability distribution of $u_{t_{k-1} + m\delta t_k}$ given $Y_0 = y_0, \dots, Y_{k-1} = y_{k-1}$ is given by the posterior $\pi_{k-1}$ pushed forward by the corresponding Markov transition kernel for $m \delta t_k$ iterations.

We just presented a brief sketch of the filtering problem for data-assimilation in geophysical models. It is easy to see that computing the filtering distribution directly is not feasible: both the transition kernel \eqref{eq:transition_kernel} and the recurrence relation \eqref{eq:filter_recurrence} involve integrals over very high dimensional spaces which would be extremely costly to approximate  numerically. This is the reason why most of the current research in stochastic filtering applied to high-dimensional problems is focused on finding computationally efficient numerical methods to approximate as precisely as possible the (optimal) filtering measure $\pi$. Our choice of methodology in this paper is Particle Filters (PF), or Sequential Monte Carlo (SMC) methods, which we now proceed to introduce.

\subsection{Particle filters}\label{sec:PF}

The main idea behind particle filters is to approximate the filtering distribution $\pi$ by an empirical measure $ \pi^N$ such that
\begin{equation}\label{eq:particle_filter}
	\pi^N_k = \sum_{i = 1}^N w^i_k \delta_{u^i_{t_k}}, \quad \forall k = 0, 1, \dots, M,
\end{equation}
where the weights $w^i_k$ sum to $1$ for each $k$, and $u^i$ are $N$ i.i.d. copies of the signal $u$. In other words, $u^i$ are $N$ i.i.d. numerical realizations of the SPDE \eqref{eq:SPDE_general}, with initial conditions $u^i_0$ sampled from the law of $u_0$ for all $i = 1, \dots, N$. We call $u^i$ the \textit{particles}, and their weighted distribution on the space $\R^{N_u}$ at each assimilation time gives the \textit{particle filter} \eqref{eq:particle_filter}.
At $k = 0$ all the particles have equal weight, i.e. $w^i_0 = 1/N$. At the first assimilation time $t_1$, the prior $q_1 = P_1 \pi_0$ is approximated by
\begin{equation*}
	q_1 \approx  \sum_{i = 1}^N w^i_0 \delta_{u^i_{t_1}} = \frac{1}{N} \sum_{i = 1}^N \delta_{u^i_{t_1}},
\end{equation*}
obtained by evolving all the particles according to the SPDE \eqref{eq:SPDE_general} from time $t_0 = 0$ to  $t_1$. Now consider the recurrence relation \eqref{eq:filter_recurrence}. Inserting the empirical measure approximation of $q_1$ into \eqref{eq:filter_recurrence} we get that for all $S \in \mathcal{B}(\R^{N_u})$
\begin{equation*}
	\pi_1(S) = \frac{\int_S g_{1}(u) q_{1} (\di u)}{\int_{\R^{N_u}}  g_{1}(u)  q_1 (\di u) } \approx
	\sum_{i = 1}^N \frac{g_1(u^i_{t_1})}{\sum_{j=1}^N g_1(u^j_{t_1})} \delta_{u^i_{t_1}}(S)  =
	\sum_{i = 1}^N w^i_1 \delta_{u^i_{t_1}}(S)
	= \pi^N_1(S),
\end{equation*}
where the weights $w^i_1$ are given by the likelihood of $Y_1$ given each particle $u^i_{t_1}$ (and normalized to sum to 1).
In the most basic instance of particle filter, this process is then repeated recursively so that at time $t_k$ we have the estimate
\begin{equation}\label{eq:basic_PF_weights_rec}
	\pi^N_k = \sum_{i=1}^N w^i_k \delta_{u^i_{t_k}}, \qquad w^i_k = \frac{w^i_{k-1} g_k(u^i_{t_k})}{\sum_{i = 1}^N w^i_{k-1} g_k(u^i_{t_k})}.
\end{equation}
Theoretically, it has been proven that the above particle filter approximation $ \pi^N_k$ does in fact converge to the true filter $\pi_k$ asymptotically as $N \to \infty$ (see e.g.~\cite[Ch.~10]{bain_crisan} for convergence in the weak topology almost surely). 

However, in practice we can only run a finite number of particles, and this simple scheme degenerates quickly due to weight-collapse: after very few assimilation steps, the weights of all particles but one collapse to 0, so that the approximate filter essentially reduces to a Dirac mass at the position of the one surviving particle, and looses all statistical meaning. As we remarked in the introduction, steps can be undertaken to prevent the degeneracy of the particle filter. In this paper we add the procedures of \textit{resampling} (the so-called \textit{bootstrap particle filter}), \textit{jittering} and \textit{tempering} to the basic particle filter to improve its stability. For the sake of keeping our exposition somewhat self-contained, we provide a brief introduction to these algorithms (with pseudocodes) in Appendix~\ref{app:basic_filt}. A pseudocode for a PF with tempering and jittering can be found in Algorithm~\ref{alg:PF}.

As we will see in Section~\ref{sec:numerics}, the particle filter can be effective in high dimensions, even with a relatively low number of particles, but, especially when a large amount of observations is available at each assimilation step, the tempering algorithm needs many iterations to prevent degeneracy of the ensemble, and thus a Algorithm~\ref{alg:PF} is generally computationally expensive. One way to address this issue is localization: by ``localizing'' one breaks up the update step of the particle filter into smaller problems, each dealing with a fraction of the total number of observations. By parallelizing, it is possible to achieve a considerable speed up in the computations. Finally, in situations when the particle filter happens to degenerate, localization can help make the algorithm more stable.

\section{The localization algorithm}\label{sec:localization}

For a clear description of the localization algorithm, throughout this section and for the rest of the paper, we assume that the spatial domain $D$ of the SPDE \eqref{eq:SPDE_general} is a $[0,1] \times [0,1] \subset \R^2$ square (discretized). Recall the notation $N_u$ for the number of grid-points in $D$ that determine the dimension of the numerical approximation $u \in \R^{N_u}$ of the solution to \eqref{eq:SPDE_general}. The localization method that we propose can be easily applied to other types of domain in $\R^2$ (not necessarily square), and similarly it can be extended to deal with $d$-dimensional problems, although some work will be needed to adapt the decomposition strategy we describe to an $d$-dimensional domain. Recall that, if $z \in D$ is a grid-point, and $v$ a numerical field on $D$, by $v(z)$ we indicate the value of $v$ at grid-point $z$. Since $D$ is 2-dimensional, we might also write $v(x,y)$ where $x$ and $y$ are the coordinates of a grid-point $z = (x,y) \in D$.

From now on, we also assume that the observations $Y_k$ at each assimilation time $t_k$ are located at some prescribed grid-points in $D$, and, crucially, only depend on the value of the signal $u$ at those points. More specifically, for every $k \ge 1$ let $N_{\mathrm{obs}_k}$ be the number of observations available at assimilation time $t_k$, and let $ \{z_k^1, \dots, z_k^{ N_{\mathrm{obs}_k}} \}$, $z^s_k \in D$, be a collection of grid-points in $D$. Then we take the observations $Y$ to be given by
\begin{equation}\label{eq:discrete_obs}
	Y_{k} = 
	\left( \begin{array}{l}
		Y^1_k \\
		\vdots \\
		Y_k^{N_{\mathrm{obs}_k}}
	\end{array}
	\right)	
	=
	\left( \begin{array}{c}
		h_k^1(u_{t_k}(z_k^1)) + B^1_{k} \\
		\vdots \\
		h_k^{N_{\mathrm{obs}_k}}(u_{t_k}(z_k^{N_{\mathrm{obs}_k}})) + B^{N_{\mathrm{obs}_k}}_{k}
		\end{array}
	\right)	
	= h_k(u_{t_k}) + B_k,
\end{equation}
where $h^s_k : \R \to \R^{\tilde{d_y}}$ for $s = 1, \dots, N_{\mathrm{obs}_k}$, and the total dimension of the observations is $d_y(k) = \tilde{d_y}\times N_{\mathrm{obs}_k}$ (possibly time-dependent if $N_{\mathrm{obs}_k}$ changes with $k$). As usual, $B^s_k \iid N(0, \Sigma_k^s)$ for all $s$, with $\Sigma_k^s \in \R^{\tilde{d_y} \times \tilde{d_y}}$ positive-definite, so in particular $B_k \sim N(0, \Sigma_k)$ with
\begin{equation*}
	\Sigma_k =
	\left( \begin{array}{@{}cc@{}}
		\begin{matrix}
		\Sigma_k^1 &   \, \\
		\, & \Sigma_k^2
		\end{matrix}
		& \bigzero \\
		\bigzero &
		\begin{matrix}
		\ddots & \, \\
		\, &\Sigma_k^{N_{\mathrm{obs}_k}}
		\end{matrix}
		\end{array}\right).
\end{equation*}

	For all $k \ge 1$, we call the points $\{z^s_k\}_{s = 1}^{N_{\mathrm{obs}_k}} \in D$ the \textit{locations of the observations $Y_k$}.

The main idea behind our algorithm is as follows: we divide $D$ into subdomains, and in each subdomain we compute separately a particle filter of the type of Algorithm~\ref{alg:PF}. The observations will be weighted, so that the further away they are from a region, the less impact they will have in the computation of the weights of the local particles. Briefly, the algorithm can be explained as follows:
\begin{enumerate}
	\item At time $t_k$ start with the usual ensemble of $N$ particles $\{u^i_{t_k}\}_{i=1}^N$. Evolve it forward according to the dynamics in \eqref{eq:SPDE_general} up to the next assimilation time $t_{k+1}$.
	\item Construct local particle ensembles in each subregion by restricting the global ensemble to the subregions' domains.
	\item Compute the particle filter described in Algorithm~\ref{alg:PF} in each subregion, using a \textit{local loglikelihood function} that limits the impact of observations far away from the subregion on the update of the local particles' positions. We make use of (a version of) the Gaspari--Cohn function \cite{gaspari_cohn} in this step.
	\item Merge the particles back together using spatial interpolation to recover the global ensemble.
\end{enumerate}

Fundamentally, we want to define a localization algorithm that is \textit{scalable} with the number of subdomains. In what follows, we will take the shape and the number of the subdomains to be fixed to simplify the exposition, but it is straightforward to extend the algorithm to a varying number of subdomains of different (convex) shapes.

\subsection{Domain decompositions and local subregions}

\begin{figure}[ht]
	\centering
	\includegraphics[width=0.7\linewidth]{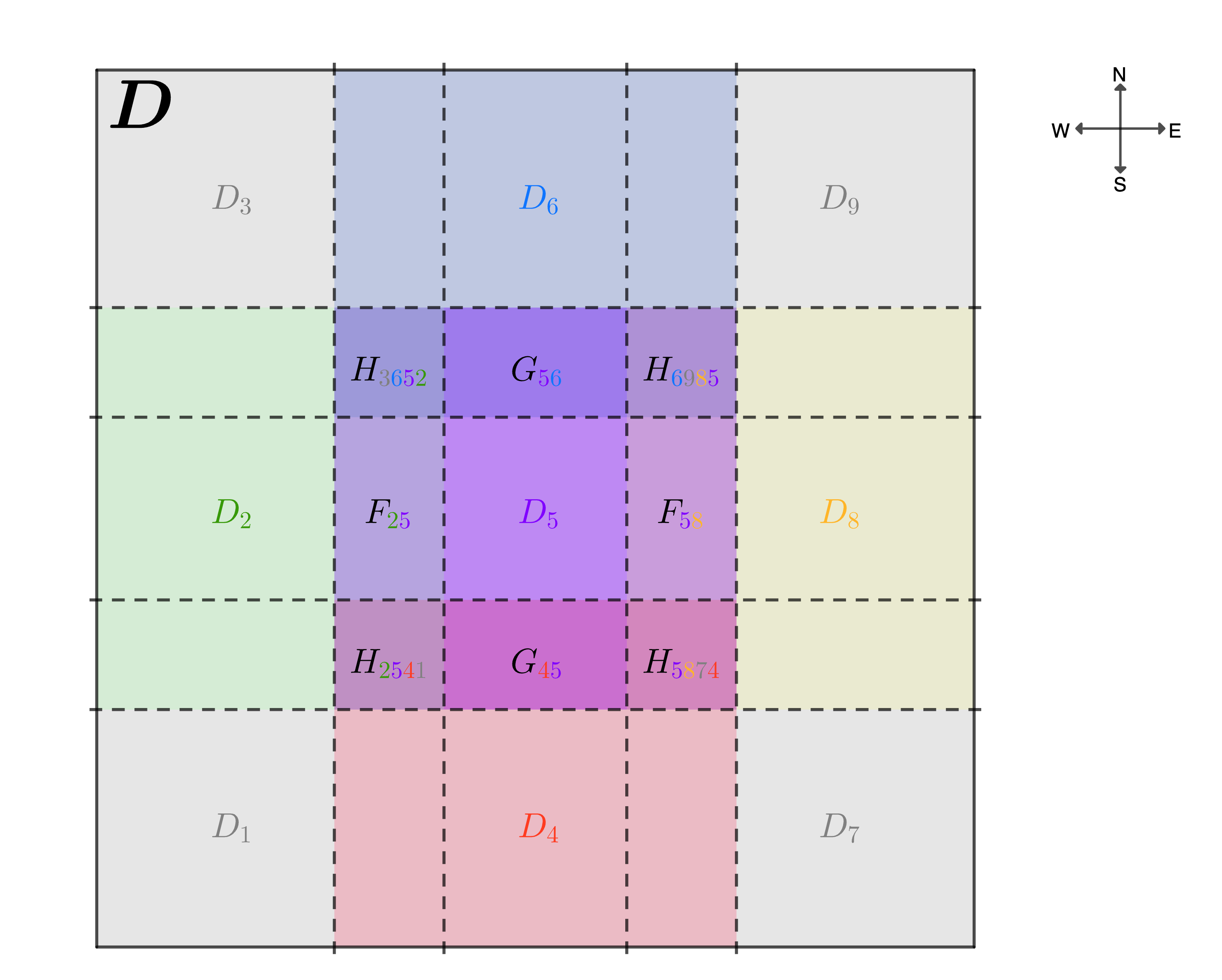}
	\caption{Square domain $D$ divided into 9 subregions $D_1, \dots, D_9$. We highlight $D_5$ (purple) and the regions adjacent to it at the north ($D_6$, blue), east ($D_8$, yellow), south ($D_4$, red) and west ($D_2$, green). Within $D_5$, we highlight the WE-overlap regions $F_{25} = D_5 \cap D_2$ and $F_{58} = D_5 \cap D_8$, and the SN-overlap regions $G_{45} = D_5 \cap D_4$ and $G_{56} = D_5 \cap D_6$. The corner-overlap $H_{3652}$ at the $NW$-corner of $D_5$ is the intersection of 4 subregions (i.e. $H_{3652} = D_5 \cap D_2 \cap D_3 \cap D_6$), and similarly $H_{6985}$, $H_{5874}$ and $H_{2541}$. }
	\label{fig:center_region_overlap}
\end{figure}

Let us start by introducing some notation. Let $D_1, \dots, D_{N_{\mathrm{loc}}} \subset D$ be \textit{convex subsets} of $D$ such that $\cup_j D_j = D$, and $N_{\mathrm{loc}} \in \N$ is the number of subregions. In what follows, we take $D_j$ to be squares, and $N_{\mathrm{loc}} = c^2$ for some $c \in \N$ for simplicity, but any number of any type of quadrilateral shapes would also work, as long as they cover the whole of $D$. Note that the squares $D_j$ are chosen not to be disjoint: in fact, every $D_j$ partly overlaps to the north (N), east (E), south (S) and west (W) with its neighbouring squares, as shown in Figure~\ref{fig:center_region_overlap}. We denote the WE-overlap regions by $F_{we} = D_w \cap D_e$, where $w$ is the index of the region to the west and $e$ is the index of the one to the east. Similarly, we denote by $G_{sn} = D_s \cap D_n$ the SN-overlap regions, with region $s$ to the south and $n$ to the north. Again looking at Figure~\ref{fig:center_region_overlap}, we see that there are some regions in the domain, which we denote by $H_{abcd} = D_a \cap D_b \cap D_c \cap D_d $, where 4 regions overlap, region $D_a$ to the NW-corner, $D_b$ to the NE, $D_c$ to the SE and $D_d$ to the SW-corner.

\begin{figure}[ht]
	\centering
	\includegraphics[width=0.7\linewidth]{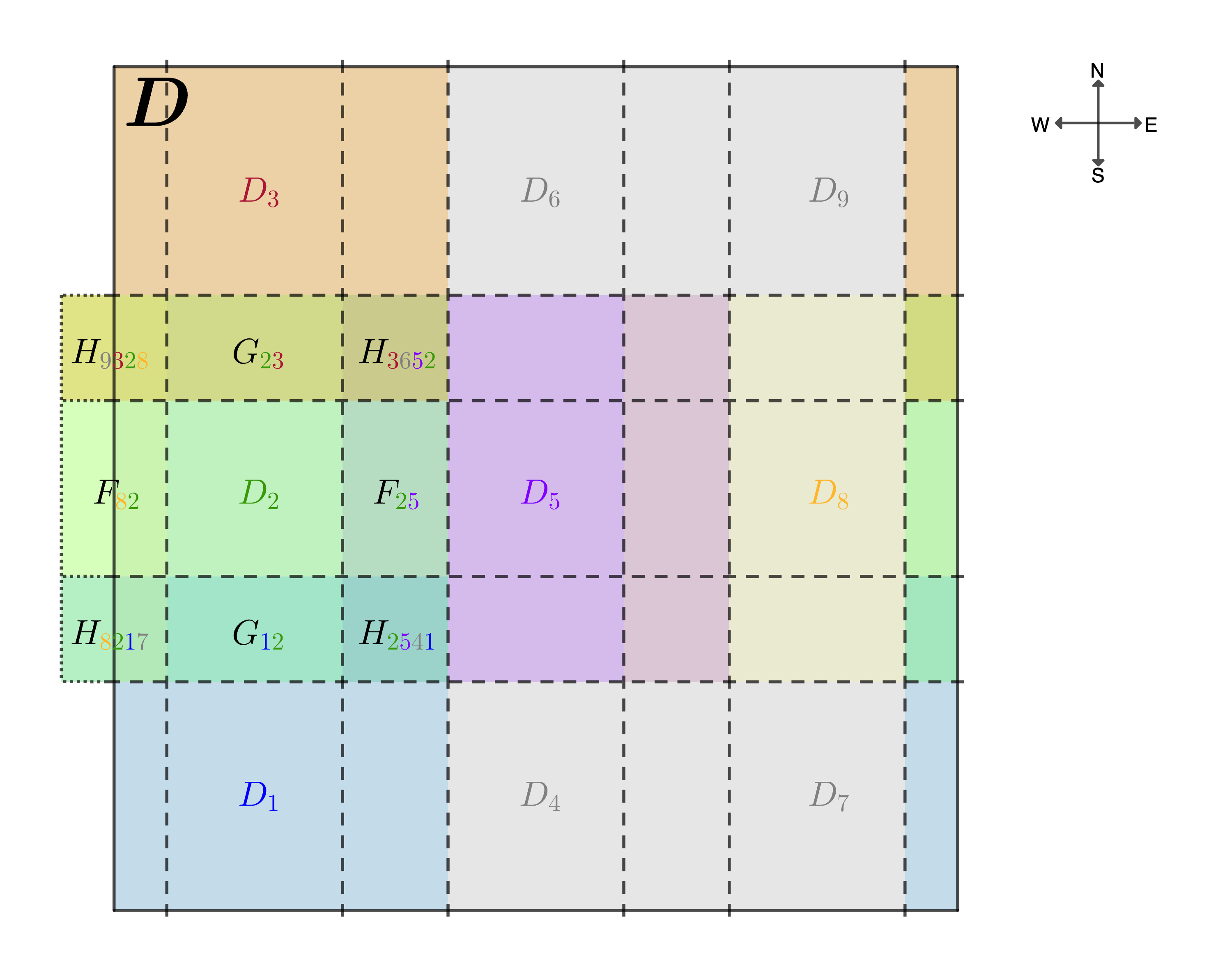}
	\caption{Square domain $D$ divided into 9 subregions $D_1, \dots, D_9$. We assume periodic $EW$-boundary conditions. We highlight the boundary region $D_2$ (green) and the regions adjacent to it at the north ($D_3$, orange), east ($D_5$, purple), south ($D_1$, blue) and west ($D_8$, yellow). Within $D_2$, we highlight the WE-overlap regions $F_{82} = D_2 \cap D_8$ and $F_{25} = D_2 \cap D_5$, the SN-overlap regions $G_{12} = D_2 \cap D_1$ and $G_{23} = D_2 \cap D_4$, and the corner-overlaps $H_{9328} = D_2 \cap D_8 \cap D_9 \cap D_3 $ at the NW-corner of $D_5$, $H_{3652}$ at the NE-corner, $H_{2541}$ at the SE-corner and $H_{8217}$ at the SW-corner. Note that $D_2$ ``wraps-around'' from the east to the west boundary of $D$, and so do the overlap-regions $F_{82}, H_{9328}$ and $H_{8217}$.}
	\label{fig:boundary_region_overlap}
\end{figure}

\begin{rmk}
	We will make the purpose of the overlap regions clear when we discuss our strategy for merging the local particles back into global ones in Section~\ref{sec:merging}. At this stage, we simply remark that they provide a buffer at the interface between two (or more) adjacent subregions to make spatial interpolation easier. For this reason, if we impose periodic boundary conditions on $D$, we need to be particularly careful when constructing the subregions at the boundary: we should think of the regions at a periodic boundary as ``wrapping around", so they overlap with the regions adjacent to them on the opposite side of the domain. We refer to Figure~\ref{fig:boundary_region_overlap} for a visual explanation.
\end{rmk}

For each region $D_j$, denote the local particle ensemble at assimilation time $t_k$ by $\{u^{i,(j)}_{t_k}\}_{i=1}^N$, for all $k \ge 1$. In particular, given the global ensemble $\{u^i_{t_k}\}$, the local particles are defined by the restriction of the global particles to $D_j$, i.e.
\begin{equation}\label{eq:local_parts}
	u^{i,(j)}_{t_k} = u^{i}_{t_k} |_{D_j}, \quad \forall i = 1, \dots, N, \, \forall j = 1, \dots, N_{\mathrm{loc}}.
\end{equation}
\begin{rmk}
	The subregions $\{ D_j \}_{j=1}^{N_{\mathrm{loc}}}$ naturally inherit the same discretization of $D$, so the restriction of a numerical field $v$ on $D$ to a region $D_j$ (denoted by $v |_{D_j}$ as in \eqref{eq:local_parts}) can be understood as selecting the entries of the $\R^{N_u}$-dimensional vector $v$ corresponding to the grid-points in $D_j \subset D$. In particular, the number of grid-points in $D_j$ determines the dimension of the local numerical field $v |_{D_j}$. As before, the notation $v |_{D_j}(z)$ for $z \in D_j$ indicates the value of $v$ at grid-point $z$.
\end{rmk}
Finally, consider the observations $Y$ defined in \eqref{eq:discrete_obs}, so that at each assimilation time $t_k$ the observations $Y_k$ are found at the locations given by the grid-points $\{z^s_k\}_{s=1}^{N_{\mathrm{obs}_k}}, \, z^s_k \in D$. For all $k \ge 1$, we denote the set of indices of the observations belonging to region $D_j$ by $\mathcal{I}^j_{k}$, i.e.
\begin{equation}\label{eq:obs_indices_per_region}
	\mathcal{I}^j_{k} = \left\{ s \, : \, z^s_k \in D_j, \, s \in \{ 1, \dots, N_{\mathrm{obs}_k} \} \right\}.
\end{equation}
 Note that when $z^s_k$ belongs to an overlap region, $s$ belongs to $\mathcal{I}^j_k$ for more than one $j$. Thus, for all $k \ge 1$ and all $j = 1, \dots, N_{\mathrm{loc}}$, we also define the functions \[f^j_k: \{1, \dots, N_{\mathrm{obs}_k} \} \to \{1, \dots, N_{\mathrm{loc}} \}\] such that
 \begin{equation}\label{eq:selection_func}
 	f^j_k(s) =
 	\left\{  \begin{array}{ll}
 		 j & \quad \mathrm{if}\, s \in \mathcal{I}^j_k, \\
 		\min \{ r \, : \,  s \in \mathcal{I}^r_k \} &\quad \mathrm{if}\, s \notin \mathcal{I}^j_k,
 	 \end{array} \right.
 \end{equation}
 which assigns the index $r$ of a subregion $D_r$ to (the index $s$ of) each observation's location $z^s_k$. If $z^s_k \in D_j$, the index $j$ is always assigned (even when $z^s_k$ belongs to multiple regions).  If $z^s_k \notin D_j$, by convention we select the smallest among the indices of the subregions which $z^s_k$ belongs to.
 
 \subsection{Local likelihood}

We are now ready to introduce a local likelihood function for the computation of the (local) weights of the local ensembles. Let us start with a version of the Gaspari--Cohn localization function, as introduced in \cite{gaspari_cohn}.
\begin{defn}[Gaspari-Cohn localization function]\label{def:gaspari-cohn}
	Let $D_1, \dots, D_{N_{\mathrm{loc}}} \subset D$ be $N_{\mathrm{loc}}$ subregions of $D$,  let the observations $Y$ be defined as in \eqref{eq:discrete_obs}, so that at each assimilation time $t_k$ the observations $Y_k = (Y^1_k, \dots, Y^{N_{\mathrm{obs}_k}}_k)$ are located at the points $\{z^s_k\}_{s=1}^{N_{\mathrm{obs}_k}}$, $z^s_k \in D$. Let $\alpha \in \R^{+}$. For all $k \ge 1$, let $\rho^{\alpha}_k : \{ 1, \dots, N_{\mathrm{loc}}\} \times \{1, \dots, N_{\mathrm{obs}_k}\} \to [0,1] $ be the \textit{Gaspari--Cohn localization function}, defined by
	\begin{equation}\label{eq:gaspari-cohn}
		\rho_k^{\alpha}(j,s) = \left\{  \begin{array}{ll}
			1 & \quad \mathrm{if}\, z^s_k \in D_j, \\
			\exp\{ - \alpha d(D_j, z^s_k) \} &\quad \mathrm{if}\, z^s_k \notin D_j,
			\end{array}\right.
	\end{equation}
where $d(A, z)$ is a distance function between a convex region $A \subset D$ and a point $z \in D\setminus A$.
\end{defn}
\begin{defn}[Local likelihood function]\label{def:local_likelihood}
	Let $v$ be a numerical field on $D$, and let $D_1, \dots, D_{N_{\mathrm{loc}}}$, $Y$ and $\alpha$ be as in Definition~\ref{def:gaspari-cohn}. Recall the definitions of $\mathcal{I}^j_k$ and $f^j_k$ for $j = 1, \dots, N_{\mathrm{loc}}$ and $k \ge 1$ given in \eqref{eq:obs_indices_per_region} and \eqref{eq:selection_func}.
	For all $j = 1, \dots, N_{\mathrm{loc}}$ and all $k \ge 1$, denote by $g^{\alpha, j}_k : \R^{N_u} \to \R$ the \textit{local likelihood function} for region $D_j$, and let it be defined by
	\begin{align}\label{eq:loc_likelihood}
		g^{ \alpha, j}_k (v) \approx  \exp \Bigg\{
		&-\frac{1}{2}  \sum_{s = 1}^{N_{\mathrm{obs}_k}} \tilde g_{j, k , s} (v)^{\top} (\Sigma^{s}_k)^{-1} \tilde g_{j, k , s} (v) \rho_k^{\alpha}(j,s) \Bigg\},
	\end{align}
	where $\rho^{\alpha}_k$ is the Gaspari--Cohn localization function defined in \eqref{eq:gaspari-cohn}, and $\tilde g_{j, k , s} : \R^{N_u} \to \R^{\tilde{d_y}}$ is given by
	\begin{equation*}
		\tilde g_{j, k , s} (v) := Y^{s}_k - h^s_k(v|_{D_{f^j_k(s)}}(z^s_k)).
	\end{equation*}
\end{defn}
\begin{rmk}
	Note that if we evaluate \eqref{eq:loc_likelihood} at $u^{i}_{t_k}$, where $u^i$ is a particle of the (global) ensemble and $t_k$ is an assimilation time, we have
	\begin{align}\label{eq:loc_likelihood_loc_ensemble}
		g^{ \alpha, j}_k (u^i_{t_k}) 
%		&\approx  \exp \Bigg\{
%		-\frac{1}{2}  \sum_{s \in \mathcal{I}^j_k}  \big( Y^{s}_k - h^s_k(u^i_{t_k}|_{D_j}(z^s_k)) \big)^{\top} (\Sigma^{s}_k)^{-1} \big( Y^{s}_k - h^s_k(u^i_{t_k}|_{D_j}(z^s_k)) \big) \nonumber \\
%		&\qquad \quad -\frac{1}{2}  \sum_{s \notin \mathcal{I}^j_k}  \big( Y^{s}_k - h^s_k(u^i_{t_k}|_{D\setminus D_j}(z_k^s)) \big)^{\top} (\Sigma^{s}_k)^{-1}   \big( Y^{s}_k - h^s_k(u^i_{t_k}|_{D\setminus D_j}(z_k^s)) \big) \rho_k^{\alpha}(j,s) \Bigg\} \nonumber \\
		&\approx  \exp \Bigg\{
		-\frac{1}{2}  \sum_{s \in \mathcal{I}^j_k}  \big( Y^{s}_k - h^s_k(u^{i,(j)}_{t_k}(z^s_k)) \big)^{\top} (\Sigma^{s}_k)^{-1} \big( Y^{s}_k - h^s_k(u^{i,(j)}_{t_k}(z^s_k)) \big) \nonumber \\
		&\qquad \quad -\frac{1}{2}  \sum_{s \notin \mathcal{I}^j_k}  \big( Y^{s}_k - h^s_k(u^{i, (f^j_k(s))}_{t_k}(z_k^s)) \big)^{\top} (\Sigma^{s}_k)^{-1}   \big( Y^{s}_k - h^s_k(u^{i, (f^j_k(s))}_{t_k}(z_k^s)) \big) \rho_k^{\alpha}(j,s) \Bigg\},
	\end{align}
	where the function $f^j_k$ is defined in \eqref{eq:selection_func}, and we recognize the local particle $u^{i, (j)}$ for the computation of the part of the likelihood dealing with observations $Y^s_k$ located at grid-points $z^s_k \in D_j$, and the local particles $u^{i, (f^j_k(s))}$ for the observations $Y^s_k$ such that $z^s_k \notin D_j$ and $z^s_k \in D_{f^j_k(s)}$.
\end{rmk}

\begin{rmk}
	For simplicity, we took the observation noise in \eqref{eq:discrete_obs} to be Gaussian, but the definition of the local likelihood function \eqref{eq:loc_likelihood} can be adapted to any other noise distribution.
\end{rmk}

\begin{defn}[Local weights]\label{def:local_weights}
	Let $D_1, \dots, D_{N_{\mathrm{loc}}}$, $Y$, $\alpha$ and $\mathcal{I}^j_k$ for $j = 1, \dots, N_{\mathrm{loc}}$ be as in Definition~\ref{def:local_likelihood}. For all $j = 1, \dots, N_{\mathrm{loc}}$, $k \ge 1$, and $i = 1, \dots, N$, we denote by $w^{i, \alpha, (j)}_k$ the \textit{local weight} in subregion $D_j$ of particle $u^i$ at time $t_k$, and define it by
	\begin{equation}\label{eq:local_weights}
		w^{i, \alpha, (j)}_k = \frac{g^{\alpha, j}_k(u^i_{t_k})}{\sum_{j=1}^N g^{\alpha, j}_k(u^i_{t_k})},
	\end{equation}
	where $g^{\alpha, j}_k$ evaluated at particle $u^i_{t_k}$ is to be understood as a function of the local particles $u^{i, (l)}_{t_k}$ for $l = 1, \dots, N_{\mathrm{loc}}$ as in \eqref{eq:loc_likelihood_loc_ensemble}.
\end{defn}

We compute the weights of the local ensembles using Definition~\ref{def:local_weights}, and then proceed to apply tempering, jittering and resampling as in Algorithm~\ref{alg:tempering}, but now locally. However, because of the dependence of the local weights on the global particles (or, equivalently, the dependence of the local weights for region $D_j$ on all of the local ensembles $\{u^{i, (l)}\}_{i = 1}^N$, for $l = 1, \dots, N_{\mathrm{loc}}$), we come to an important choice in our implementation: whether or not to allow local updates of particles (due to tempering and jittering) outside of region $D_j$ to influence the local particle filter in region $D_j$. Let us expand briefly on this problem.

Assume that at time $t_{k-1}$ we have the usual global ensemble $\{u_{t_{k-1}}^i\}_{i=1}^N$, and that observation $Y_{k-1}$ has been assimilated. We evolve the ensemble forward according to \eqref{eq:SPDE_general} up to the next assimilation time $t_{k}$, and we split the resulting global ensemble $\{u_{t_{k}}^i\}_{i=1}^N$ in its local counterparts $\{u_{t_{k}}^{i, (j)}\}_{i=1}^N$ for $j = 1, \dots, N_{\mathrm{loc}}$. We now want to apply the tempering Algorithm~\ref{alg:tempering} in each subregion $D_j$ for $j = 1, \dots, N_{\mathrm{loc}}$. For the purpose of clarity, let us fix $j$ for the moment and focus on region $D_j$. At the start of the tempering iterations, the weights for the local ensemble $\{u^{i, (j)}_{t_{k}}\}$ computed using \eqref{eq:local_weights} are simply given by evaluating the global ensemble at the locations $z^s_k$ for $s = 1, \dots, N_{\mathrm{obs}_k}$ of the observations $Y_{k}$. At the end of the first tempering iteration, the local particles $\{u^{i, (j)}_{t_{k}}\}$ have been modified (due to resampling and jittering). For all subsequent tempering iterations, we have two choices:
\begin{enumerate}
	\item In \eqref{eq:local_weights}, only update the part of the likelihood function \eqref{eq:loc_likelihood} which deals with observations at locations $z^s_k \in D_j$, determined by the ensemble $\{u^{i, (j)}_{t_{k}}\}$, and consider the other local ensembles $\{u^{i, (l)}_{t_{k}}\}$ for $l \neq j$, which are needed to compute the part of the likelihood that deals with observations at $z^s_k \notin D_j$, as \textit{fixed}.
	\item Let the local ensembles $\{u^{i, (l)}_{t_{k}}\}$ for $l \neq j$ vary according to the tempering algorithm applied locally in each region: for example, before proceeding with the second tempering iteration in region $D_j$, wait until the first tempering iteration has completed in all the other regions as well, communicate the new local particles positions and update the likelihood accordingly for the computations of the local weights.
\end{enumerate}

\begin{rmk}
	Recall that we are looking for a localization algorithm which scales with the number of subdomains $N_{\mathrm{loc}}$: in other words, we want to be able to run the tempering algorithm in parallel in each region. If one is satisfied with running the data-assimilation sequentially, the issue raised here is not relevant, since in the computation of the weights for each subregion $D_j$ one should simply use the local ensembles, some of which will have already gone through the tempering procedure (those in $D_l$ for $l < j$), and some of which will only be updated after the data-assimilation in $D_j$ has been completed.
\end{rmk}

Clearly, from an algorithmic point of view, between the two options above the second one is the hardest to implement, since, when writing the parallelization scheme, the cores which deal with the computations in each subdomain need to be made to communicate information to each other throughout the data-assimilation process. The task is made more complex by the fact that the number of tempering iterations needn't be equal across the subdomains: in fact, they will most likely be different, since one of the advantages of localizing the data-assimilation procedure is that (almost) no update to the particles should be needed in the subregions where there are no observations to be found. The implementation of such an algorithm goes beyond the scope of this paper. In terms of performance of the particle filter, the choice between the first and second option matters only if $\alpha$ in \eqref{eq:gaspari-cohn} is small, or if the observations are located in neighbouring subregions. Operating under the assumption that the observations outside a subdomain $D_j$ should have very little impact on the weights of the local ensemble, the first option suggested above should yield acceptable results, and is the one that we choose to implement (see Algorithm~\ref{alg:loc_PF}).

\subsection{Merging the local particles and spatial interpolation}\label{sec:merging}
At the end of each assimilation step, once the tempering iterations have concluded in all the local subregions, we are left with $N_{\mathrm{loc}}$ local ensembles $\{u^{i, (j)}_{t_k}\}_{i=1}^N$ for $j = 1, \dots, N_{\mathrm{loc}}$ of $N$ local particles each. These local ensembles are generally \textit{discontinuous} at the subregions' boundaries. We briefly expand on this.

Fix an index $i$ (and a time-step $t_k$) and consider the local particles $\{u^{i, (j)}_{t_k}\}_{j=1}^{N_{\mathrm{loc}}}$. When we construct the local particles at the start of the assimilation step (line~\ref{line:local_ens_construction} of Algorithm~\ref{alg:loc_PF}), they coincide on all the overlap regions, i.e.~$u^{i, (e)}_{t_k}(F_{ew}) = u^{i, (w)}_{t_k}(F_{ew})$ for all EW-overlap regions $F_{ew}$, $u^{i, (s)}_{t_k}(G_{sn}) = u^{i, (n)}_{t_k}(G_{sn})$ for all SN-overlap regions $G_{sn}$, and $u^{i, (a)}_{t_k}(H_{abcd}) = u^{i, (b)}_{t_k}(H_{abcd}) = u^{i, (c)}_{t_k}(H_{abcd}) = u^{i, (d)}_{t_k}(H_{abcd})$ for all corner-overlap regions $H_{abcd}$. By the end of the local assimilation step, for each $j = 1, \dots, N_{\mathrm{loc}}$, the local particle $u^{i,(j)}_{t_k}$ has gone through independent tempering and jittering iterations: for the same index $i$, particle  $u^{i,(j)}_{t_k}$ might have been left invariant in some sub-regions $D_j$, while being resampled and modified in others. Since the algorithm operates independently in each subregion, the local particles $u^{i,(j)}_{t_k}$ do not necessarily agree on the overlap regions $F_{ew}, G_{sn}$ and $H_{abcd}$ any more. Thus, in order to reconstruct a global particle $u^i_{t_k}$ on the whole domain $D$, we need to take care of these spatial discontinuities.

\begin{rmk}
	The use of the Gaspari--Cohn function and the introduction of the overlap regions give us some hope that the local weights of particles $u^{i,(j)}_{t_k}$ (for a fixed index $i$) in neighbouring regions are not too dissimilar. However, while these features mitigate excessive local weights discontinuities across neighbouring regions, the tempering, resampling and jittering algorithms ultimately rely on the generation of random variables, so significant spatial discontinuities can still, in principle, happen.
\end{rmk}

The idea behind the algorithm for merging local particles back together is as follows. By defining the overlap regions, at the end of the local assimilation step we end up with two (in EW- and NS-overlap regions) or four (in corner-overlap regions) independent copies of each local particle $u^{i, (j)}_{t_k}$ in each overlap region. Define the following interpolation functions \footnote{Note that for simplicity we use linear interpolation, but other types of interpolation are equally suitable.}.
\begin{defn}[Interpolation functions]
	Let $u$ and $v$ be numerical fields defined on the same quadrilateral grid $L = [x_e, x_w] \times [y_s, y_n]$ consisting of $m_1 \times m_2$ grid-points $(x,y)$, where $x_e$ and $x_w$ are the $x$-coordinates of the eastern- and western-most grid points, and $y_s$ and $y_n$ are the $y$-coordinates of the southern- and northern-most grid-points. Define the \textit{horizontal} and \textit{vertical interpolation functions} $\mathrm{Interp}_{ew}, \mathrm{Interp}_{sn} : \R^{m_1 \times m_2} \times \R^{m_1 \times m_2} \to \R^{m_1 \times m_2}$ as
	\begin{align*}
		\mathrm{Interp}_{ew}(u,v)(x,y) &:= u(x,y) \frac{x_w - x}{x_w - x_e} + v(x,y) \frac{x - x_e}{x_w - x_e}, \quad \forall (x,y) \in L, \\
		\mathrm{Interp}_{sn}(u,v)(x,y) &:= u(x,y) \frac{y_n - y}{y_n - y_s} + v(x,y) \frac{y - y_s}{y_n - y_s}, \quad \forall (x,y) \in L.
	\end{align*}
\end{defn}

Using Figure~\ref{fig:center_region_overlap} for visual reference as needed, for each $i = 1, \dots, N$, we construct the global particle $u^i_{t_k}$ by merging the local particles $\{u^{i, (j)}_{t_k} \}_{j = 1}^{N_{\mathrm{loc}}}$ as follows:
\begin{itemize}
	\item For each $j$, on the subregion $\mathring{D}_{j} = D_j \setminus (F_{ej} \cup F_{jw} \cup G_{sj} \cup G_{jn})$, where $e, w, s$ and $n$ are the indices of the subregions neighbouring $D_j$ to the east, west, south and north respectively, we set $u^i_{t_k} |_{\mathring{D}_j} = u^{i, (j)}_{t_k} |_{\mathring{D}_j}$.
	\item On each EW-overlap region $F_{ew}$, we define $u^i_{t_k} |_{{F}_{ew}}$ as the (horizontal) interpolation of $u^{i, (e)}_{t_k} |_{{F}_{ew}}$ to the east and $u^{i, (w)}_{t_k} |_{{F}_{ew}}$ to the west, i.e.~$u^i_{t_k}(x,y) = \mathrm{Interp}_{ew}(u^{i, (e)}_{t_k}, u^{i, (w)}_{t_k} )(x,y)$ for all $(x,y) \in F_{ew}$.
	\item On each SN-overlap region $G_{sn}$, we define $u^i_{t_k} |_{{G}_{sn}}$ as the (vertical) interpolation of $u^{i, (s)}_{t_k} |_{{G}_{sn}}$ to the south and $u^{i, (n)}_{t_k} |_{{G}_{sn}}$ to the north, i.e.~$u^i_{t_k}(x,y) = \mathrm{Interp}_{sn}(u^{i, (s)}_{t_k}, u^{i, (n)}_{t_k} )(x,y)$ for all $(x,y) \in G_{sn}$ .
	\item On each corner-overlap region $H_{abcd}$, we define $u^i_{t_k} |_{H_{abcd}}$ by repeated interpolation of $u^{i, (a)}_{t_k} |_{H_{abcd}}$, $u^{i, (b)}_{t_k} |_{H_{abcd}}$, $u^{i, (c)}_{t_k} |_{H_{abcd}}$ and $u^{i, (d)}_{t_k} |_{H_{abcd}}$. We first interpolate horizontally $u^{i, (a)}_{t_k} |_{H_{abcd}}$ to the east and $u^{i, (b)}_{t_k} |_{H_{abcd}}$ to the west, and $u^{i, (d)}_{t_k} |_{H_{abcd}}$ to the east and $u^{i, (c)}_{t_k} |_{H_{abcd}}$ to the west, obtaining $\mathrm{Interp}_{ew}(u^{i, (a)}_{t_k}, u^{i, (b)}_{t_k} )$ and $\mathrm{Interp}_{ew}(u^{i, (d)}_{t_k}, u^{i, (c)}_{t_k} )$. We then interpolate the two resulting numerical fields vertically, the first to the north and the second to the south, and get, for all $(x,y) \in H_{abcd}$,
	\begin{equation*}
		u^{i}_{t_k}(x,y) =\mathrm{Interp}_{sn} \Big( \mathrm{Interp}_{ew}(u^{i, (d)}_{t_k}, u^{i, (c)}_{t_k} ), \mathrm{Interp}_{ew}(u^{i, (a)}_{t_k}, u^{i, (b)}_{t_k} ) \Big) (x,y).
	\end{equation*}
\end{itemize}
	
	\begin{rmk}
		For clarity of exposition, in this section we assumed the evaluation of the various numerical fields at \textit{fixed reference grid-points}, denoted by $(x,y)$ in $D$. In practice, since the local numerical fields are defined on different subregions, and they only overlap on subsets of these, when defining the interpolation algorithms one needs to pay close attention to matching the indices of the arrays correctly.
	\end{rmk}
	
\subsection{On the error due to localization}
In this section we briefly give a heuristic understanding of the error due to the localization procedure, in comparison to the PF with resampling, tempering and jittering of Algorithm~\ref{alg:PF}. There are two errors to consider: the first is due to the modification of the likelihood function, and the second is due to interpolation. 

The first error is somewhat easier to understand. Consider the local likelihood \eqref{eq:loc_likelihood}. Since taking the limit as $\alpha \to 0$ in the Gaspari--Cohn function \eqref{eq:gaspari-cohn} yields $\rho_k^{\alpha}(j,s) \to 1$ for all $k \ge 1$, $j \in \{1, \dots, N_{\mathrm{loc}}\}$ and $s \in \{1, \dots, N_{\mathrm{obs}_k}\}$, we have that the local likelihood $g^{ \alpha, j}_k$ converges to the likelihood function $g_k$ as $\alpha \to 0$. Thus, as $\alpha \to 0$, the local ensembles in each subregion $D_j$ are weighted equally: if one considers a specific particle $u^i_{t_k}$ at assimilation time $t_k$, its local weights $w^{i, \alpha, (j)}_k$ for all $j = 1, \dots, N_{\mathrm{loc}}$ computed according to \eqref{eq:local_weights} converge to its global weight $w^i_k$ computed using \eqref{eq:basic_PF_weights_rec}, as $\alpha \to 0$. Now, if one forgoes resampling, tempering and jittering, and proceeds with a basic PF (described in \eqref{eq:basic_PF_weights_rec}) in each subregion, there are no resulting discontinuities (so no interpolation is necessary). Thus, in this simple setting, the LPF coincides with the basic PF described by the recursive relation \eqref{eq:basic_PF_weights_rec} as $\alpha \to 0$ (and should similarly converge to the filtering distribution as $N \to \infty$). Intuitively, then, one could be tempted to claim that as $\alpha \to 0$, the LPF converges to the PF, and so the error between the two might be somewhat quantified by looking at the difference between the standard and local likelihood functions. However, already by introducing resampling (not to mention tempering and jittering), the matter becomes much more complicated. 

Consider the same situation as before, with $\alpha \to 0$ so the local and global weights agree, although this time we allow resampling: in effect, we want to compare a \textit{local bootstrap PF} and a \textit{boostrap PF}. Clearly, due to the randomness within the resampling algorithms, even if the weights agree across subregions, we cannot discard the possibility that local particles might differ on the overlap regions. This leads us necessarily to the error due to interpolation.

To shed some light on this, introduce notation $\pi^{N, PF}$ for the bootstrap PF and $\pi^{N,LPF}$ for the bootstrap LPF. Recall that $\{ \mathring{D}_j \}_j \cup \{ F_{ew} \}_{(e,w)} \cup \{ G_{sn}\}_{(s,n)} \cup \{ H_{abcd}\}_{(a,b,c,d)}$, where $\mathring{D}_j = D_j \setminus (F_{ej} \cup F_{jw} \cup G_{jn} \cup G_{sj})$, form a partition of $D$. As $D$ is discretized in $N_u$ grid-points, for $A \subset D$ let $X_A$ denote the subset of grid-points of $D$ also in $A$, and $|X_A|$ the number of grid-points in $X_A$, so in particular
\begin{equation*}
	\sum_j |X_{\mathring{D}_j} | + \sum_{(e,w)} |X_{F_{(ew)}}| + \sum_{(s,n)} |X_{G_{(sn)}}| + \sum_{(a,b,c,d)} |X_{H_{(abcd)}}| = N_u.
\end{equation*}
Then  $\{ X_{\mathring{D}_j} \}_j \cup \{ X_{F_{ew}} \}_{(e,w)} \cup \{ X_{G_{sn}}\}_{(s,n)} \cup \{ X_{H_{abcd}}\}_{(a,b,c,d)}$ is a partition of $\R^{N_u}$. In particular, any function $u$ on $D$ can be identified with a vector in $\R^{N_u}$, and we can decompose it as
\begin{equation}\label{eq:vector_decomp}
	u = (\dots, u|_{\mathring{D}_j}, \dots, u|_{F_{(ew)}}, \dots, u|_{G_{(sn)}}, \dots, u|_ {H_{(abcd)}}, \dots),
\end{equation}
where $u|_A \in X_{A}$ is the restriction of $u$ on $A \subset D$, discretized as a vector in $\R^{|X_A|}$. Similarly, given functions on $\mathring{D}_j, F_{(ew)}, G_{(sn)}$ and $H_{(abcd)}$, identified as vectors in $\R^{|X_{\mathring{D}_j}|}, \R^{|X_{F_{(ew)}}|}, \R^{|X_{G_{(sn)}}|}$ and $\R^{|X_{H_{(abcd)}}|}$ respectively, one can define an element of $\R^{N_u}$ by concatenating the vectors as in \eqref{eq:vector_decomp}.

Denote by $\pi^{N, D_j}_{k}$ the local empirical measures on $\mathring{D}_j$ at assimilation time $t_k$ given by the local ensembles $\{u^{i,(j)}_{t_k}\}_{i = 1}^N$, i.e.~after resampling,
\begin{equation*}
	\pi^{N, \mathring{D}_j}_{k} = \frac{1}{N} \sum_{i = 1}^N \delta_{u^{i, (j)}_{t_k}|_{\mathring{D}_j}}, \quad \forall j = 1, \dots, N_{\mathrm{loc}},
\end{equation*}
and by $\pi^{N, F_{ew}}_{k}$ the local empirical measures on the EW-overlap regions, i.e.
\begin{equation*}
	\pi^{N, F_{ew}}_{k} = \frac{1}{N} \sum_{i = 1}^N \delta_{ \mathrm{Interp}_{ew}(u^{i, (e)}_{t_k}|_{F_{(ew)}}, u^{i, (w)}_{t_k}|_{F_{(ew)}})}, \quad \forall (e,w) \in \{ (e,w) \, : \, F_{ew} \text{ is EW-overlap in }D\},
\end{equation*}
and similarly by $\pi^{N, G_{sn}}_{k}$ and $\pi^{N, H_{abcd}}_{k}$ the local empirical measures on the SN-overlap and corner-overlap regions. For all $S \in \mathcal{B}(\R^{N_u})$, the bootstrap LPF can be written as
\begin{align*}
	\pi^{N, LPF}_{k}(S) 
	&= \sum_{j = 1}^{N_{\mathrm{loc}}}  \omega_{\mathring{D}_j}	\pi^{N, \mathring{D}_j}_{k}(S \cap X_{\mathring{D}_j}) + \sum_{(e,w)} \omega_{F_{ew}} \pi^{N, F_{ew}}_{k}(S \cap X_{F_{ew}}) + \sum_{(s,n)}\omega_{G_{sn}} \pi^{N, G_{sn}}_{k}(S \cap X_{G_{sn}}) \\
	&\quad+ \sum_{(a,b,c,d)} \omega_{H_{abcd}} \pi^{N, H_{abcd}}_{k}(S \cap X_{H_{abcd}}),
\end{align*}
where $\omega_{A} = |X_A|/N_u$ for $A \subset D$.

Now, after the first assimilation step we can expect $\pi^{N, LPF}_{1}(S) \approx \pi^{N, PF}_{1}(S)$ for $S \in \mathcal{B}(\cup_j X_{\mathring{D}_j})$ since $\alpha = 0$ and global and local weights agree (and they should be equal if we let $N \to \infty$). This is of course not the case when $S \in \mathcal{B}(\R^{N_u} \setminus \cup_j X_{\mathring{D}_j})$, and quantifying the difference between $\pi^{N, LPF}_{1}(S)$ and $\pi^{N, PF}_{1}(S)$ remains problematic. However, if the overlap regions are small, and the weights are smooth enough across subregions, one can hope that the overall difference between $\pi^{N, LPF}_{1}$ and $\pi^{N, PF}_{1}$ is small too. Sadly, this reasoning only holds up to the first assimilation time $t_1$. As the system moves forward in time, the errors in the bootstrap LPF due to interpolation in the overlap-regions propagate in space according to the model dynamics \eqref{eq:SPDE_general}, which, in a geophysical setting, are most likely nonlinear. Thus, at later assimilation times, $\pi^{N, LPF}_{1k}(S)$ and $\pi^{N, PF}_{k}(S)$ might not agree even for $S \in \mathcal{B}(\cup_j X_{\mathring{D}_j})$.

All these considerations should not discourage us. When developing PF in practice, we already allow for an approximation error (compared to the optimal filtering distribution) due to the finite-ness of $N$. The further we modify the basic PF, the bigger the error we expect in our estimate: the resampling, jittering and tempering procedures all introduce further approximation errors, necessarily diverging further from the optimal filter. Following along this reasoning, we do not expect the added error due to localization to have significant repercussions. In fact, when working with small $N$, the greater flexibility allowed by the local tempering procedures (compared to a global one) can even improve on the performance of the filter. We plan to explore the theoretical aspects of the localization error in more detail in future works. For the moment, we present some numerical results comparing the PF of Algorithm~\ref{alg:PF} and the LPF of Algorithm~\ref{alg:loc_PF} which show that the difference in performance between the two, when the parameters are properly calibrated, is indeed small.

\section{Numerical results: data-assimilation for the (viscous) stochastic RSW equations}\label{sec:numerics}

In this section we test the particle filter with tempering and jittering (comparing results from both the localized version of Algorithm~\ref{alg:loc_PF} and the non-localized version of Algorithm~\ref{alg:PF}) on a geophysical model from the SALT (Stochastic Advection by Lie Transport) framework \cite{Holm_SALT_15}. We will perform what in the data-assimilation literature is often referred to as a ``twin experiment", which is in fact simply a filtering problem: the dynamics of the signal, the sensor function and the distribution of the noise are all known, thus errors due to the misspecification of model parameters are avoided. The data is generated by fixing a realization of an SPDE as the signal, and perturbing it to obtain the observations; the particles of the (L)PF are evolved forward with the same SPDE model used to generate the signal, and the distribution of the noise and the sensor functions are known and used in the computation of the likelihood weights.
\begin{notation}
	For the purpose of clarity, in what follows we identify vectors or vector-valued functions with boldface characters.
\end{notation}
We consider the 2-dimensional SALT (viscous) Rotating Shallow Water equations, given by
\begin{equation}\label{eq:SRSW}
\begin{aligned}
  \mathrm{d} \mathbf{u}
	 &+ \bigg[ (\mathbf{u} \cdot \nabla) \mathbf{u} + \frac{f}{\mathrm{Ro}} \hat{\mathbf{z}} \times \mathbf{u} \bigg] \dt \\
	 &\begin{aligned} \, + \sum_n \left[ \left(\boldsymbol{\xi}_n \cdot \nabla \right) \mathbf{u}  + \nabla \boldsymbol{\xi}_n \cdot \mathbf{u} +\frac{f}{\mathrm{Ro}} \hat{\mathbf{z}} \times \boldsymbol{\xi}_n \right] \circ \mathrm{d} W_t^n &= \nu \Delta \mathbf{u} - \frac{1}{\mathrm{Fr}^2} \nabla \eta \dt, \\
	 \di \eta + \nabla \cdot(\eta \mathbf{u}) \dt  +\sum_n \nabla \cdot\left(\eta \boldsymbol{\xi}_n \right) \circ \mathrm{d} W_t^n 
	 &=0,
	 \end{aligned}
\end{aligned}
\end{equation}
where $\mathbf{u}(t, \mathbf{x}) = (u(t,\mathbf{x}),v(t,\mathbf{x}))$ is the flow velocity field and $\eta(t,\mathbf{x})$ is the height of the fluid column, $\hat{\mathbf{z}}$ is the unit vector pointing outwards and $f$ is the Coriolis parameter. The equations are stated in non-dimensional form, with domain $D = [0,1]^{\times 2}$ and $\mathbf{x} \in D$, so that the typical velocity, length and height scales are encoded in the parameters $\mathrm{Ro}$ and $\mathrm{Fr}$. In particular, if $U$ is the typical velocity-scale, $L$ the length-scale and $H$ the height-scale, the \textit{Rossby number} is defined as $\mathrm{Ro} =  U/fL$ and the \textit{Froude number} as $\mathrm{Fr} = U/ \sqrt{gH}$, where $g$ is the gravitational acceleration constant. Finally, the driving noise is given by a spatial Gaussian noise of the form
\begin{equation}\label{eq:noise_SRSW}
	\sum_{n \ge 1} \boldsymbol{\xi}_n(\mathbf{x}) \di W_t^n 
	= \sum_{n \ge 1} \left( \begin{matrix}
		\xi_n^u(\mathbf{x}) \\
		\xi_n^v(\mathbf{x})
	\end{matrix} \right) \di W_t^n, 
\end{equation}
where $\boldsymbol{\xi}_n(\mathbf{x}) = (\xi_n^u(\mathbf{x}), \xi_n^v(\mathbf{x}))$, $n \ge 1$ is a smooth vector field defined on the domain $D$.

Note that, compared to our general setup in Section~\ref{sec:set-up}, the noise term in the SPDE \eqref{eq:SRSW} is expressed in Stratonovich form. One can recover the It\^o formulation by using the classical conversion formula. As regards the infinite-dimensional noise \eqref{eq:noise_SRSW} itself, this corresponds to  \eqref{eq:noise_SPDE} with $\mathcal{K}=L^2(D)$ in each of the two coordinates $u$ and $v$.

The well-posedness of the solution of equation \eqref{eq:SRSW} on the 2-dimensional torus $\mathbb{T}^2$ has been established, in the viscous case, in Crisan~and~Lang~\cite{crisan_lang_RSW}. In particular, the authors show that for initial data $u_0$, $v_0$ and $\eta_0$ in $W^{1,2}(\mathbb{T}^2)$, for all $t$ up to a positive stopping time there exists a unique strong solution to \eqref{eq:SRSW} with $u_t, v_t$ and $\eta_t$ continuous in $W^{1,2}(\mathbb{T}^2)$ and in $L^2([0,T], W^{2,2}(\mathbb{T}^2))$. For further details, we refer to \cite[Theorem~3.1]{crisan_lang_RSW}. Using the existence and pathwise uniqueness proved in \cite{crisan_lang_RSW}, the Markov property can also be derived, along the lines of Remark~4.2.11 and Proposition~4.3.3 in \cite{rockner_spdes}. Thus, the conditions for the signal we stated in Section~\ref{sec:set-up} are satisfied, and we can take the (numerical) solution to \eqref{eq:SRSW} driven by a realization of the noise \eqref{eq:noise_SRSW} to be our signal process.

\begin{rmk}
	In the numerical experiments in this section we set a very small viscosity coefficient $\nu \ll 1$, so our simulations are in fact closer to the inviscid setting. To the best of our knowledge, well-posedness of the the inviscid RSW equations (with or without transport noise) is still an open problem.
\end{rmk}

\subsection{Experiments set-up}
We start by discretizing space and time: we denote by $\delta x$ and $\delta t$ the units of space and time respectively. We discretize the domain $D = [0,1]^{\times 2}$ into a $d \times d$ grid (with $d = 1/\delta x$), and fix $\delta t \ll 1$, so that $t \in \{0, \delta t, 2\delta t, \dots, m \delta t, \dots \}$. We set up our filtering experiment as follows: let $(u_t, v_t, \eta_t)$ be a (fixed) numerical realization of the SRSW equations \eqref{eq:SRSW}. We take the signal to be given by the height of the fluid $\eta_t$; in other words, we do \textit{not} observe at all the velocities. Any numerical solution $(u_t, v_t, \eta_t)$ to \eqref{eq:SRSW} belongs to $(\R^{d\times d})^{\times 2} \times \R^{d \times d}_{+}$, so, in particular, the signal $\eta_t \in \R^{d\times d}_{+}$, and it is a $d^2$-dimensional vector. For the observations, let the assimilation times $t_k = k r \delta t$, for some fixed $r \in \N$ (we will experiment with different values of $r$ in what follows) and all $k \ge 1$ (we assume no observation at time $t=0$), and recall expression \eqref{eq:discrete_obs}.
%\begin{notation}
%	 Throughout this section, let the grid-points of the discretized domain be denoted by $x^s$ with $s = 1, \dots, d^2$, and similarly think of (numerical) solutions $(u_t, v_t, \eta_t)$ to \eqref{eq:SRSW} as vectors, with components $1, \dots, d^2$ denoted with a superscript, so that for all $s$, $(u^s_t, v^s_t, \eta^s_t)$ is the solution to \eqref{eq:SRSW} at grid-point $x^s$. 
%\end{notation}

 For a set of indices $\mathcal{J}$ with $|\mathcal{J}| = N_{\mathrm{obs}}$, let $h_{\mathcal{J}}: \R^{d^2} \to \R^{N_{\mathrm{obs}}}$ be the projection operator that takes a vector in $\R^{d^2}$ and returns a vector in $\R^{N_{\mathrm{obs}}}$ by selecting only the components corresponding the indices in $\mathcal{J}$. Define the observations $Y$ as
\begin{equation}\label{eq:obs_numerics_vec}
	Y_{k} = h_{\mathcal{J}}(\eta_{t_k}) + \sigma B_{k}, \quad \forall k \ge 1,
\end{equation}
where $B_k \iid N(0, \mathrm{Id})$ are independent $N_{\mathrm{obs}}$-dimensional scaled multivariate standard normal random variables, and $\sigma \in \R^+.$

\begin{rmk}
	One can write \eqref{eq:obs_numerics_vec} in the form of \eqref{eq:discrete_obs} as follows: for a collection of $N_{\mathrm{obs}}$ grid-points $ \{z^1, \dots, z^{ N_{\mathrm{obs}}} \} \in D$, take the observations $Y$ to be given by
	\begin{equation*}
		Y_{k} = 
		\left( \begin{array}{l}
			Y^1_k \\
			\vdots \\
			Y_k^{N_{\mathrm{obs}}}
		\end{array}
		\right)	
		=
		\left( \begin{array}{c}
			\eta_{t_k}(z^1) + \sigma B^1_{k} \\
			\vdots \\
			\eta_{t_k}(z^{N_{\mathrm{obs}}}) + \sigma B^{N_{\mathrm{obs}}}_{k}
		\end{array}
		\right),	
	\end{equation*}
	where by $\eta_{t_k}(z)$ we mean the value of the signal $\eta$ at time $t_k = k r \delta t$ and ``evaluated" at grid-point $z$, and $B^i_k$ are i.i.d. normal random variables (1-dimensional) for $i  = 1, \dots, N_{\mathrm{obs}}$ and all $k \ge 1$.
\end{rmk}
\begin{figure}[H]
	\centering
	\includegraphics[width=0.315\linewidth]{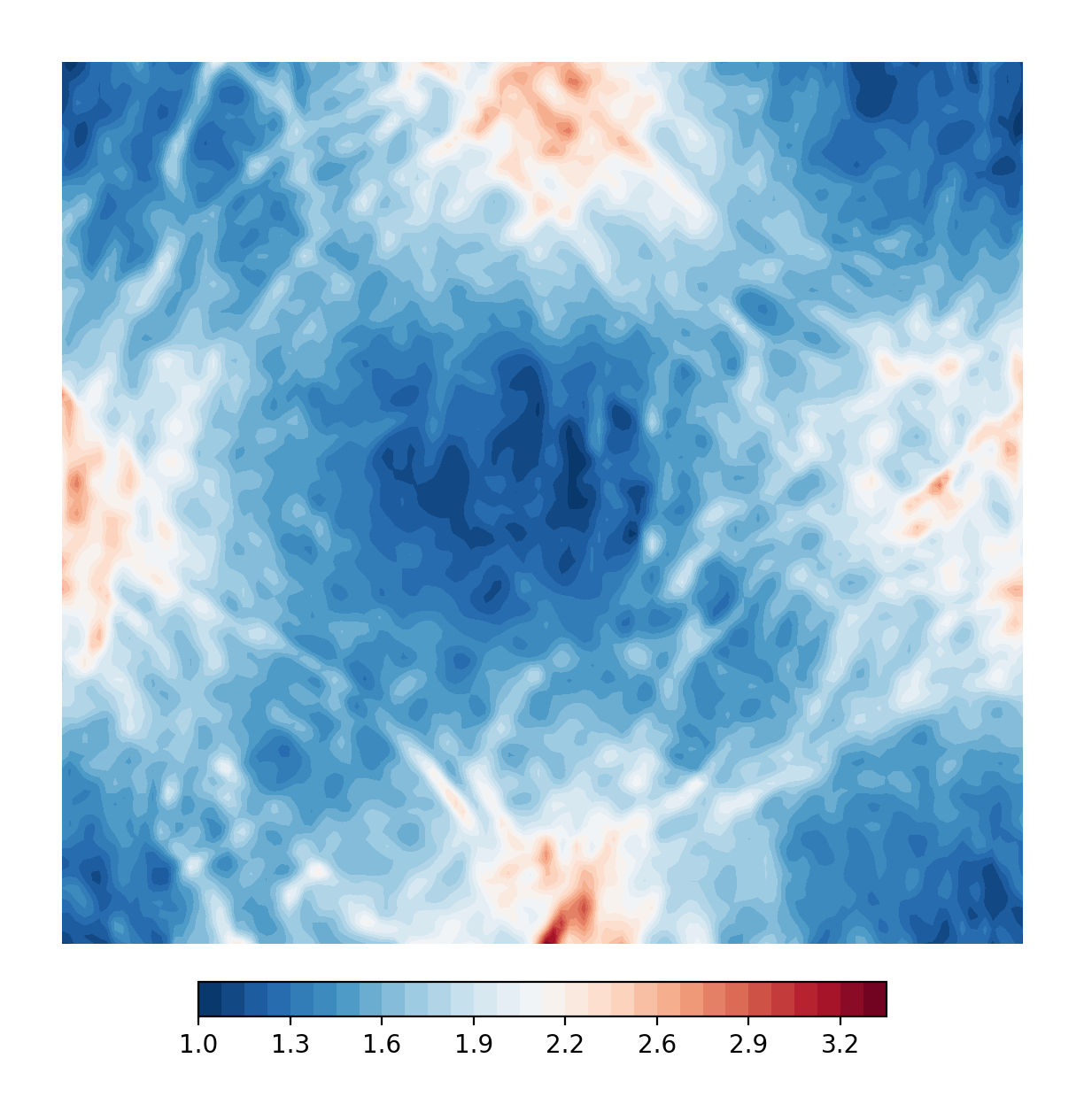}
	\includegraphics[width=0.315\linewidth]{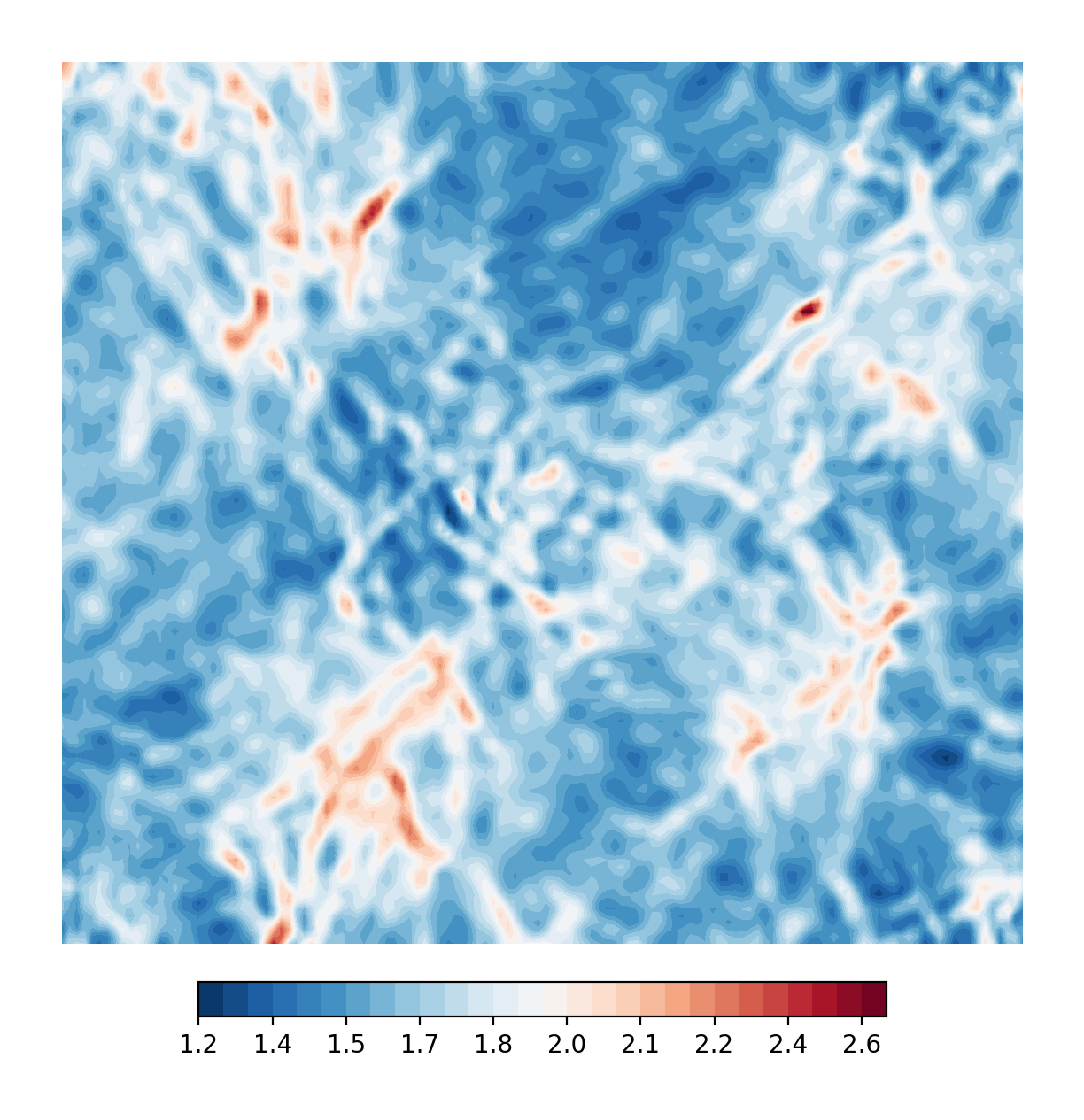}
	\includegraphics[width=0.315\linewidth]{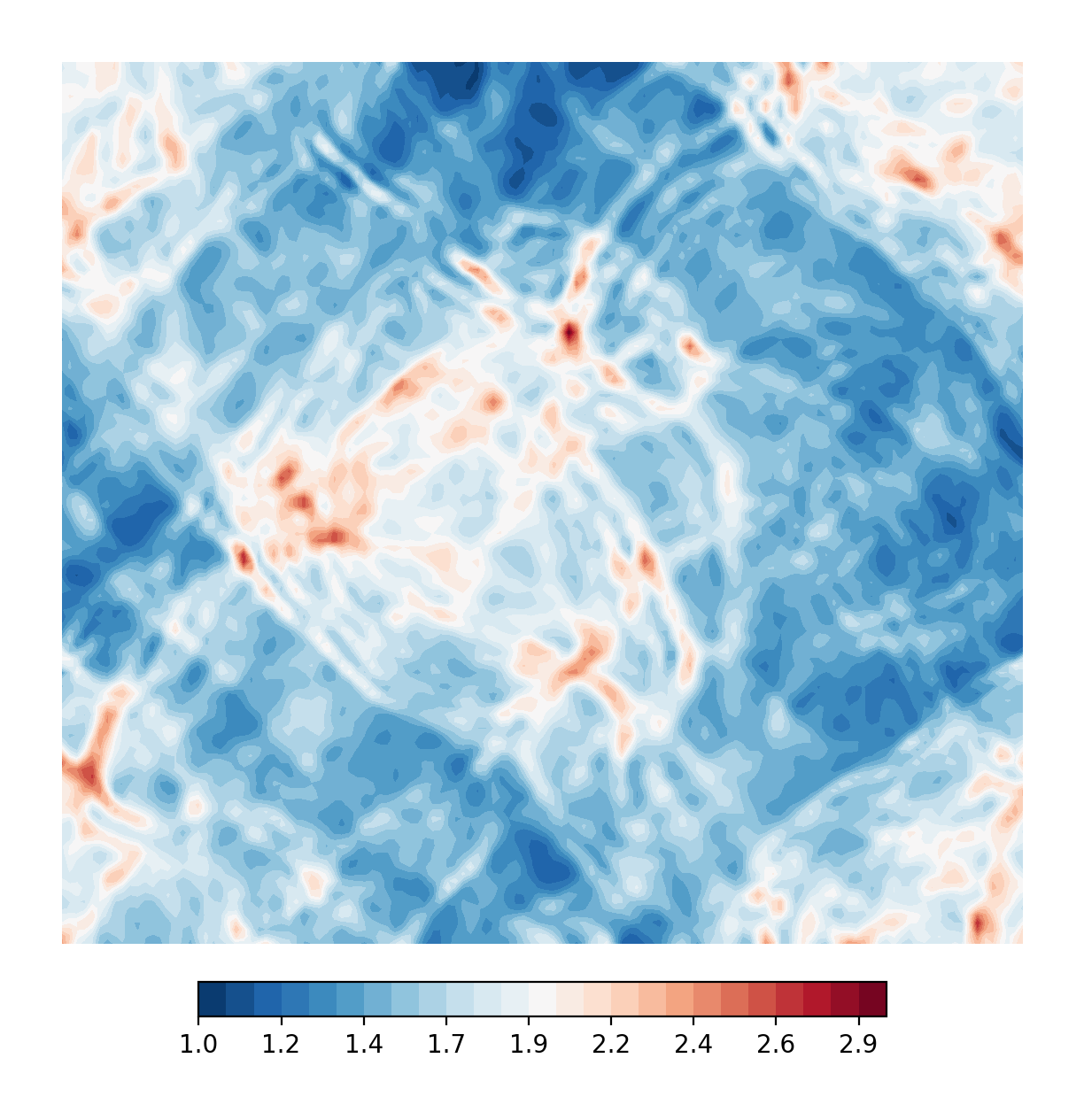} \\
	\includegraphics[width=0.315\linewidth]{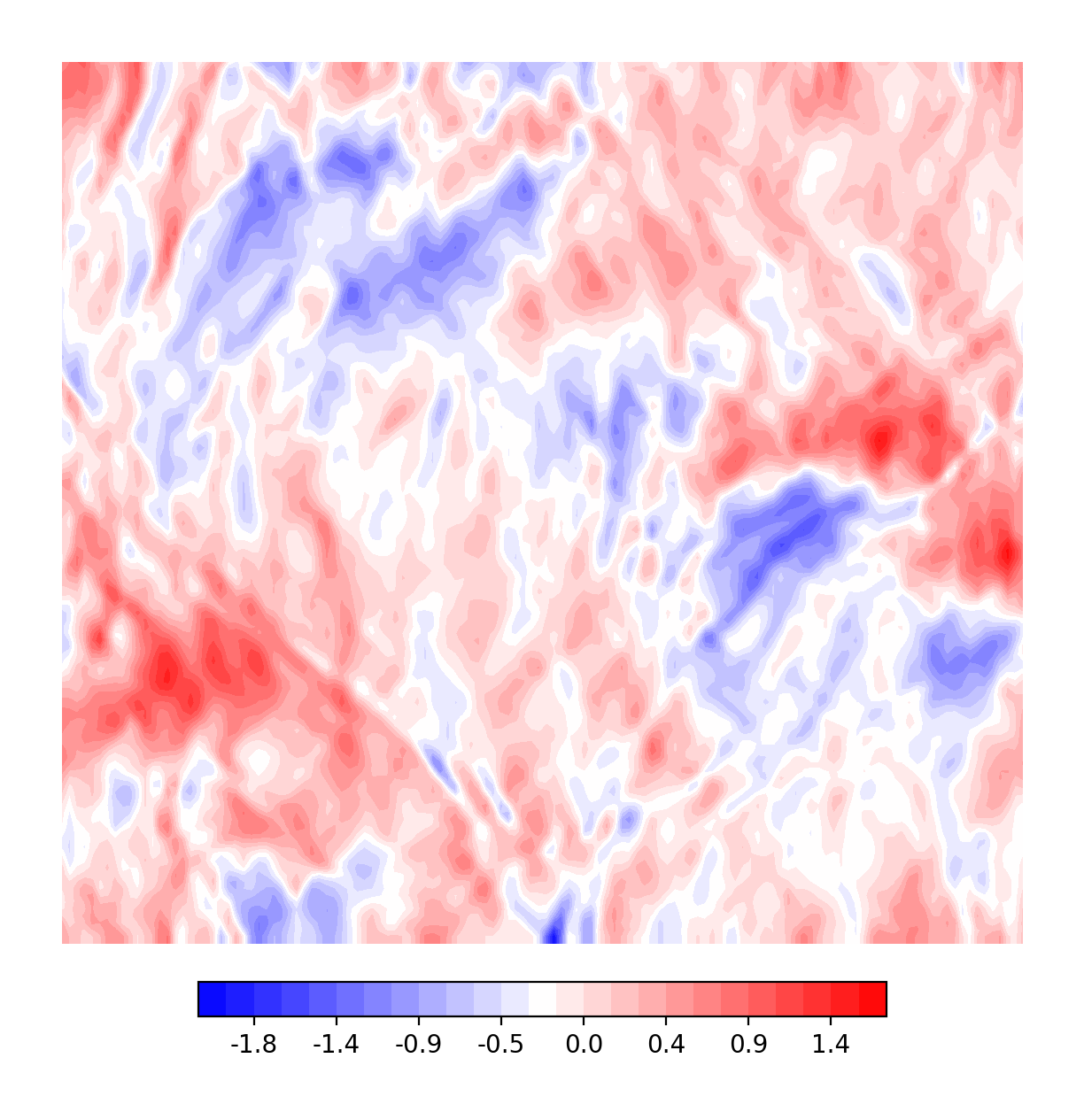}
	\includegraphics[width=0.315\linewidth]{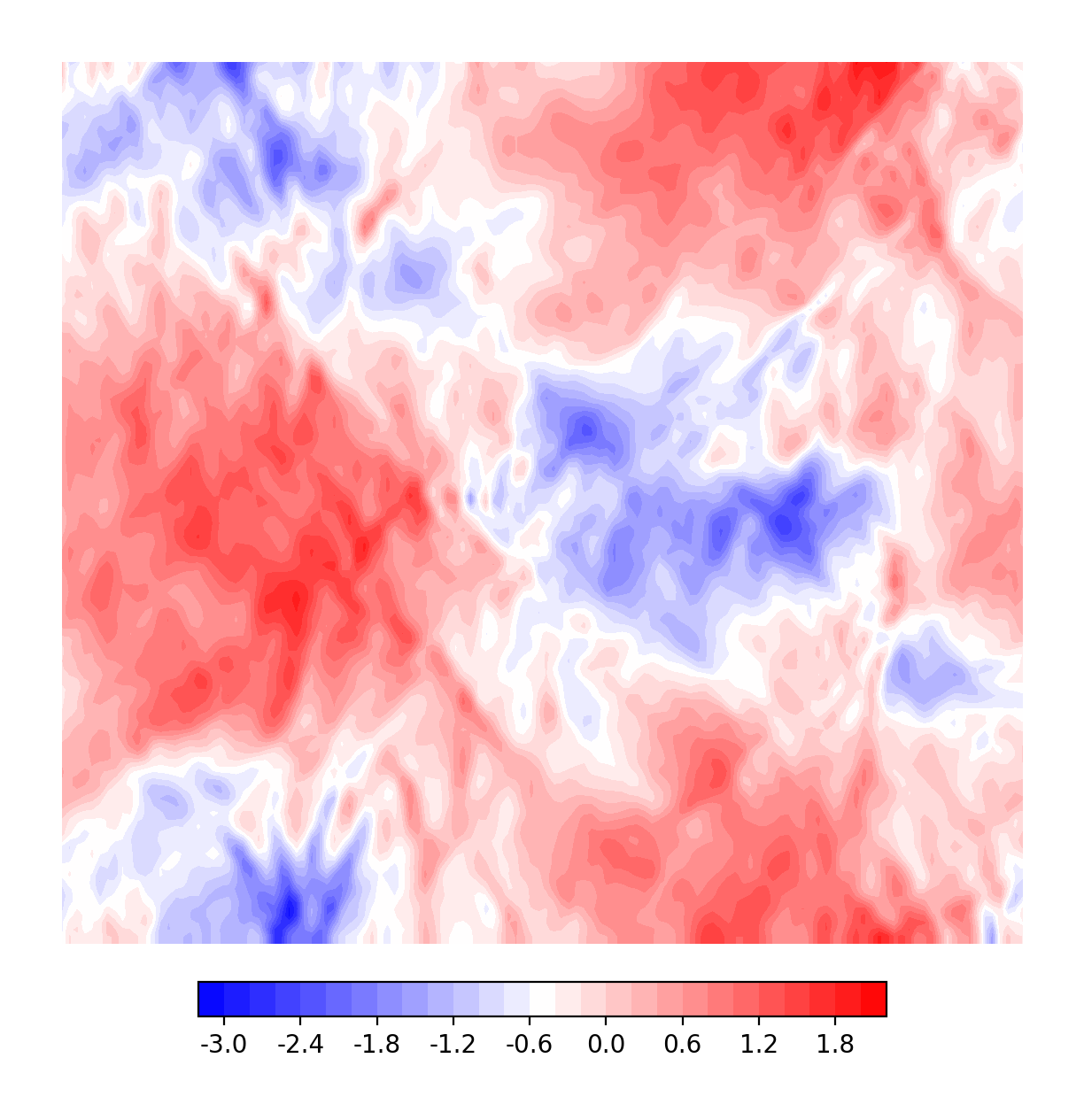}
	\includegraphics[width=0.315\linewidth]{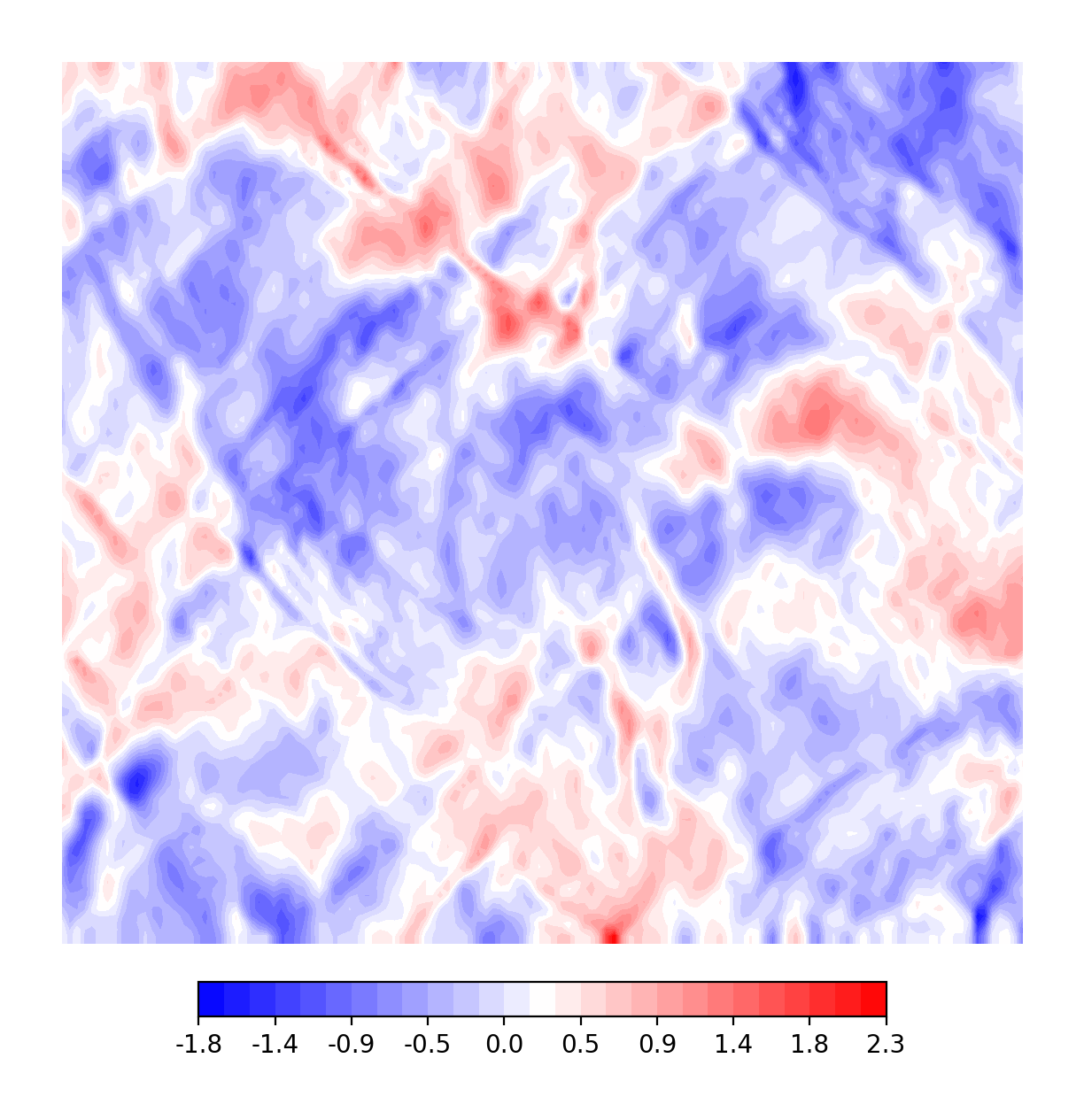} \\
	\includegraphics[width=0.315\linewidth]{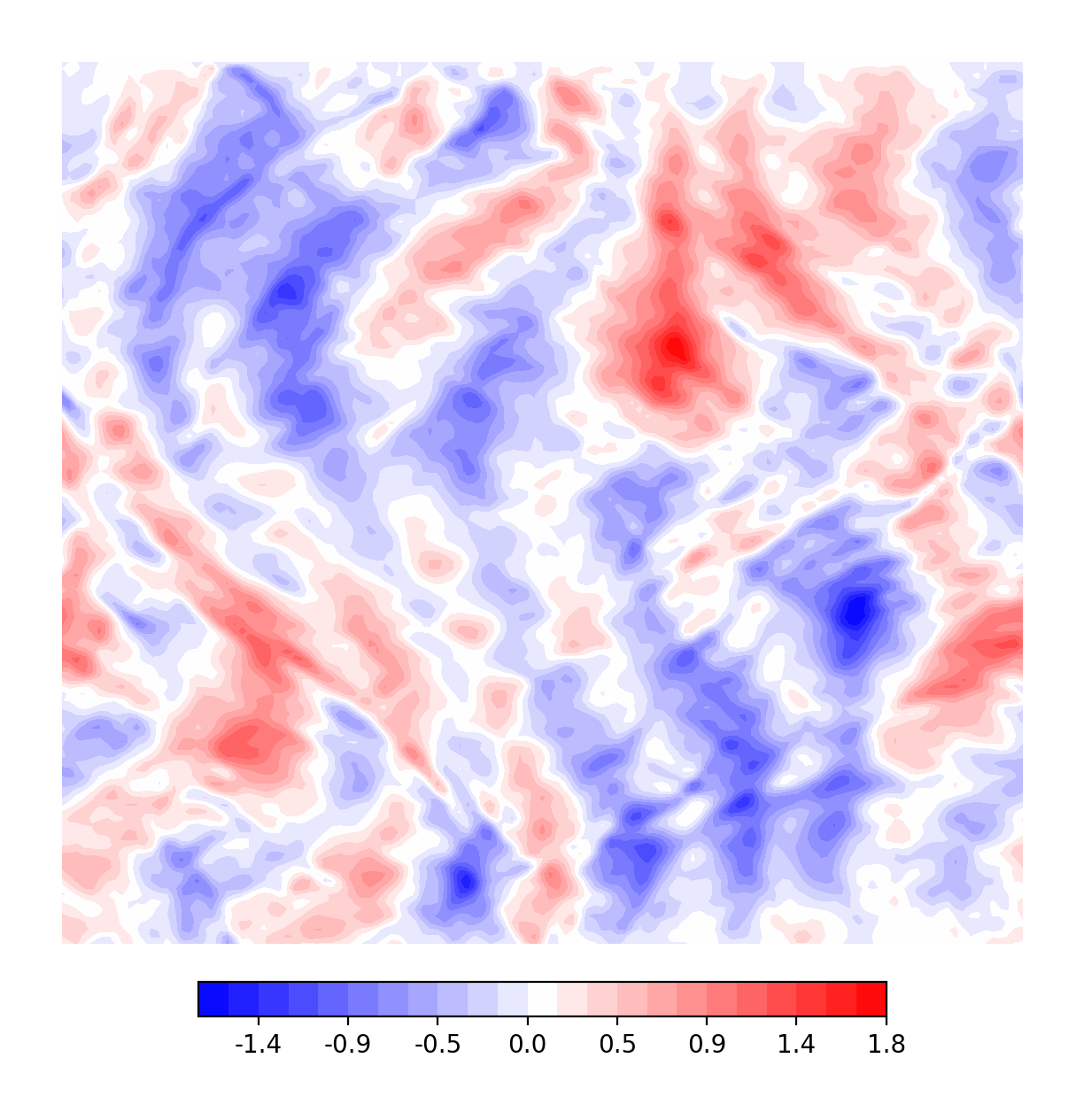}
	\includegraphics[width=0.315\linewidth]{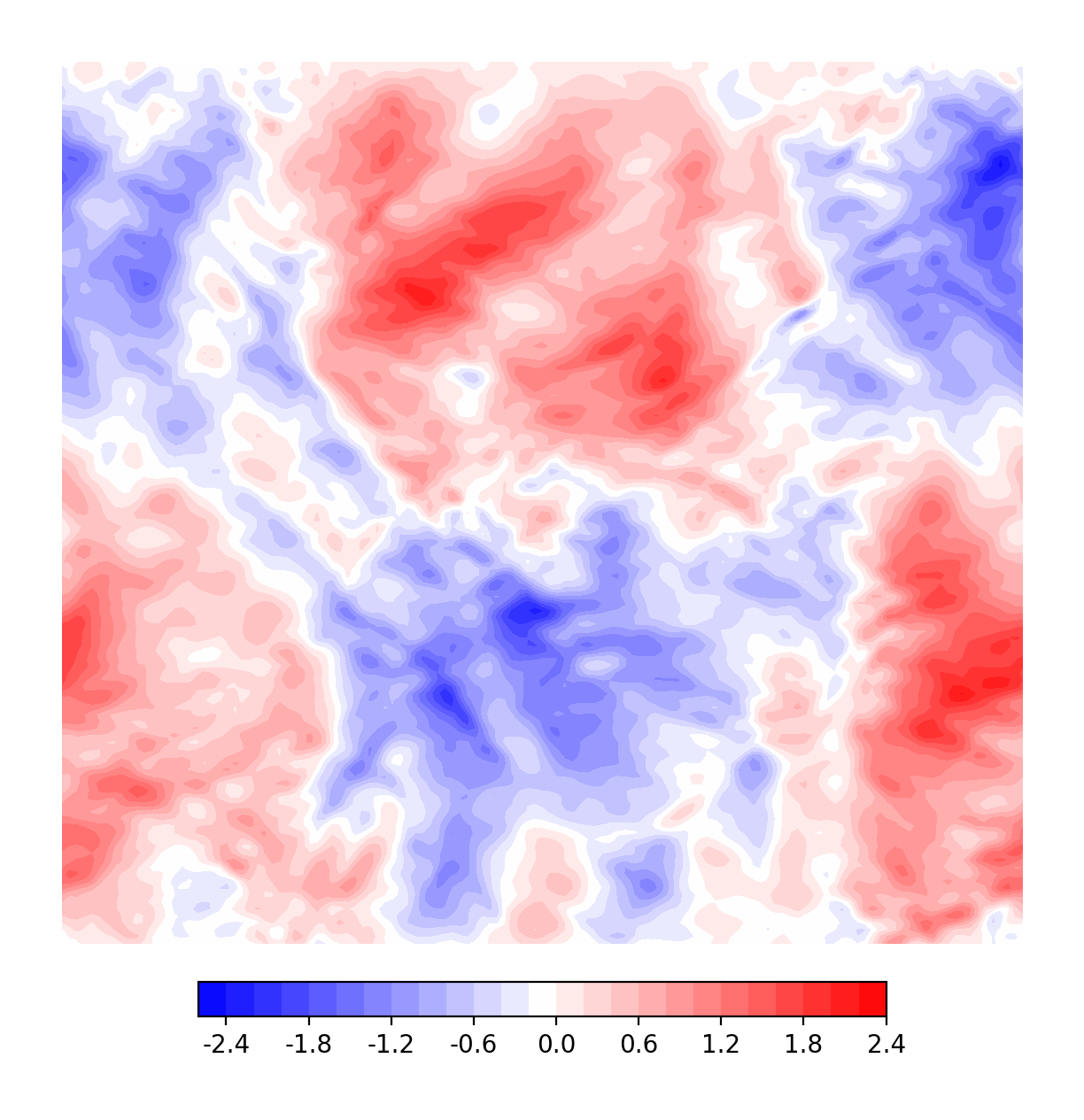}
	\includegraphics[width=0.315\linewidth]{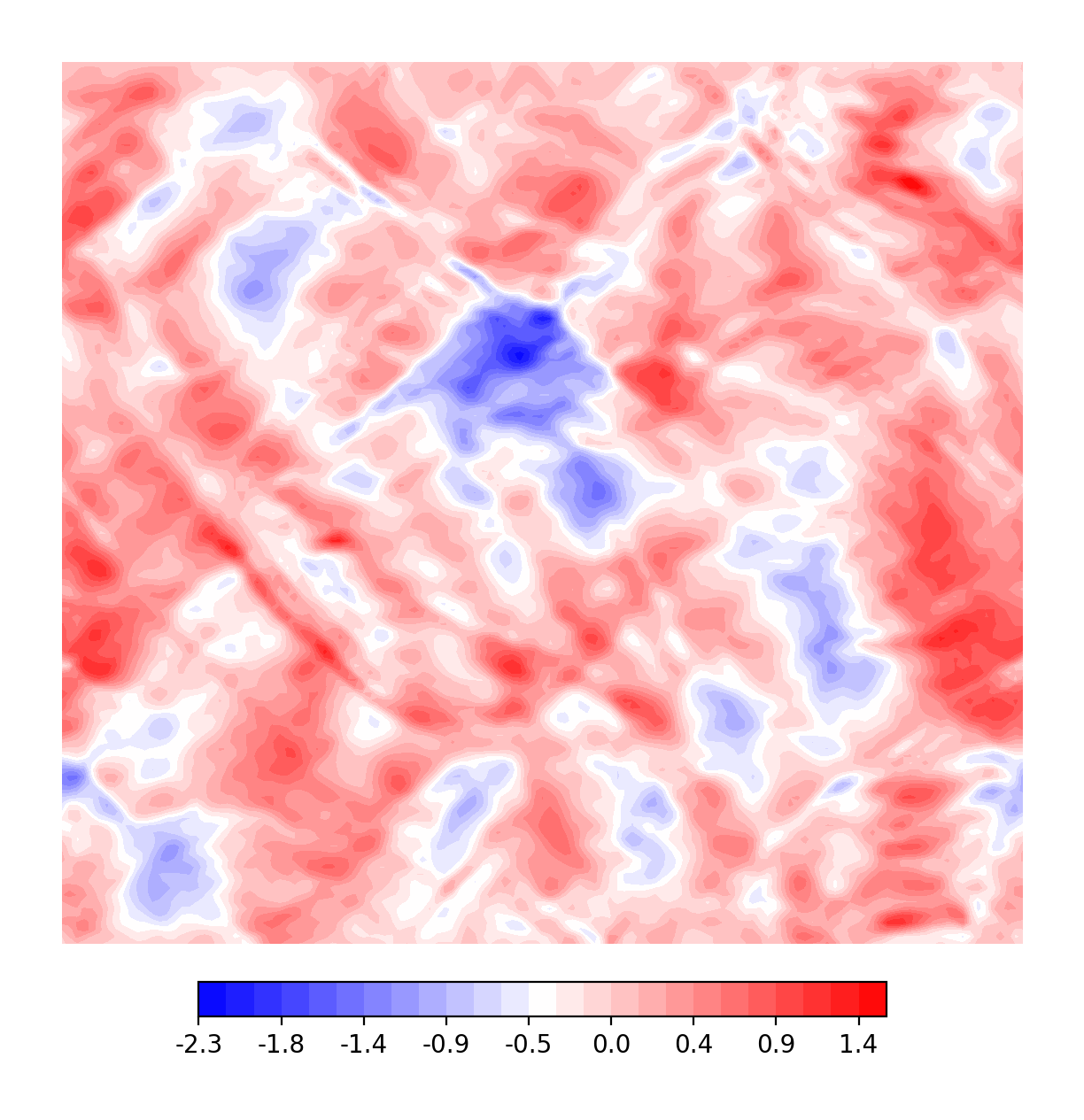}
	\caption{Plots of the signal process (a realization of the RSW SPDE \eqref{eq:SRSW}) in the \textit{S} regime on a $128 \times 128$ square grid. On the top row, the height $\eta$ at time-steps 0, 50 and 100. On the second row, the associated zonal component $u$ of the velocity field, at the same time-steps, and on the last row the meridional velocity $v$.}
	\label{fig:signal_subM}
\end{figure}

The SPDE \eqref{eq:SRSW} exhibits different behaviour depending on the choice of parameters $\mathrm{Ro}$ and $\mathrm {Fr}$. We test our particle filters in two different oceanic settings: the sub-mesoscale  (\textit{S}) regime and the mesoscale (\textit{M}) regime (see Appendix~\ref{app:params}). We will present most of our results for the \textit{S} dynamics---indeed, if not specified, it has to be assumed that this is the regime we are working in---but we will also include a few for the \textit{M} dynamics, to demonstrate that our methodology works independently of the choice of $\mathrm{Ro}$ and $\mathrm{Fr}$. As we can see in the simulations, the magnitude of the transport noise (given by a combination of sines and cosines, see \eqref{eq:transport_noise_form}) plays an important role in the dynamics of  \eqref{eq:SRSW}. While a lower noise magnitude makes the system more realistic (and indeed the models from the SALT framework were developed in this spirit, with the transport noise modelling the small-scale dynamics, see \cite{Holm_SALT_15}), we chose to run \eqref{eq:SRSW} with a large amount of noise (see Remark~\ref{rmk:noise_magnitude}): the nonlinear nature of the problem will cause small errors to grow and propagate rapidly, thus increasing the difficulty of the filtering problem.

In Figure~\ref{fig:signal_subM} we provide plots of a realization of \eqref{eq:SRSW} (in the regime \textit{S}) at different time-steps: we take this to be the signal process. In Figure~\ref{fig:signal_M} we do the same but for regime \textit{M}. The signal is simulated on a $128 \times 128$ space grid. For details on the numerical simulations, including the numerical scheme, the initial conditions $(u_0, v_0, \eta_0)$, the boundary conditions, the choice of parameters $f, \mathrm{Ro}$ and $\mathrm{Fr}$, and the basis $\boldsymbol{\xi}_n$ for the noise, we refer to Appendix~\ref{app:numerics_details}.

\begin{figure}[H]
	\centering
	\includegraphics[width=0.325\linewidth]{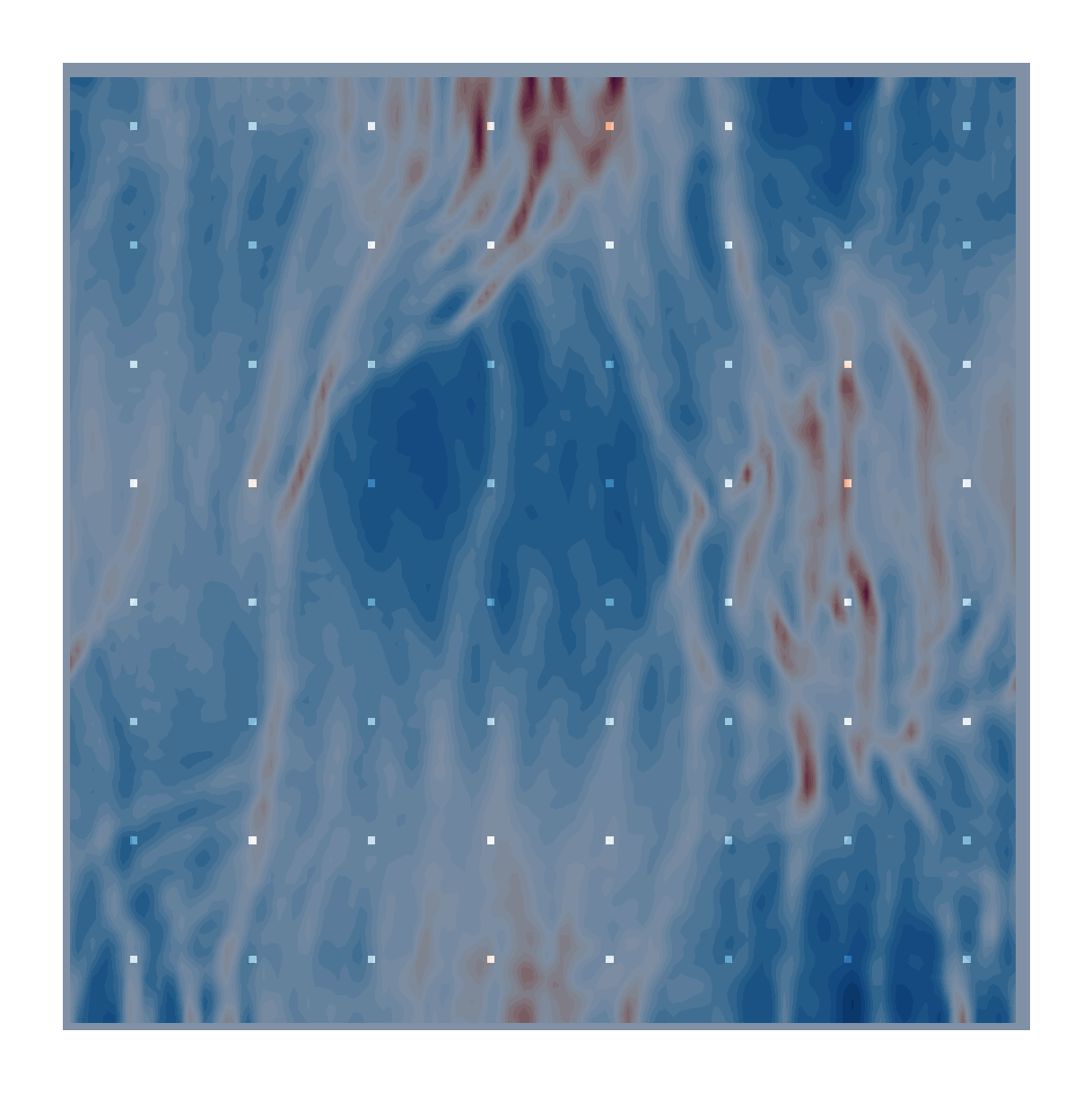}
	\includegraphics[width=0.325\linewidth]{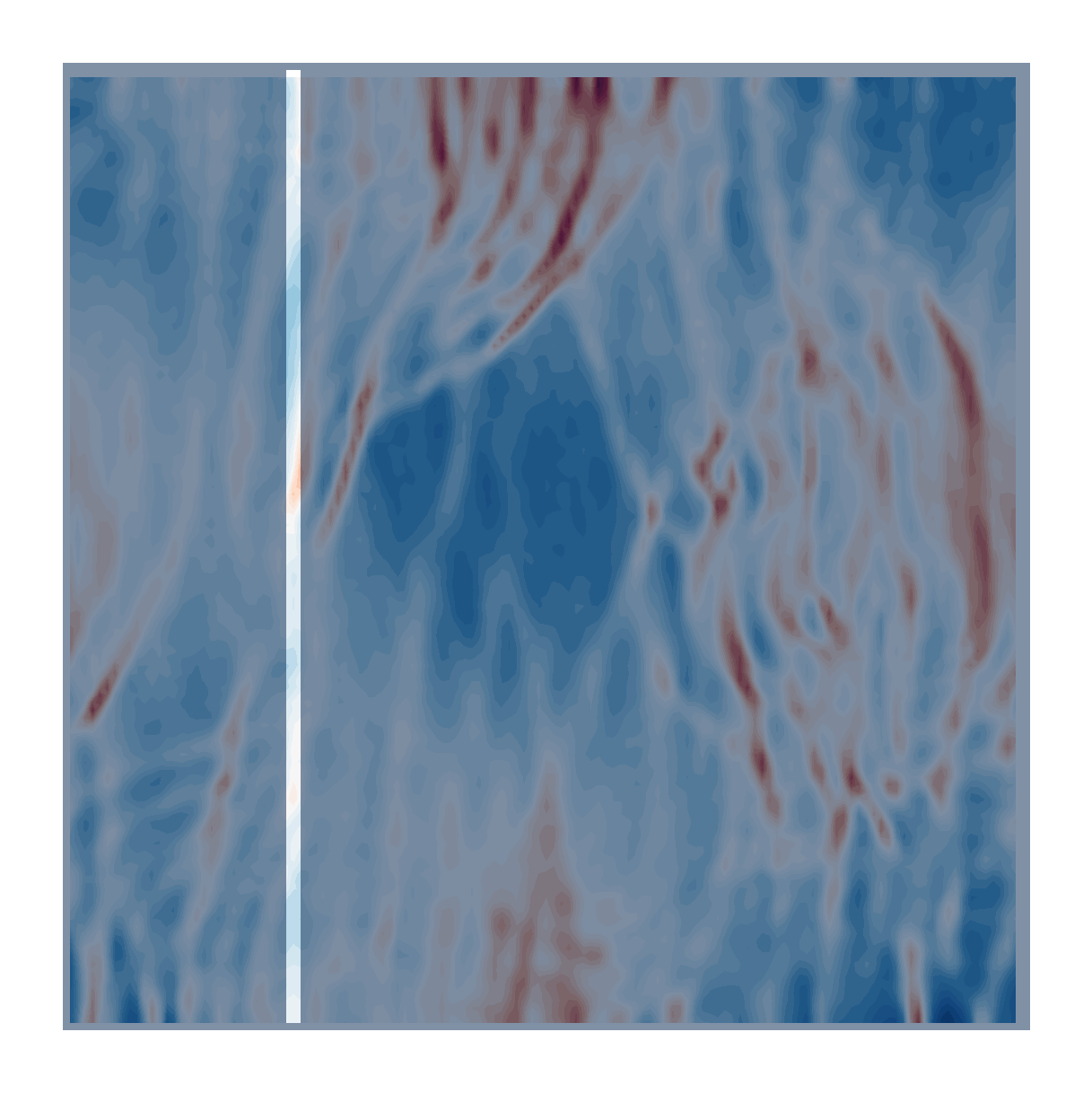}
	\caption{Locations for the observations superimposed over the signal process $\eta_t$. On the left, a $16 \times 16$ observations-grid of type (a). On the right, a strip of observations of type (b).}
	\label{fig:obs_types}
\end{figure}

\begin{figure}[ht!]
	\centering
	\includegraphics[width=0.315\linewidth]{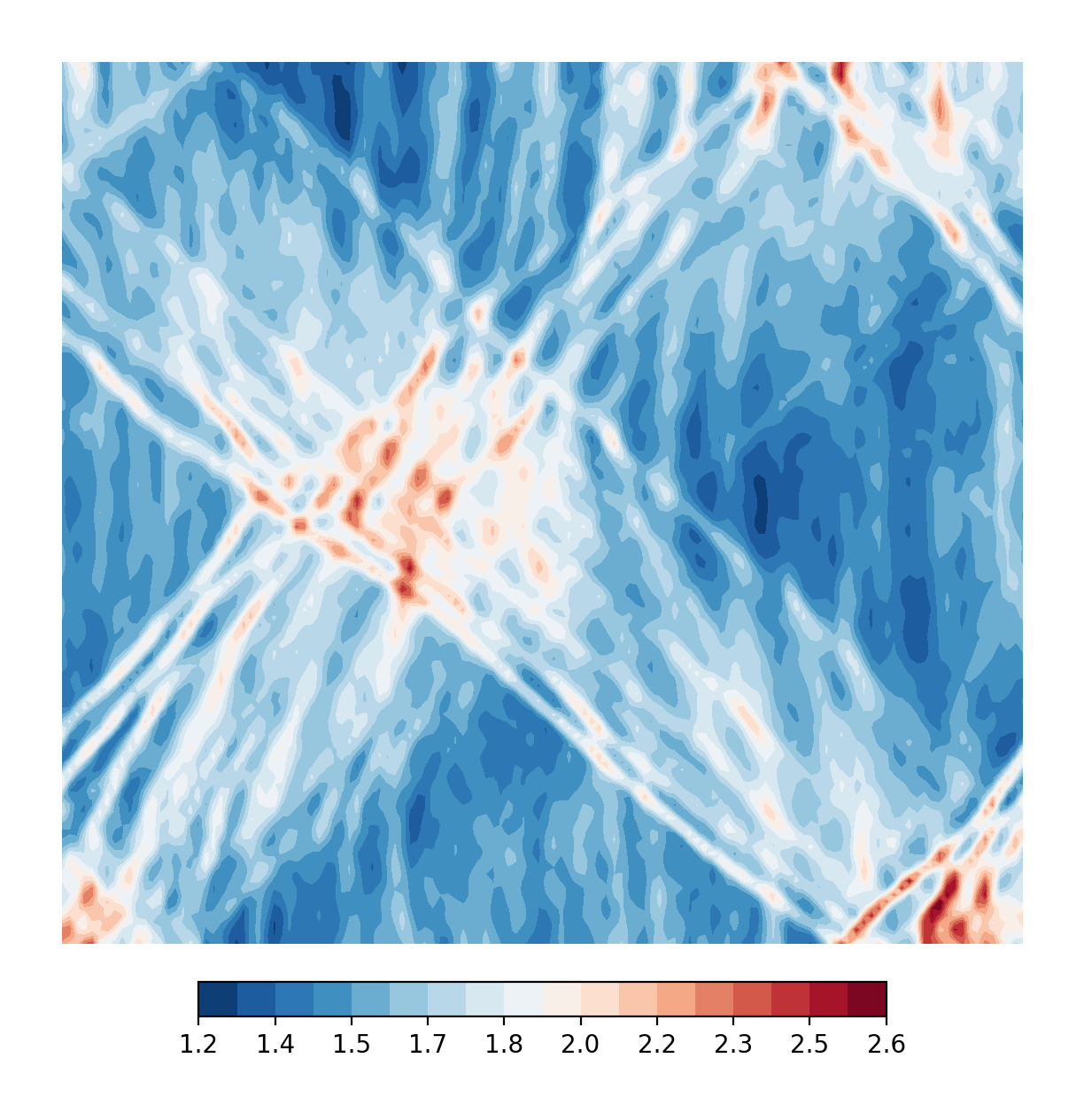}
	\includegraphics[width=0.315\linewidth]{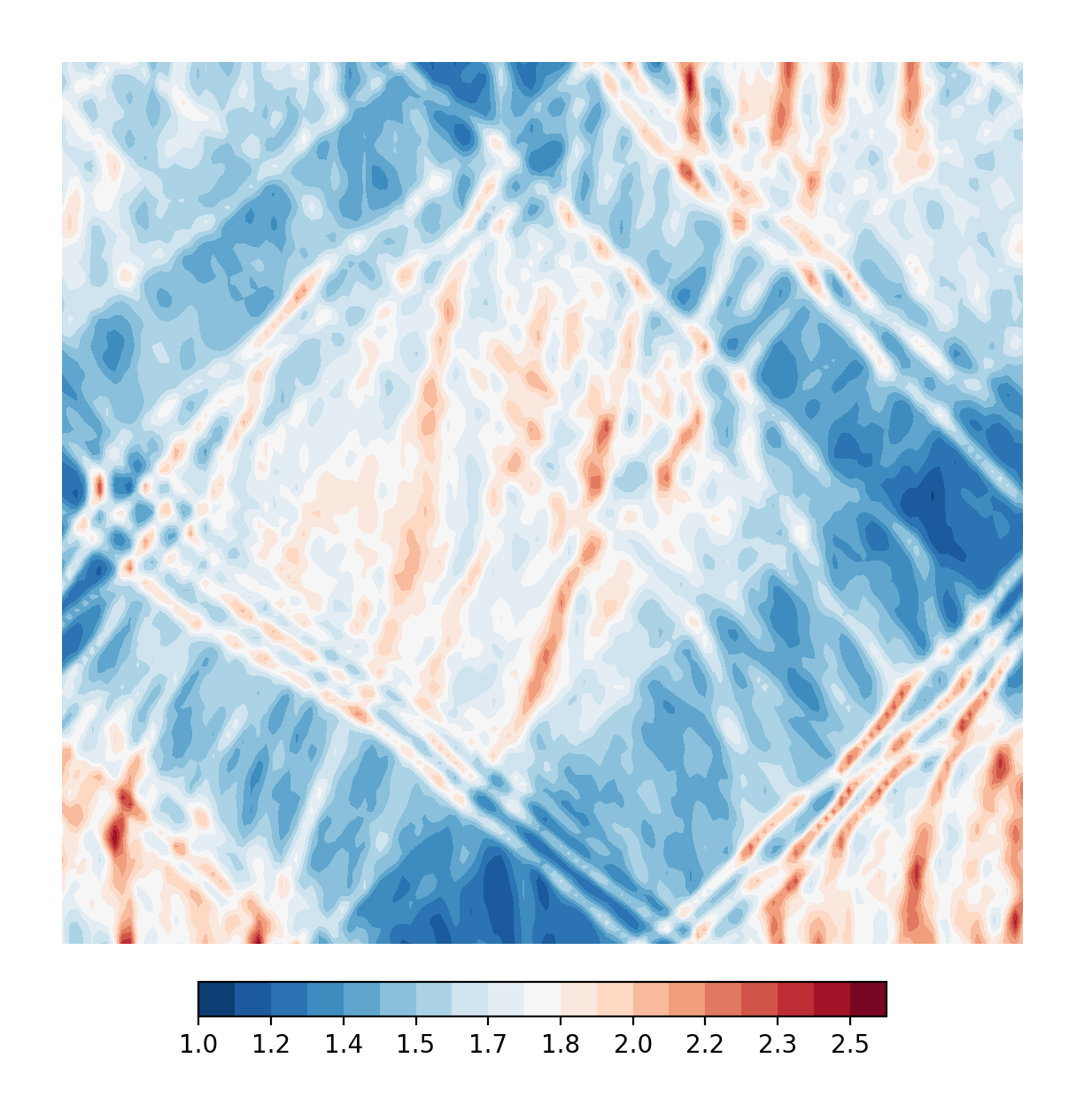}
	\includegraphics[width=0.315\linewidth]{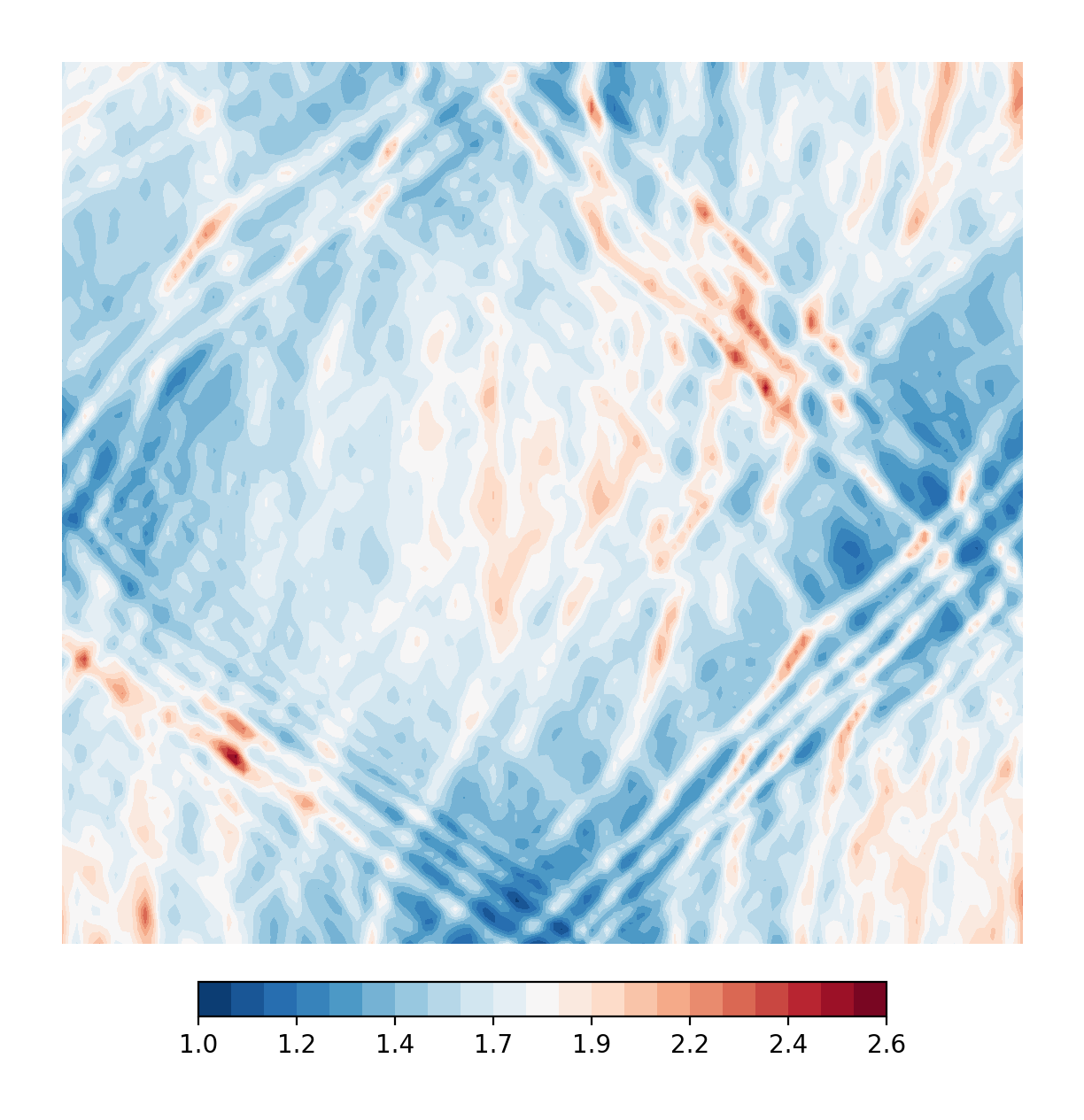} \\
	\includegraphics[width=0.315\linewidth]{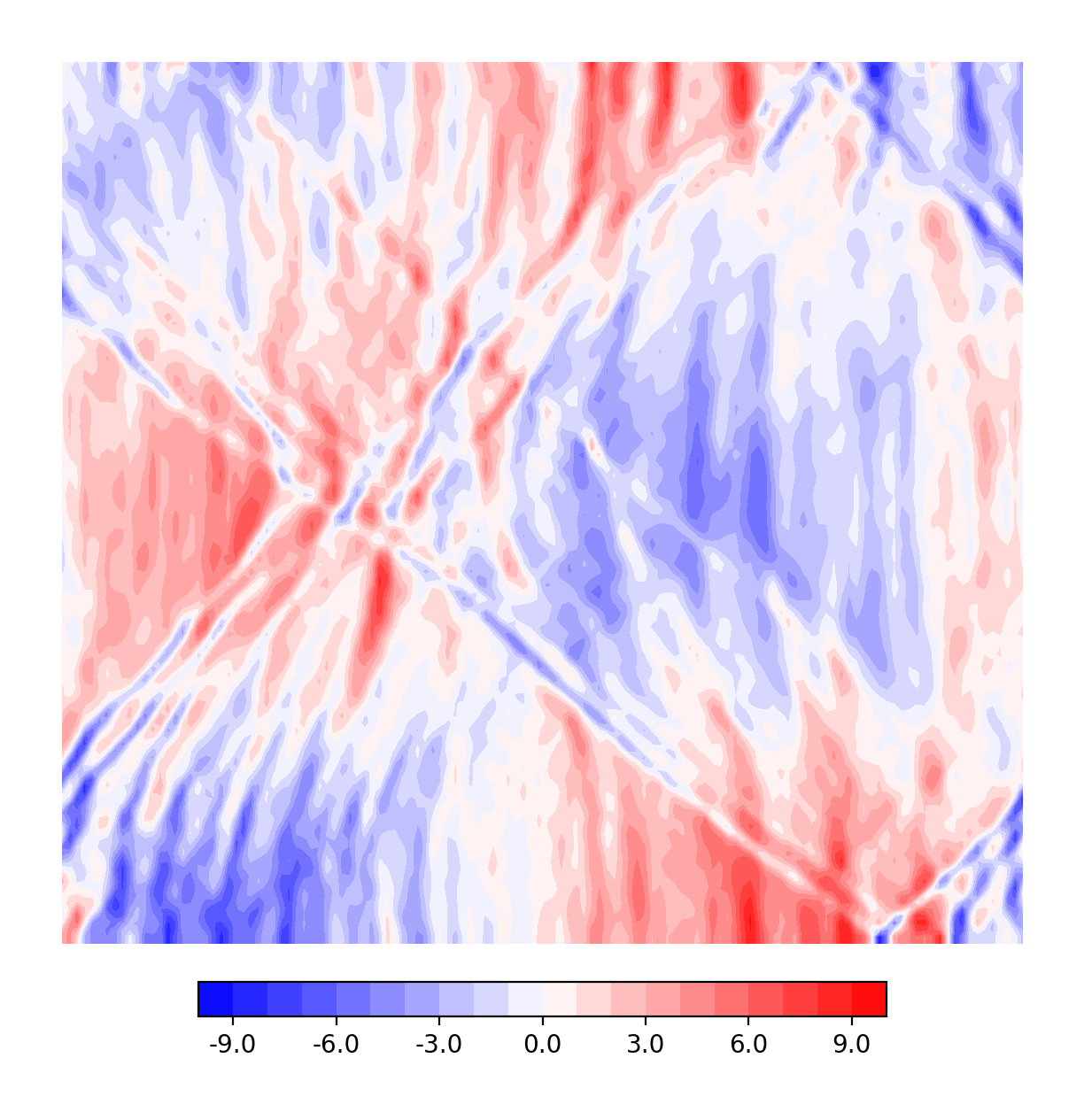}
	\includegraphics[width=0.315\linewidth]{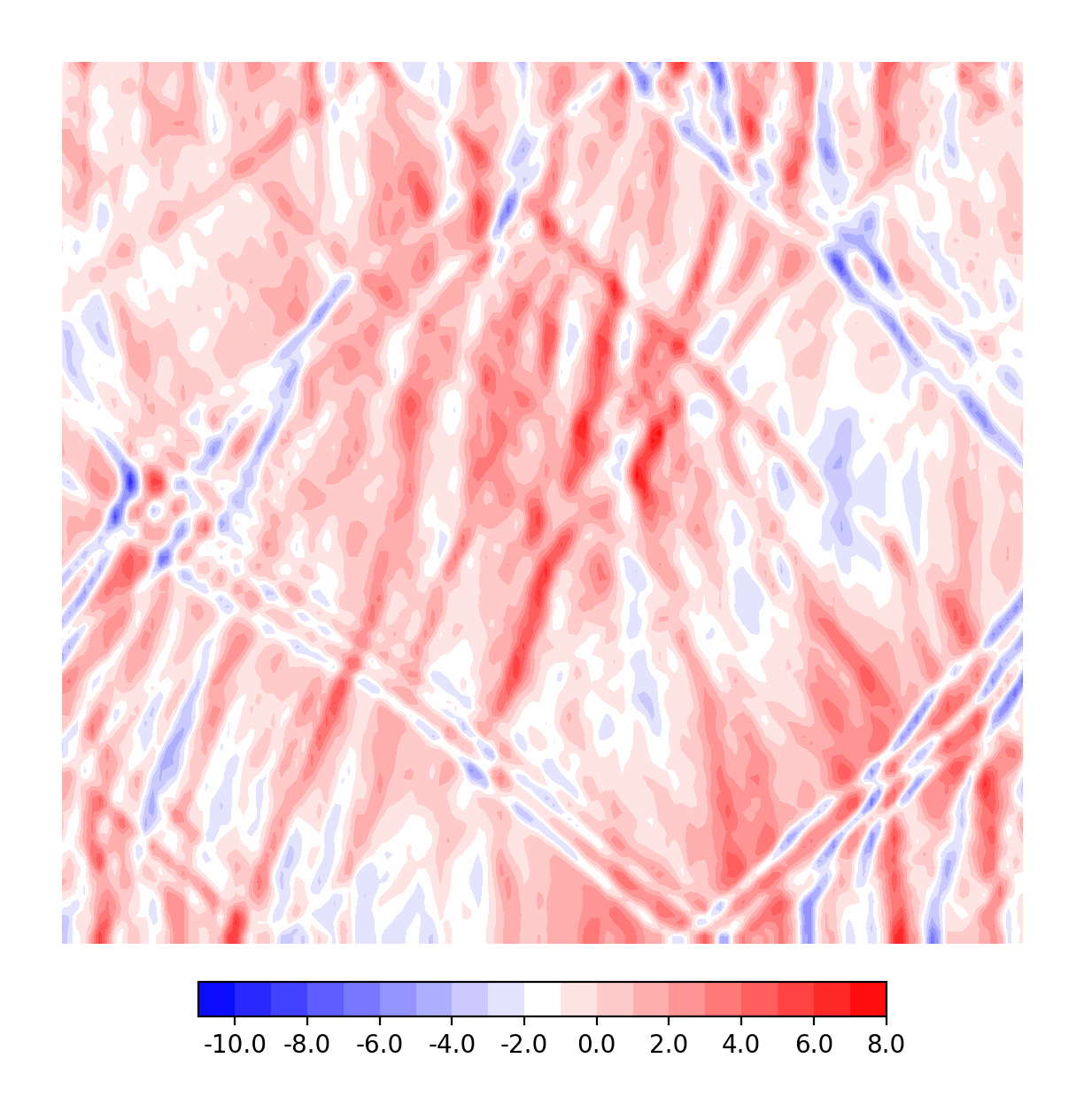}
	\includegraphics[width=0.315\linewidth]{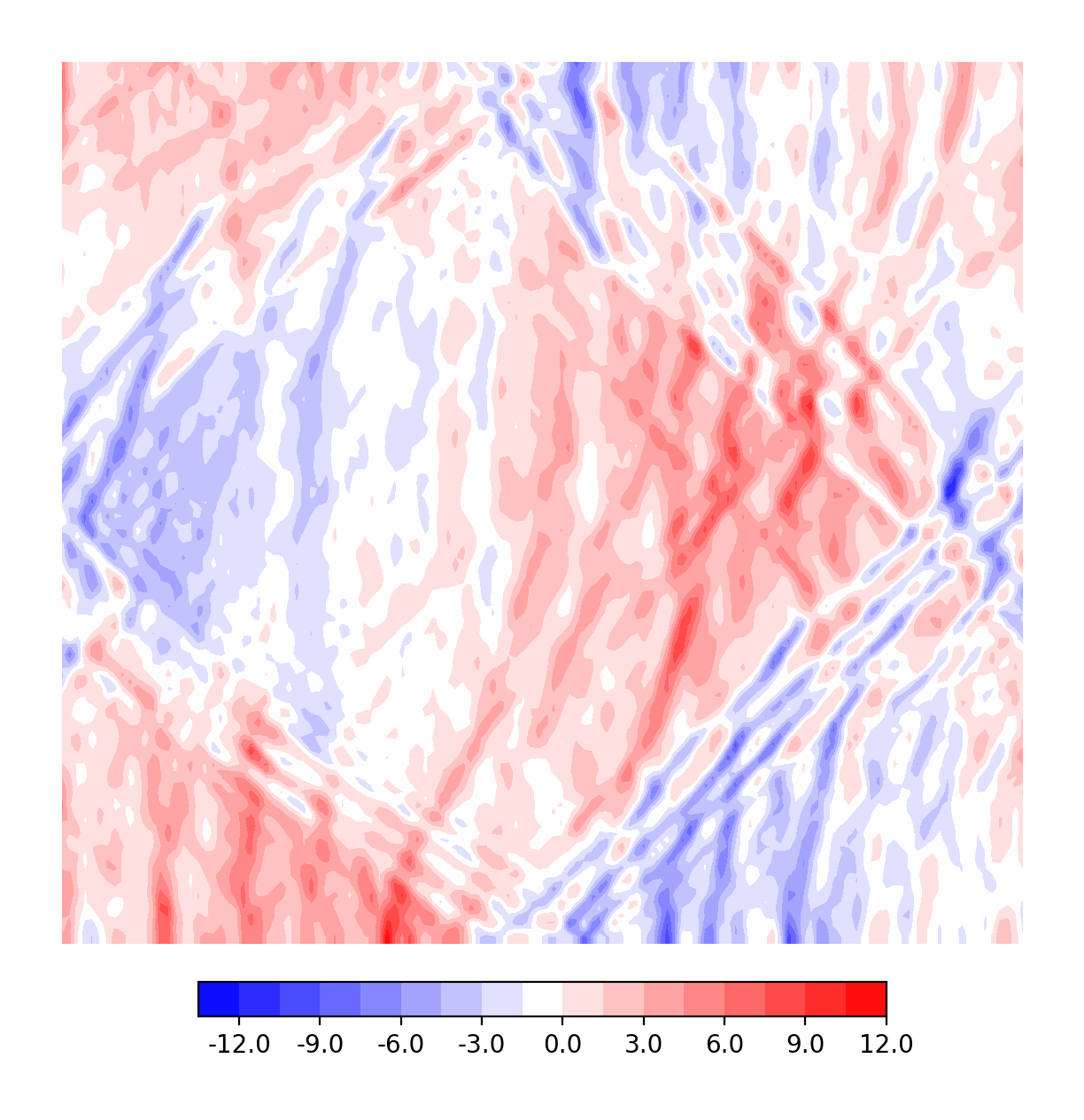} \\
	\includegraphics[width=0.315\linewidth]{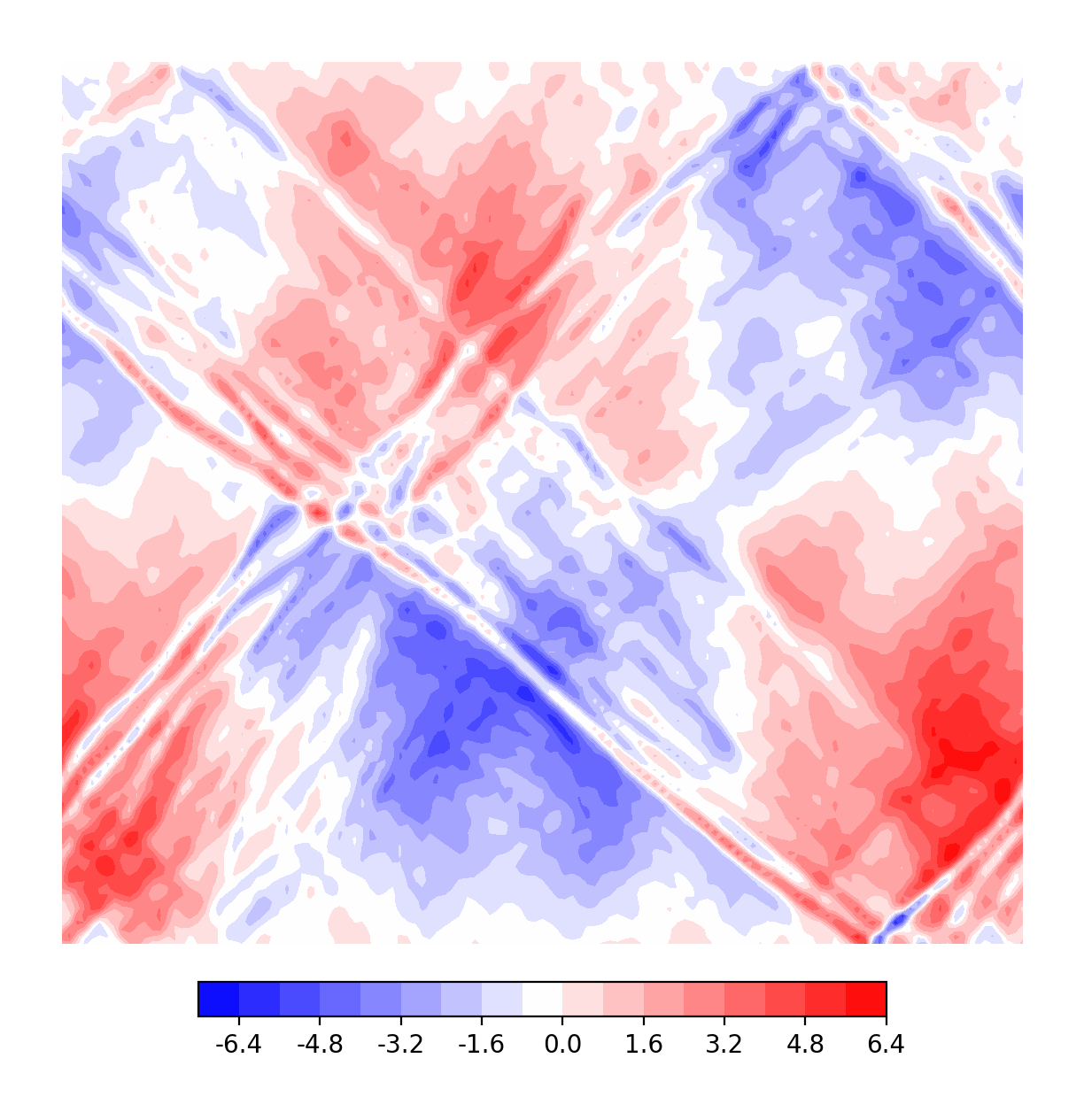}
	\includegraphics[width=0.315\linewidth]{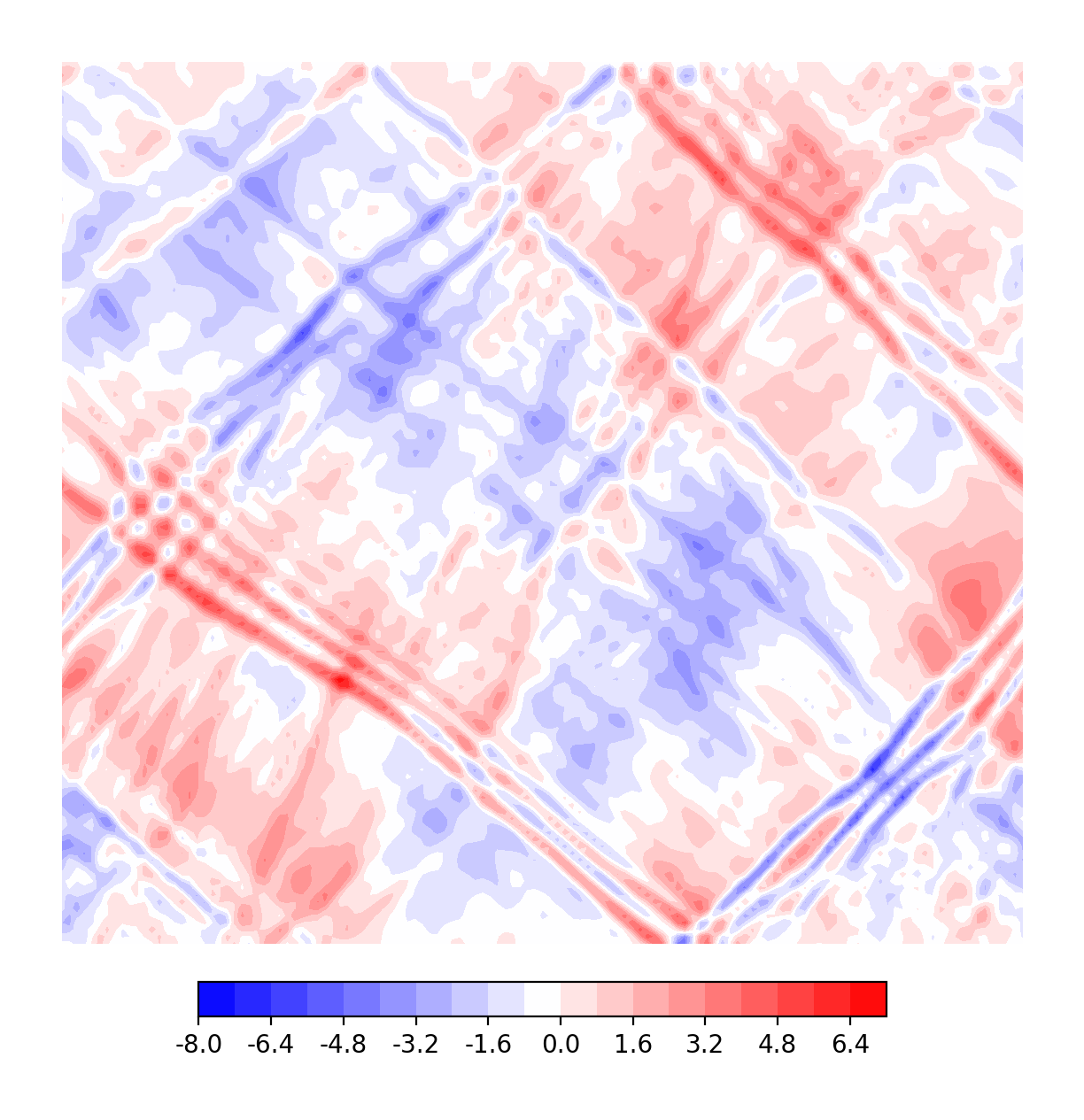}
	\includegraphics[width=0.315\linewidth]{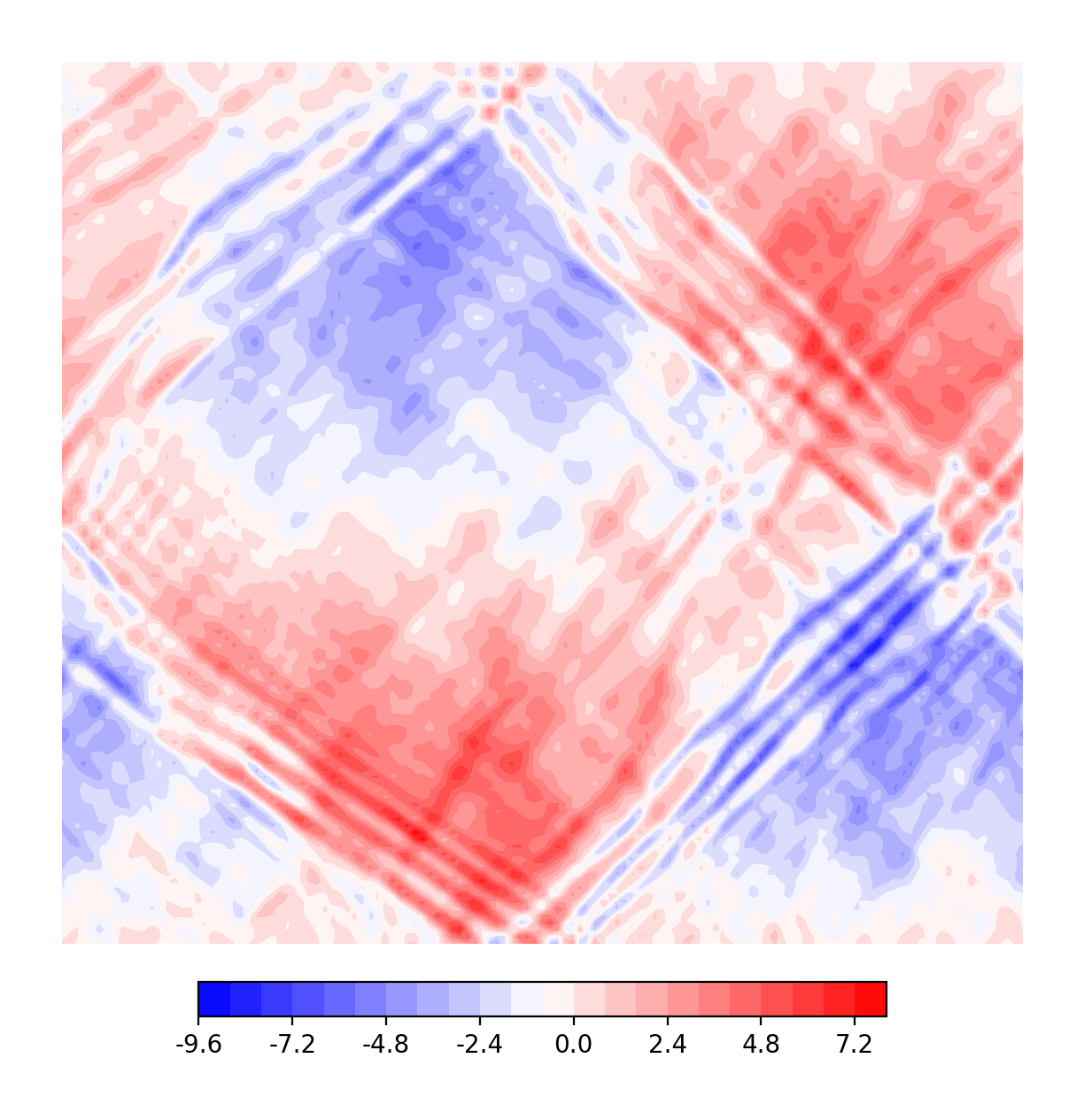}
	\caption{Plots of the signal process (a realization of the RSW SPDE \eqref{eq:SRSW}) in the \textit{M} regime on a $128 \times 128$ square grid. On the top row, the height $\eta$ at time-steps 0, 50 and 100. On the second row, the associated zonal component $u$ of the velocity field, at the same time-steps, and on the last row the meridional velocity $v$.}
	\label{fig:signal_M}
\end{figure}

Given the signal, we create our observations as in \eqref{eq:obs_numerics_vec}. To test our particle filters, we experiment with two different choices of $\mathcal{J}$ in \eqref{eq:obs_numerics_vec}:
\begin{enumerate}
	\item[(a)] \textit{fixed grid observations}: we take a grid of $d_{\mathrm{obs}} \times d_{\mathrm{obs}}$ equispaced points in $D$, with $d_{\mathrm{obs}}$ varying from 2 to $d = 128$ (and $d_{\mathrm{obs}} = d$ corresponds to observations on the whole domain);
	\item[(b)] \textit{moving strip observations}: we consider a set of points that form a vertical line in $D$ and cyclically vary its position in the domain. 
\end{enumerate}
In Figure~\ref{fig:obs_types} we plot the locations of the observations $Y$ for these two choices of $\mathcal{J}$. Note that in the case of observations of type (a) we have (partial and noisy) information spread evenly throughout the domain $D$, while in case (b) the information is sparser, and much more localized. We expect this feature to reflect in the performance of the particle filter. Locations of type (b) are meant to replicate (in a simplified way) the form of the satellite data from the SWOT mission \cite{swot}.

\subsubsection{Testing the particle filters}\label{subsec:test_PF}

 We now test the local particle filter (LPF), comparing it with the PF, in the experiment settings defined above. We asses the performance of the filtering algorithms by computing the following quantities after each assimilation step: the Ensemble Mean $L^2$-norm Relative Error (EMRE), the Relative Bias (RB), and the forecast Relative Ensemble Spread (RES). Recall that the assimilation times are given by $t_k = kr \delta t$, where $r \in \N$ is fixed and determines the frequency of the observations, and $k\ge0$. For $f \in \{u,v, \eta\}$, the error metrics are defined as follows: 
\begin{align*}
	EMRE (f_{t_k}) &: = \tfrac{1}{N} \sum_{i = 1}^N \frac{\| f_{t_k} - f_{t_k}^i  \|_{L^2} }{\| f_{t_k}  \|_{L^2}}, \\
	RB(f_{t_k}) &:= \frac{\left\| f_{t_k} - \frac{1}{N}\sum_{i = 1}^N f^i_{t_k}  \right\|_{L^2}}{\| f_{t_k}  \|_{L^2}}, \\
	RES(f_{t_k})  &:= \tfrac{1}{N-1} \sum_{i = 1}^N \frac{\left\| f_{t_k}^i - \frac{1}{N} \sum_{j=1}^N f^j_{t_k} \right\|_{L^2} }{\| f_{t_k}  \|_{L^2}}.
\end{align*}
Note that the errors above are computed over the entire domain, regardless of how many points in the domain are effectively observed. Pointwise at each observation location we compute the Root Mean Square Error (RMSE) and the Continuous Ranked Probability Score (CRPS). We then average over the number of observation points to obtain
\begin{align*}
	RMSE(f_{t_k}) := \frac{1}{N_{\mathrm{obs}}} \sum_{j = 1}^{N_{\mathrm{obs}}} rmse(f_{t_k}(z^j)), \quad CRPS(f_{t_k}) := \frac{1}{N_{\mathrm{obs}}} \sum_{j = 1}^{N_{\mathrm{obs}}} crps(f_{t_k}(z^j)),
\end{align*}
where the $rmse$ and $crps$ function are defined as usual, i.e.
\begin{equation*}
	rmse(f_{t_k}(z^j)) = \sqrt{\tfrac{1}{N}\sum_{i = 1}^N \big( f_{t_k}(z^j) - f^i_{t_k}(z^j) \big)^2},
	\quad
	crps(f_{t_k}(z^j)) = \int_{\R} \big( F_{E, z^j}(x) - H(x-f_{t_k}(z^j)) \big) \dx,
\end{equation*}
where $F_{E, z^j}(x)$ is the empirical CDF of the ensemble at point $z^j$.

Let us now test the particle filters. We recall from Section~\ref{sec:PF} and Appendix~\ref{app:basic_filt} that the PF depends on several parameters:
\begin{itemize}
	\item $N$, the number of particles that constitute our ensemble;
	\item $N_{\mathrm{ess}}$, the threshold value for the $ESS$ which determines when the tempering, jittering and resampling procedures are applied;
	\item the spatial noise process \eqref{eq:pert_jittering_noise} for the jittering procedure (Algorithm~\ref{alg:perturbation_jittering}). In particular, the choice of spatial coefficients and the scaling factor $\sigma^{jitt}$.
\end{itemize}
Other secondary parameters appear in the algorithm (for example, we need to set a tolerance level for the bisection search algorithm within the tempering step in Algorithm~\ref{alg:tempering}). All these require careful tuning for optimality. While we have spent some time searching for reasonable parameters for the PF and LPF algorithm, with further tuning for the parameters (and a higher number of particles), there is always room for improvement.

In the following experiments we set the number of particles $\mathbf{N=50}$ and $\mathbf{N_{\mathrm{ess}} = 0.8 N}$, and we fix the perturbation for the jittering algorithm to be given by
\begin{equation}\label{eq:jitt_perturbation}
	\sigma^{jit} \sum_{n = 1}^{50} \frac{ \boldsymbol{\zeta}_n(\boldsymbol{x})}{n^2} Z_n, \quad Z_n \iid N(0,1),
	\end{equation}
	where we tune $\sigma^{jit}$ individually in each experiment, and the spatial coefficients are given by
\begin{equation*}
 \boldsymbol{\zeta}_n(\boldsymbol{x}) = \left( \begin{matrix}
	\zeta_n^u(x,y) \\
	\zeta_n^v(x,y) \\
	\zeta_n^{\eta}(x,y)
\end{matrix} \right)
=  \left( \begin{matrix} ( \cos (2 \pi n  y) + \sin (2 \pi n  y + \tfrac{\pi}{2}) ) (
\sin(2 \pi n  x) +  \cos(2 \pi n x) )\\
(\cos (2 \pi  n y) + \sin (2 \pi n  y + \tfrac{\pi}{2}) ) (
\sin(2 \pi n  x) +  \cos(2 \pi n x) )\\
	(\sin (2 \pi n y) - \cos (2 \pi n  y + \tfrac{\pi}{2}) ) ( \sin(2 \pi n x) + \cos(2 \pi n x) )
\end{matrix} \right).
\end{equation*}
Note that the coefficients $\boldsymbol{\zeta}_n$ satisfy the boundary conditions for $(u_t, v_t, \eta_t)$, and are similar to the choice of basis $\boldsymbol{\xi}_n$ for the driving noise of the SPDE in \eqref{eq:noise_SRSW} (see also Appendix~\ref{app:params}).
We introduced further parameters in the LPF:
\begin{itemize}
	\item $N_{\mathrm{loc}}$, the number of subregions in $D$;
	\item $\alpha$, which regulates the amount of influence that observations external to a subregion $D_j$ have on the weights of the local ensemble $\{ (u_t,v_t, \eta_t)^{i, (j)} \}_{i=1}^N$;
	\item the size of the overlap areas: we define the variable $p_{\mathrm{ov}} \in [0,1]$, with the convention that an overlap size of $p_{\mathrm{ov}}$ for a subregion of size $\R^{m \times m}$ corresponds to the area given by $[m p_{\mathrm{ov}}]$ grid points counting outwards from the N, S, W and E boundaries of the subregion, and $[m p_{\mathrm{ov}}]$ grid-points counting inwards.
\end{itemize}
We discuss the sensitivity of the LPF algorithm to its parameters in Section~\ref{subsec:LPF_params}.

Finally, we should mention how we construct our initial particle ensemble. The way one initializes the particle ensemble plays a crucial role in the behaviour of the filter. First and foremost, the initial ensemble needs to have a good spread around the signal process. For example, one can simply add a (spatial) Gaussian perturbation to the initial state $(u_0, v_0, \eta_0)$ of the signal to create the initial conditions for each particle, and this gives a viable spread. However, a Gaussian ensemble centred around the signal process does not give a faithful representation of its prior distribution, which is surely non-Gaussian. With this in mind, we proceed as follows: we first compute $N+1$ different realizations of the SPDE \eqref{eq:SRSW} starting from a common (deterministic) initial condition. Once, after a few iterations, the realizations are different enough one from the other, we pick one of them to be the initial condition for the signal process, and use the others to initialize the ensemble. Note that this procedure gives i.i.d. samples, also i.i.d. with respect to the signal process, and, crucially, the signal is not necessarily close to the sample mean. We explain our methodology in more detail in Appendix~\ref{app:ic}. In Figure~\ref{fig:ic_ensemble} we plot the initial conditions for the PF ensemble (for the \textit{S} regime) over a section of the domain. 

\begin{figure}[ht!]
	\centering
	\includegraphics[width=0.325\linewidth]{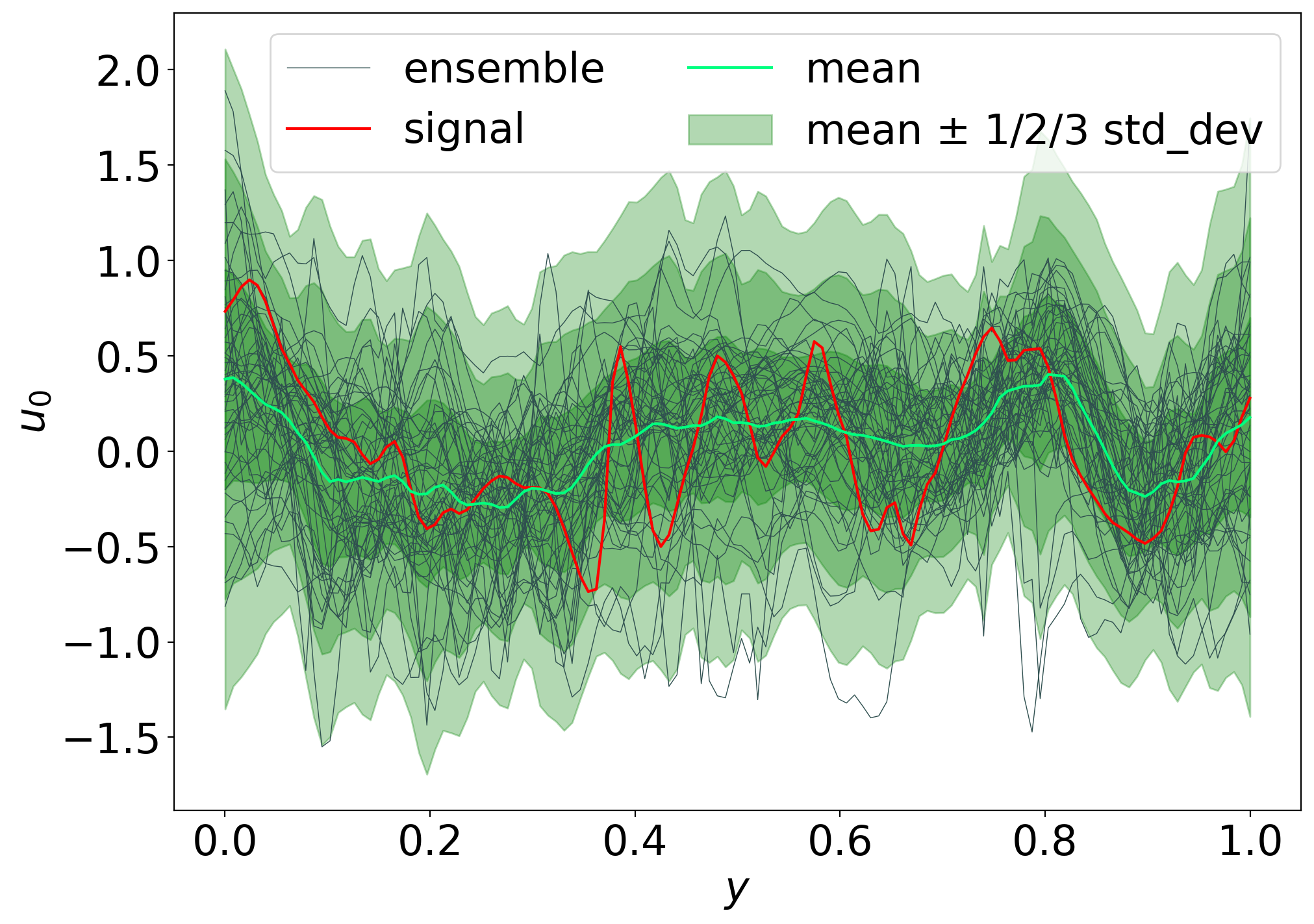}
	\includegraphics[width=0.325\linewidth]{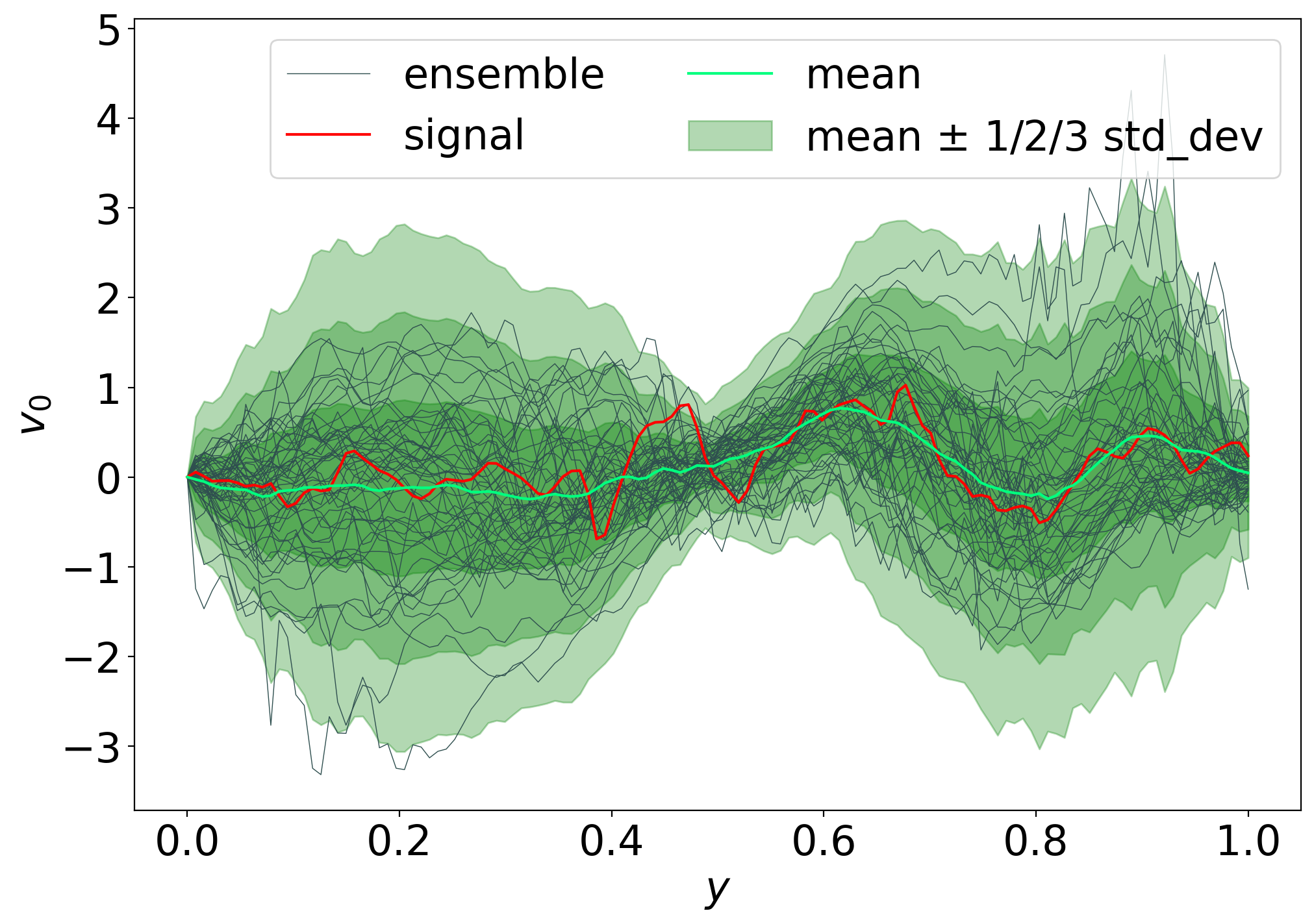}
	\includegraphics[width=0.325\linewidth]{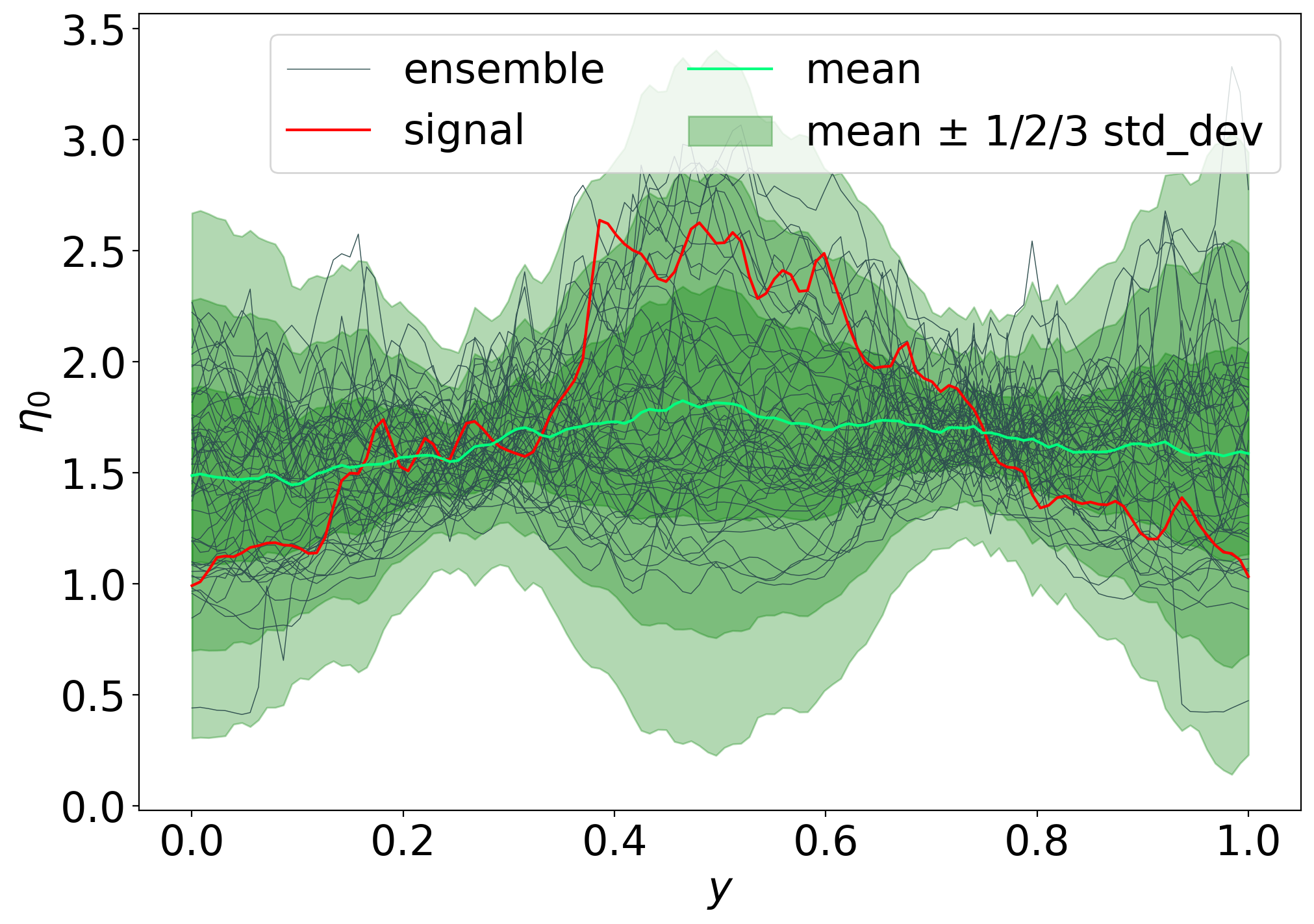}
	\caption{Plots of the initial conditions for a PF ensemble with 50 particles (in green), with $\{u^i_0\}_i$ on the left, $\{v^i_0\}_i$ in the center and $\{\eta^i_0\}_i$ on the right. In red the signal process at time $t=0$ and in lighter green the ensemble mean. We see that the signal is distinct from the ensemble mean. For this plots we selected a (vertical) section of the domain, fixed the $x$-coordinate and plotted $(u_0, v_0, \eta_0)$ as a function of $y$.}
	\label{fig:ic_ensemble}
\end{figure}

\subsection{The (L)PF with fixed grid-like observations}\label{subsec:LPF_grid}
In our first experiment we test the sensitivity of the (L)PF to the number of observations, when these are distributed in an equi-spaced grid, evenly spread around the domain $D$. We consider \textit{grid-like observations} of type (a), and grids of sizes $2^n\times 2^n$ for $n = 1, \dots, 7$. Note that when $n=7$ we are effectively observing a perturbation of the whole signal. As specified in Appendix~\ref{app:params}, for the \textit{S} regime we take $\delta t = 10^{-3}$. We take the frequency of the observations to be given by $\mathbf{r = 10}$, so we assimilate every 10 time-steps. We set the scale of the noise for the observations to be $\mathbf{\boldsymbol{\sigma} = 0.05}$ in \eqref{eq:obs_numerics_vec}. For each of the observations-grid's sizes, we run a filtering experiment for both a PF (with tempering and jittering) and an LPF with 4 localization regions ($\mathbf{N_{\mathrm{loc}}=4}$).

\begin{figure}[ht!]
	\centering
	\includegraphics[width=0.325\linewidth]{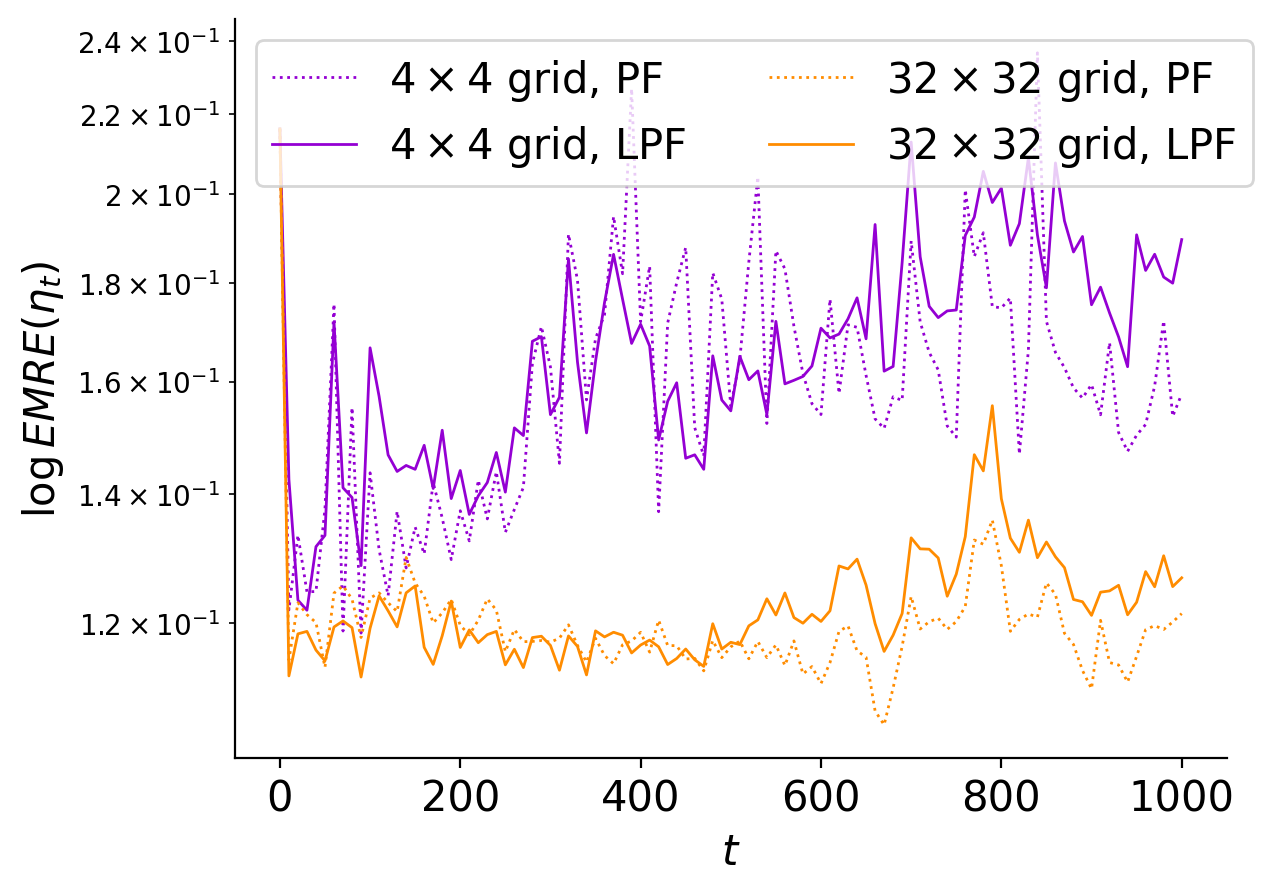}
	\includegraphics[width=0.325\linewidth]{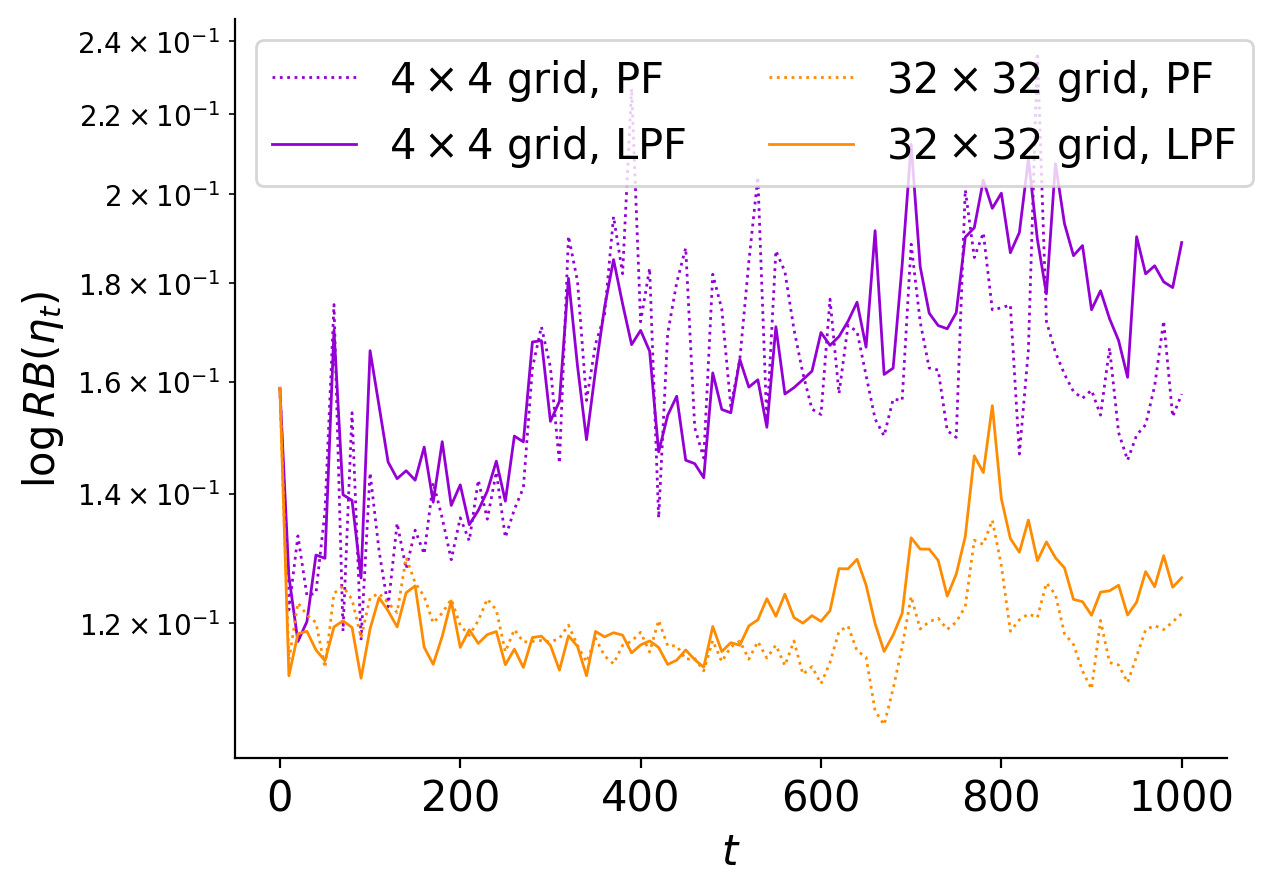}
	\includegraphics[width=0.325\linewidth]{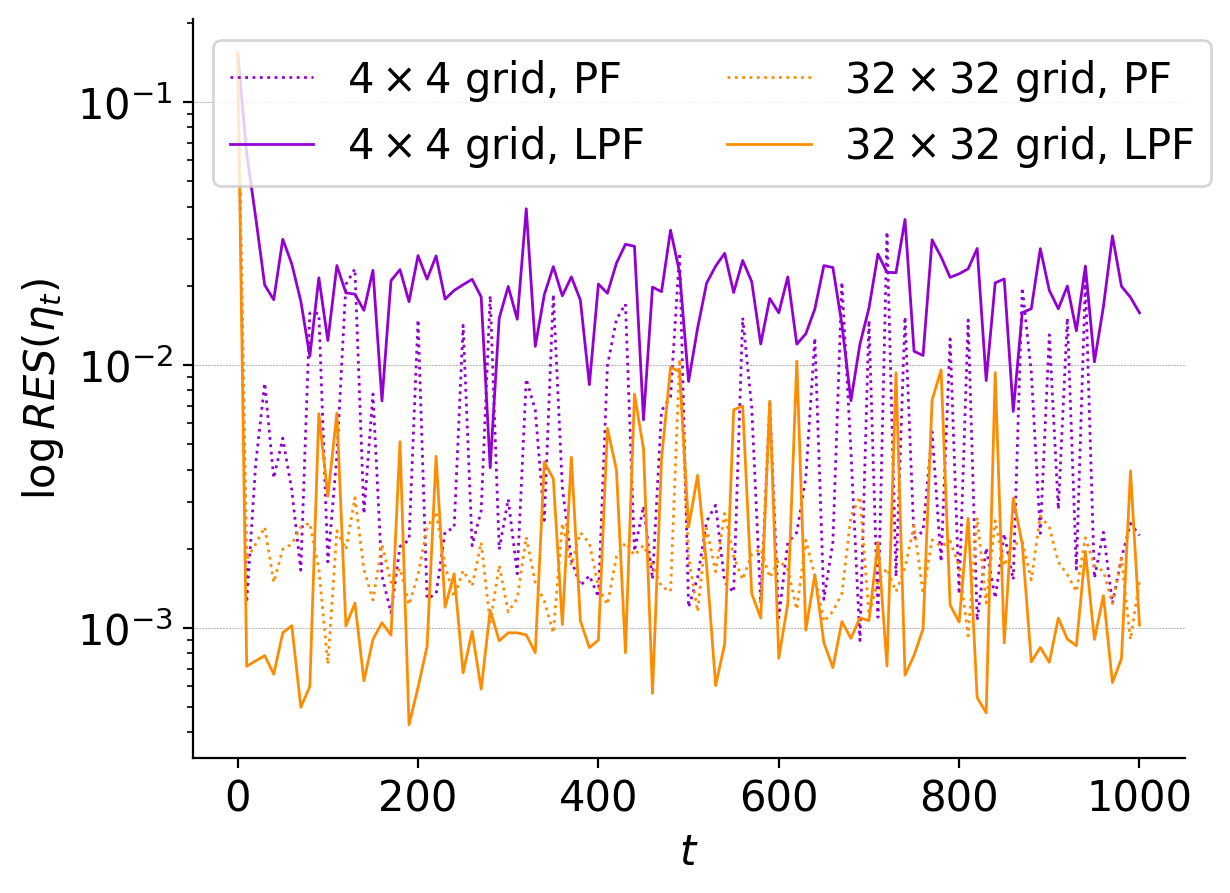} \\
	\includegraphics[width=0.325\linewidth]{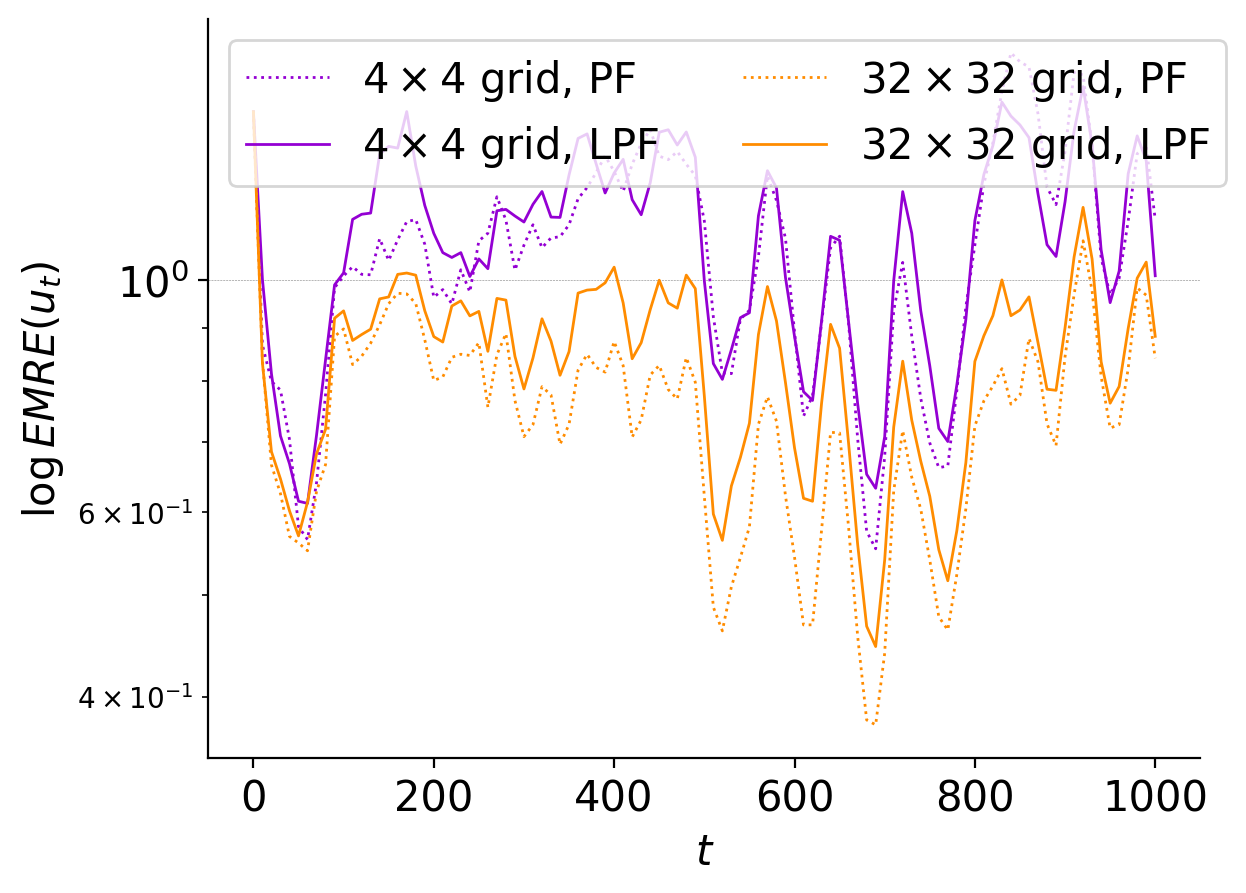}
	\includegraphics[width=0.325\linewidth]{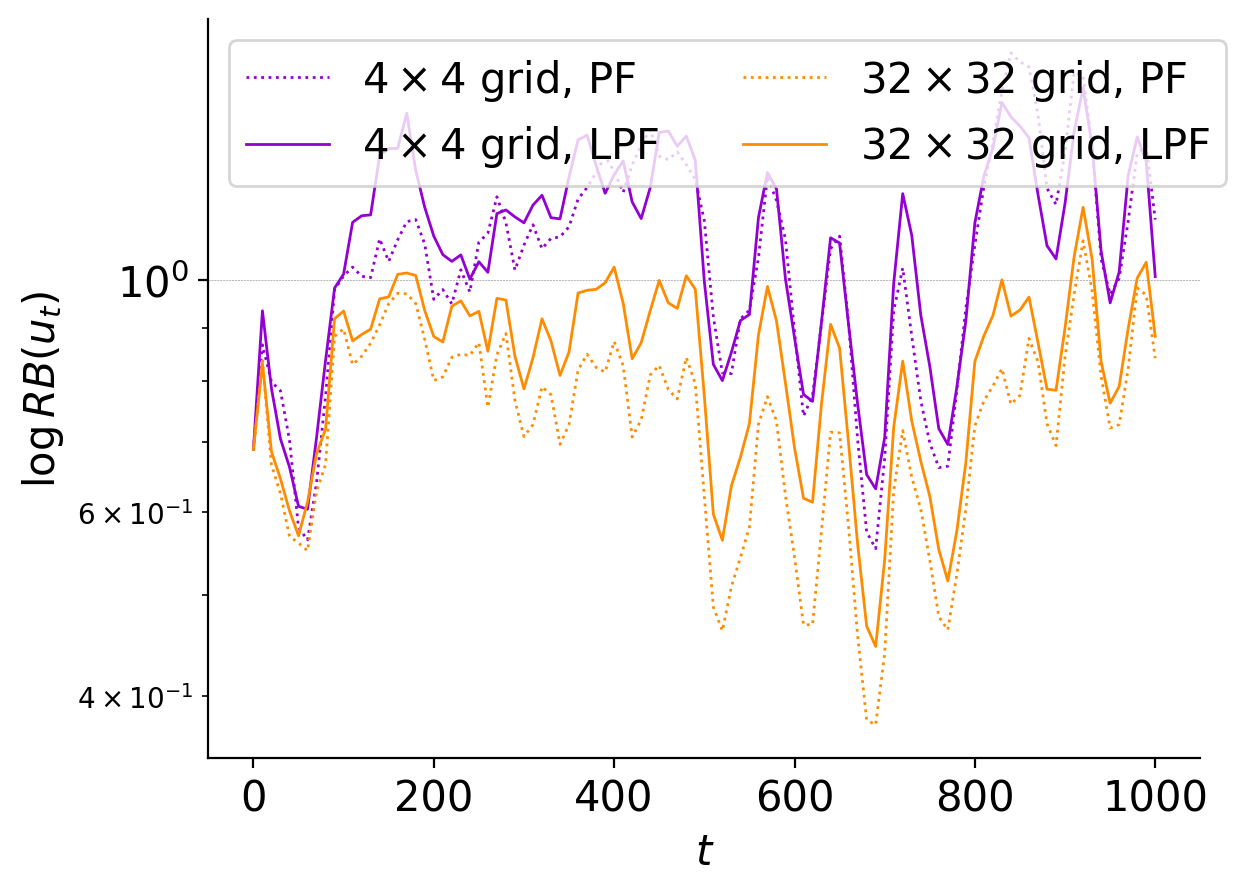}
	\includegraphics[width=0.325\linewidth]{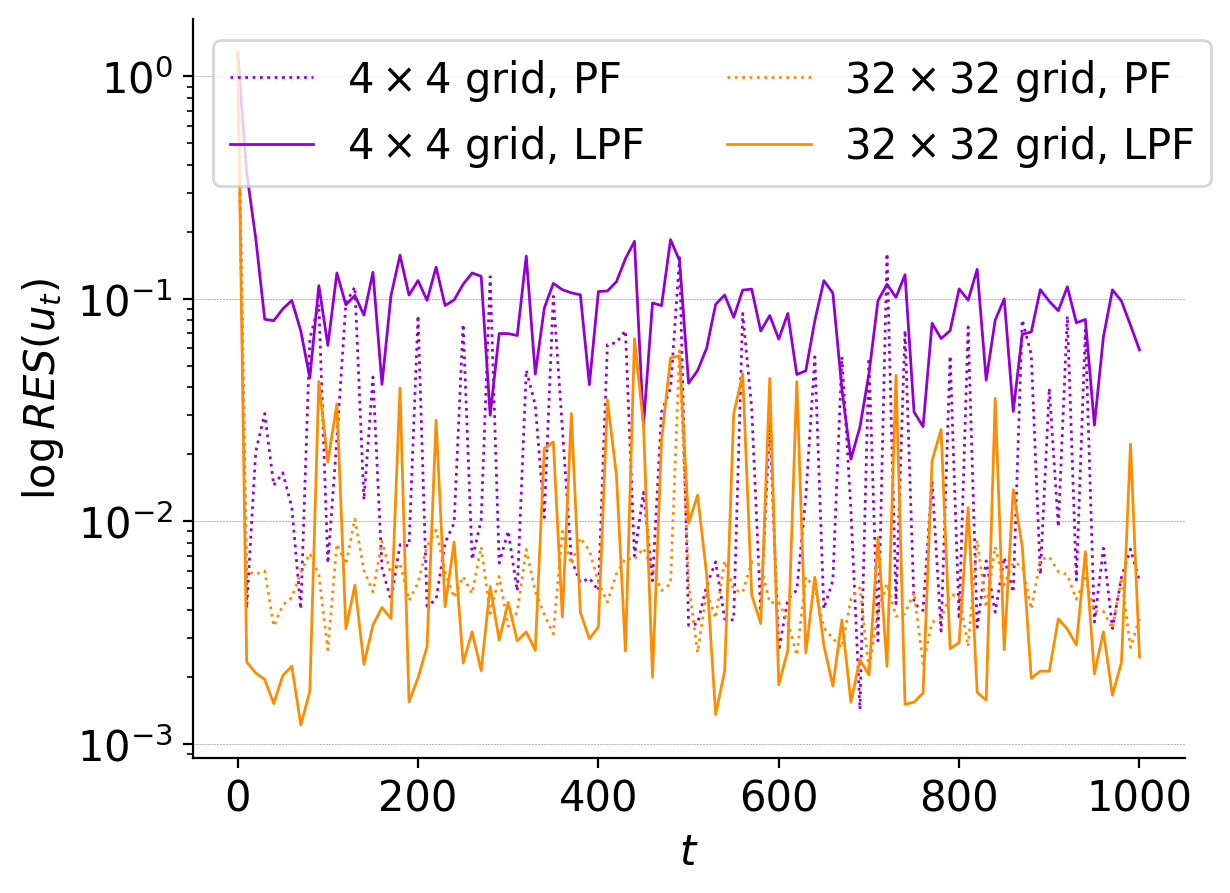} \\
	\includegraphics[width=0.323\linewidth]{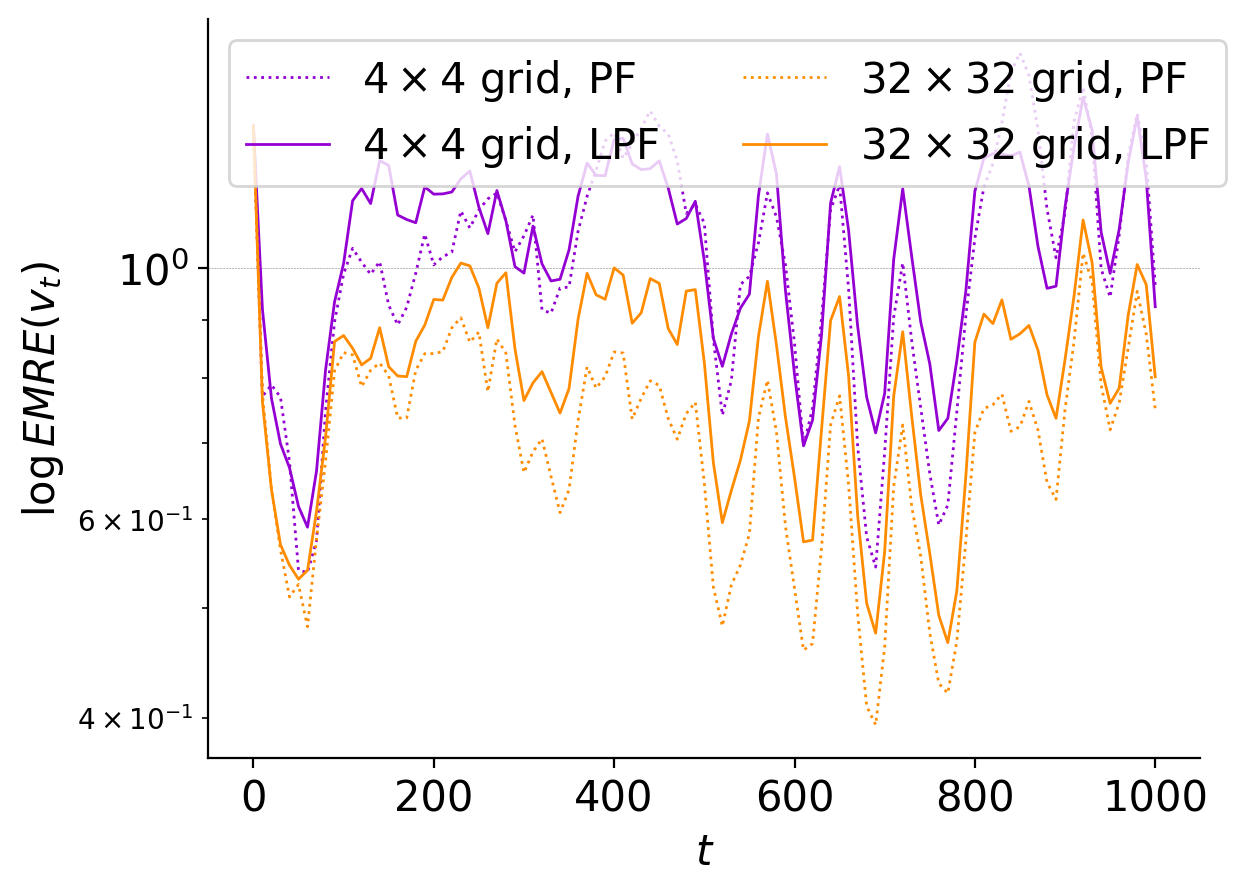}
	\includegraphics[width=0.323\linewidth]{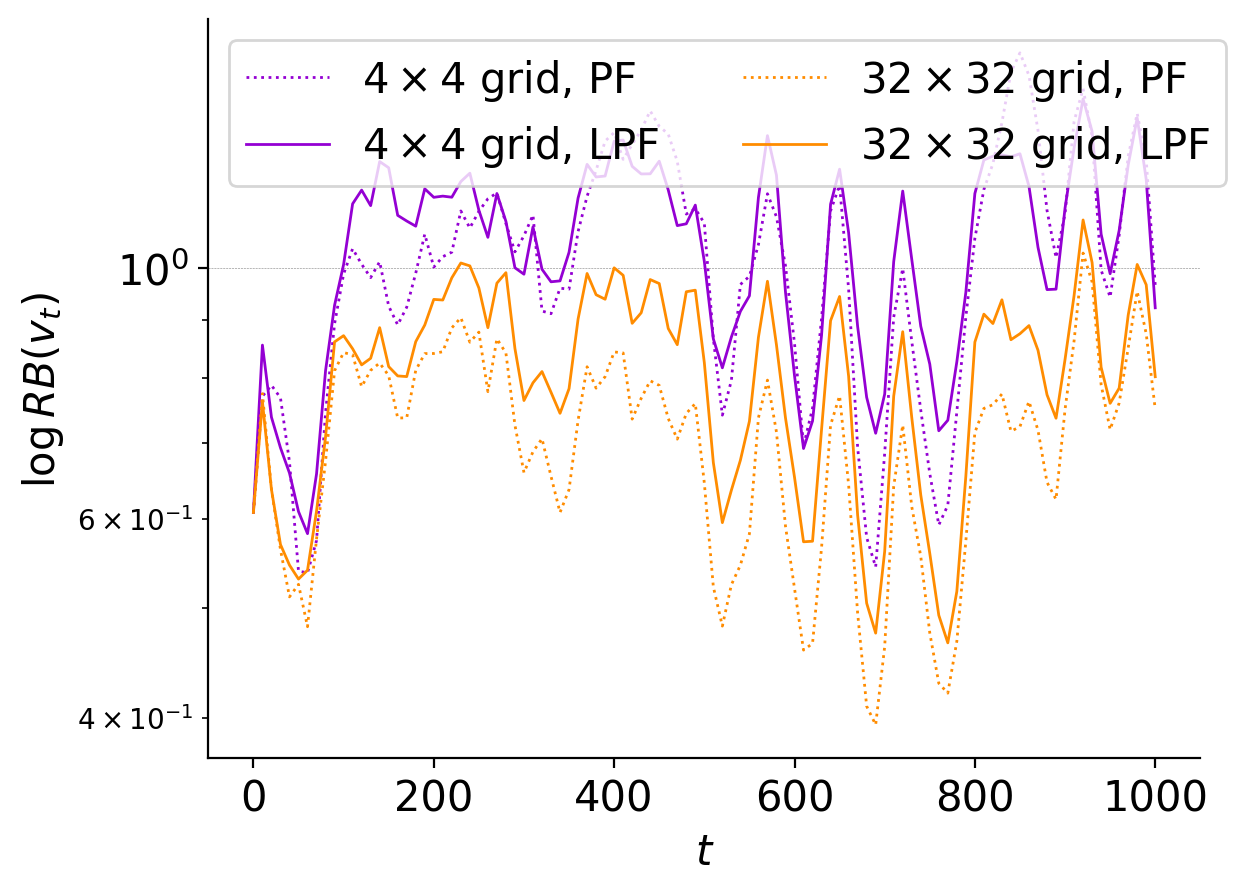}
	\includegraphics[width=0.323\linewidth]{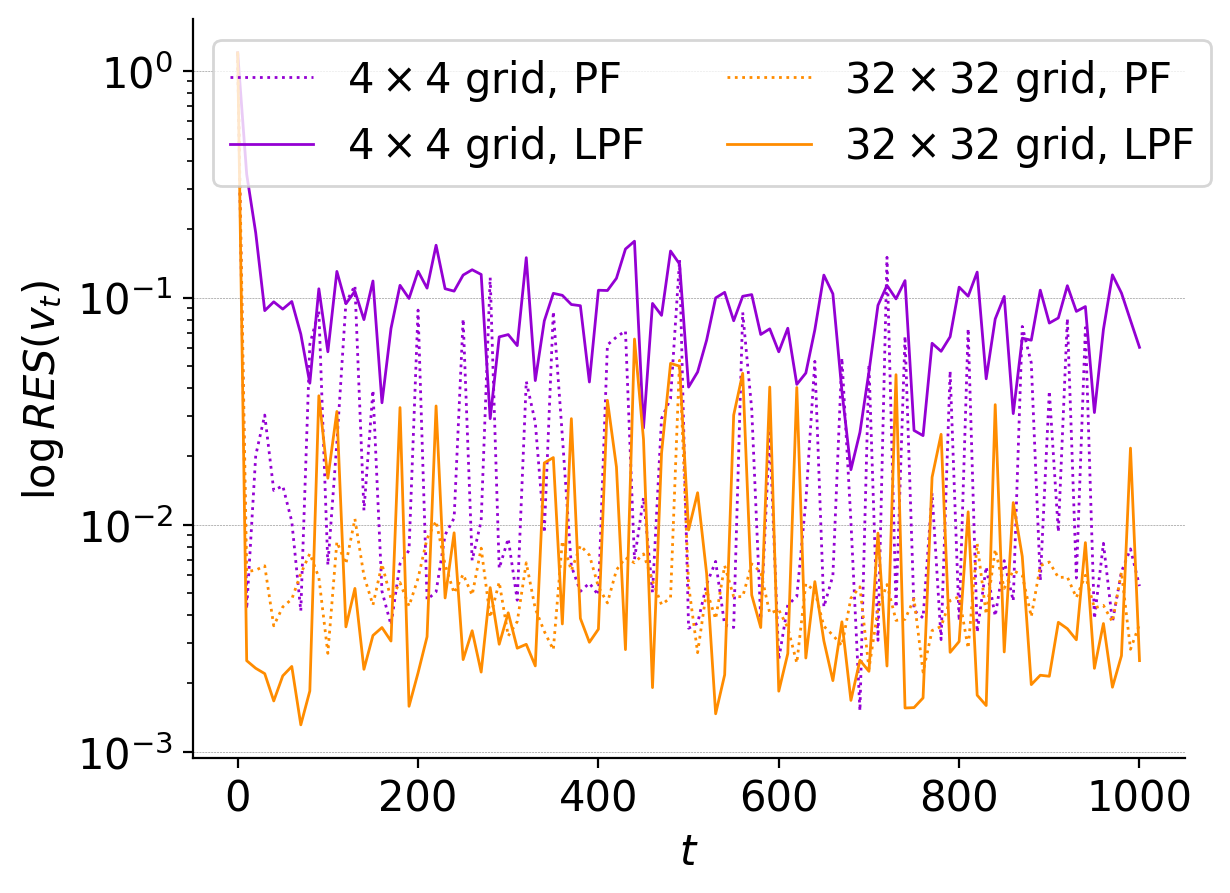}
	\caption{Plots of the $EMRE$, $RB$ and $RES$ of $\eta_t$, $u_t$ and $v_t$ for a PF (dotted lines) and a LPF ($N_{\mathrm{loc}}=4$) (solid lines) with $N=50$ particles (the $y$-axis is in log-scale). We compare these errors for observation grids of sizes $4 \times 4$ and $32 \times 32$. Observations are taken every 10 time-steps. On the top row we plot the errors for $\eta_t$, on the second row for $u_t$ and on the last one for $v_t$. The logarithm of the $EMRE$ for each field is in the first column, of the $RB$ in the second and of the $RES$ in the third. The LPF exhibits errors of a similar magnitude and behaviour to the PF.}
	\label{fig:PF_obs_grid_test_2_grids}
\end{figure}

\begin{figure}[ht!]
	\centering
	\includegraphics[width=0.45\linewidth]{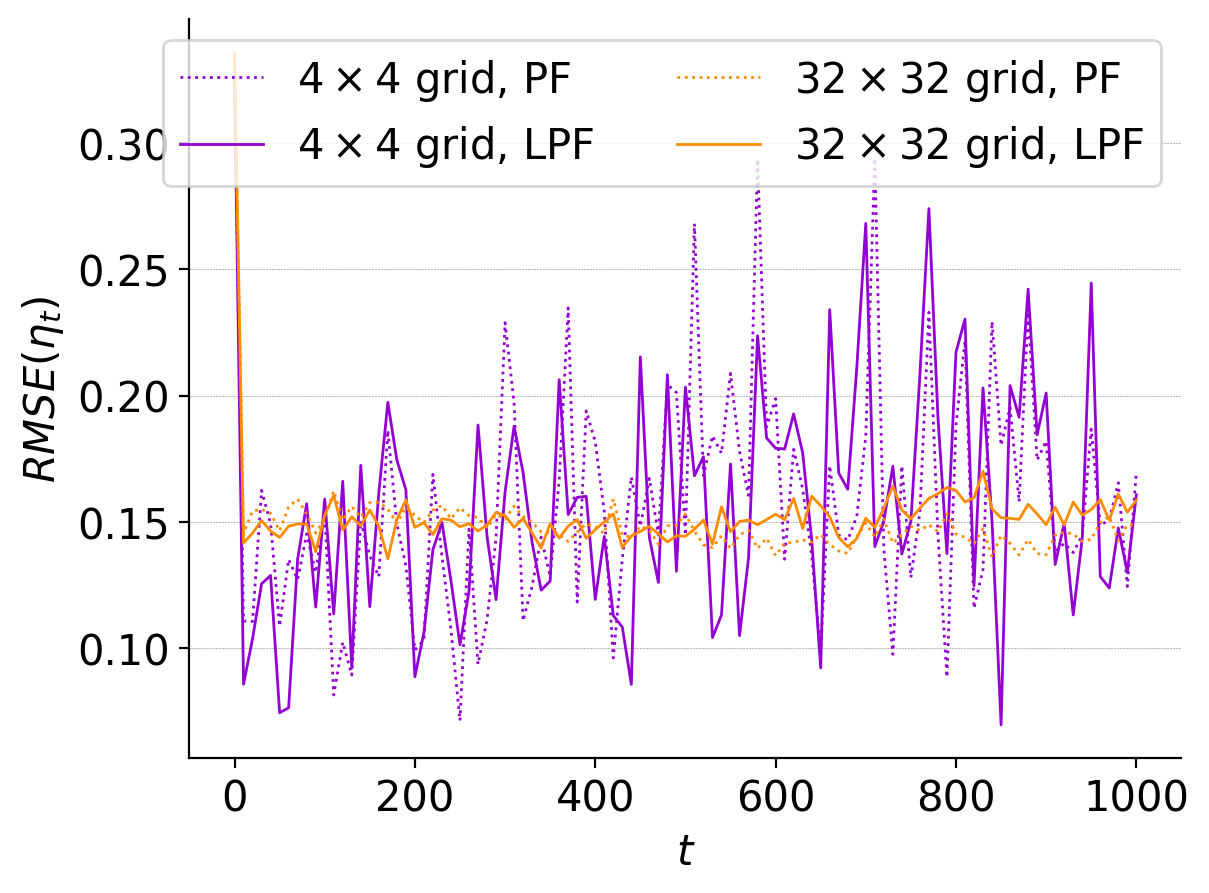}
	\includegraphics[width=0.45\linewidth]{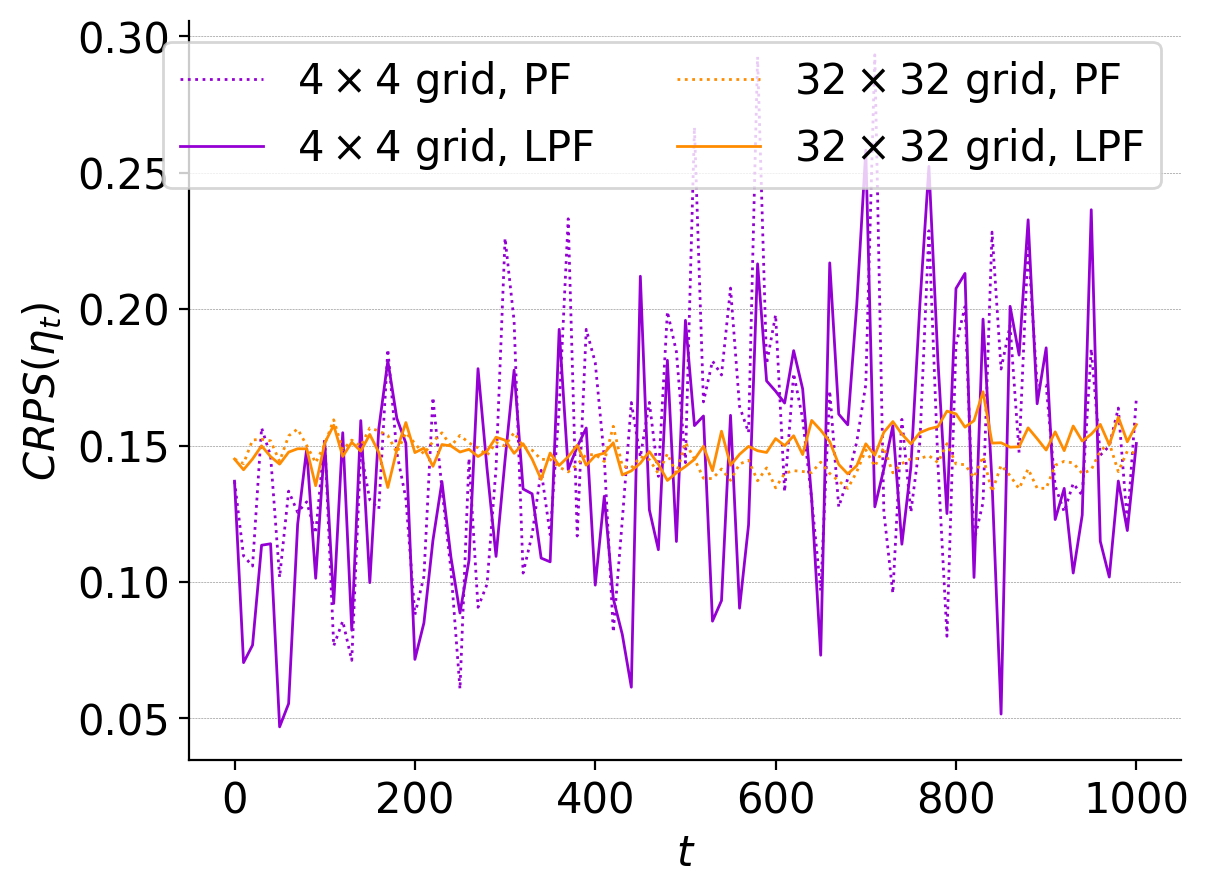}
	\caption{Plots of $RMSE(\eta_t)$ and $CRPS(\eta_t)$ for a PF (dotted lines) and a LPF (solid lines) of $N=50$ particles. We compare these errors for observation grids of sizes $4 \times 4$ and $32 \times 32$. Observations are taken every 10 time-steps. The LPF exhibits errors of a similar magnitude and behaviour to the PF.}
	\label{fig:PF_obs_grid_test_rmse_2_grids}
\end{figure}

Our first and most important concern is to show that the localization procedure does not impact excessively the efficacy of the particle filter. In Figure~\ref{fig:PF_obs_grid_test_2_grids} we plot the $EMRE$, $RB$ and $RES$ of $\eta_t$, $u_t$ and $v_t$ for 100 assimilation time-steps, in log-scale. We plot the errors of the LPF as solid lines, and those of the PF as dotted lines, for observation grids of size $4\times4$ and $32\times32$. We see that our localization procedure is working: with the same number of particles, the errors of the LPF exhibit roughly the same behaviours of the errors of the PF. This implies that our procedure for splitting the global particles into local particles, performing the filtering algorithm locally, and merging the particles back together is not creating excessive discontinuities in the system, so that both the filtering algorithm and the numerical scheme to solve \eqref{eq:SRSW} remain stable. This claim is further supported by the plots in Figure~\ref{fig:PF_obs_grid_test_rmse_2_grids}, where we plot $RMSE(\eta_t)$ and $CRPS(\eta_t)$. While in the case of the $4\times4$ observation grid the errors oscillate, we see that in the $32\times32$ grid case the errors of both the PF and the LPF stabilize at around the same level ($\approx0.15$).

In passing, we note that clearly a $4 \times 4$-grid of observations (only 16 locations!) is not enough to ensure a stable assimilation procedure: $RMSE(\eta_t)$ and $CRPS(\eta_t)$ oscillate wildly in Figure~\ref{fig:PF_obs_grid_test_rmse_2_grids}, and in Figure~\ref{fig:PF_obs_grid_test_2_grids} $EMRE(\eta_t)$ and $RB(\eta_t)$ exhibit a growing trend that betrays the instability of the filtering algorithm when so few observations are available. This leads us to wonder what amount of observations might be sufficient for the errors of the PF and the LPF to reach some kind of optimal level (conditional, of course, on the number of particles $N$, the observation frequency $r$, and all the other parameters of this particular filtering experiment).

\begin{figure}[ht!]
 	\centering
 	\includegraphics[width=0.45\linewidth]{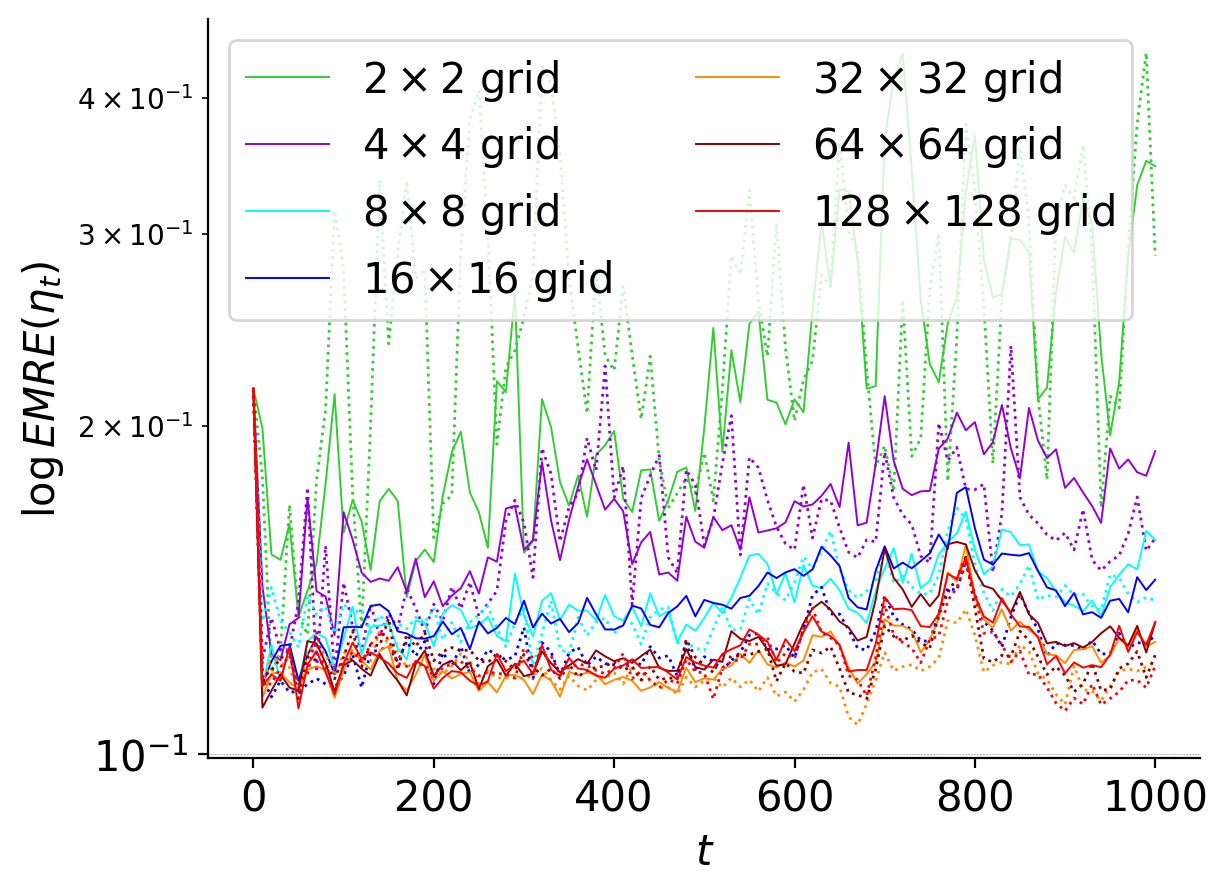}
 	\includegraphics[width=0.45\linewidth]{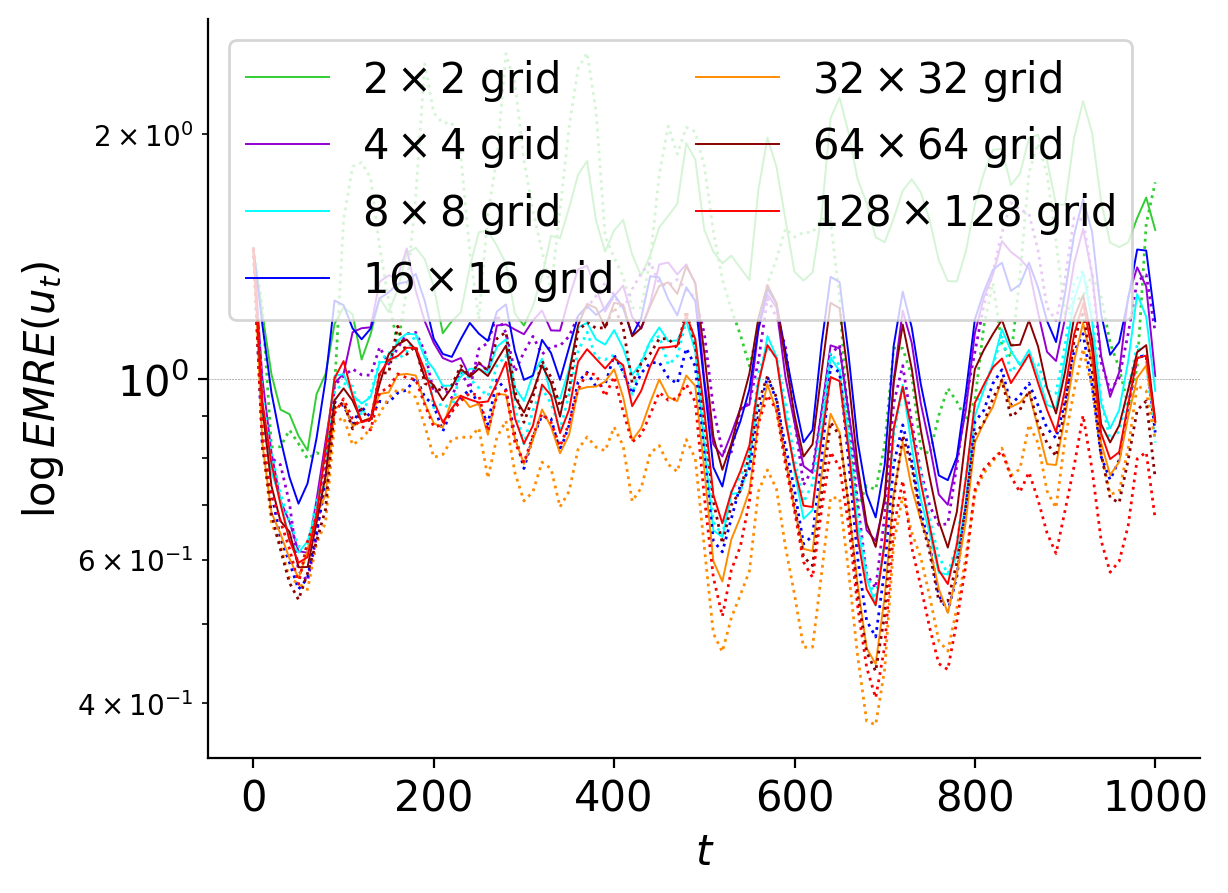}
 	\caption{Plots of $EMRE(\eta_t)$ and $EMRE(u_t)$ for a PF (dotted lines) and a LPF ($N_{\mathrm{loc}}=4$) (solid lines) with $N=50$ particles (the $y$-axis is in log-scale). We compare these errors for observation grids of size $2^n \times 2^n$ for $n = 1, \dots 7$. The more grid-points are observed, the smaller the errors.}
 	\label{fig:PF_obs_grid_test_2}
 \end{figure}

\begin{figure}[ht!]
	\centering
	\includegraphics[width=0.45\linewidth]{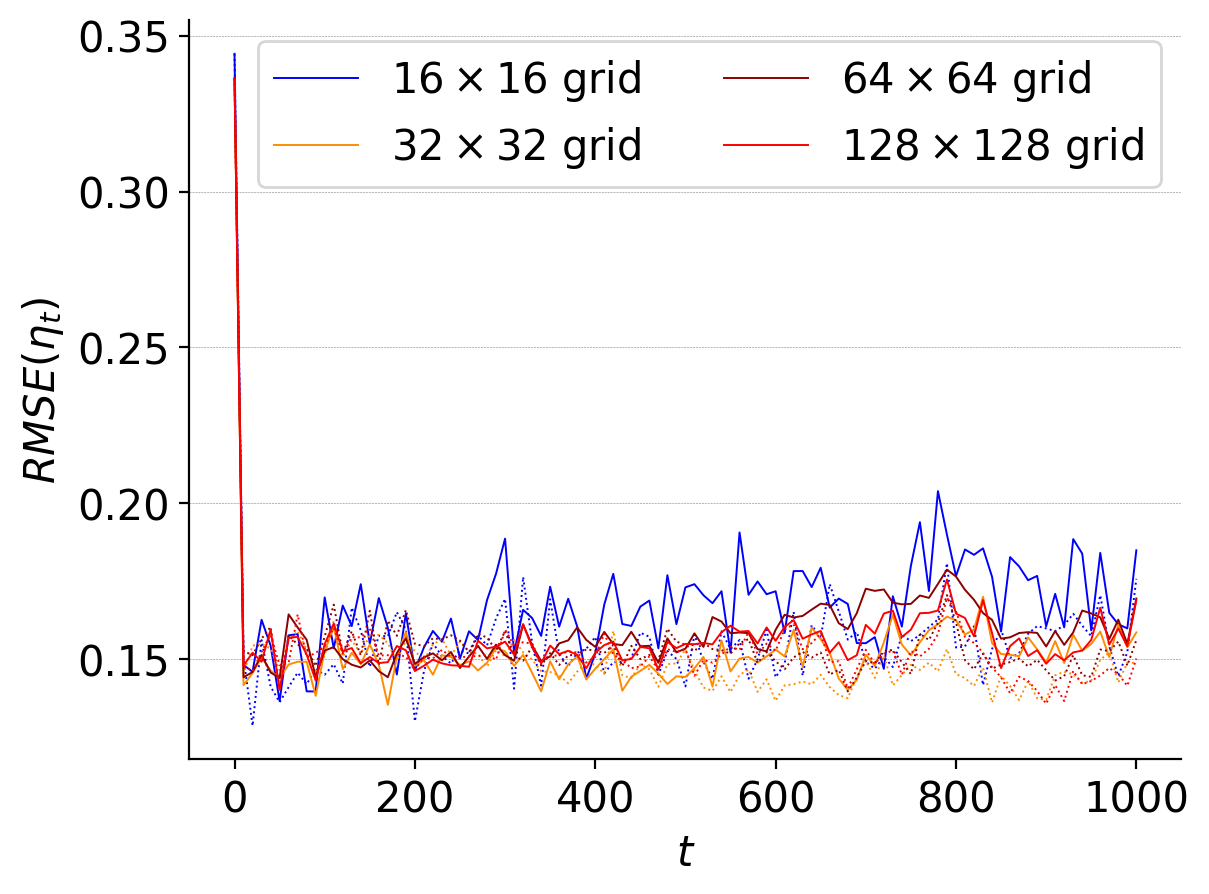}
	\includegraphics[width=0.45\linewidth]{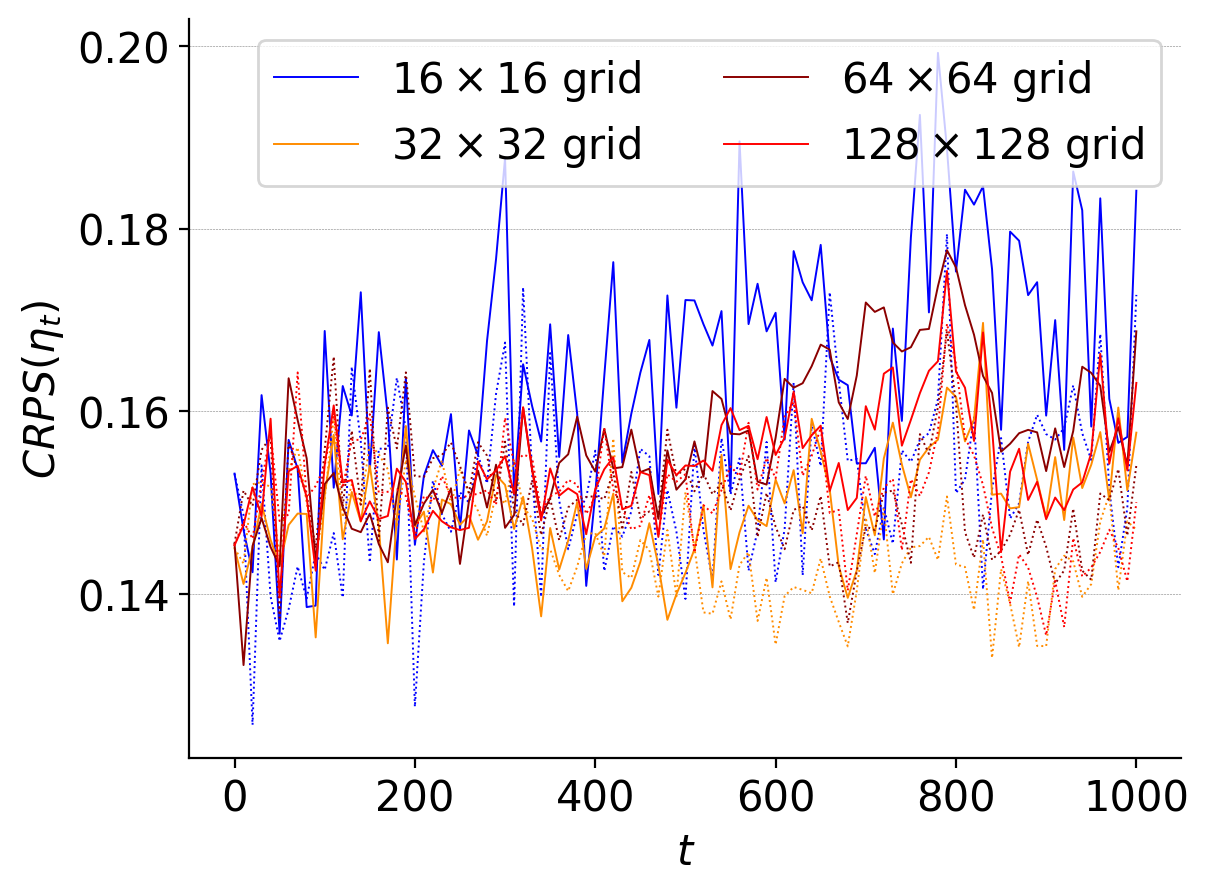}
	\caption{Plots of $RMSE(\eta_t)$ and $CRPS(\eta_t)$ for a PF (dotted lines) and a LPF (solid lines) of $N=50$ particles. We compare these errors for observations-grids of size $2^n$ for $n = 4, \dots 7$. Observations are taken every 10 time-steps.}
	\label{fig:PF_obs_grid_test_rmse_2}
\end{figure}

 We investigate this question further in Figures~\ref{fig:PF_obs_grid_test_2} and~\ref{fig:PF_obs_grid_test_rmse_2}. In Figure~\ref{fig:PF_obs_grid_test_2} we plot $EMRE(\eta_t)$ and $EMRE(u_t)$ for 100 assimilation time-steps, in log-scale, for all observation grid's sizes $2^n \times 2^n$ for $n = 1, \dots, 7$ ($EMRE(v_t)$ has similar behaviour and magnitude to $EMRE(u_t)$ and we omit its plot). In Figure~\ref{fig:PF_obs_grid_test_rmse_2}, for the same experiments, we plot the $RMSE$ and $CRPS$ of $\eta_t$ (we omit the plots for the smaller observation grids $2\times2, 4\times 4$ and $8\times8$ for clarity, since the errors oscillate a lot and cloud the figures). Again, we plot the errors of the LPF as solid lines, and those of the PF as dotted lines. As expected, we see that as the number of observations increase, the errors decrease.We also note that already an observation grid of size $16 \times 16$ produces reasonable results, compared to experiments with a larger number of observed locations. Once we increase the number of observations to a $32\times32$ grid, the errors are almost indistinguishable from the $64 \times 64$ and $128 \times 128$ cases. Note that we are only observing $\eta_t$, but the errors in the velocity fields $(u_t, v_t)$, while oscillatory, also do not grow, with a maximum $EMRE(u_t)$ for the larger observation grid around 1 in Figure~\ref{fig:PF_obs_grid_test_2}. 
 
From the above we can draw a couple of conclusions related to the dimension of our problem: if one considers the signal process to be the full numerical solution of \eqref{eq:SRSW} (so not only the height-field $\eta_t$ but also the zonal and meridional velocities $u_t$ and $v_t$), one could naively assume to be dealing with a system of dimension $3 \times 128^2 \approx 5 \times 10^{5}$. Being able to give an estimate for such a system with only $16^2$ data-points, which amount to less than 1\% of the total dimension of the system, and only using 50 particles for the ensemble, could seem impossible. Our results supply evidence to the hypothesis that the true dimension of the system is, in fact, much lower than $5 \times 10^{5}$. Not only the dependencies between $\eta_t$, $u_t$ and $v_t$ make the system highly constrained, so that by observing one we can also improve our estimates of the others; but also, the RSW model \eqref{eq:SRSW} is clearly very correlated in space, so that by observing only a few points we can get an accurate enough estimate on the whole space.

\subsection{LPF: sensitivity to the parameters}\label{subsec:LPF_params}
Our experiments in the previous subsection yielded that the LPF does approximately as well as the PF for an equi-spaced grid of observations. They also yielded that having observations in a $32 \times 32$ grid equi-spaced in the domain is ``enough'': adding further observations points does not significantly improve the errors. In this subsection we experiment with changing the settings of the filtering problem and the parameters in our LPF algorithm, and discuss the results. We fix the observation noise to have standard deviation $\mathbf{\boldsymbol{\sigma} = 0.05}$ and the location of the observations to be the $\mathbf{32\times32}$ \textbf{grid} throughout.

\begin{figure}[ht!]
	\centering
	\includegraphics[width=0.43\linewidth]{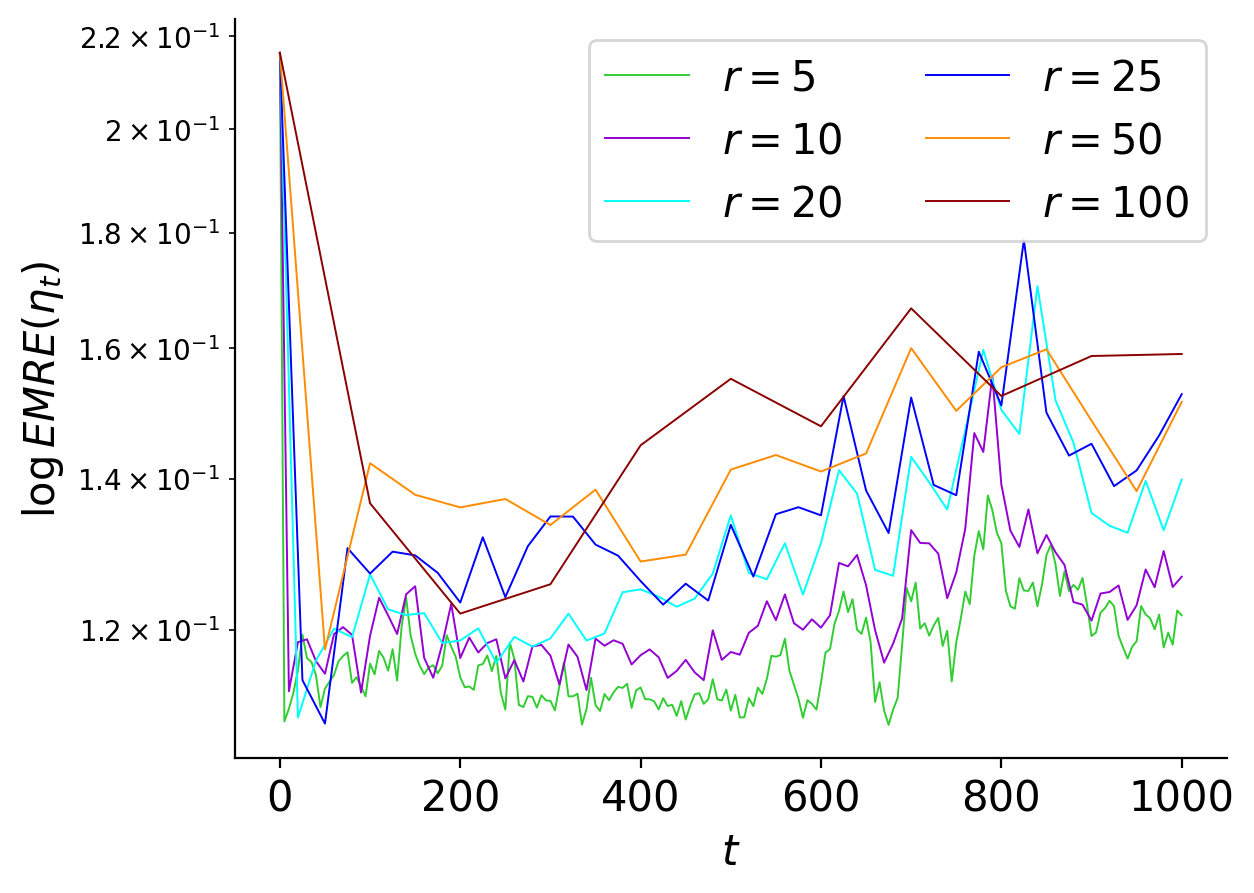}
	\includegraphics[width=0.43\linewidth]{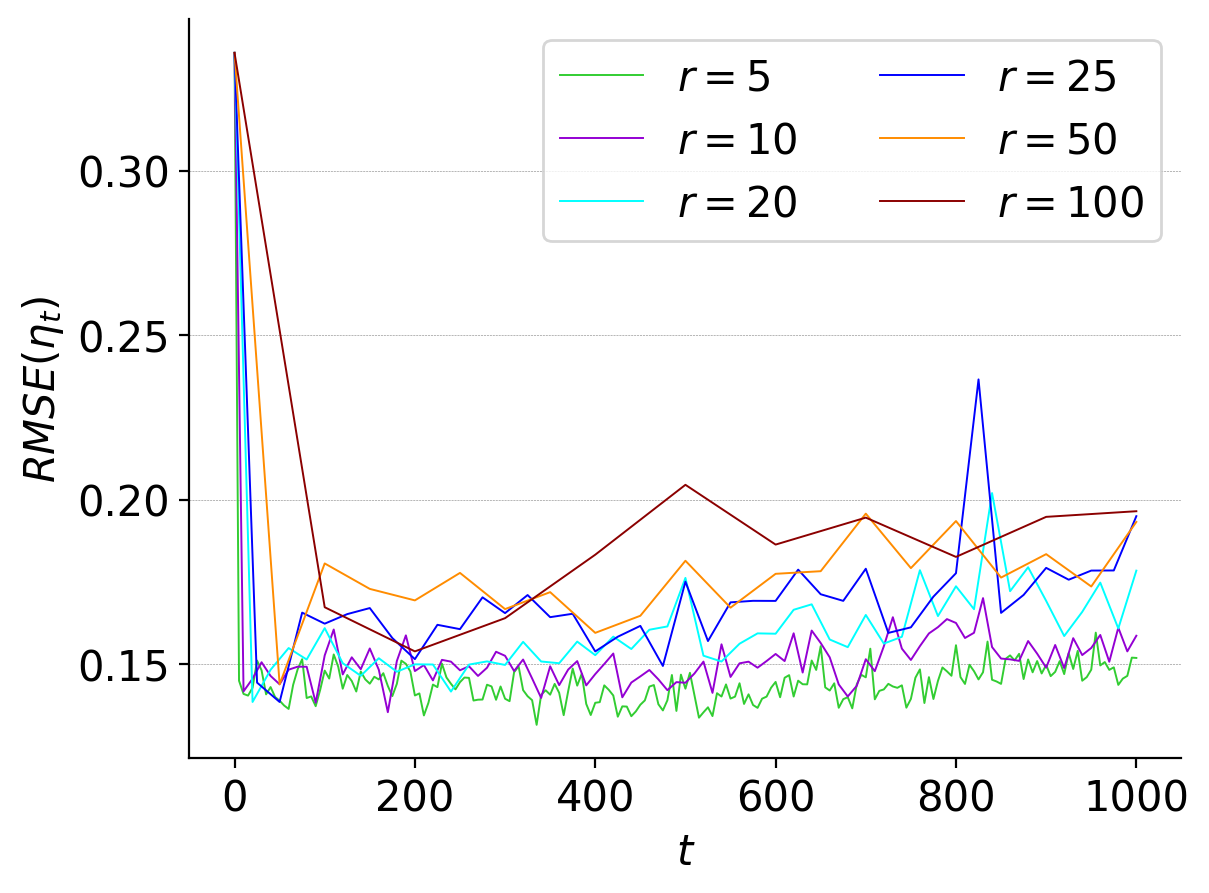}
	\caption{Plots of $EMRE(\eta_t)$ (in log-scale) and $RMSE(\eta_t)$ for a LPF with $N_{\mathrm{loc}}=4$ and $N=50$, with observations located on a $32 \times 32$ equi-spaced grid within $D$. As we increase the observation frequency by varying $r \in \{5, 10, 20, 25, 50, 100\}$ (smaller $r$ indicates more frequent observations), the errors decrease.}
	\label{fig:LPF_test_freq}
\end{figure}

\begin{figure}[ht!]
	\centering
	\includegraphics[width=0.43\linewidth]{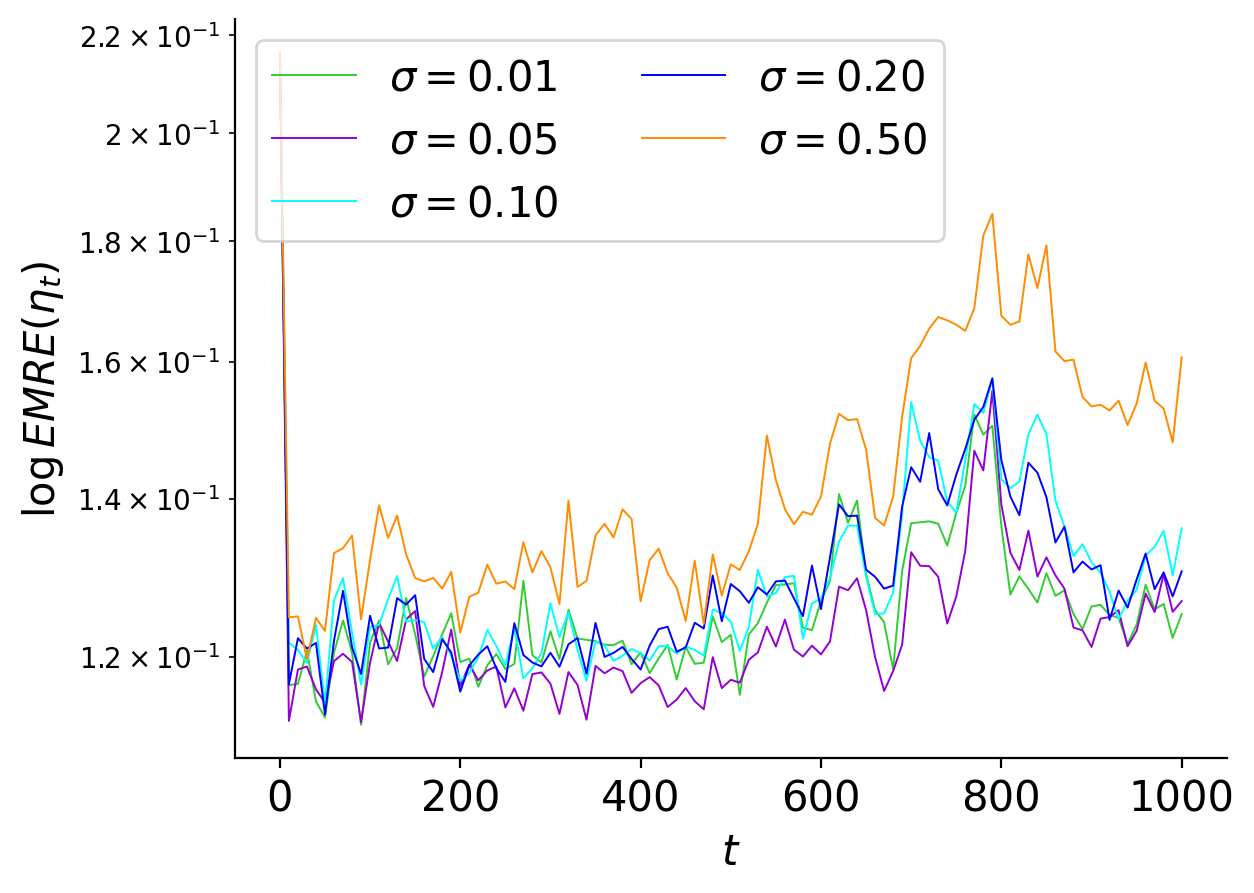}
	\includegraphics[width=0.43\linewidth]{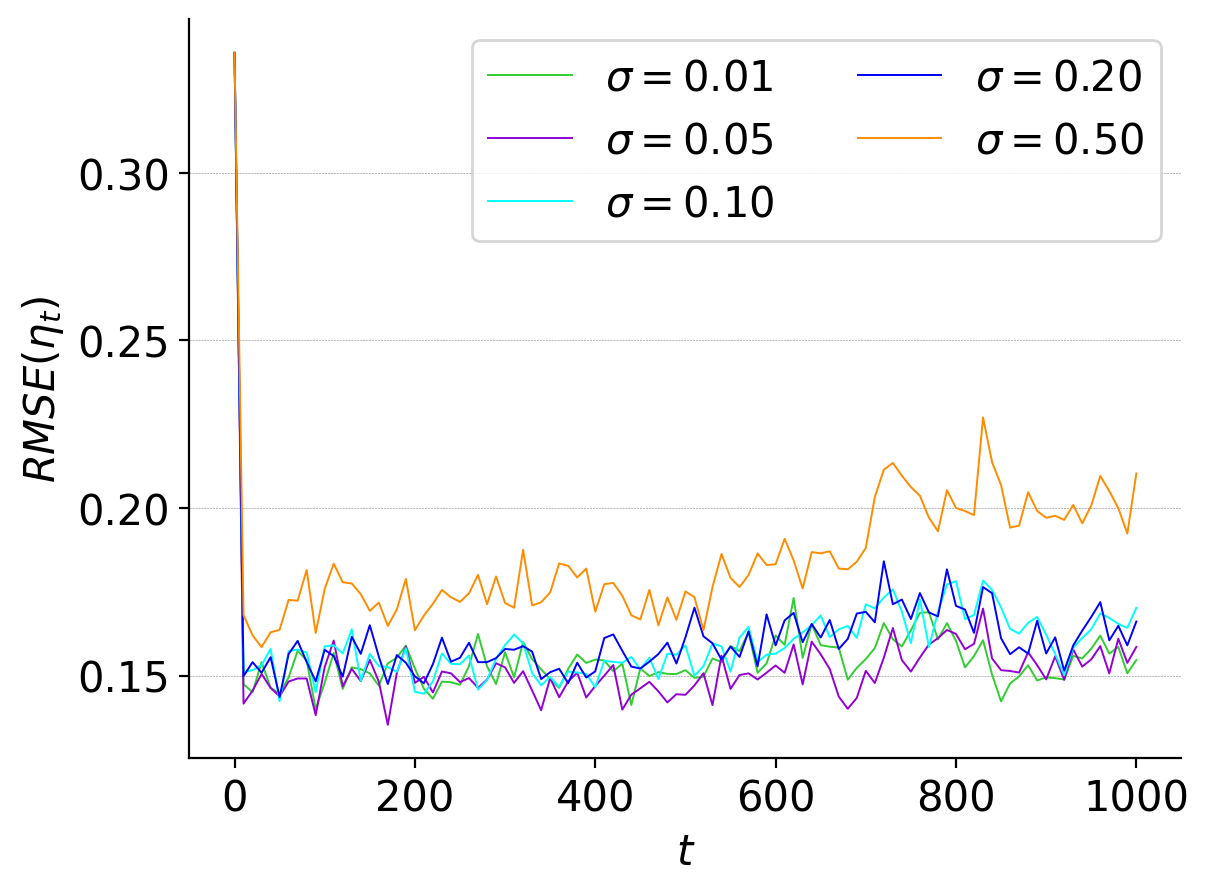}
	\caption{Plots of $EMRE(\eta_t)$ (in log-scale) and $RMSE(\eta_t)$ for a LPF with $N_{\mathrm{loc}}=4$ and $N=50$, with observations located on a $32 \times 32$ equi-spaced grid within $D$. As we decrease the observation noise by varying $\sigma \in \{0.01, 0.05, 0.1, 0.2, 0.5\}$, the errors decrease.}
	\label{fig:LPF_test_obs_noise}
\end{figure}

\begin{figure}[ht!]
	\centering
	\includegraphics[width=0.325\linewidth]{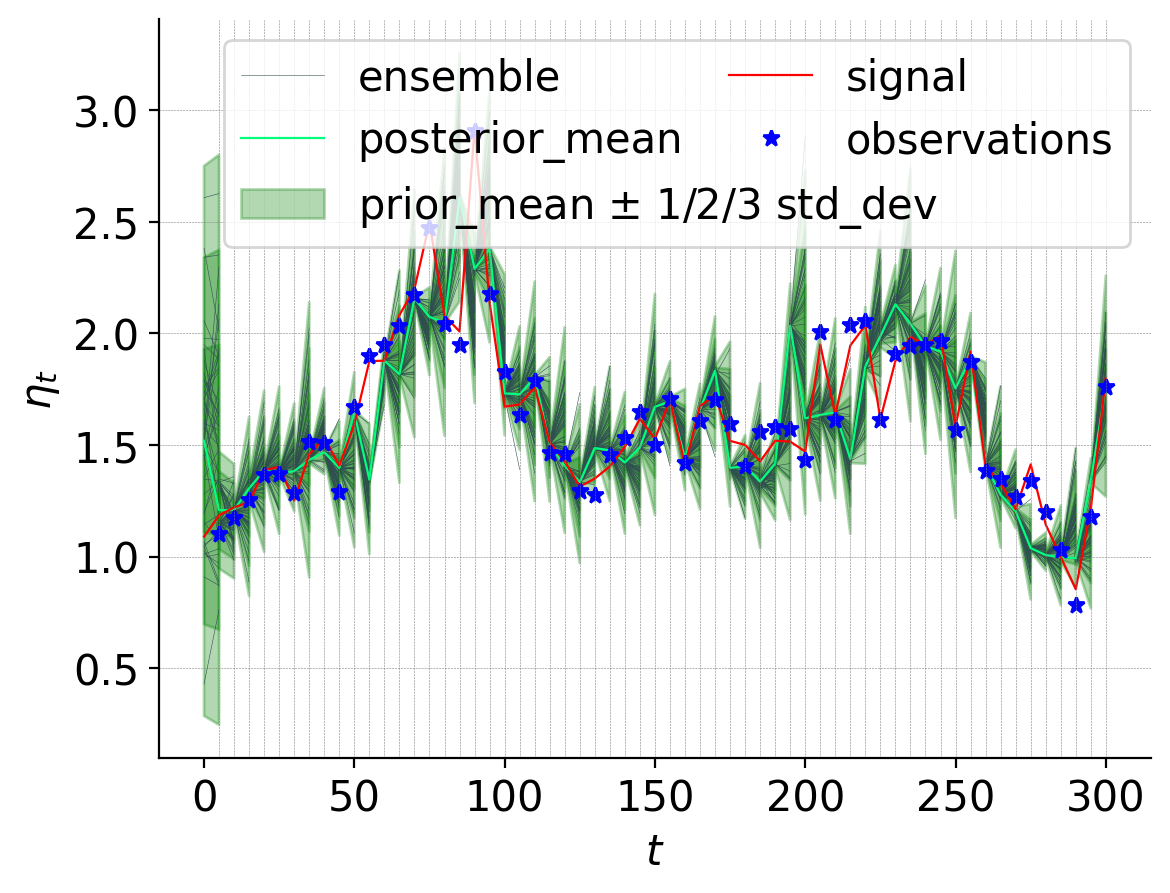}
	\includegraphics[width=0.325\linewidth]{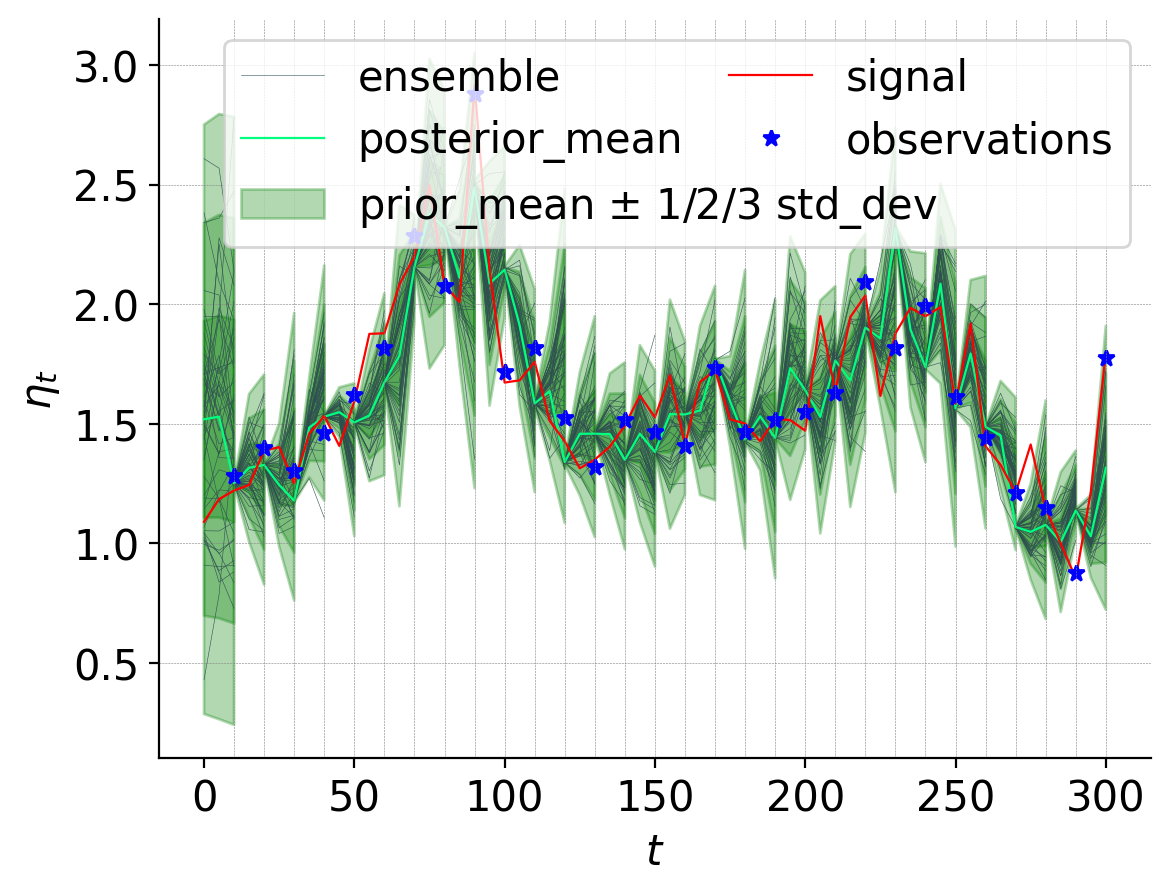}
	\includegraphics[width=0.325\linewidth]{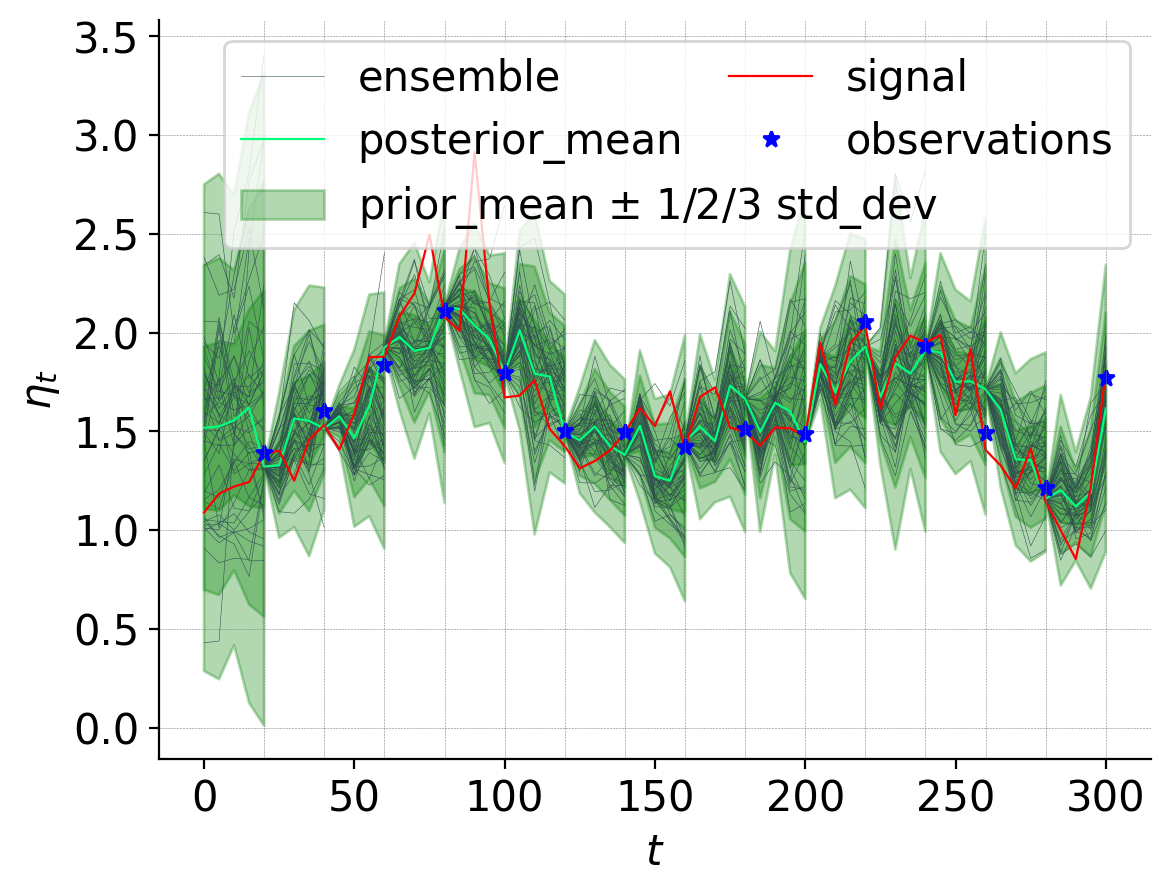} \\
	\includegraphics[width=0.325\linewidth]{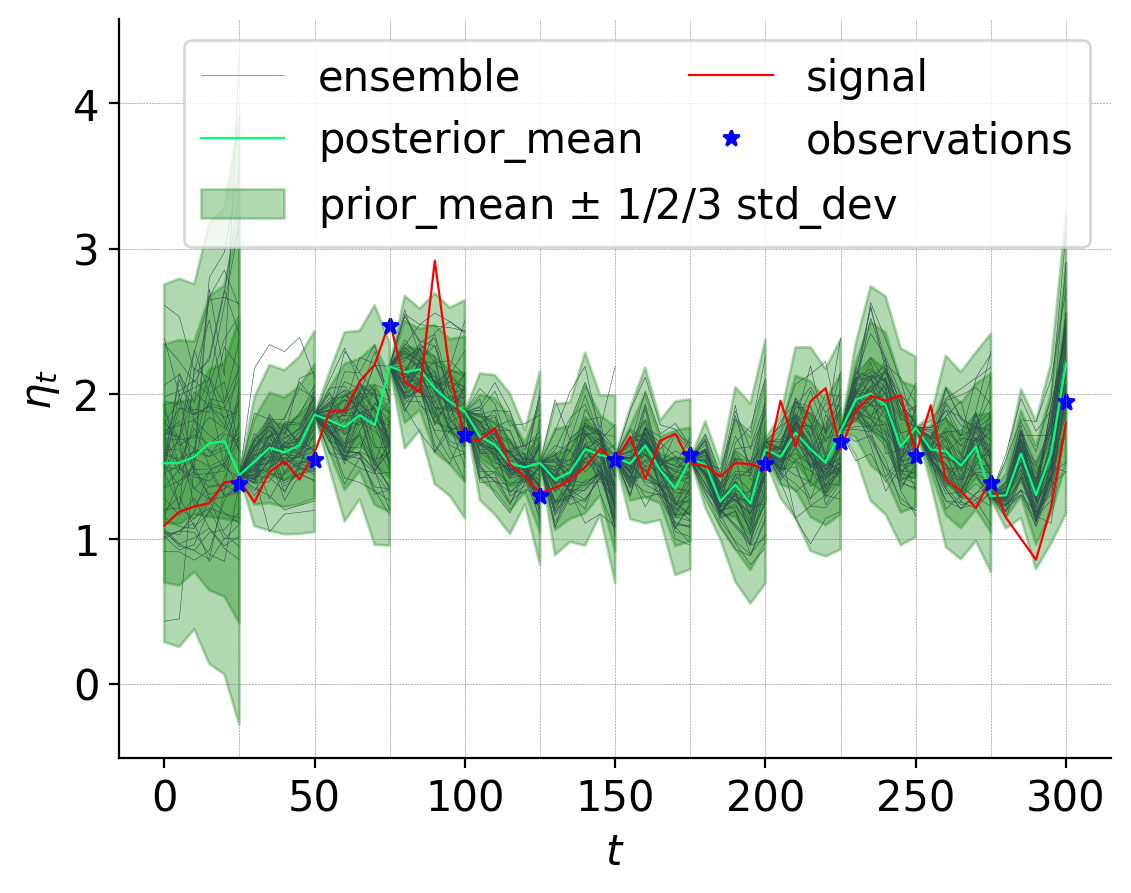}
	\includegraphics[width=0.325\linewidth]{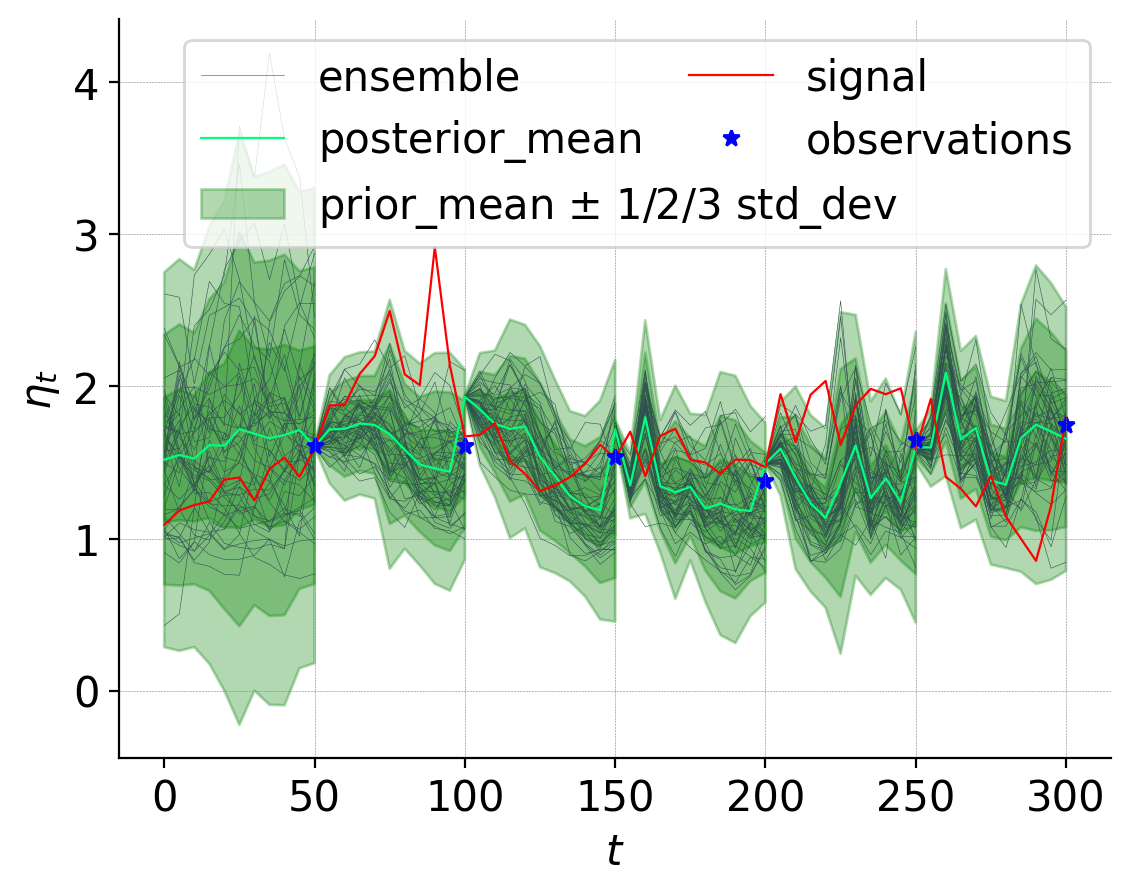}
	\includegraphics[width=0.325\linewidth]{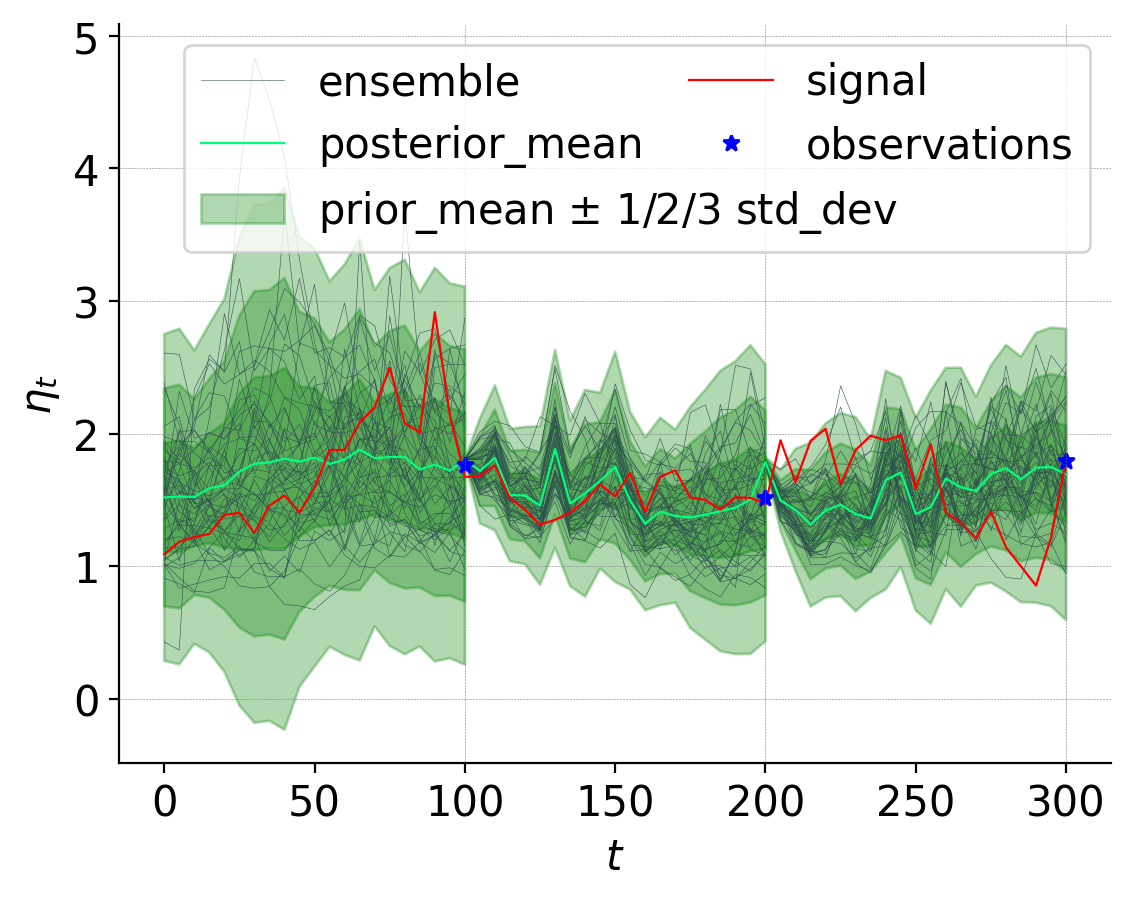}
	\caption{Plots of the evolution in time, for the first 300 time-steps, of the signal (red) and the ensemble (green) of the LPF ($N_{\mathrm{loc}}=4$) at a fixed point in the domain, for different observation frequencies. In order from the top-left, assimilation happens every 5, 10, 20, 25, 50 and 100 time-steps. The posterior mean of the ensemble is highlighted in lighter green, and the observations are shown in blue. The more frequent the assimilations, the more accurately the ensemble is able to capture the movement of the signal.}
	\label{fig:LPF_freq_evol}
\end{figure}

We first perform a couple of consistency-checks. For a PF, we expect the errors to increase as the frequency of the observations decreases, or when the noise of the observations increases. We expect the LPF to perform similarly. To check the first, we run a LPF with $\mathbf{N_{\mathrm{loc}} = 4}$ and $\mathbf{N = 50}$, and observation frequency $r \in \{5,10,20,25,50,100\}$, so we assimilate every 5, 10, 20, 25, 50 and 100 time-steps respectively.  To check the second, we run a LPF (again with $\mathbf{N_{\mathrm{loc}} = 4}$ and $\mathbf{N = 50}$) with fixed observation frequency $\mathbf{r =10}$, and vary the observation noise $\sigma \in \{0.01, 0.05, 0.1, 0.2, 0.5\}$. The results of these experiments, which confirm that indeed the errors grow conversely to the observation frequency, and proportionally to the observation noise, are found in Figure~\ref{fig:LPF_test_freq} and Figure~\ref{fig:LPF_test_obs_noise}. Finally, in Figure~\ref{fig:LPF_freq_evol} we plot the evolution in time of the signal $\eta_t$ and the ensemble of the LPF, at a fixed point in the domain $D$. We choose the point $(x,y) = (67,67)$, which is roughly in the middle of $D$. We observe the effect of the assimilation on the ensemble for the different observations-frequencies: the more frequent the assimilations, the more accurately the ensemble is able to capture the movement of the signal.

We now move on to discussing the parameters $N_{\mathrm{loc}}$, $\alpha$ and $p_{\mathrm{ov}}$, which are specific to the LPF.

We start by understanding the impact of increasing the number of localization regions $N_{\mathrm{loc}}$. There are both clear benefits and clear disadvantages of increasing $N_{\mathrm{loc}}$. From a computational perspective, if one has a large number of CPU/GPU cores at their disposal, one can argue that the more localization regions, the better, since the LPF can run in parallel on each subregion, and thus the computational time decreases proportionally to $N_{\mathrm{loc}}$. Another advantage is that, especially when working with a limited number of particles, the update procedure in the LPF (especially with large $\alpha$) makes the filter less sensitive about the observations outside each specific localization regions: thus, the particles have fewer constrains during the tempering/jittering steps, and can move more easily in each subregion to try and capture the signal dynamics.

On the other hand, the larger $N_{\mathrm{loc}}$, the higher the amount of interpolation needed at the end of the local update steps to reconstruct the global particles. Interpolation not only introduces errors, but also makes the resulting global particles less `physical': since our model is highly nonlinear, interpolation (even linear) of two solutions does not yield another solution to the SPDE \eqref{eq:SRSW}. This also impacts the stability of the filtering algorithm, since this is linked to the stability of the numerical scheme to solve \eqref{eq:SRSW}, which has higher chances of breaking the more our localization procedure introduces potential `non-physical' behaviours and discontinuities. It then becomes very important to tune carefully $\alpha$ and $p_{\mathrm{ov}}$ to avoid such instability.

\begin{figure}[ht!]
	\centering
	\includegraphics[width=0.325\linewidth]{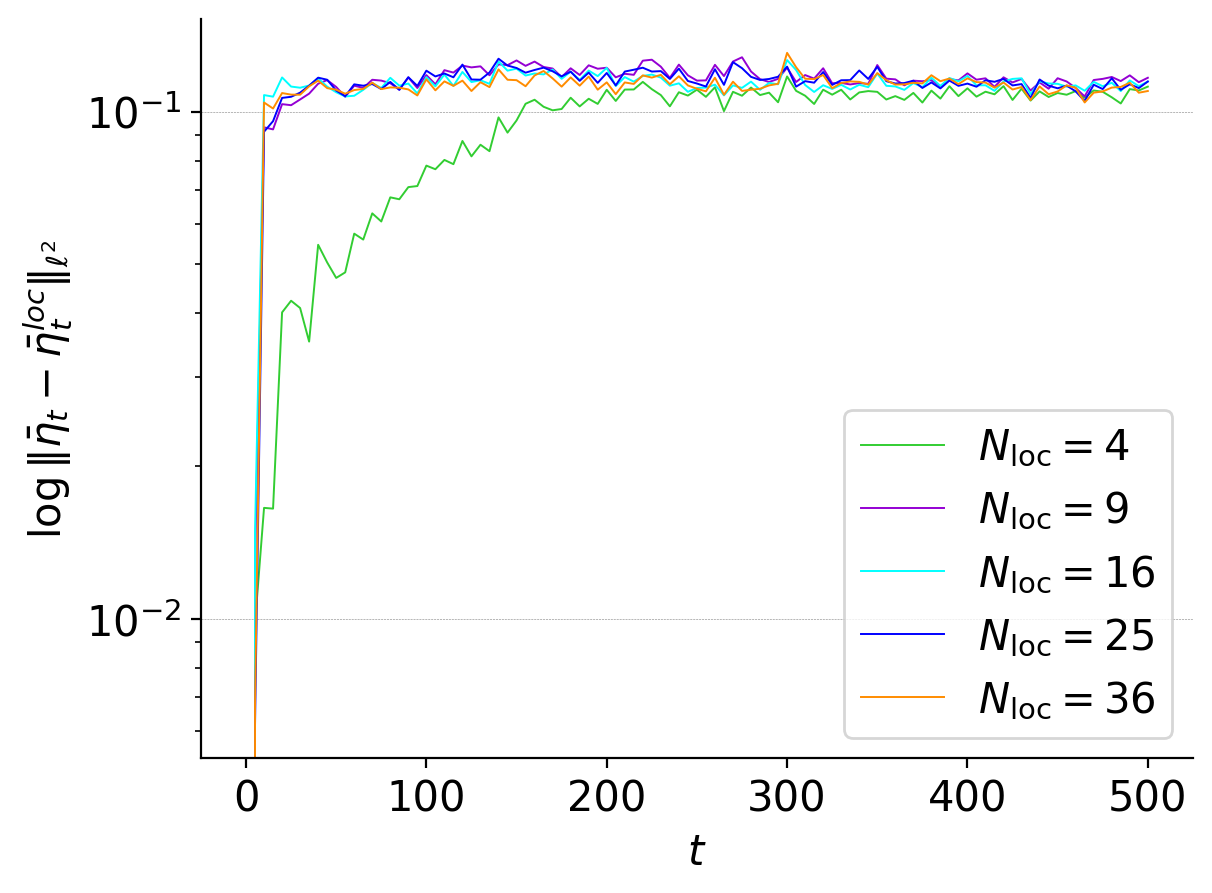}
	\includegraphics[width=0.325\linewidth]{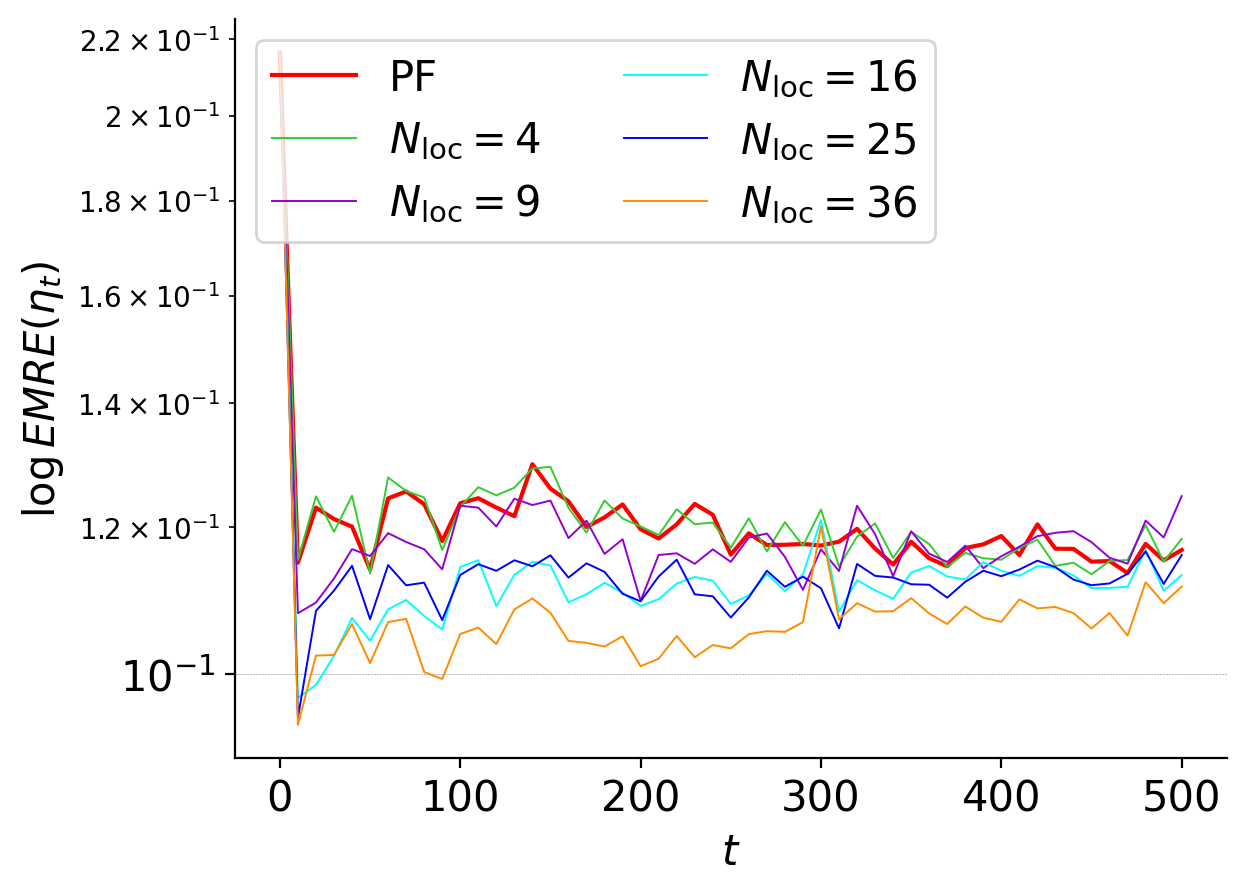}
	\includegraphics[width=0.325\linewidth]{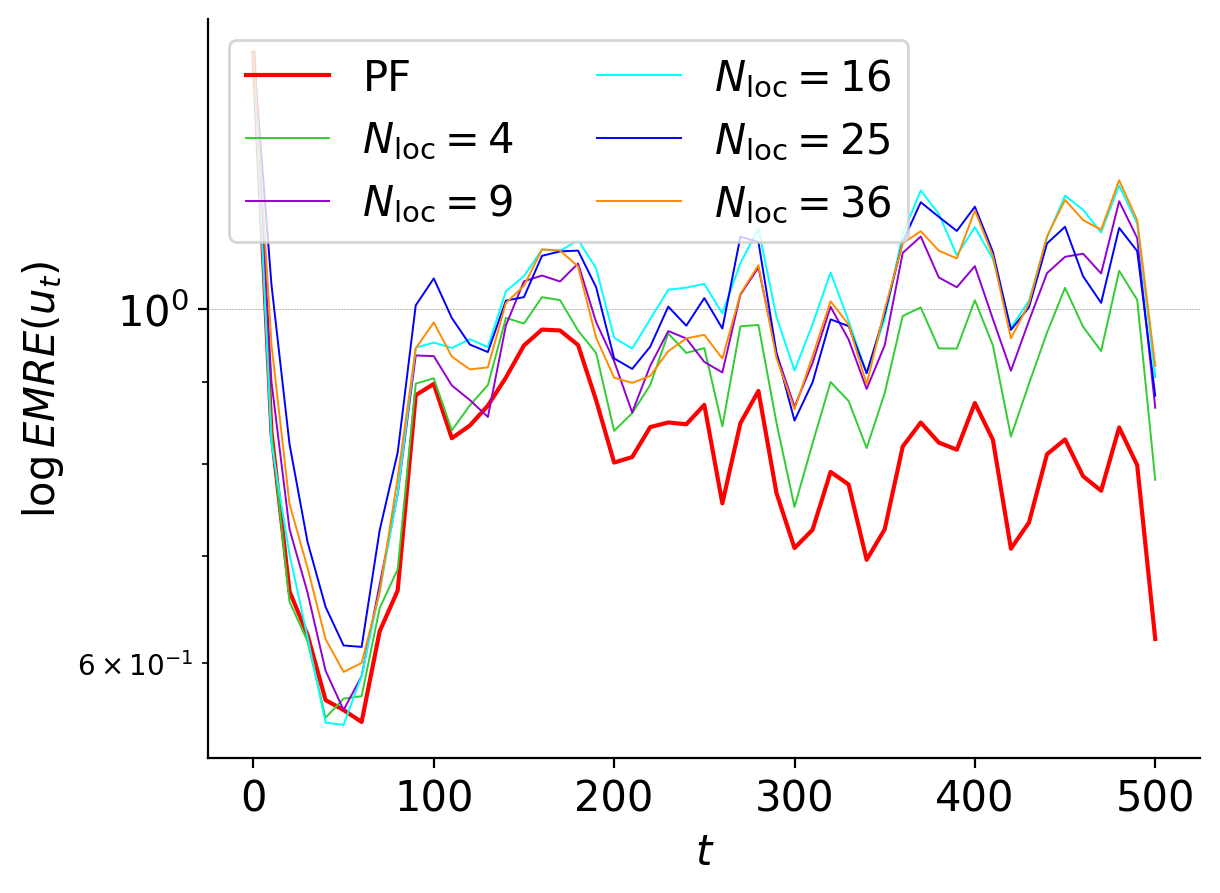} 
	\caption{Comparison between LPFs run with different numbers of localization regions $N_{\mathrm{loc}} \in \{4,9,16,25,36\}$. In the first plot we show the $\ell^2$ error between the mean of the PF and the mean of each of the LPFs. The second and the third plot show the EMRE of $\eta_t$ and $u_t$ respectively, for each of the LPFs; for reference, the same errors for the PF are plotted in red. Observations are taken every 10 time-steps on a equi-spaced grid of $32 \times 32$ points in $D$. The $y$-axis is in log-scale. A larger number of localization regions decreases the error for $\eta_t$ (which is observed), but increases the error for $u_t$ (unobserved).}
	\label{fig:LPF_test_nregions}
\end{figure}

In Figure~\ref{fig:LPF_test_nregions} we show some plots that compare LPFs run with different numbers of subregions. We vary $N_{\mathrm{loc}} \in \{4,9,16,25,36\}$. For each of the LPFs, we choose $p_{\mathrm{ov}}$ so that the width of the overlap regions is fixed at $12$ grid-points. For example, when $N_{\mathrm{loc}} =4$, each region starts out as a square of $64 \times 64$ grid-points, to which we add $6$ points to the N, S, E and W of the boundary, making them into $76 \times 76$ squares instead: this corresponds to $p_{\mathrm{ov}} \approx 0.1$, and a total width of the overlap of 12 grid-points all around the subregion. When $N_{\mathrm{loc}} =36$, each region starts out as $22 \times 22$ grid-points, so adding 6 points for the overlap to the N, S, E and W corresponds to $p_{\mathrm{ov}} \approx 0.3$. As for $\alpha$, it is tuned individually in each experiment, but we note that, due to the observations being spread around all of $D$, we should choose $\alpha$ relatively large ($\ge500$): $\alpha$ too small in not only unnecessary, but could slow down the algorithm significantly due to the dependencies of the local weights on the global particles (which are considered as fixed during the local tempering iterations). With a $32\times 32$ grid of observations, in each subregions there is enough data to correct the ensemble, and we know from our experiments in Section~\ref{subsec:LPF_grid} that the system is highly correlated in space, so local ensembles should match closely after assimilation, even with $\alpha$ very large.

In the first plot of Figure~\ref{fig:LPF_test_nregions} we compute the $\ell_2$ difference between the mean of the PF (without localization) and the mean of each of the LPFs with different $N_{\mathrm{loc}}$. We see that the error stabilizes around 0.1 for all the LPFs, with the error of the LPF with only 4 localization regions growing more slowly than the others initially, but then stabilizing at the same level. In the second and third plot we compare the $EMRE$ of $\eta_t$ and $u_t$ for the LPFs, using the $EMRE$ of the PF for reference. We see that $EMRE(\eta_t)$ decreases as we increase $N_{\mathrm{loc}}$. However, we also see that $EMRE(u_t)$ displays exactly the opposite behaviour: the bigger $N_{\mathrm{loc}}$, the bigger the error (the behaviour of $EMRE(v_t)$ is similar, and we omit the plot). This is to be expected: recall that we are observing only $\eta_t$, so, while the localization procedure allows $\eta_t$ to be matched more closely by the ensemble in each subregion, the effect of the errors introduced by the interpolation is felt more keenly in the velocities fields $u_t$ and $v_t$, since we have no control over their update, and by localizing we also partly disrupt the dependencies between $u_t, v_t$ and $\eta_t$.

\begin{figure}[ht!]
	\centering
	\includegraphics[width=0.43\linewidth]{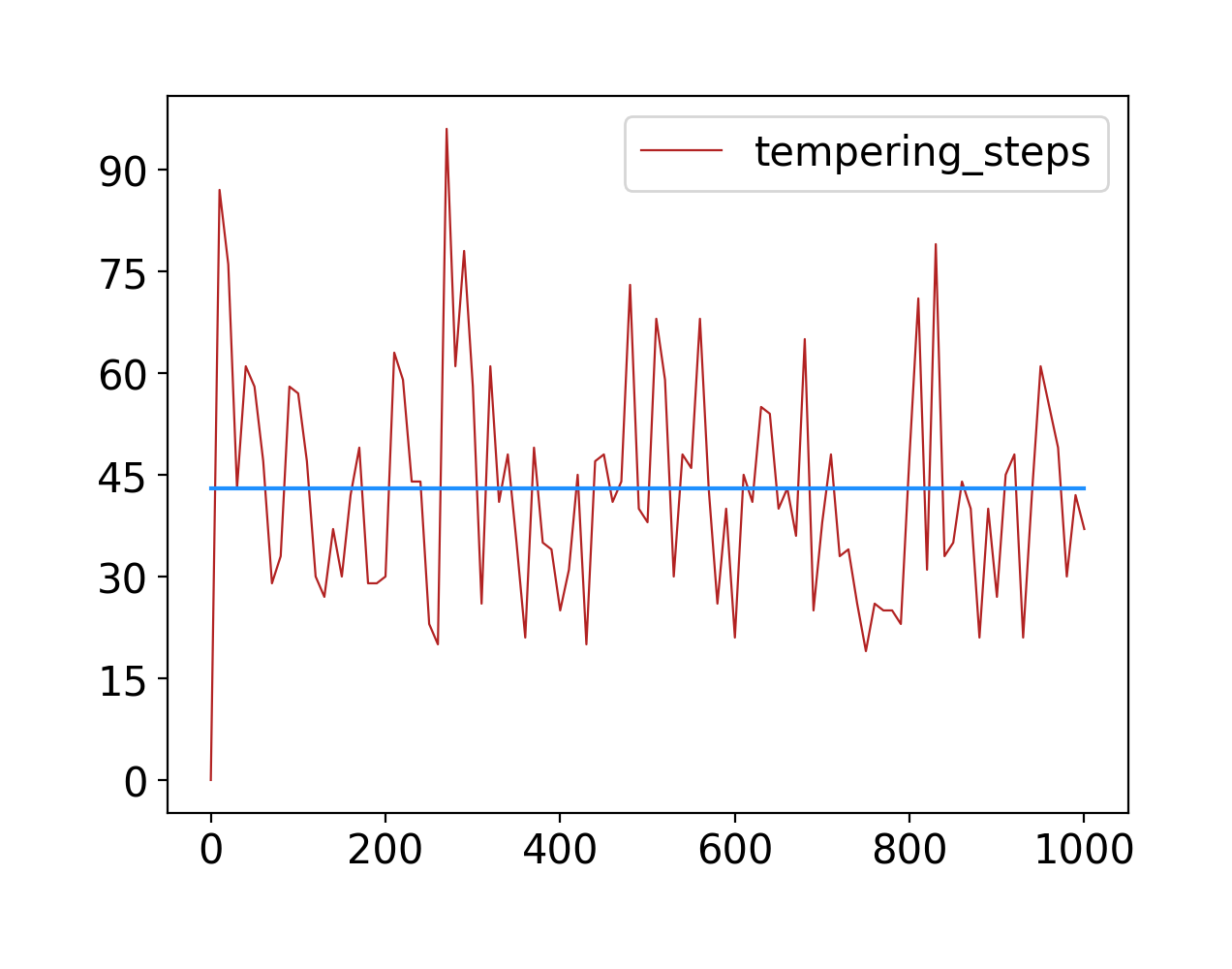}
	\includegraphics[width=0.43\linewidth]{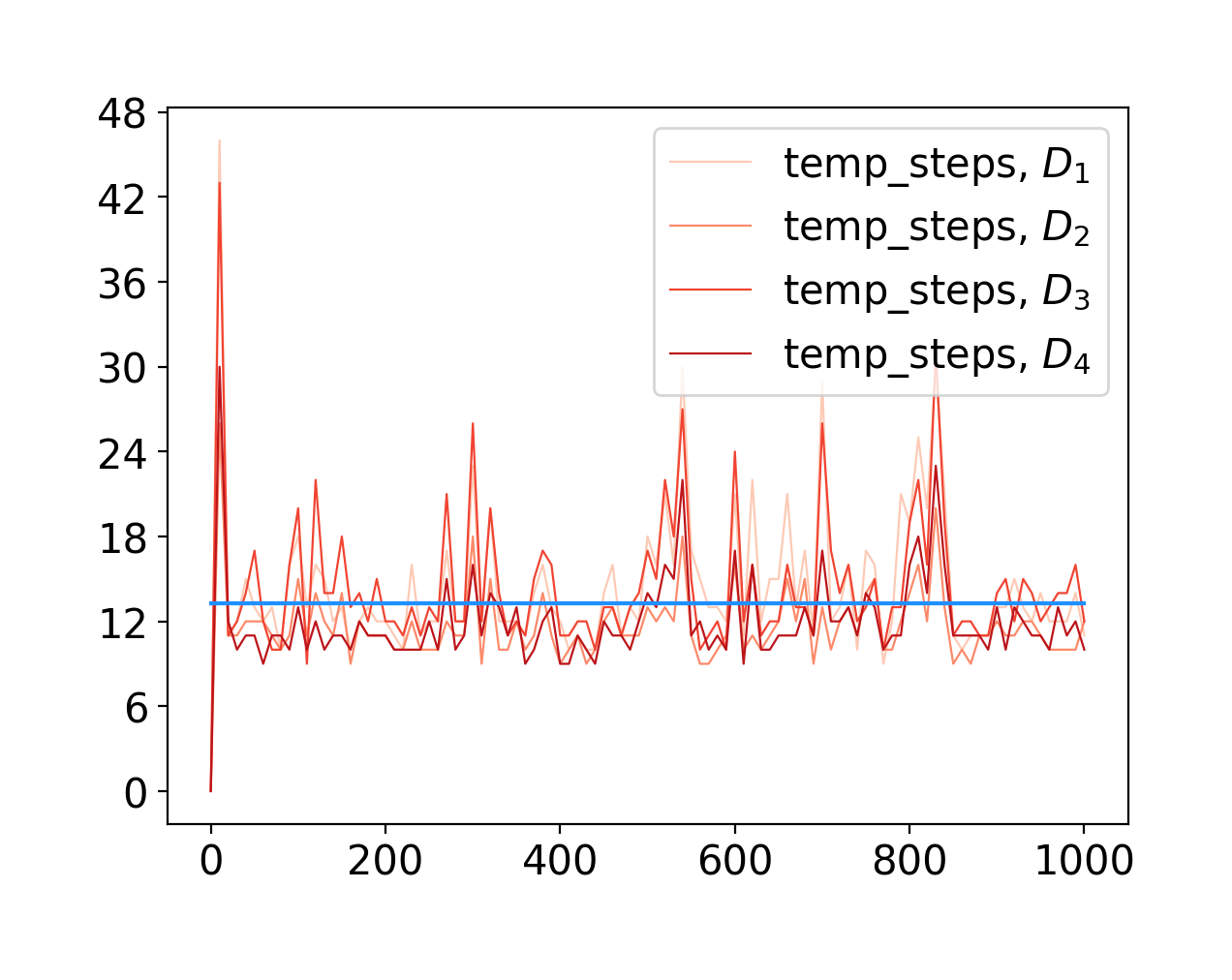}
	\caption{On the left, graph showing the number of tempering steps required for a PF with a fixed equi-spaced $32\times32$ observation grid. In red we plot the tempering steps at each assimilation time, and in blue their average. On the right, the number of tempering steps required in each subregion of a LPF with 4 localization region, for the same $32 \times 32$ grid of observations (in shades of red we plot the tempering steps for each subregion, and in blue their average). We see that the average number of tempering steps for the LPF is around 13, against around 44 steps for the PF.}
	\label{fig:tempering_steps}
\end{figure}

Finally, we iterate once more that localization gives a substantial speed-up to the particle filtering algorithm, not only through its parallelization capabilities, but also due to a significant reduction of the dimensionality of the filtering problem. This computational advantage is reinforced by the reduced tempering requirements in the localized approach: as can be seen in Figure~\ref{fig:tempering_steps}, on average, each region in the LPF requires fewer tempering steps to achieve effective particle mixing compared to the global tempering process in the standard PF. The lower-dimensional state space within each localized region enables more efficient exploration of the posterior distribution, reducing the number of intermediate needed in the tempering sequence. Consequently, while the PF must perform extensive tempering across the full high-dimensional space, the LPF achieves comparable sampling quality with reduced computational overhead per region, amplifying the overall speed advantage when combined with parallel processing across multiple regions.

\subsection{The LPF with moving strip-like observations}
We now test the particle filter in the case of \textit{moving strip-like observations} of type (b). Note that in this setting, where the observation locations vary periodically in time, $h_{\mathcal{J}}$ in \eqref{eq:obs_numerics_vec} must be time-dependent. We present results for both the sub-mesoscale dynamics \textit{S} and the mesoscale dynamics \textit{M}.

We fix the scale of the observation noise $\mathbf{\boldsymbol{\sigma} = 0.05}$, set $\mathbf{N_{\mathrm{loc}} = 4}$ and $\mathbf{N = 50}$. For each of the two regimes, we run a filtering experiment in which a vertical strip of observations (of width 2 grid-points) is observed at a different position every 10 time-steps (so the observations frequency is $\mathbf{r = 10}$): the position of the strip cycles through the $x$-coordinates $[10, 90, 40, 120, 70, 20, 100, 50]$ (as grid-points). To evaluate the performance of the LPF with this type of observations, we compare it with two LPF run with a fixed $32\times32$ grid of observations, respectively with frequency $\mathbf{r=10}$ (experiment \textit{A}) and $\mathbf{r=100}$ (experiment \textit{B}). We expect the performance of the LPF with strip-like observations to fall roughly between the two: observations are as frequent as in experiment \textit{A}, but we only observe enough of the space (a whole cycle in the strip-positions) roughly every 80 time-steps, which should bring us closer in performance to experiment \textit{B}.

\begin{figure}[ht]
	\centering
	\includegraphics[width=0.325\linewidth]{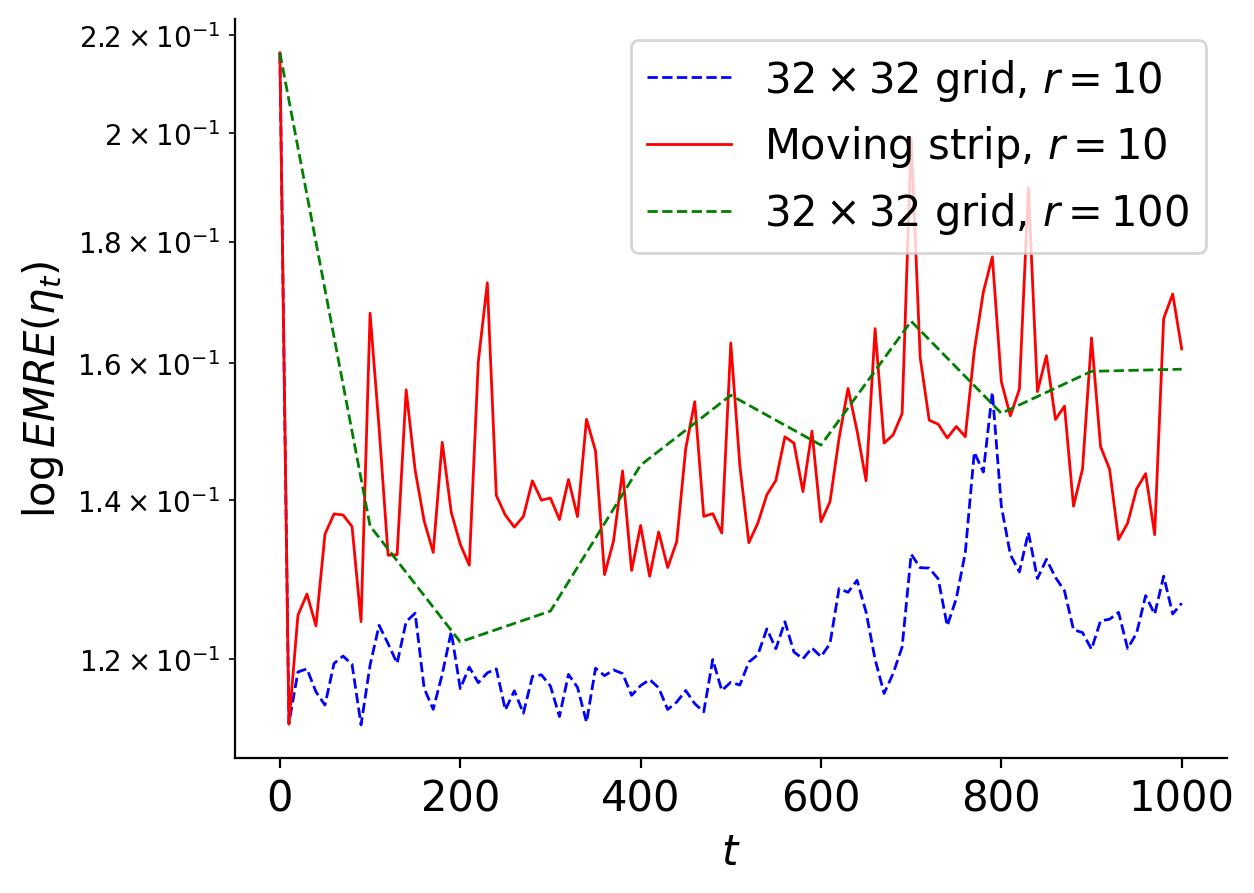}
	\includegraphics[width=0.325\linewidth]{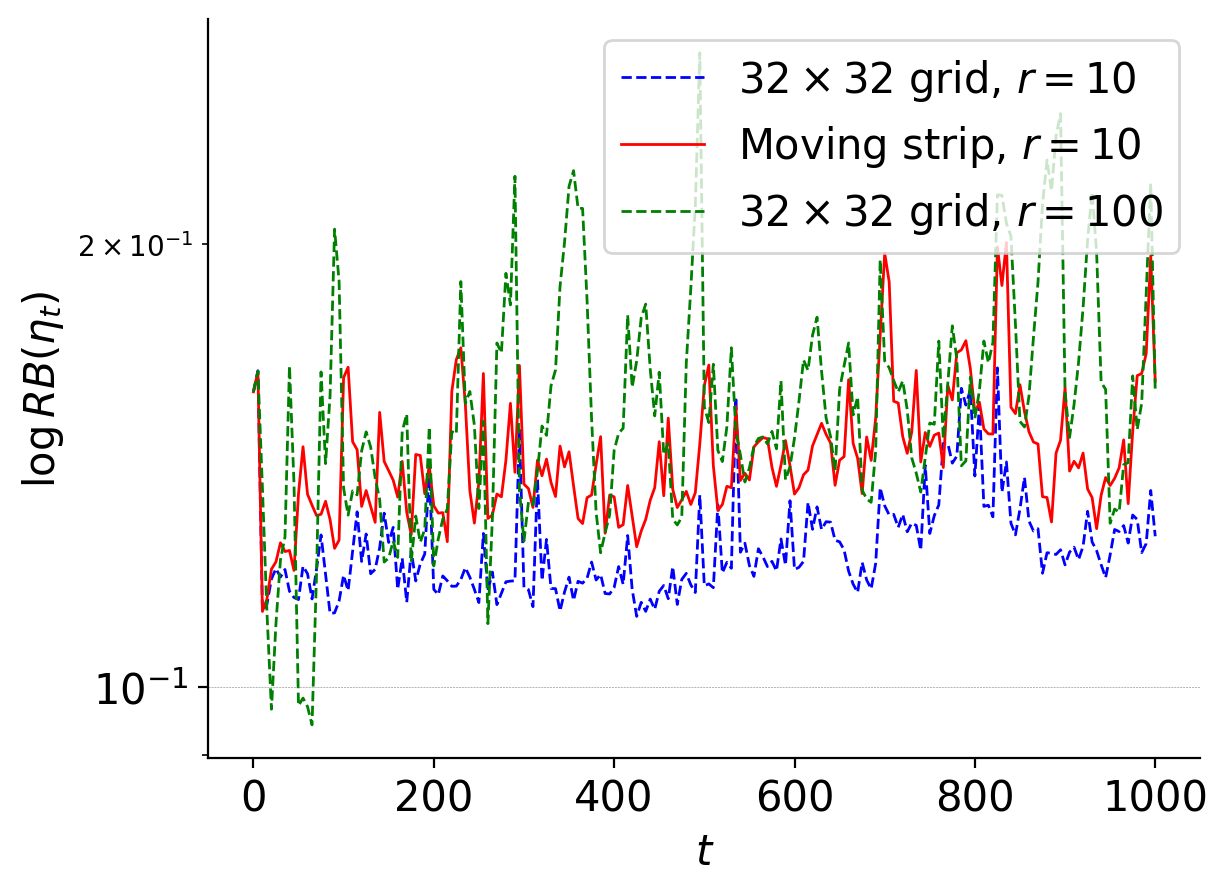}
	\includegraphics[width=0.325\linewidth]{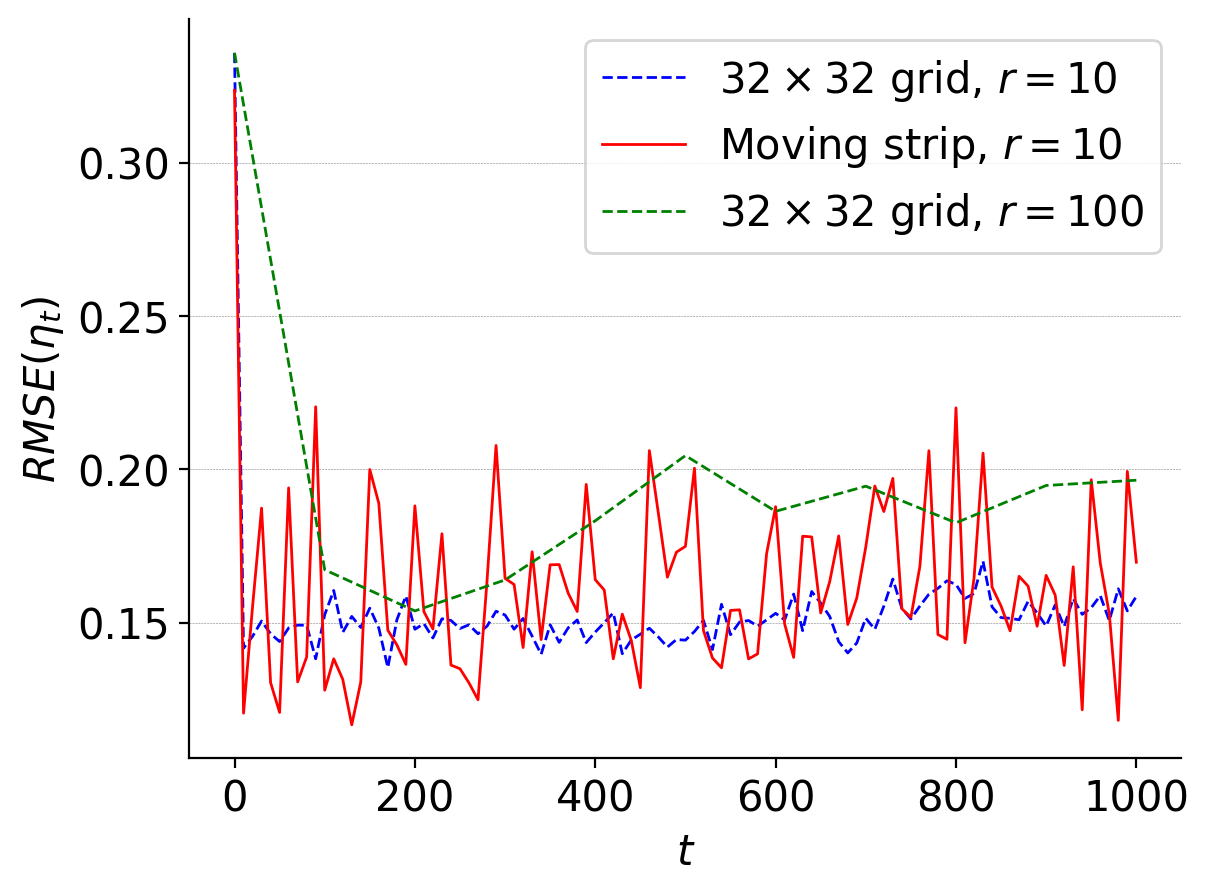} 
	\caption{Comparison between LPFs run with different types of observations (for the \textit{S} dynamics). In solid red we plot the error of a LPF with moving strip-like observations that change locations every 10 time-steps, in dotted green those of a LPF with a fixed $32 \times 32$ grid of observations observed every 10 time-steps (experiment \emph{A}), and in dotted blue those for a LPF with the same observation grid, but observed every 100 time-steps (experiment \emph{B}). The left plot shows $EMRE(\eta)$ (computed at assimilation times) in log-scale, the center plot $RB(\eta)$ (computed every 10 time-steps regardless of observations frequency) in log-scale, and the right plot $RMSE(\eta_t)$ (computed at assimilation times). The errors for the filtering experiment with moving strip observations are roughly in-between the errors for experiment \emph{A} and those for experiment \emph{B}.}
	\label{fig:strip_subM}
\end{figure}

\begin{figure}[ht]
	\centering
	\includegraphics[width=0.325\linewidth]{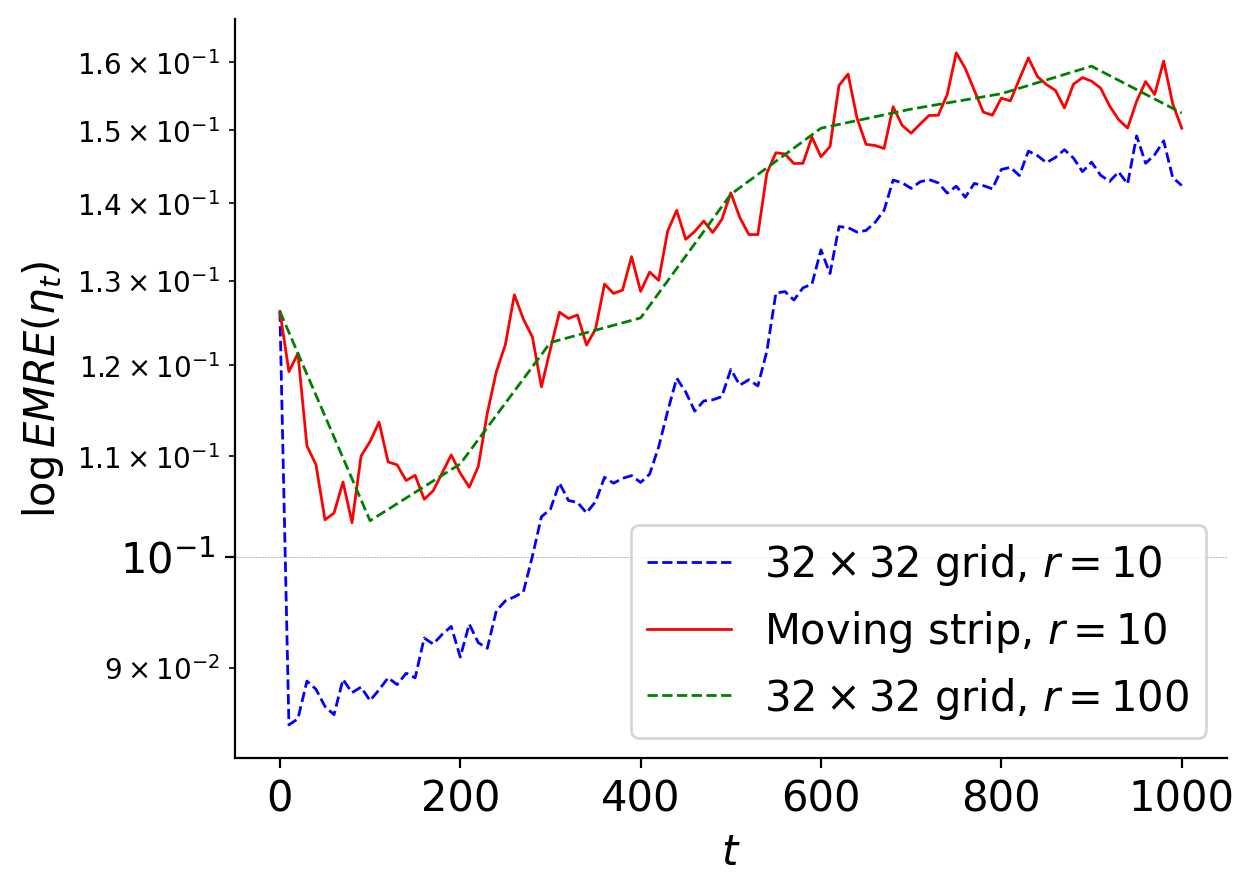}
	\includegraphics[width=0.325\linewidth]{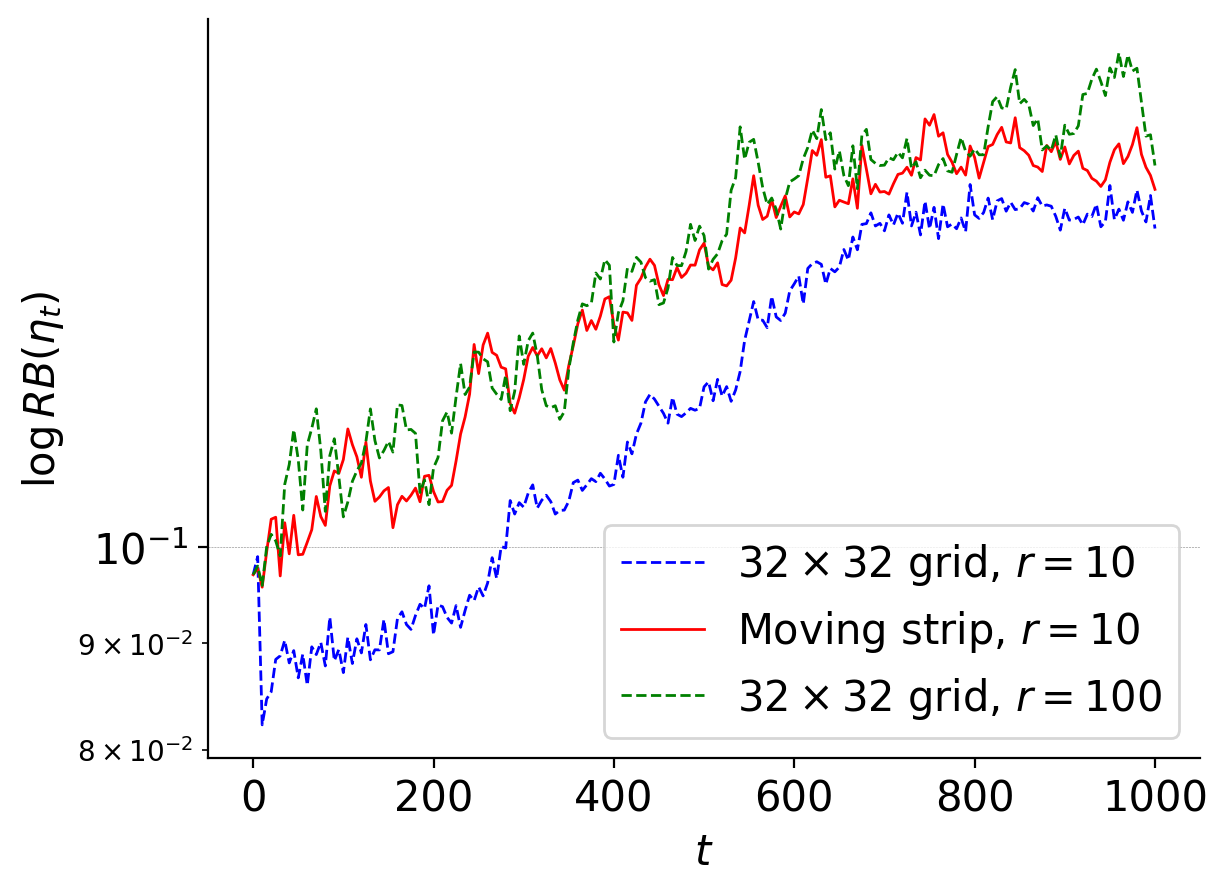}
	\includegraphics[width=0.325\linewidth]{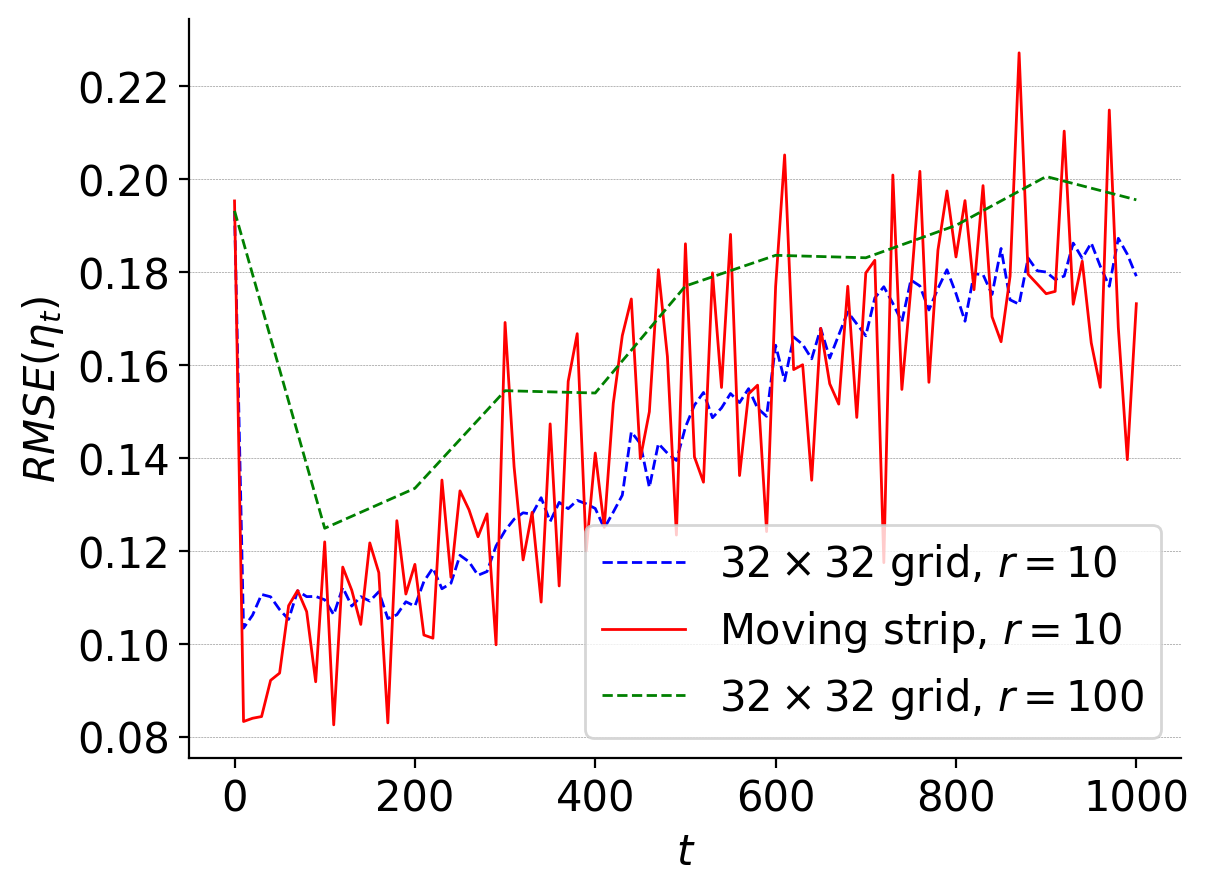} 
	\caption{Same as Figure~\ref{fig:strip_subM}, but for the $M$ dynamics} 
	\label{fig:strip_M}
\end{figure}

The results for the \textit{S}-regime are plotted in Figure~\ref{fig:strip_subM}, and for the \textit{M}-regime in Figure~\ref{fig:strip_M}. We see that $EMRE(\eta_t)$, $RB(\eta_t)$ and $RMSE(\eta_t)$ behave as we expected, with the errors for the strip-like observations oscillating more (due to the intermittent nature of the observations), but staying roughly between the results for experiment \textit{A} and \textit{B}. We also note that the errors tend to stabilize around the same level for the two regimes, $\approx0.16$ for $EMRE(\eta_t)$ and $RB(\eta_t)$ and $\approx0.18$ for $RMSE(\eta_t)$.

\section{Conclusion}

In this work we presented a localized particle filter (LPF) incorporating localized tempering and jittering procedures designed to address the computational challenges of high-dimensional nonlinear data assimilation. We compared its performance with that of a particle filter (PF) with tempering and jittering to evaluate both accuracy and computational efficiency in realistic geophysical scenarios.

We tested both algorithms on filtering experiments for a stochastic rotating shallow water model, \textit{observing only the sea surface height} while \textit{leaving the velocity fields unobserved}. This experimental design better mimics satellite altimetric data, such as that from the SWOT (Surface Water and Ocean Topography) mission, where direct velocity measurements are unavailable and must be inferred from surface height observations. The filtering problem was formulated as estimating the full state of the shallow water system from these indirect observations, representing a challenging test case for both algorithms.

Our findings demonstrate that the localized particle filter maintains stability throughout the assimilation process and produces results comparable in accuracy to the standard particle filter. The LPF successfully captures the essential dynamics of the rotating shallow water system while effectively handling the observational constraints. Importantly, the localized approach achieves this performance with the potential for significant computational speedup due to two key advantages: the inherent parallelization capabilities across localized regions, and the dimensionality reduction of the state-space data assimilation problem within each localization region, as demonstrated by the reduced tempering requirements in the lower-dimensional local spaces.

These results suggest that the localized particle filter represents a promising approach for operational data assimilation in high-dimensional geophysical systems, offering a practical pathway to scale particle filtering methods to realistic oceanographic and atmospheric applications. Future work will focus on extending the methodology to more complex models and exploring the integration of additional observational datasets to further validate the robustness of the approach and its computational advantages.

\paragraph{Acknowledgements} This research has been supported by the European Research Council (ERC) under the European Union’s Horizon 2020 Research and Innovation Programme (Sinergy grant STUOD – DLV-856408). EF would like to thank Darryl Holm and Etienne M\'emin for the many useful discussions about the stochastic RSW model used in this paper, and Alexander Lobbe for his valuable input regarding the numerics and the JAX Python library.

%%-------------------------------------------------------------------------------
\clearpage
\appendix
\section{Numerical simulations}\label{app:numerics_details}
In this appendix we provide some details about the numerical simulations for the results presented in Section~\ref{sec:numerics}. Let us start from the numerical scheme for simulating the solution of the 2-d SALT viscous Rotating Shallow Water equation \eqref{eq:SRSW} on $D = [0,1]^{\times 2}$. Our starting point is the code used in \cite{lobbe_RSW_23}, which is described in details in the appendix therein, and which the authors shared with us at the start of this research project. The chosen numerical solver for \eqref{eq:SRSW} is a Runge--Kutta scheme of order 4, which automatically takes care of the second-order correction term due to the conversion from Stratonovich to It\^o's integral. The spatial discretization is implemented through a staggered Arakawa C-grid (Arakawa~and~Lamb~\cite{arakawa77}) for numerical accuracy (see also \cite[Fig.~11]{lobbe_RSW_23}). Conceptually, we do not make any change to the numerical scheme described in \cite{lobbe_RSW_23}; we do however vary slightly the implementation. The two main changes in this sense are:
\begin{itemize}
	\item the placement of the ghost cells/points and the grid-size. Ghost cells are extra cells at the boundary of the grid which are used to implement the boundary conditions. In our implementation, we place ghost cells at each boundary of $D$, including the E (East: $x = 0$) boundary, while in \cite{lobbe_RSW_23} ghost cells are only at the W (West: $x=1$), S (South: $y=0$) and N (North: $y=1$) boundaries (again, see \cite[Fig.~11]{lobbe_RSW_23}). Since in \cite{lobbe_RSW_23} (and in this paper as well, see the next bullet point) the EW boundary conditions are taken to be periodic, ghost points to the east of the grid are not strictly necessary (once we have them at the west boundary, at least). However, for the sake of symmetry, which simplifies the implementation of the domain decomposition necessary for the localization algorithm, we prefer to add ghost points to the east of the grid as well. This makes the total dimension of the spatial grid $(d+2) \times (d+2)$, where $d = 1/\delta x$, instead of $(d+1) \times (d+1)$ as in \cite{lobbe_RSW_23}, and we need to adapt the numerical scheme appropriately. Recall that we take $d = 128$ in our simulations.
	\item the enforcing of the boundary conditions. As in \cite{lobbe_RSW_23}, we impose the following conditions at the E, W, S and N boundaries.
	\begin{itemize}
		\item EW boundary conditions: periodic, i.e.
		\begin{equation*}
			\eta(0, y) = \eta(1, y), \quad u(0, y) = u(1, y), \quad v(0, y) = v(1, y).
		\end{equation*}
		\item SN boundary conditions: no meridional velocity, so Dirichlet boundary conditions for $v$, and free-slip for the zonal velocity, so Neumann boundary conditions for the zonal flux $u \eta$, i.e.
		\begin{equation*}
			v(x,0) = v(x,1) = 0, \quad \frac{\partial (u \eta)}{\partial y}(x,0) = \frac{\partial (u \eta)}{\partial y}(x,1) =0.
		\end{equation*}
		To enforce the free-slip boundary conditions, we set Neumann boundary conditions for both $u$ and $\eta$, so that the $y$-derivative of the flux $u \eta$ is indeed 0, that is
		\begin{equation*}
			\frac{\partial \eta}{\partial y}(x,0) =	\frac{\partial \eta}{\partial y}(x,1) = 0, \quad
			\frac{\partial u}{\partial y}(x,0) =	\frac{\partial u}{\partial y}(x,1) = 0.
		\end{equation*}	
		We enforce these numerically in the Euler steps within the Runge--Kutta scheme.
	\end{itemize}
\end{itemize}

\subsection{Choice of parameters and noise coefficients}\label{app:params}
We consider two different settings for our simulations:
\begin{enumerate}
	\item Sub-mesoscale \textit{(S)} dynamics: we set the Rossby number $\mathrm{Ro} = 0.9$ and the Froude number $\mathrm{Fr} = 0.3$;
	\item Mesoscale \textit{(M)} dynamics: we set $\mathrm{Ro} = \mathrm{Fr}  = 0.05$.
\end{enumerate}
Throughout, we fix the Coriolis parameter $f$ to be constant $1$ for simplicity, and the viscosity coefficient $\nu = 10^{-5}$. Note that the \textit{M}-dynamics seem much faster than the \textit{S}-dynamics (when compared with the same $\delta t$ in this dimensionless setting), therefore we take $\delta t = \times 10^{-3}$ in the time-discretization for system \textit{S}, and $\delta t = 10^{-4}$ for system \textit{M}.

For the construction of the driving noise \eqref{eq:noise_SRSW}, we let
\begin{equation}\label{eq:transport_noise_form}
	\boldsymbol{\xi}_n(\mathbf{x})
	=  \left( \begin{matrix}
		\xi_n^u(x,y) \\
		\xi_n^v(x,y)
	\end{matrix} \right)
	= \frac{\sigma_{\mathrm{noise}} }{n^p } \left( \begin{matrix} \cos (2 \pi n  y) (
		\alpha_{n} \sin(2 \pi n  x) + \beta_{n} \cos(2 \pi n x) )\\
		\sin (2 \pi n y) (\gamma_{n} \sin(2 \pi n x) + \delta_{n} \cos(2 \pi n x) )
	\end{matrix} \right),
\end{equation}
for $n = 1, \dots, N_{\mathrm{noise}}$, where  $N_{\mathrm{noise}}$ is the number of noise components, $\sigma_{\mathrm{noise}}$ is a scaling factor, $p$ is a positive constant, and $\alpha_n, \beta_n, \gamma_n$ and $\delta_n$ are scalar coefficients. Note that $\xi_n^u$ and $\xi_n^v$ satisfy the boundary conditions for $u$ and $v$. The quadratic variation of the driving noise \eqref{eq:noise_SRSW} as an element of $L^2(D)$ with such a choice of spatial coefficients is given by
\begin{equation}\label{eq:quad_var_noise}
	\sum_{n = 1}^{N_{\mathrm{noise}} } \|  \boldsymbol{\xi}_n \|_{L^2}^2 t =
	\frac{\sigma_{\mathrm{noise}}^2}{4} \sum_{n=1}^{N_{\mathrm{noise}} } \frac{\alpha_n^2 + \beta_n^2 + \gamma_n^2 + \delta_n^2}{n^{2p}} t.
\end{equation} 
Given the typical scale of the velocity field in each of the sub-mesoscale and mesoscale regimes described above, we fix the parameters $\sigma_{\mathrm{noise}}, \alpha_n, \beta_n, \gamma_n$ and $\delta_n$ such that the magnitude of the noise is roughly $5-10 \%$ of the magnitude of the velocity. Specifically,
\begin{itemize}
	\item we fix $N_{\mathrm{noise}} = 25$ and $p = 2$, and for all $n$ we draw $\alpha_n, \beta_n. \gamma_n$ and $\delta_n$ independently from a Uniform distribution. Since $\mathbb{E}[U^2] = 1/3$ for $U\sim U(0,1)$, we can approximate $\tfrac{\alpha_n^2 + \beta_n^2 + \gamma_n^2 + \delta_n^2}{4} \approx 1/3$, so a rough estimate of the magnitude of the noise is given by
	\begin{equation*}
		\frac{ \sigma_{\mathrm{noise}}^2}{4} \sum_{n=1}^{N_{\mathrm{noise}} } \frac{\alpha_n^2 + \beta_n^2 + \gamma_n^2 + \delta_n^2}{n^{2p}} \approx \frac{ \sigma_{\mathrm{noise}}^2}{3} \sum_{n = 1}^{N_{\mathrm{noise}}} \frac{1}{n^{2p}};
	\end{equation*}
	\item to determine the typical magnitude of the velocity in the sub-mesoscale and mesoscale regimes, we run the \textit{deterministic RSW equations} in each of the regimes (starting from the initial conditions described in the next subsection) for a large number of time-steps. At each time, we compute the magnitude of the velocity as
	\begin{equation*}
		U_t = \sqrt{\| u_t \|_{L^2}^2 + \| v_t \|_{L^2}^2},
	\end{equation*}
	and then average over the number of time-steps to obtain $\bar{U}$;
	\item equating $10^{-1}\bar{U} \approx \sqrt{ \tfrac{ \sigma_{\mathrm{noise}}^2 }{3} \sum_{n = 1}^{N_{\mathrm{noise}}} \frac{1}{n^{2p}}}$, we get
	\begin{equation*}
		\sigma_{\mathrm{noise}} \approx \mathcal{O} \Big(10^{-1} \bar{U} \sqrt{3 }\big(\sum \tfrac{1}{n^{2p}}\big)^{-1/2}\Big).
	\end{equation*}
	Since for the \textit{S}-dynamics, $\bar{U} \approx 0.7$, and $\bar{U} \approx 1.5$ for the  \textit{M}-dynamics, we finally pick $\sigma_{\mathrm{noise}} \hspace{-2.5pt} =\hspace{-1pt}  0.1$.
\end{itemize}

\begin{rmk}\label{rmk:noise_magnitude}
	Note that at each time-step, the increment due to the (deterministic) velocity $\bar{U}$ is of order $\delta t$, while the one due to the stochastic velocity perturbation of order $\sqrt{\delta t}$. In view of this, the influence of the transport noise on the system dynamics is indeed high. 
\end{rmk}

\subsection{Initial conditions}\label{app:ic}

In this section we describe the procedure we followed to produce the initial conditions for the signal process and the particle ensemble. We start with some artificial initial conditions for $\eta$ by taking a combination of $\arctan$, $\sin$, $\cos$, and negative exponentials, given by
\begin{gather*}
	\bar{\eta}(x,y) = 1.5+ 0.1 \arctan(y - 0.5) - 0.05 \arctan(0.5 y) + 0.03 \arctan(0.9 x(1-x)) \\
	+ 0.2 \sin (2 \pi x) + 0.03 \sin(4 \pi x) \sin^4(y) - 0.05 e^{-0.5 x (1-x)} + 0.05 e^{-0.5 y} \\
	+ 0.1 \sin(2 \pi y) \cos(2 \pi x) + 0.3 \cos^2(6 \pi x).
\end{gather*}
Once we fix $\bar\eta$, we derive the initial condition for the velocity field using the geostrophic balance condition
\begin{equation}\label{eq:geostrophic_balance}
	\mathbf{\bar u}(x,y) = - \frac{1}{f} \frac{\mathrm{Ro}}{\mathrm{Fr}^2} \nabla^{\perp} \bar\eta_0(x,y),
\end{equation}
and scale the velocity components by $1/\max (\bar u)$ and $1/\max (\bar v)$ so they are roughly of order $\mathcal{O}(1)$.
Note that $\bar \eta$ satisfies the periodic EW boundary conditions for $\eta$, but not the Neumann NS boundary conditions. This is unimportant, however, since as soon as we start running the numerical scheme, the boundary conditions are enforced automatically.

Consider the timeline below:

\begin{tikzpicture}[line cap=round,line join=round,>=triangle 45,x=1cm,y=1cm]
	\clip(2.3,6.5) rectangle (16.4,9.3);
	\draw [->,line width=1.5pt] (3,8) -- (16,8);
	\draw [line width=1.5pt] (3,8.1)-- (3,7.9);
	\draw [line width=1.5pt] (9,8.1)-- (9,7.9);
	\draw (8.76,7.9) node[anchor=north west] {$0$};
	\draw (15.3,7.9) node[anchor=north west] {$t$};
	\draw (2.6,7.9) node[anchor=north west] {$-T_1$};
	\draw [line width=1.5pt] (6,8.1)-- (6,7.9);
	\draw (5.6,7.9) node[anchor=north west] {$-T_0$};
	\draw [line width=1pt] (3.03,8.25)-- (3.03,8.4);
	\draw [line width=1pt] (5.97,8.25)-- (5.97,8.4);
	\draw [line width=1pt] (6.03,8.25)-- (6.03,8.4);
	\draw [line width=1pt] (8.97,8.25)-- (8.97,8.4);
	\draw [line width=1pt] (3.03,8.4)-- (5.97,8.4);
	\draw [line width=1pt] (6.03,8.4)-- (8.97,8.4);
	\draw (3.5,8.9) node[anchor=north west] {$\textit{pde burn-in}$};
	\draw (6.3,8.9) node[anchor=north west] {$\textit{spde burn-in}$};
	\draw [line width=1.5pt] (11,8.1)-- (11,7.9);
	\draw [line width=1.5pt] (13,8.1)-- (13,7.9);
	\draw [line width=1pt] (9.03,8.25)-- (9.03,8.4);
	\draw [line width=1pt] (11.03,8.25)-- (11.03,8.4);
	\draw (10.8,7.9) node[anchor=north west] {$t_1$};
	\draw (12.8,7.9) node[anchor=north west] {$t_2$};
	\draw [line width=1pt] (10.97,8.25)-- (10.97,8.4);
	\draw [line width=1pt] (9.03,8.4)-- (10.97,8.4);
	\draw [line width=1.5pt] (9.4,8.1)-- (9.4,7.9);
	\draw (9.13,7.9) node[anchor=north west] {$\delta t$};
	\draw [line width=1pt] (12.97,8.25)-- (12.97,8.4);
	\draw [line width=1pt] (12.97,8.4)-- (11.03,8.4);
	\draw (11.63,8.9) node[anchor=north west] {$r\delta t$};
	\draw (9.53,8.9) node[anchor=north west] {$r\delta t$};
	%	\draw [line width=1.5pt] (9.8,8.1)-- (9.8,7.9);
	\draw [line width=1pt] (3.03,7.25)-- (3.03,7.4);
	\draw [line width=1pt] (8.97,7.25)-- (8.97,7.4);
	\draw [line width=1pt] (3.03,7.25)-- (8.97,7.25);
	\draw [line width=1pt] (9.03,7.25)-- (9.03,7.4);
	\draw [line width=1pt] (9.03,7.25)-- (14.95,7.25);
	\draw [line width=1pt] (15.05,7.25)-- (15.15,7.25);
	\draw [line width=1pt] (15.25,7.25)-- (15.35,7.25);
	\draw [line width=1pt] (15.45,7.25)-- (15.55,7.25);
	\draw (6,6.7) node[anchor=south] {$\textit{set-up}$};
	\draw (12.3,6.7) node[anchor=south] {$\textit{filtering experiment}$};
\end{tikzpicture}

Let $t=0$ be the start of our filtering experiment. Consider negative times $-T_1 < -T_0 < 0$. At time $t = -T_1$ we initialize the \textit{deterministic RSW equations} with $\eta_{-T_1} = \bar \eta$ and $\mathbf{u}_{-T_1} = \mathbf{\bar u}$ as above, and run them forward up until $t = -T_0$, when the fields have evolved sufficiently for the artificial structure of the initial data to be no longer noticeable. We stop the PDE simulations and take the resulting $ \eta_{-T_0}$ and $\mathbf{ u}_{-T_0}$ (numerical) fields as (deterministic) initial conditions for the SPDE \eqref{eq:SRSW} at time $t = -T_0$.

\begin{figure}[H]
	\centering
	\includegraphics[width=0.325\linewidth]{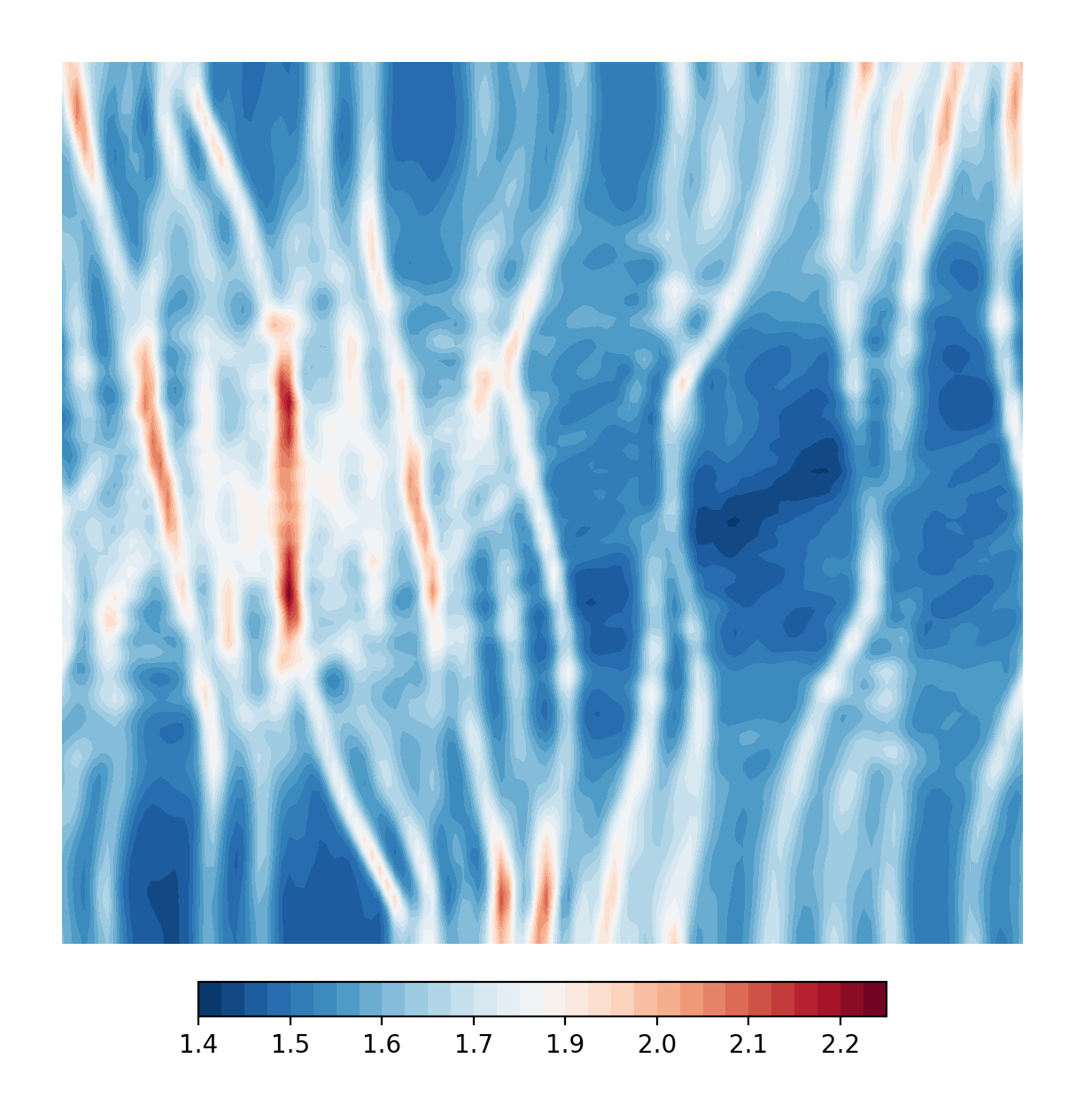}
	\includegraphics[width=0.325\linewidth]{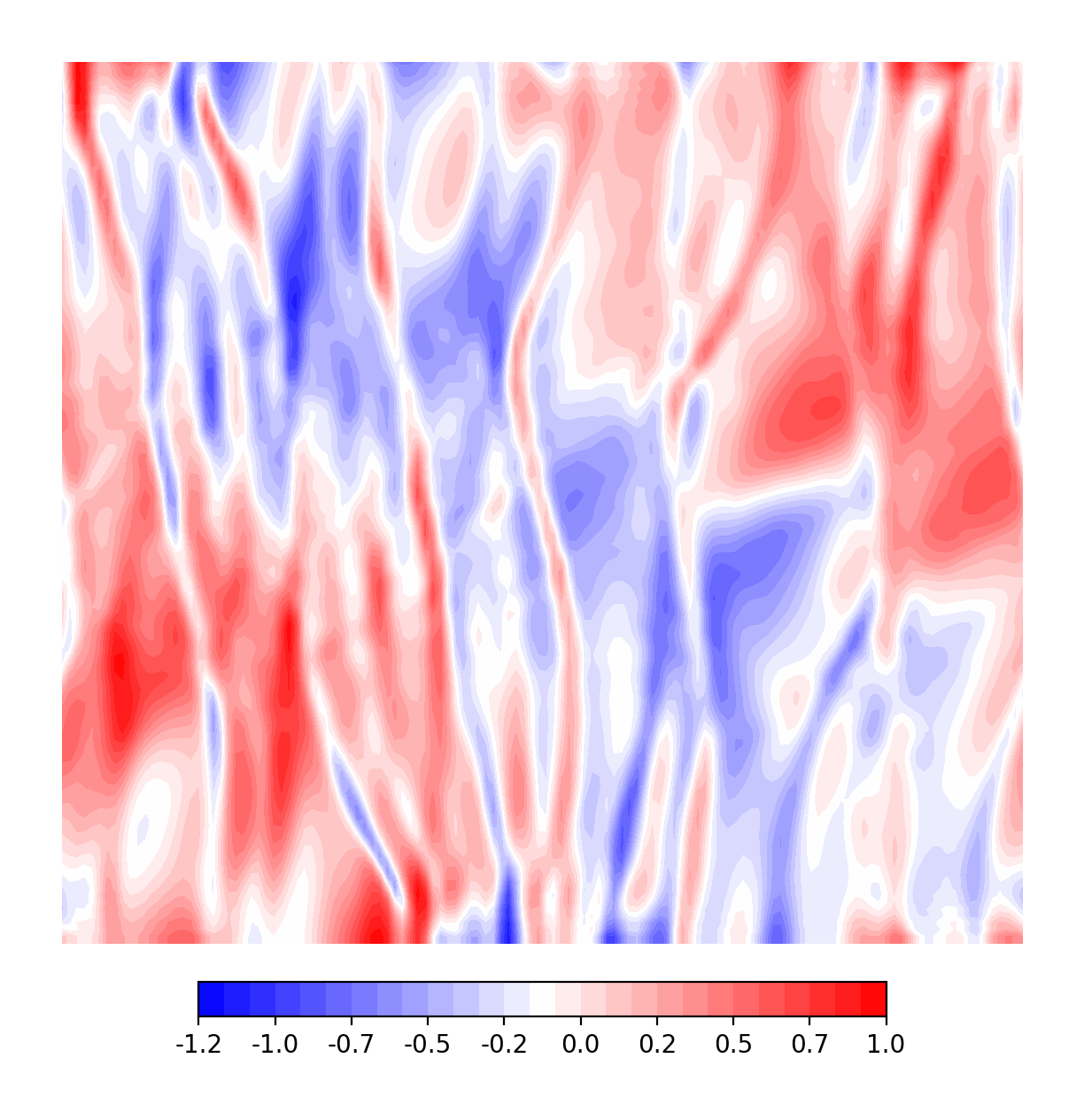}
	\includegraphics[width=0.325\linewidth]{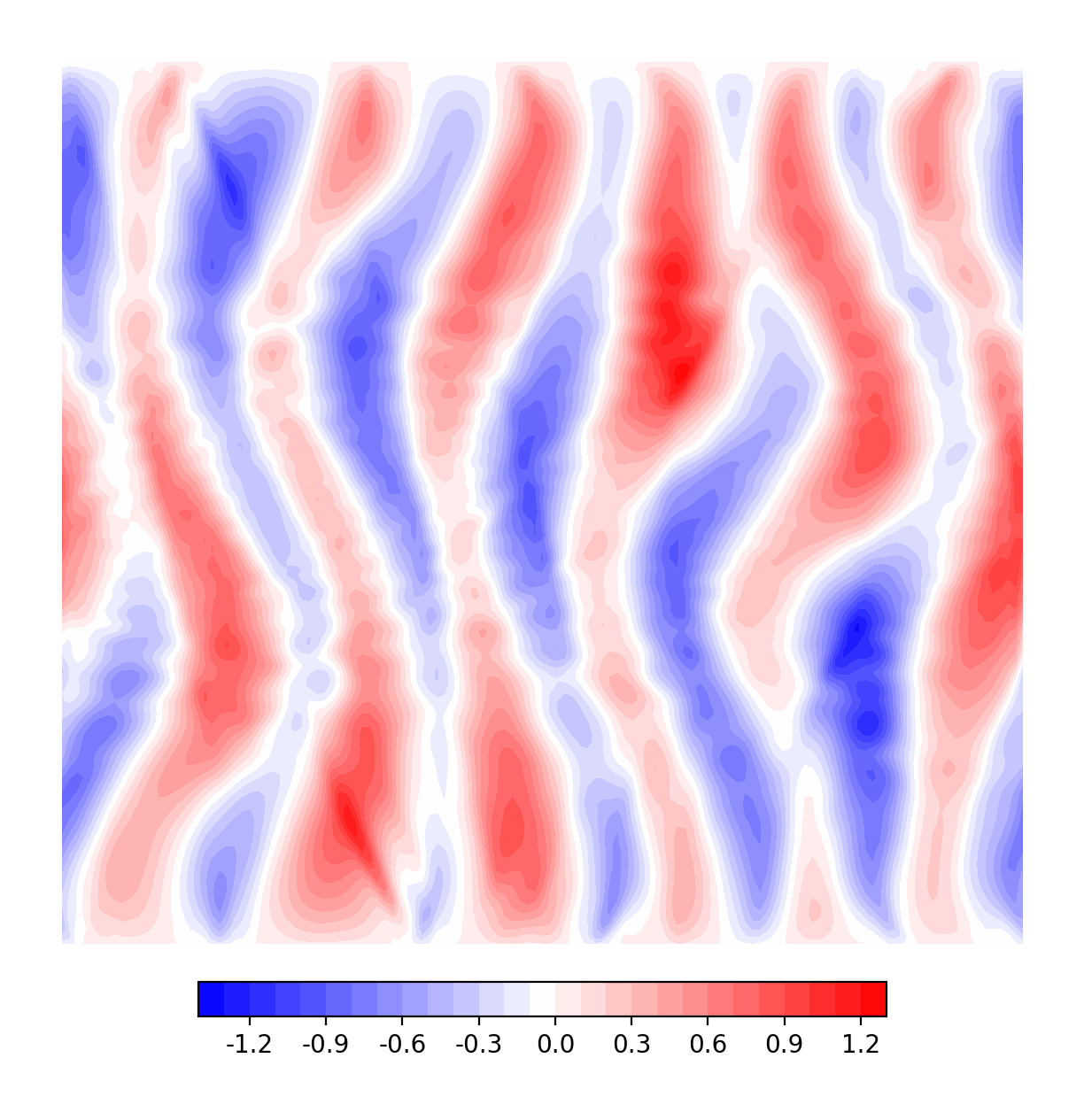}
	\caption{Plots of the fields $\eta_{-T_0}$ (left), $u_{-T_0}$ (centre) and $v_{-T_0}$ (right) at the final time-step $t = -T_0$ of a deterministic run of the RSW equations in regime \textit{S}. We use these to initialize the stochastic runs of \eqref{eq:SRSW} which yield the initial conditions at time $t=0$ for the signal process and the particle ensemble for the numerical experiments in regime \textit{S} in Section~\ref{sec:numerics}.}
	\label{fig:ic_regime_subM}
\end{figure}

\begin{figure}[H]
	\centering
	\includegraphics[width=0.325\linewidth]{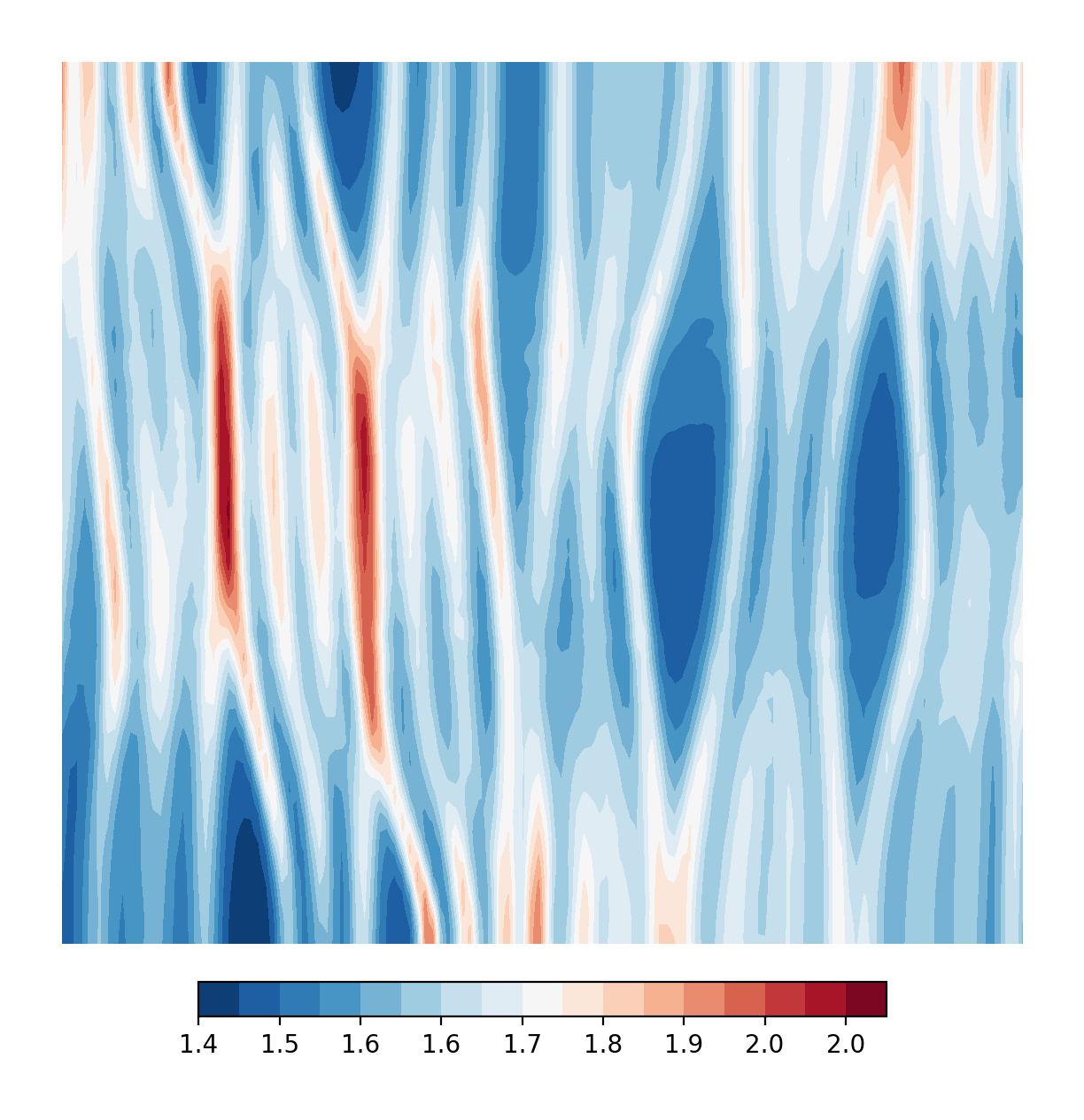}
	\includegraphics[width=0.325\linewidth]{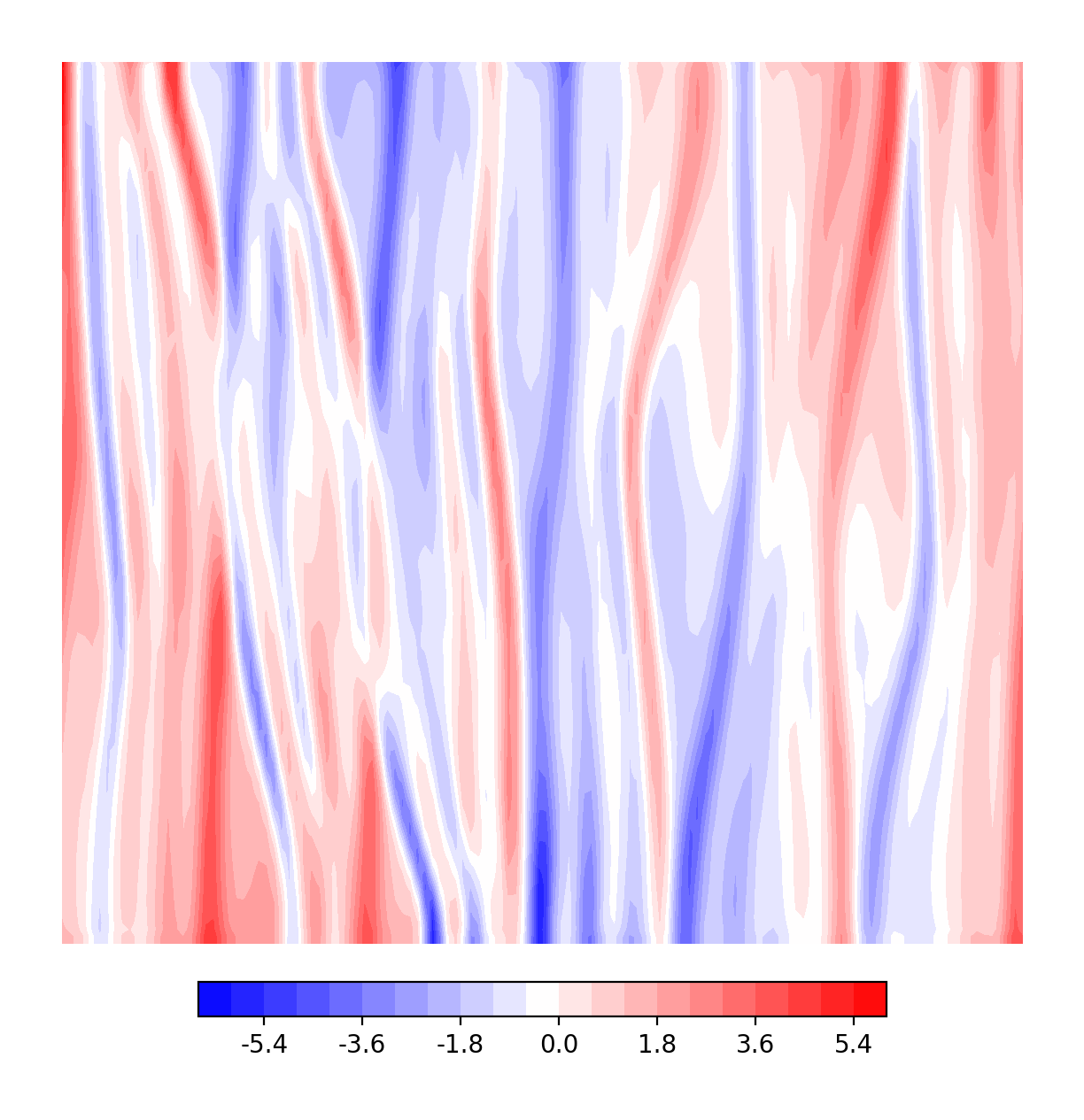}
	\includegraphics[width=0.325\linewidth]{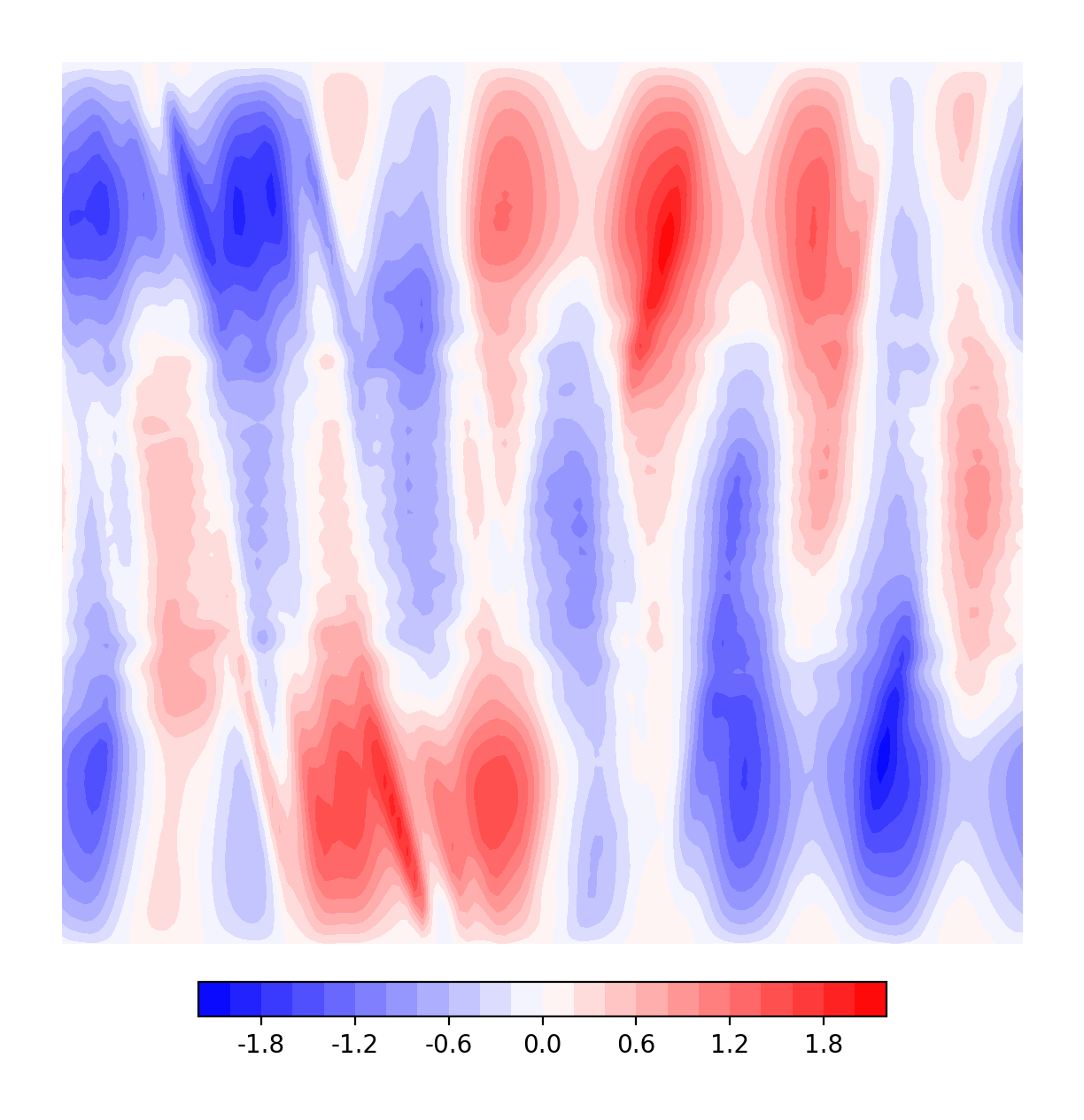}
	\caption{Plots of the fields $\eta_{-T_0}$ (left), $u_{-T_0}$ (centre) and $v_{-T_0}$ (right) at the final time-step $t = -T_0$ of a deterministic run of the RSW equations in regime \textit{M}. We use these to initialize the stochastic runs of \eqref{eq:SRSW} which yield the initial conditions at time $t=0$ for the signal process and the particle ensemble for the numerical experiments in regime \textit{M} in Section~\ref{sec:numerics}.}
	\label{fig:ic_regime_M}
\end{figure}

In Figure~\ref{fig:ic_regime_subM} and Figure~\ref{fig:ic_regime_M} we plot the deterministic solutions $ \eta_{-T_0}$ and $\mathbf{ u}_{-T_0}$ of the RSW PDE that we take as initial conditions for the stochastic runs, respectively for regime \textit{S} and regime \textit{M}. Finally, to ensure a varied initial spread for the ensemble of the particle filter, we proceed as follows: starting from the same i.c. $\eta_{-T_0}$ and $\mathbf{ u}_{-T_0}$, we produce $N+1$ realizations of the SPDE \eqref{eq:SRSW}, running our numerical scheme forward until $t=0$. Of these $N+1$ realizations, we select one at random to be the initial conditions for the signal, and use the others to initialize the $N$ particles of the ensemble. Note that by proceeding in this manner, we have that the signal is not the average of the ensemble, i.e. $\eta_0 \neq \tfrac{1}{N} \sum_i \eta_0^i$, and that $\eta_0$ and $\eta^i_0$ for $i = 1, \dots, N$ are all (independent and) distributed according to the same law.

\section{Pseudocodes for particle filtering algorithms}\label{app:pseudo-codes}
This appendix contains pseudocodes for the particle filtering algorithms (localized and not-localized) implemented for this work.

\subsection{Bootstrap, tempering and jittering}\label{app:basic_filt}

This section consists of a brief, self-contained introduction to the procedures of resampling, jittering and tempering, which have been developed to increase the stability of the particle filter. Note that these procedures do not guarantee non-degeneracy of the PF algorithm, but can nonetheless be very effective, as we have seen in our numerical experiments in Section~\ref{sec:numerics}.

Recall the notation from Section~\ref{sec:PF}. A first step towards improving the degeneracy of the basic particle filter is the so-called \textit{bootstrap particle filter} (Gordon~et~al.~\cite{gordon_bootstrap_93}), or sequential importance resampling (SIR) algorithm. In this case, after computing the weights for the posterior, one resamples the particles such that they return to being equally weighted at $1/N$: in other words, particles with large weights are multiplied, while particles with small weights are discarded. From the point of view of genetic algorithms, the duplicated particles are \textit{parent particles}, and they each get assigned a number of discarded particles based on their weight (the bigger the weight, the higher the number of discarded particles assigned). The discarded particles are then \textit{child particles}, which are `reborn' as copies of their assigned parent. Note that there are many ways to resample. In our implementation we make use of stochastic universal sampling, which is computationally cheap and has been shown to introduce the least amount of sampling noise (Kitagawa~\cite{Kitagawa96}). In Algorithms~\ref{alg:bootstrap} and \ref{alg:univ_resampling} we provide pseudocodes for the bootstrap particle filter and the resampling algorithm. Finally, we should remark that the bootstrap filter has also been shown to converge to the true filter asymptotically (see \cite[Ch.~10]{bain_crisan} and reference therein).

\begin{algorithm}[ht]
	\caption{Bootstrap particle filter}\label{alg:bootstrap}
	\begin{algorithmic}[1]
		\State {Assume assimilation times at $t_1, t_2, \dots, t_M$}
		\State {Initialize particle ensemble $\{u^i_0\}_{i=1}^N$ by drawing i.i.d. samples from $\pi_0 = \mathrm{law}(u_0)$}
		\State {Initialize weights $w^i_0 = 1/N$ for all $i = 1, \dots, N$}
		\For{$k = 1,2,  \dots, M$}
		\State {Evolve particles forward by solving the SPDE \eqref{eq:SPDE_general} for $t \in [t_{k-1}, t_{k}]$}
		\State \parbox[t]{408pt}{
			Gather observations $Y_k$, compute the weights $w^i_k = g(Y_k - h_k(u^i_{t_k}))$ for all $i = 1, \dots, N$, and normalize by dividing them by $\mathbf{c} = \sum_{i = 1}^N w^i_k $ \strut}
		\State {Resample according to Algorithm~\ref{alg:univ_resampling} }
		\State {Reset all weights to $1/N$}
		\EndFor
	\end{algorithmic}
\end{algorithm}

\begin{algorithm}[ht]
	\caption{Stochastic universal sampling}\label{alg:univ_resampling}
	\begin{algorithmic}[1]
		\Require {$t = t_k$}
		\State {Partition $[0,1]$ into $N$ intervals of length $w^i_k$; each interval represents the `bin' of the corresponding particle $i$}
		\State {Draw a uniform r.v. $U_0 \in [0, 1/N]$; consider the points $U_r = U_0 + r/N$ for $r = 0, \dots, N-1$ }
		\For{i = 1, \dots, N}
		\State {Count how many $U_r$ fall into the bin of particle $i$; let the number of $U_r$ in bin $i$ be $c_i$}
		\If {$c_i = 0$}
		\State {Discard particle $i$}
		\Else
		\State {Make $c_i-1$ copies of particle $i$}
		\EndIf
		\EndFor
	\end{algorithmic}
\end{algorithm}

The bootstrap particle filter is again unable to to avoid degeneracy of the ensemble when the dimension of the state-space is high. This can be partly due to the lack of diversity in the ensemble after the resampling step: as we mentioned before, weight-collapse in high-dimensional problems generally causes all particles but one to be discarded. The surviving particle, of which we have N copies after resampling, is the one with the best weight (relative to the other particles), but it might have lost track of the signal nonetheless. In such situations it is unlikely that the particle filter will recover, since the ensemble will have limited forecast capabilities, due to being concentrated at a single point.

One way to reintroduce some diversity in the ensemble is to perturb the particles slightly after resampling: this process is called \textit{rejuvenation}, or \textit{jittering}. A common type of jittering is to use a Markov Chain Monte Carlo (MCMC) method with the posterior as the target distribution. In the context of geophysical data assimilation, this form of jittering was implemented for example in \cite{kantas_14, cotter_et_al_PF_2D_euler_20}. In Algorithm~\ref{alg:MCMC_jittering} we present a pseudocode for a similar MCMC jittering algorithm adapted to our setting.

\begin{algorithm}[h]
	\caption{MCMC jittering}\label{alg:MCMC_jittering}
	\begin{algorithmic}[1]
		\Require {$t = t_k$}
		\Require {Parameter $\rho \in (0,1)$ and $M_{\mathrm{jit}} \in \N$}
		\State {After resampling, identify duplicated particles as \textit{parents} and assign them discarded particles as \textit{children}}
		\For {each parent particle $p$}
		\For {each child particle $c$ of $p$}
		\State $r \gets 0$
		\While {$r \le M_{\mathrm{jit}}$ }
		\State {Set $u^{c,r}_{t_{k-1}} \gets u^p_{t_{k-1}}$}
		\State \parbox[t]{376pt}{
			Set the noise $W^r = \rho W^p + \sqrt{1 - \rho^2} V^r$, where $W^p$ is the driving noise of particle $p$ for $t \in [t_{k-1}, t_k]$ and $V^r$ is an independent cylindrical BM \strut}
		\State {For $t \in [t_{k-1}, t_{k}]$, evolve particle $c$ forward by solving \eqref{eq:SPDE_general} with noise $W^r$ }
		\State Obtain proposal particle $u^{c,r}_{t_{k}}$ and compute its weight $w^{c,r}_k$
		\State Draw uniform r.v. $U \in (0,1)$
		\If {$U \le \frac{w^{c,r}_k}{w^{p}_k}$}  $u^c_{t_k} \gets u^{c,r}_{t_k}$
		\BREAK
		\Else
		\State $r \gets r+1$
		\If {$r = M_{\mathrm{jit}}$} $u^c_{t_k} \gets u^{p}_{t_k}$
		\EndIf
		\EndIf
		\EndWhile
		\EndFor
		\EndFor
	\end{algorithmic}
\end{algorithm}

At this point we should make a remark. An MCMC jittering procedure maintains invariant the posterior distribution, so it is clear that, at least in theory, it should not impact the convergence of the particle filter. In practice, however, there are a few drawbacks.

First of all, the parameters appearing in Algorithm~\ref{alg:MCMC_jittering} require careful tuning. If $\rho$ is chosen too large, for example, children particles will be heavily correlated to the parent particle and the spread is not going to be improved much by the jittering procedure. Conversely, if $\rho$ is chosen too small, the jittering algorithm might take a lot of iterations before a proposed particle is accepted. The parameter $M_{\mathrm{jit}}$ plays the role of a `safety stop': when the numbers of iterations in the MCMC algorithm becomes larger than $M_{\mathrm{jit}}$, we stop proposing new particles and simply assign the value of the parent particle to the child particle (essentially duplicating the parent particle, in line with the bootstrap filter). On a similar note, not only the number of iterations needed before a particle is accepted can be very large, but also each iteration in the MCMC algorithm is computationally expensive, since it involves solving (numerically) \eqref{eq:SPDE_general} repeatedly.

Secondly, the estimate of the posterior probability given by the particle filter can be quite a poor approximation of the true filtering measure if, for example, we do not have a very large number of particles. This begs the question: is it worth it to implement such an expensive algorithm to preserve the posterior distribution, when the posterior is anyway just an approximation? In \cite{crisan_miguez_18_nested} the authors suggest a jittering algorithm which consists simply in perturbing the particles using a kernel with small variance, and prove that in fact this type of jittering does not affect the asymptotic convergence of the particle filter. The variance of the jittering kernel is taken to be proportional to $1/N^2$ (to preserve $L^2$ convergence, otherwise  $1/N^{\tfrac{p+2}{p}}$ for $L^p$ convergence). We describe a jittering algorithm by perturbation that can be used in our setting in Algorithm~\ref{alg:perturbation_jittering}. We call this \textit{roughening}, in the spirit of \cite{gordon_bootstrap_93}. Note that since our model is given by \eqref{eq:SPDE_general}, it makes sense to choose a perturbation which is smooth in space. Given a set (or a basis) of smooth functions $\{f_n\}$ on $D$, we take the jittering perturbation to be given by spatial Gaussian noise of the form
\begin{equation}\label{eq:pert_jittering_noise}
	Z^{jit} = \sigma^{jit} \sum_{n = 1}^{\infty} \lambda^{jit}_n f_n Z_n, \quad Z_{n} \iid N(0,1),
\end{equation}
where the coefficients $\{\lambda^{jit}_n\}$ are some parameters fixed in time, and $\sigma^{jit}$ is a scaling factor. In Appendix~\ref{app:jittering_comp} we compare the performance of particle filters using these two types of jittering.

\begin{rmk}
	Note that we only jitter discarded (or \textit{child}) particles, and not the whole ensemble, to save computational time.
\end{rmk}

\begin{algorithm}[h]
	\caption{Roughening}\label{alg:perturbation_jittering}
	\begin{algorithmic}[1]
		\Require {$t = t_k$}
		\Require {Parameters $\sigma^{jit}$, $\{\lambda_n^{jit}\} \in \R$ and basis $\{f_n\}$ on $D$}
		\State {After resampling, identify duplicated particles as \textit{parents} and assign them discarded particles as \textit{children}}
		\For {each parent particle $p$}
		\For {each child particle $c$ of $p$}
		\State Sample realization $Z^{jit, c}$ of the noise \eqref{eq:pert_jittering_noise}
		\State  $u^c_{t_k} \gets u^{p}_{t_k} + Z^{jit, c}$
		\EndFor
		\EndFor
	\end{algorithmic}
\end{algorithm}

The final modification to the basic filtering algorithm that we consider in this paper is \textit{tempering}. While jittering targets the degeneration of the particle sample after resampling, tempering is added before resampling and combined with it to assimilate the observations obtained at each time step gradually. In particular, tempering targets the weight-collapse issue of the particle filter, by introducing intermediate steps in the assimilation procedure so that the particle weights maintain relatively low variance (as opposed to one, or very few particles, being heavily favoured compared to the rest of the ensemble). We describe the tempering procedure in Algorithm~\ref{alg:tempering}. We will need a few definitions in what follows.
\begin{defn}[ESS]
	The \textit{Effective Sample Size (ESS)} of the particle ensemble $\{u^i_{t_k}\}$ with weights $\{w^i_k\}$ at time $t_k$ is given by
	\begin{equation}\label{eq:ESS}
		ESS(\{w^i_{k}\}) = \frac{1}{\sum_{i = 1}^N (w^i_k)^2}.
	\end{equation}
\end{defn}
The \textit{Effective Sample Size (ESS)} is a commonly used statistics in SMC to track the performance of the ensemble online (Kong~et~al.~\cite{kong94}, Straka~and~\v{S}imandl~\cite{straka06}).  Often, a threshold value $N_{\mathrm{ess}}$ is set so that tempering, resampling and jittering steps are only performed at assimilation times $t_k$ if the ESS is unsatisfactory, i.e. if $ESS(\{w^i_{k}\}) < N_{\mathrm{ess}}$.

\begin{defn}[Tempered weights]
	Given a \textit{temperature} $\varphi \in (0,1)$ and the weights $\{w^i_k\}$ of the particle ensemble $\{u^i_{t_k}\}$ at time $t_k$, we define the \textit{(normalized) tempered weights} as
	\begin{equation}\label{eq:tempered_weights}
		w^{i, \varphi}_k = \frac{(w^{i}_k)^{\varphi}}{\sum_j (w^{j}_k)^{\varphi}}, \quad \forall i = 1, \dots, N.
	\end{equation}
\end{defn}

In the tempering procedure, at each assimilation step we iteratively scale the weights $\{w^n_k\}$ by exponentiating them to a power $\varphi  \in (0,1)$, where $\varphi$ is called \textit{temperature}, we resample and jitter according to the tempered weights (which are now more uniform across the ensemble due to being greatly scaled down), and then repeat the process until the cumulative sum of the temperatures reaches $1$. The temperature $\varphi$ is chosen at each iteration so that the ESS computed with the tempered weights is approximately $N_{\mathrm{ess}}$. Intuitively, we are partitioning the interval $[0,1]$ into temperatures $0< \varphi_1 < \varphi_2 < \cdots < \varphi_r = 1$. Here, 1 represents the full impact of the observations at time $t_k$, and $0$ no impact. By iteratively exponentiating the weights to the power of $\varphi_{s}-\varphi_{s-1}$, we mitigate the deteriorating effect that a big discrepancy between prior and posterior would have on the ensemble. By the end of the tempering algorithm, all the information contained in the observations has been assimilated, but the assimilation has happened in multiple small steps instead of a single large one. 

\begin{algorithm}[h]
	\caption{Tempering}\label{alg:tempering}
	\begin{algorithmic}[1]
		\Require {$t = t_k$}
		\Require {Parameter $N_{\mathrm{ess}} \in (1, N)$}
		\Require {Weights $\{w^i_k\}$ }
		\State $\varphi \gets 1$
		\While {$ESS(\{w^{i, \varphi}_{k}\}) < N_{\mathrm{ess}}$}
		\State Find $\varphi' \in (1-\varphi, 1)$ such that $ESS\big( \big\{w^{i, \varphi' - (1 - \varphi)}_{k} \big\} \big) \approx N_{\mathrm{ess}}$ \label{line:bisec}
		\State Resample using Algorithm~\ref{alg:univ_resampling} according to the weights $w^{i, \varphi' - (1 - \varphi)}_k $
		\State Jitter according to Algorithm~\ref{alg:MCMC_jittering} or \ref{alg:perturbation_jittering}
		\State Update $\{w^i_k\}$ by computing the weights of the new ensemble
		\State $\varphi \gets 1 - \varphi'$
		\EndWhile
		\State Resample using Algorithm~\ref{alg:univ_resampling} according to the weights $w^{i, \varphi}_k $
		\State Jitter according to Algorithm~\ref{alg:MCMC_jittering} or \ref{alg:perturbation_jittering}
	\end{algorithmic}
\end{algorithm}

\begin{rmk}
	The step at line~\ref{line:bisec} of Algorithm~\ref{alg:tempering} can be performed by implementing a bisection search algorithm.
\end{rmk}

In Algorithm~\ref{alg:PF} we summarize the implementation of the particle filter with tempering and jittering. 

\begin{algorithm}[H]
	\caption{Particle filter (PF)}\label{alg:PF}
	\begin{algorithmic}[1]
		\State {Assume assimilation times at $t_1, t_2, \dots, t_M$}
		\State {Initialize particle ensemble $\{u^i_0\}_{i=1}^N$ by drawing i.i.d. samples from $\pi_0 = \mathrm{law}(u_0)$}
		\State {Initialize weights $w^i_0 = 1/N$ for all $i = 1, \dots, N$}
		\For{$k = 1,2,  \dots, M$}
		\State {Evolve particles forward by solving numerically the SPDE \eqref{eq:SPDE_general} for $t \in [t_{k-1}, t_{k}]$}
		\State \parbox[t]{408pt}{
			Gather observations $Y_k$ and compute the weights $w^i_k = g(Y_k - h_k(u^i_{t_k}))$ for all $i = 1, \dots, N$, and normalize by dividing them by $\mathbf{c} = \sum_{i = 1}^N w^i_k $ \strut}
		\If {$ESS(\{w^i_k\}) < N_{\mathrm{ess}}$}
		\State {Temper according to Algorithm~\ref{alg:tempering} }
		\State {Reset all weights to $1/N$}
		\EndIf
		\EndFor
	\end{algorithmic}
\end{algorithm}

\subsection{Localized algorithms}\label{app:localized_codes}
This section contains pseudocodes for the local particle filter (LPF) algorithm (see Algorithm~\ref{alg:loc_PF}); the procedure for merging local particle into global ones described in Section~\ref{sec:merging} is summarized in Algorithm~\ref{alg:merging}.

\begin{algorithm}[H]
	\caption{Local particle filter (LPF)}\label{alg:loc_PF}
	\begin{algorithmic}[1]
		\State {Assume assimilation times at $t_1, t_2, \dots, t_M$}
		\State {Initialize particle ensemble $\{u^i_0\}_{i=1}^N$ by drawing i.i.d. samples from $\pi_0 = \mathrm{law}(u_0)$}
		\State {Initialize weights $w^i_0 = 1/N$ for all $i = 1, \dots, N$}
		\State {Divide domain $D$ into $N_{\mathrm{loc}}$ subregions $D_j$, $j = 1, \dots, N_{\mathrm{loc}}$}
		\For{$k = 1,2,  \dots, M$}
		\State {Evolve particles forward by solving the SPDE \eqref{eq:SPDE_general} for $t \in [t_{k-1}, t_{k}]$}
		\State \parbox[t]{408pt}{
			Gather observations $Y_k$ and compute the normalized weights $w^i_k = g(Y_k - h_k(u^i_{t_k}))/ \mathbf{c}$ for all $i = 1, \dots, N$, where $\mathbf{c} = \sum_{i = 1}^N w^i_k $ \strut}
		\If {$ESS(\{w^i_k\}) < N_{\mathrm{ess}}$}
		\State Construct local ensembles $\{u_{t_k}^{i,(j)}\}_{i=1}^N = \{u_{t_k}^{i} |_{D_j}\}_{i=1}^N$ for $j = 1, \dots, N_{\mathrm{loc}}$ \label{line:local_ens_construction}
		\For{$j = 1, \dots, N_{\mathrm{loc}}$}
		\State \parbox[t]{375.5pt}{
			Compute local weights $w^{i, \alpha, (j)}_{k}$ for each particle $u^i_{t_k}$ using \eqref{eq:local_weights} and the global ensemble $\{u^i_{t_k}\}_i$
			\strut}
		\If {$ESS(\{ w^{i, \alpha, (j)}_{k} \}) < N_{\mathrm{ess}}$}
		\State \parbox[t]{358.5pt}{
			Temper according to Algorithm~\ref{alg:tempering}: at each iteration compute the local weights $w^{i,\alpha, (j)}_k$ according to \eqref{eq:local_weights} using the updated local particles $u^{i,(j)}_{t_k}$ for region $D_j$, and treating the local particles $u^{i,(l)}_{t_k}$ in $D_l \neq D_j$ as fixed
			\strut} \label{line:tempering}
		\EndIf
		\EndFor
		\State \parbox[t]{391.5pt}{
			Merge local particles back together to recover the global ensemble $\{u^i_{t_k}\}_{i=1}^N$ according to Algorithm~\ref{alg:merging}
			\strut}
		\State {Reset all weights to $1/N$}
		\EndIf
		\EndFor
	\end{algorithmic}
\end{algorithm}

\begin{algorithm}[H]
	\caption{Reconstruction of global particles}\label{alg:merging}
	\begin{algorithmic}[1]
		\Require {$t = t_k$}
		\Require Assimilated local ensembles $\{u^{i, (j)}_{t_k}\}_{i=1}^N$ in each subregion $D_j$ for $j = 1, \dots, N_{\mathrm{loc}}$.
		\For{$i = 1,2,  \dots, N$}
		\State Initialize array for global particle $u^i_{t_k}$
		\For{$j = 1,2,  \dots, N_{\mathrm{loc}}$}
		\State \parbox[t]{391pt}{
			Set $u^i_{t_k}(x,y) = u^{i, (j)}_{t_k}(x,y)$ at grid-points $(x,y)$ corresponding to domain $\mathring{D}_j = D_j \setminus (F_{ej} \cup F_{jw} \cup G_{jn} \cup G_{sj})$
			\strut}
		\EndFor
		\For{$(e,w) \in \{ (e,w) \, : \, F_{ew} \text{ is EW-overlap region in }D\}$}
		\State \parbox[t]{391pt}{
			Set $u^i_{t_k}(x,y) = \mathrm{Interp}_{ew}(u^{i, (e)}_{t_k}, u^{i, (w)}_{t_k})(x,y)$ at grid-points $(x,y)$ corresponding to domain $F_{ew}$
			\strut}
		\EndFor
		\For{$(s,n) \in \{ (s,n) \, : \, G_{sn} \text{ is SN-overlap region in }D\}$}
		\State \parbox[t]{391pt}{
			Set $u^i_{t_k}(x,y) =  \mathrm{Interp}_{sn}(u^{i, (s)}_{t_k}, u^{i, (n)}_{t_k})(x,y)$ at grid-points $(x,y)$ corresponding to domain $G_{sn}$
			\strut}
		\EndFor
		\For{$(a,b,c,d) \in \{ (a,b,c,d) \, : \, H_{abcd} \text{ is corner-overlap region in }D\}$}
		\State \parbox[t]{391pt}{
			Set $u^i_{t_k}(x,y) = \mathrm{Interp}_{sn} \big( \mathrm{Interp}_{ew}( u^{i, (d)}_{t_k}, u^{i, (c)}_{t_k} ), \mathrm{Interp}_{ew}( u^{i, (a)}_{t_k}, u^{i, (b)}_{t_k} ) \big) (x,y)$ at grid-points $(x,y)$ corresponding to domain $H_{abcd}$
			\strut}
		\EndFor
		\EndFor
		\State Return global ensemble $\{u^i_{t_k}\}_{i=1}^N$
	\end{algorithmic}
\end{algorithm}

\section{MCMC jittering and roughening}\label{app:jittering_comp}
In this appendix we present the results of a couple of filtering experiments comparing the two jittering algorithms which we described in Appendix~\ref{app:basic_filt}. We run two PF with tempering and jittering (Algorithm~\ref{alg:PF}), with jittering methodology specified by Algorithm~\ref{alg:MCMC_jittering} for one, and by Algorithm~\ref{alg:perturbation_jittering} for the other. We see that the performances of the two PF are equivalent, which indicates that the extra computational effort required by Algorithm~\ref{alg:MCMC_jittering} is superfluous.

The set-up for our filtering experiments is the same as in Section~\ref{subsec:LPF_grid}: the signal is given by a realization of the Stochastic Rotating Shallow Water SPDE \eqref{eq:SRSW}, in the sub-mesoscale regime, and the observations are of the form \eqref{eq:obs_numerics_vec}, with $\mathcal{J}$ ia set of indices determining an equispaced $2^n \times 2^n$-grid of observation locations within $D$ (as in Figure~\ref{fig:obs_types}, left), for $n = 1, \dots, 7$. In terms of parameters, we fix the number of particles to be $\mathbf{N=50}$, the ESS threshold for resampling, tempering and jittering $\mathbf{N_{\mathrm{ess}} = 0.8 N}$, the frequency of the observations to be given by $\mathbf{r = 10}$, so we assimilate every 10 time-steps, and the scale of the noise for the observations to be $\mathbf{\boldsymbol{\sigma} = 0.05}$ in \eqref{eq:obs_numerics_vec}. The perturbation for the jittering Algorithm~\ref{alg:perturbation_jittering} is specified as in \eqref{eq:jitt_perturbation}, with the scaling factor $\sigma^{jit}$ tuned, while in Algorithm~\ref{alg:MCMC_jittering} we set $M_{\mathrm{jit}} = 10$ (so after 10 rejected sample, we stop the Matropolis--Hastings algorithm and let the child particle equal to the parent particle), and tune $\rho$ (generally very large, $\rho \ge 0.99$).

\begin{figure}[ht!]
	\centering
	\includegraphics[width=0.325\linewidth]{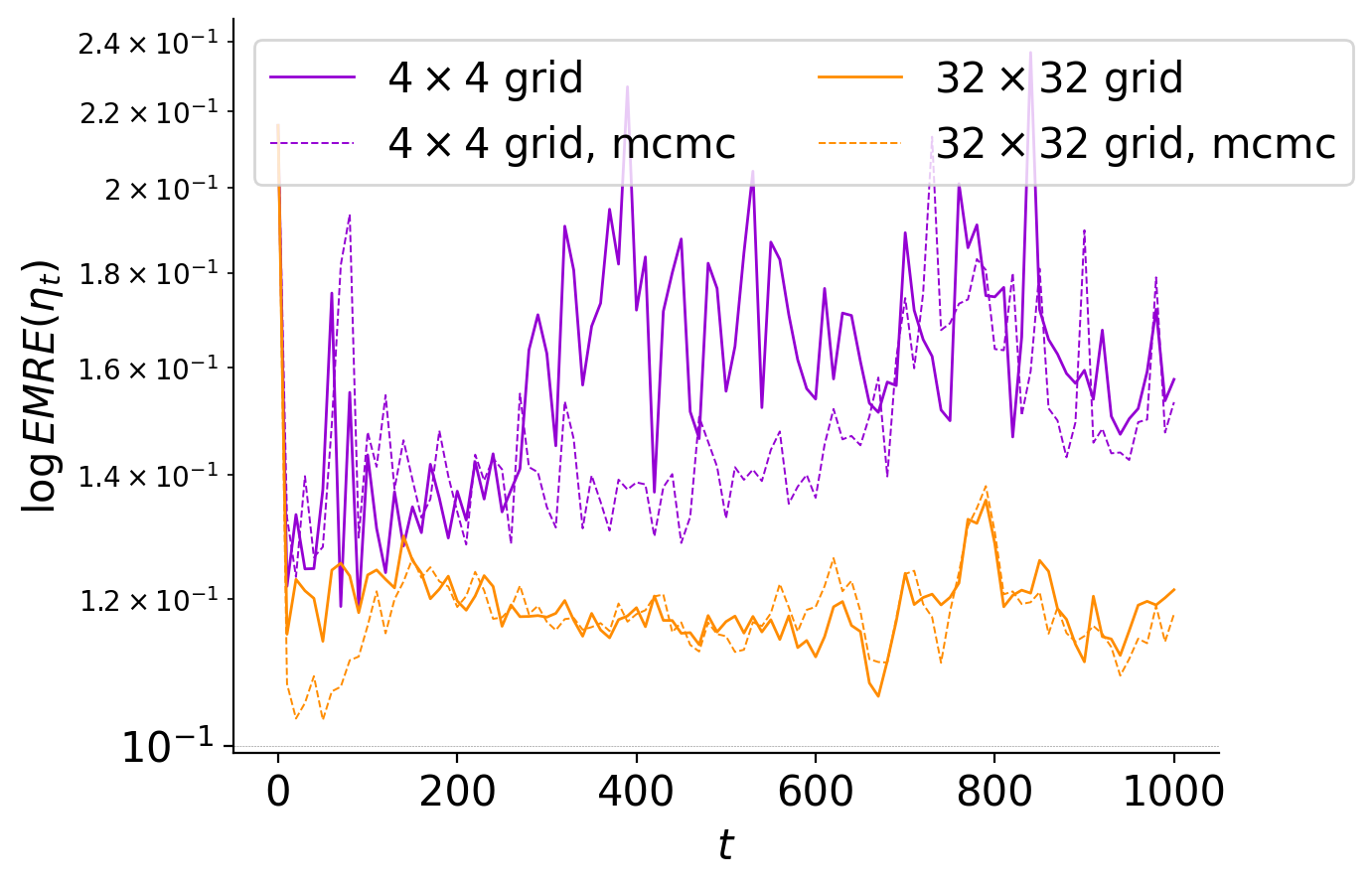}
	\includegraphics[width=0.325\linewidth]{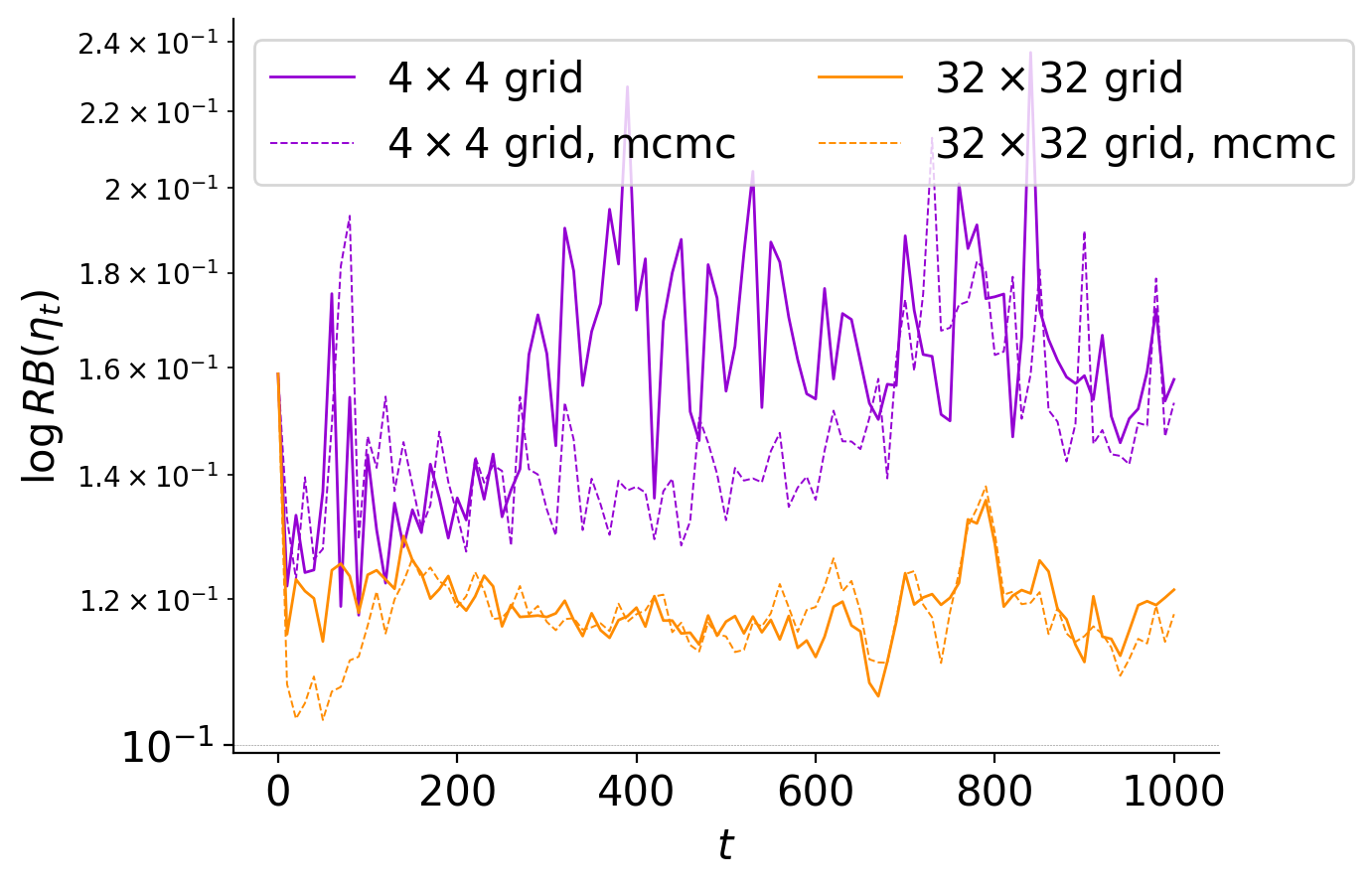}
	\includegraphics[width=0.325\linewidth]{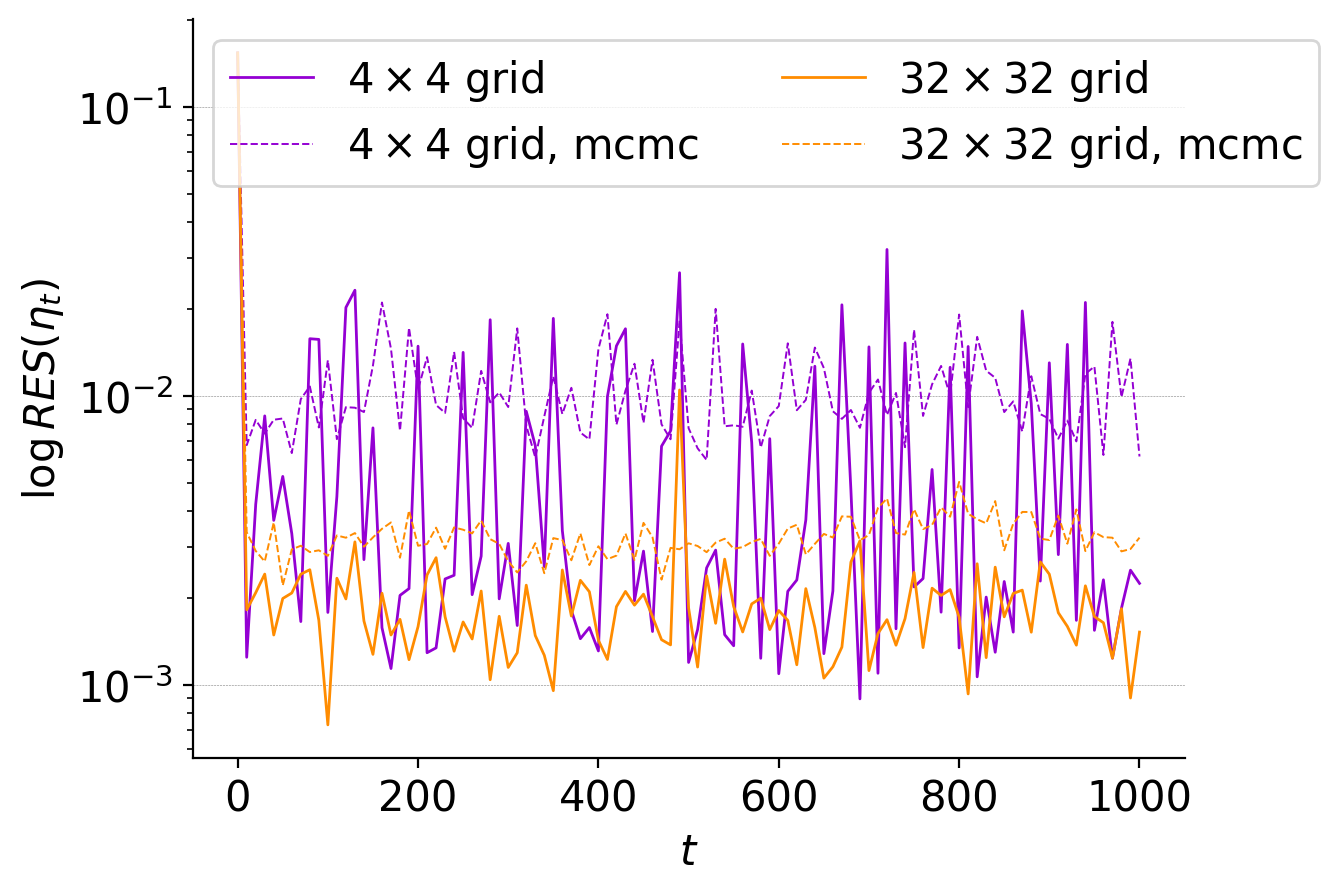} \\
	\includegraphics[width=0.325\linewidth]{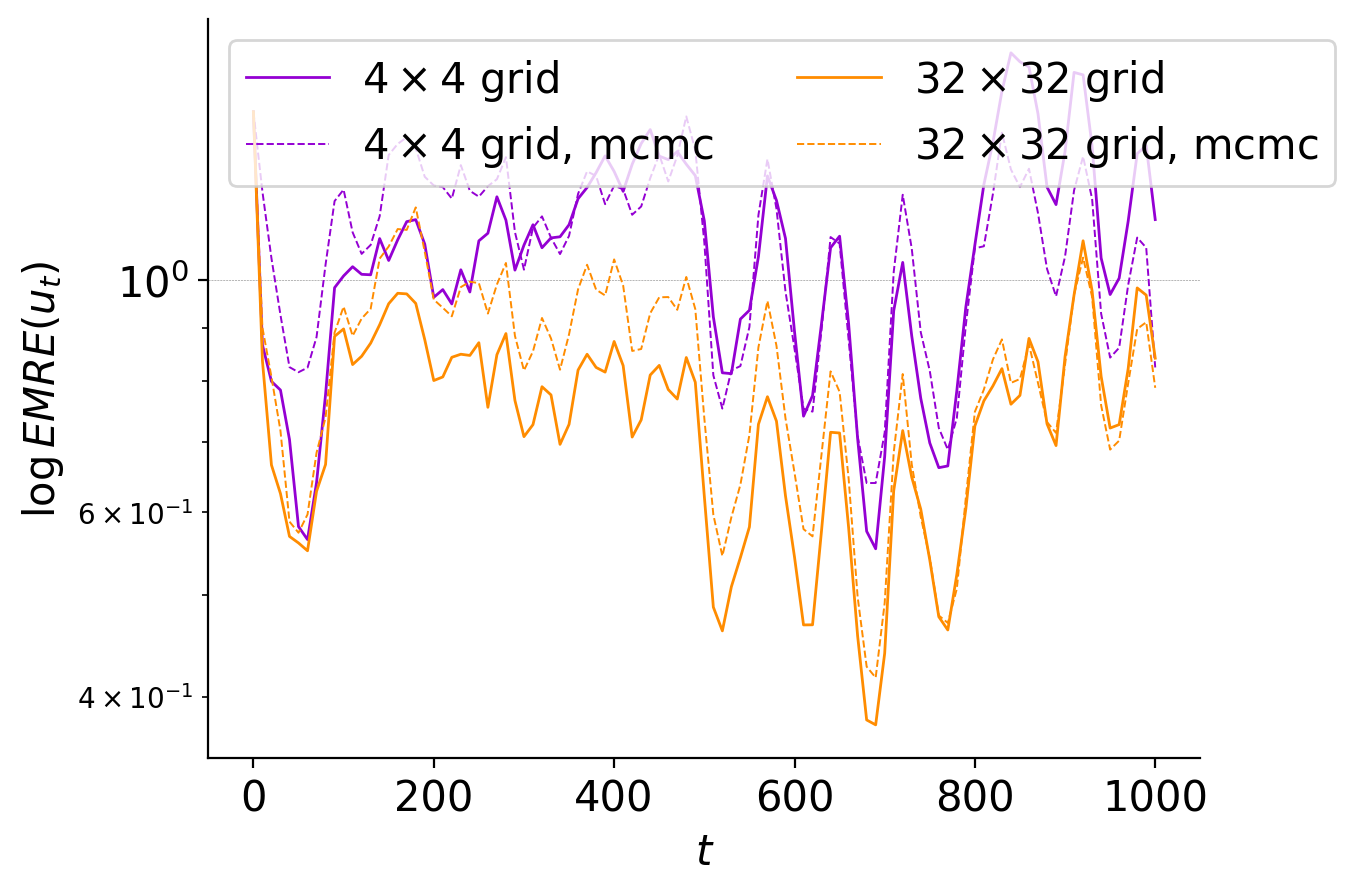}
	\includegraphics[width=0.325\linewidth]{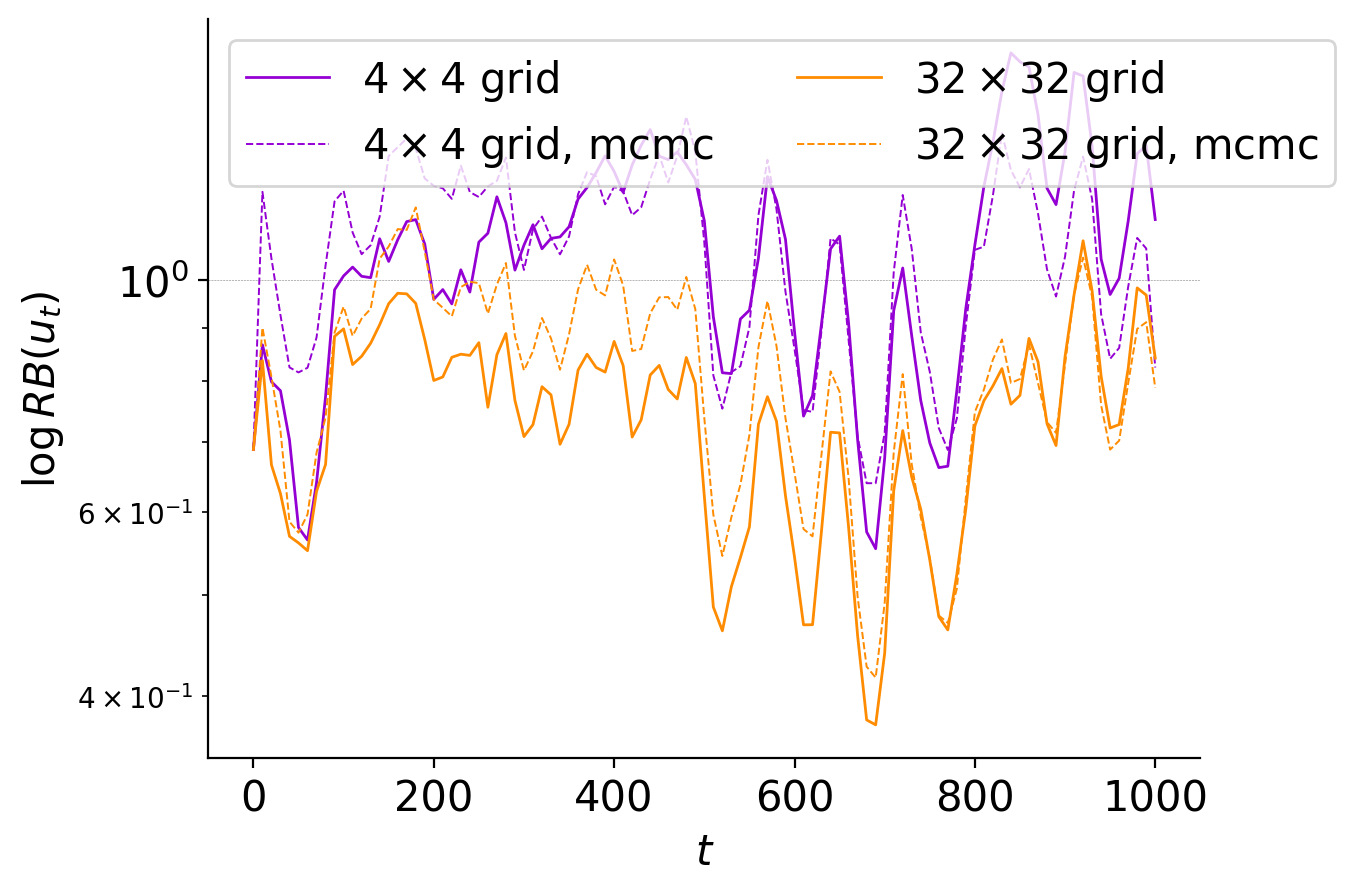}
	\includegraphics[width=0.325\linewidth]{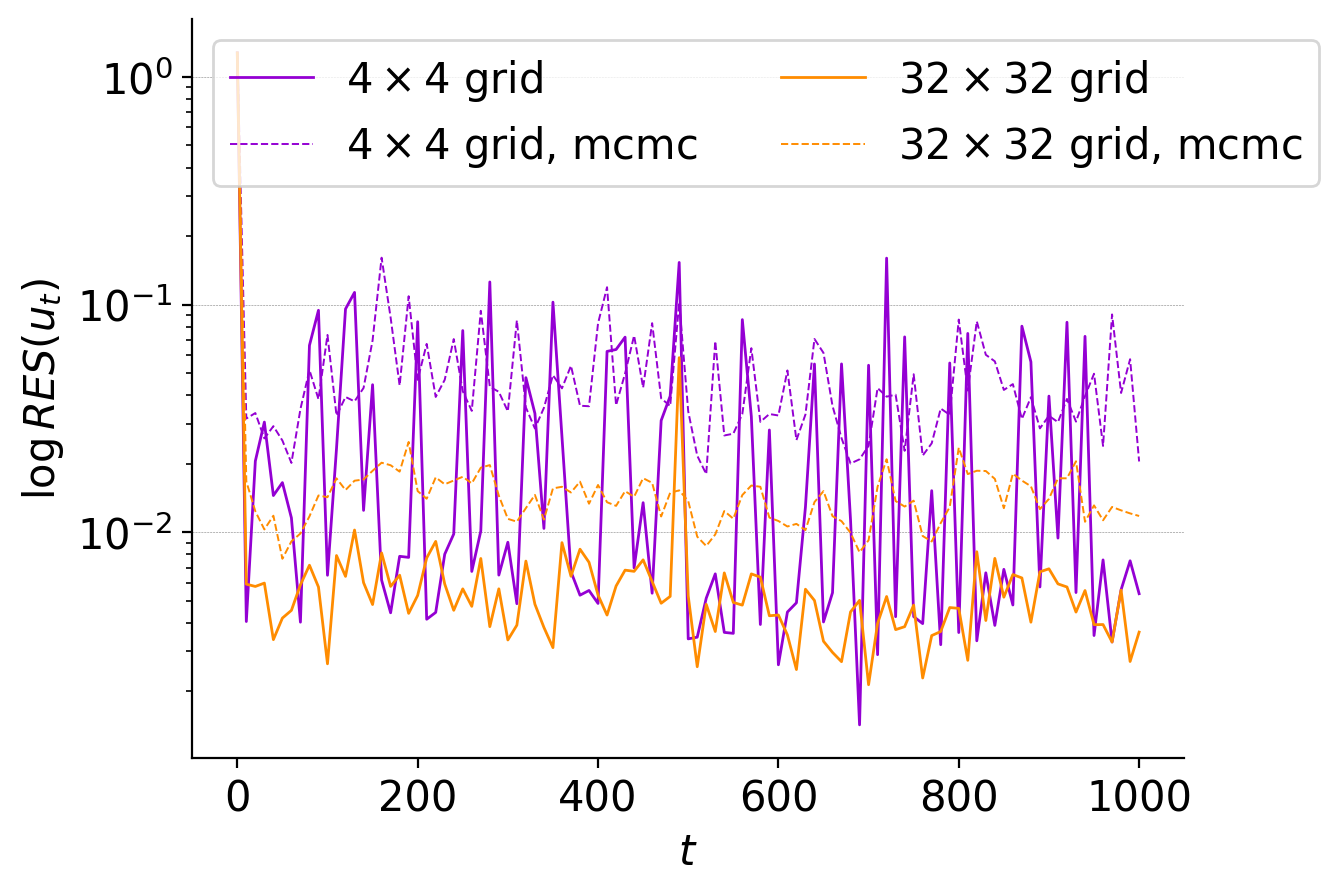} \\
	\includegraphics[width=0.323\linewidth]{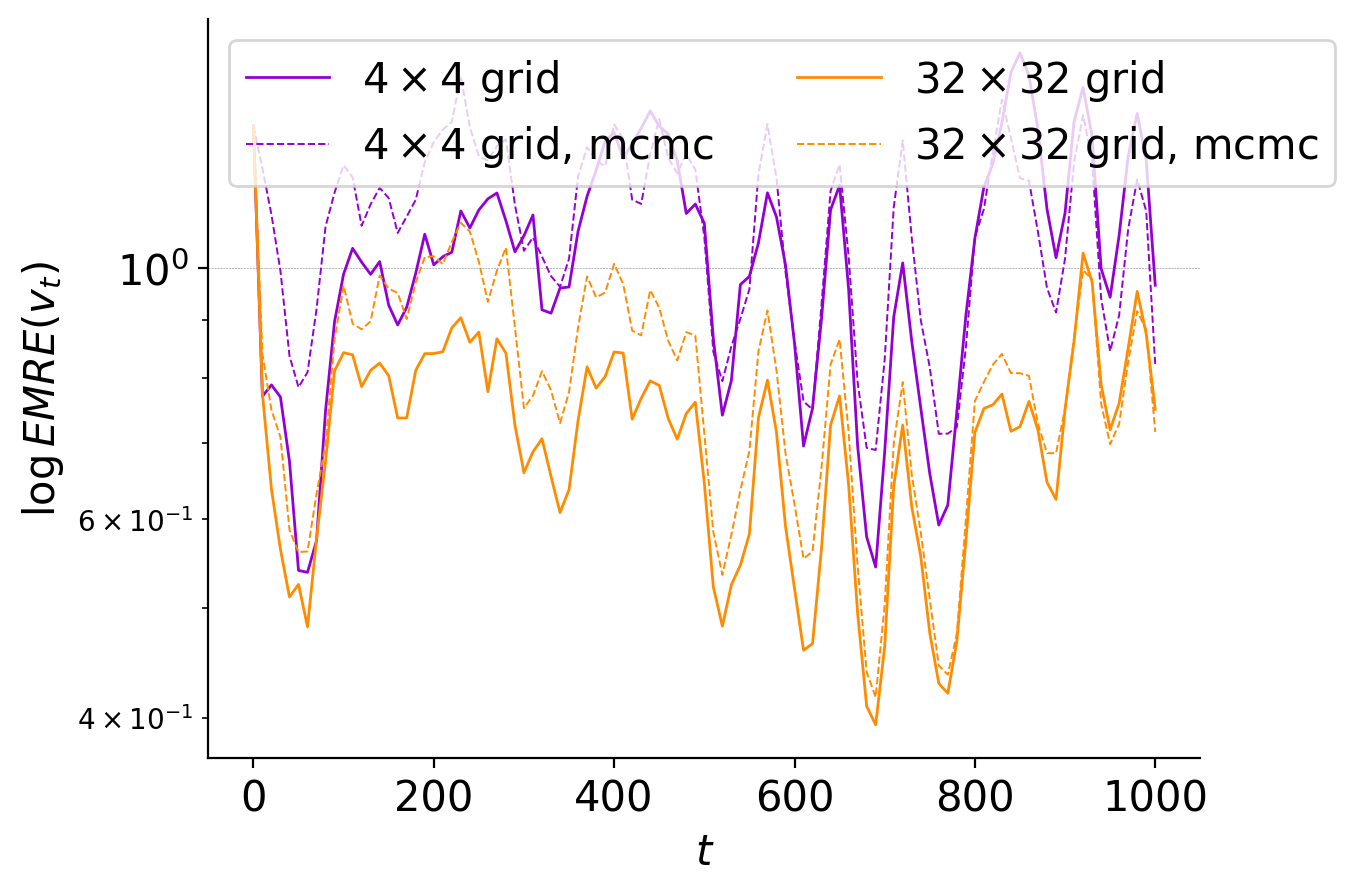}
	\includegraphics[width=0.323\linewidth]{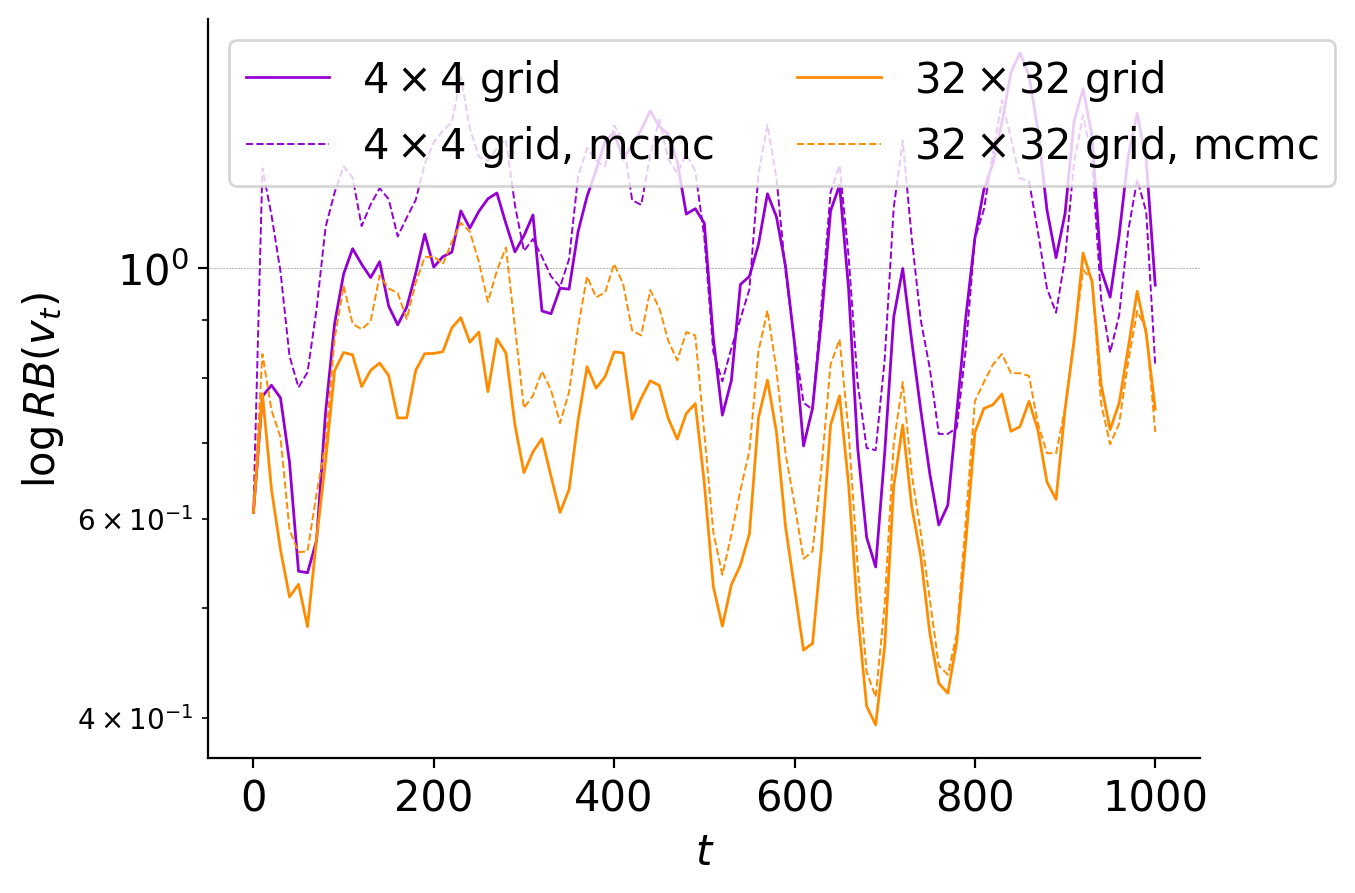}
	\includegraphics[width=0.323\linewidth]{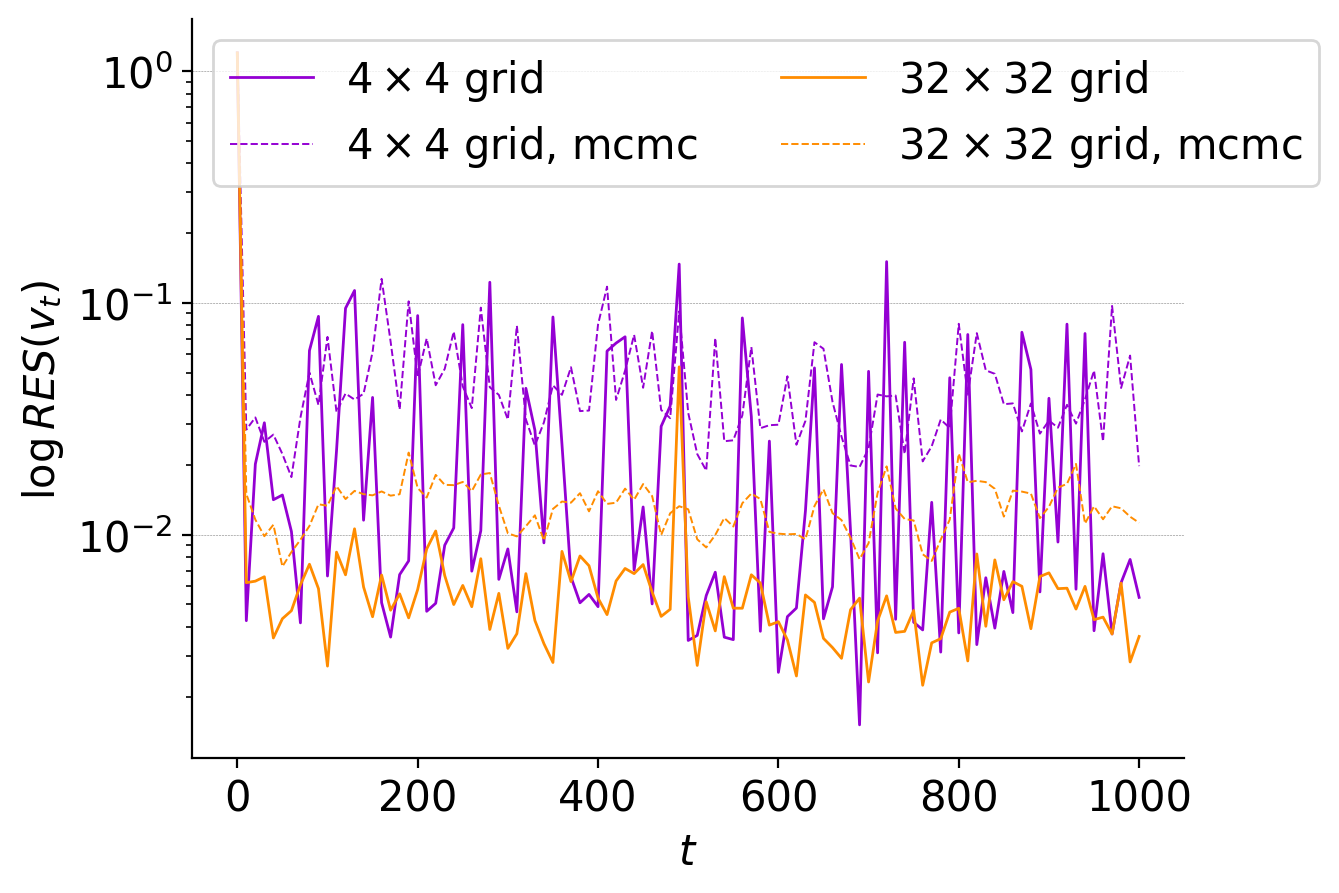}
	\caption{Plots of the $EMRE$, $RB$ and $RES$ of $\eta_t$, $u_t$ and $v_t$ for a PF with jittering Algorithm~\ref{alg:perturbation_jittering}  (solid lines) and a PF with jittering Algorithm~\ref{alg:MCMC_jittering} (dashed lines) with $N=50$ particles (the $y$-axis is in log-scale). We compare these errors for observations-grids of sizes $4 \times 4$ and $32 \times 32$. Observations are taken every 10 time-steps. On the top row we plot the errors for $\eta_t$, on the second row for $u_t$ and on the last one for $v_t$. The logarithm of the $EMRE$ for each field is in the first column, of the $RB$ in the second and of the $RES$ in the third.}
	\label{fig:PF_mcmc_pert}
\end{figure}

In Figure~\ref{fig:PF_mcmc_pert} we plot the $EMRE$, $RB$ and $RES$ for $\eta_t$, $u_t$ and $v_t$ for 100 assimilation steps, in log-scale, and two filtering experiments, one with a $4 \times 4$-grid of observations, and the other with a $32 \times 32$-grid. The errors for the PF with jittering Algorithm~\ref{alg:perturbation_jittering} are plotted as solid lines, those for the PF with jittering Algorithm~\ref{alg:MCMC_jittering} as dashed lines. We see that the magnitude and behaviour of the errors are similar.

\begin{figure}[ht!]
	\centering
	\includegraphics[width=0.325\linewidth]{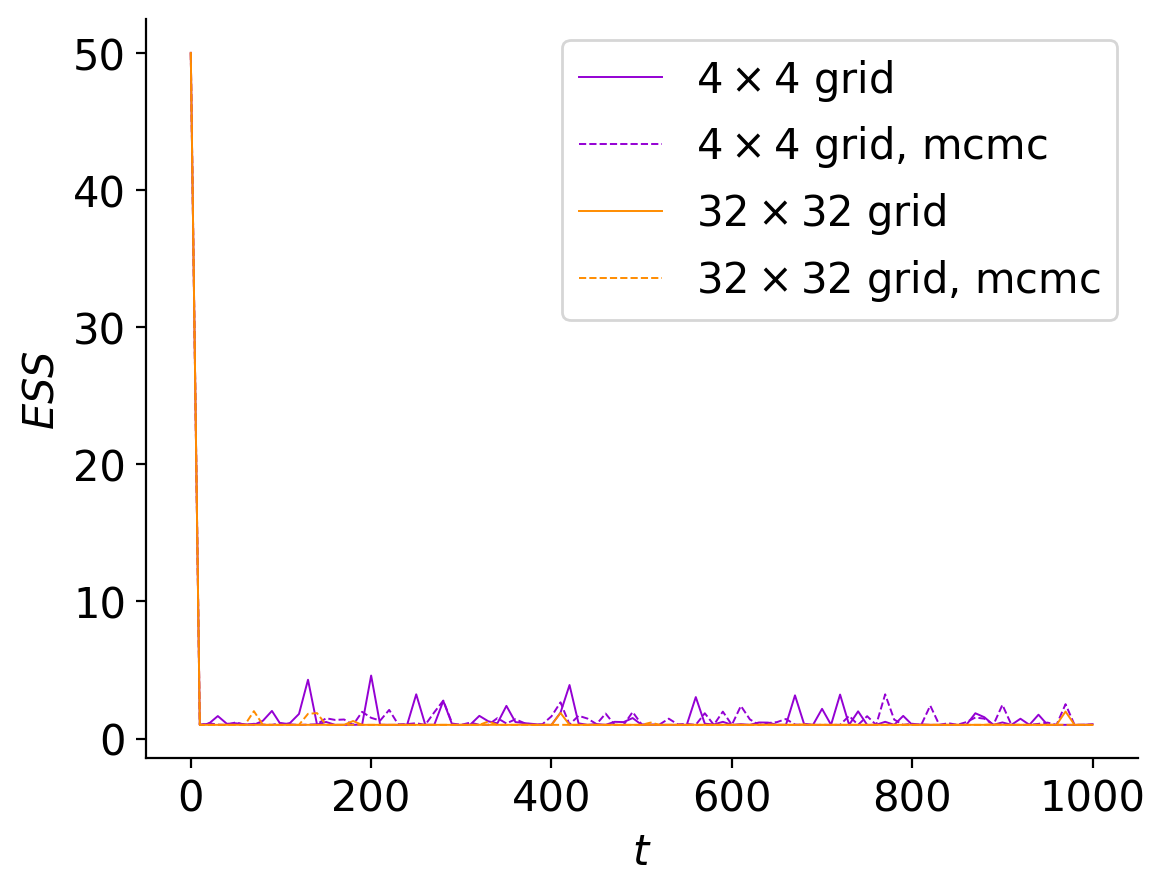}
	\includegraphics[width=0.325\linewidth]{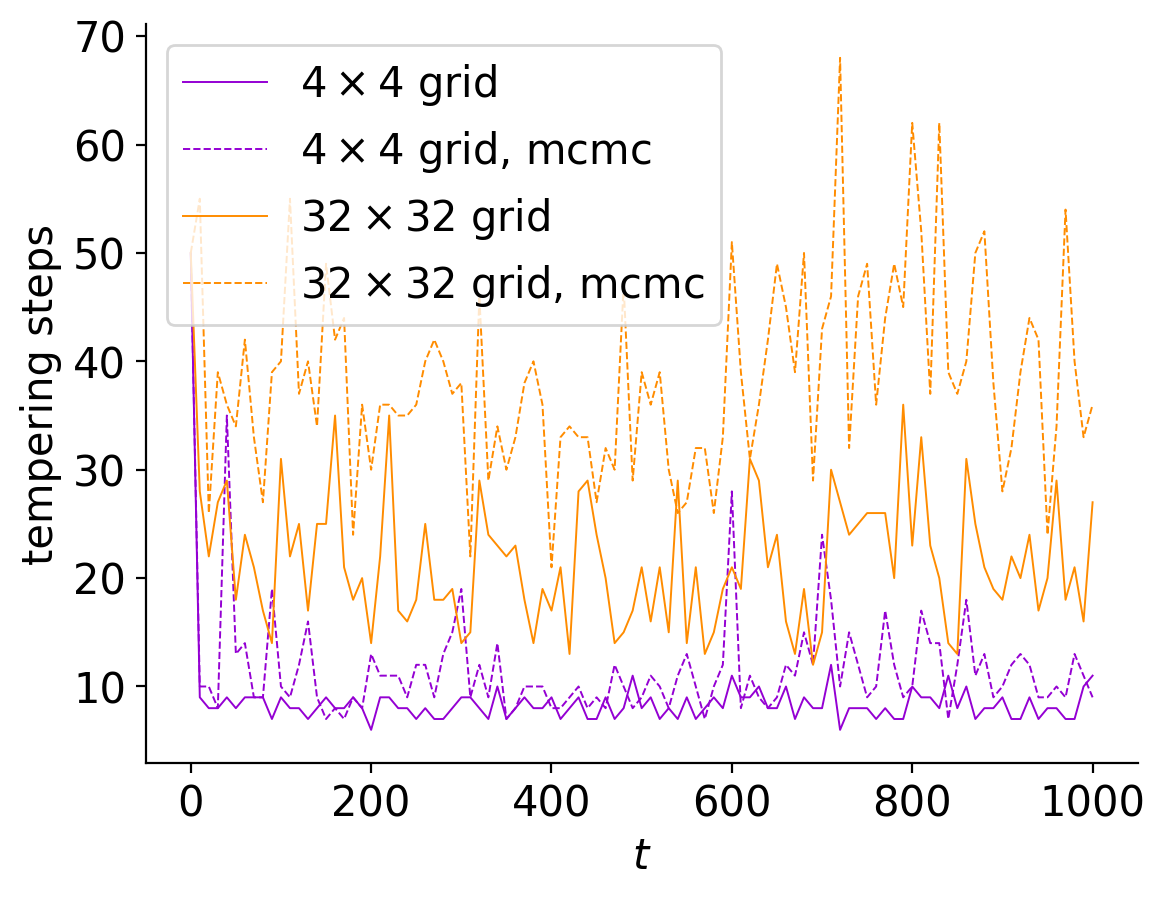}
	\includegraphics[width=0.325\linewidth]{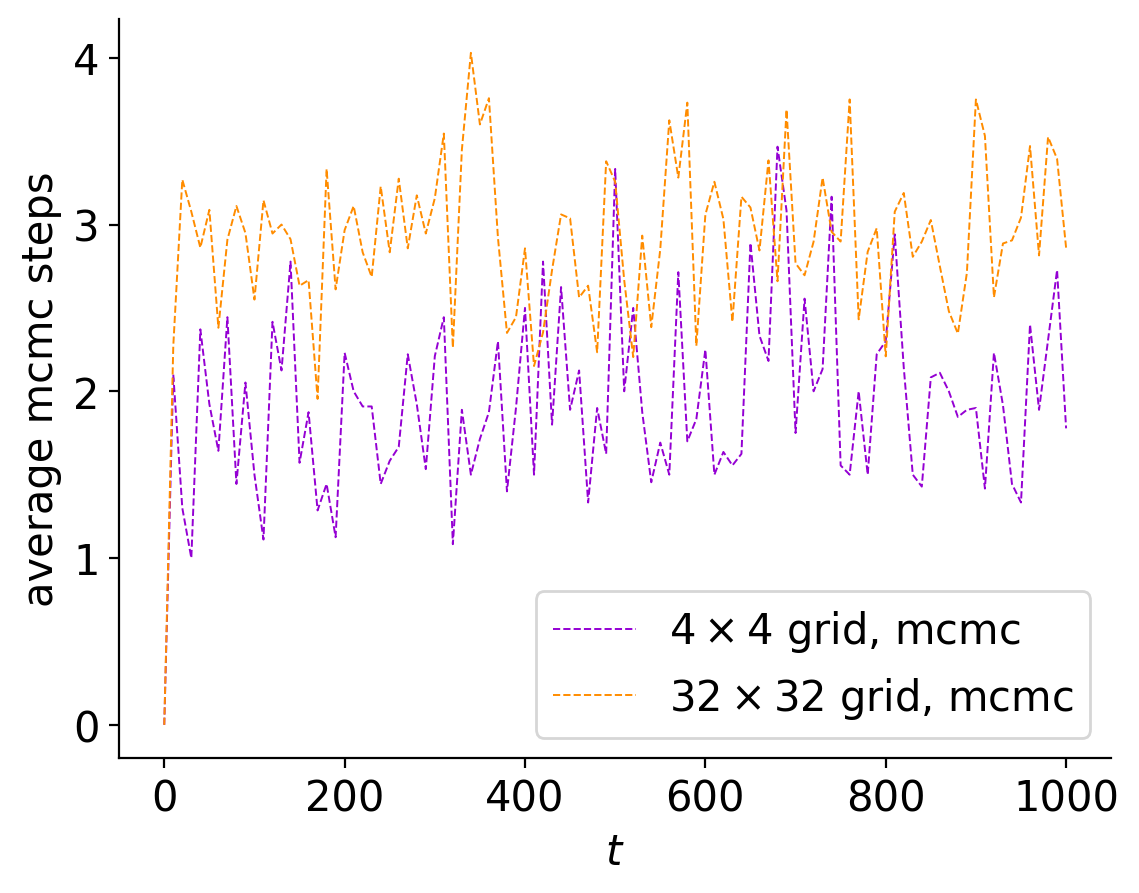} 
	\caption{Plots of the $ESS$ (left) and the number of tempering steps (centre) for a PF with jittering Algorithm~\ref{alg:perturbation_jittering}  (solid lines) and a PF with jittering Algorithm~\ref{alg:MCMC_jittering} (dashed lines) with $N=50$ particles, for observations grids of sizes $4\times 4$ and $32 \times 32$ and 100 assimilation steps. On the right, for the PF with MCMC jittering algorithm, we plot the average number of MCMC iterations per tempering step at each assimilation time.}
\label{fig:PF_mcmc_pert_stats}
\end{figure}

In Figure~\ref{fig:PF_mcmc_pert_stats} we plot the $ESS$ (before assimilation), the number of tempering steps and the average number of MCMC iterations (per tempering step) at each assimilation time. The average number of MCMC iterations is computed by collecting, at each tempering step, the maximum number of rejected samples (among all jittered particles): these are then averaged over the number of tempering steps to get a value at each assimilation time. At each tempering iterations $\approx 10$ particles get resampled and jittered (recall ~\ref{line:bisec} of Algorithm~\ref{alg:tempering} and that we set $N_{\mathrm{ess}} = 0.8 N = 40$). This means that by adapting the perturbation jittering Algorithm~\ref{alg:perturbation_jittering} instead of the MCMC Algorithm~\ref{alg:MCMC_jittering}, we save, per assimilation step, the computational time that we would need to re-run the SPDE \eqref{eq:SRSW} approximately $10 \times \textit{tempering steps} \times \textit{average MCMC iterations}$ times.

\clearpage

%%% bibliography
\bibliographystyle{plain}
\bibliography{biblio_PF}

\end{document}